\documentclass[review]{elsarticle}

\usepackage{hyperref}
\usepackage[dvipsnames]{xcolor}
\usepackage[caption = false]{subfig}

\journal{Journal of \LaTeX\ Templates}

\biboptions{authoryear}

\def\lapprox{\hbox{\lower .8ex\hbox{$\,\buildrel < \over\sim\,$}}}
\def\gapprox{\hbox{\lower .8ex\hbox{$\,\buildrel > \over\sim\,$}}}

\begin{document}

\begin{frontmatter}

\title{Synthesis of radioactive elements in novae and supernovae and their use as a diagnostic tool}






\author[a,b,c]{J. Isern\corref{J. Isern}}
\cortext[mycorresponding]{J. Isern}
\ead{isern@ice.cat}

\author[a,b]{M. Hernanz}
\author[d]{E. Bravo}
\author[e]{S. Grebenev}
\author[f]{P. Jean}
\author[g]{M. Renaud}
\author[h]{T. Siegert}
\author[i,j]{J. Vink}

\address[a]{Institut de Ci\`encies  de l'Espai (ICE,CSIC)}
\address[b]{Institut d'Estudis Espacials de Catalunya (IEEC)}
\address[c]{Reial Acad\`emia de Ciencies i Arts de Barcelona (RACAB)}
\address[d]{E.T.S. Arquitectura del Vall\`es, Universitat Polit\`ecnica de Catalunya}
\address[e]{Space Research Institute, Russian Academy of Sciences}
\address[f]{IRAP, Universit\'e de Toulouse, CNRS, CNES, UPS}
\address[g]{Laboratoire Univers et Particules de Montpellier (LUPM), Universit\'e de Montpellier, CNRS/IN2P3}
\address[h]{Center for Astrophysics and Space Sciences, University of California, San Diego}
\address[i]{ Anton Pannekoek Institute/GRAPPA, University of Amsterdam}
\address[j]{SRON, Netherlands Institute for Space Research}

\begin{abstract}
Novae and supernovae play a key role in many fields of Astrophysics and Cosmology. Despite their importance, an accurate description of which objects explode and why and how they explode is still lacking. One of the main characteristics of such explosions is that they are the main suppliers of newly synthesized chemical elements in the Galaxy. Since some of these isotopes are radioactive, it is possible to use the corresponding gamma-rays as a diagnostic tool of the explosion thanks to their independence on the thermal state of the debris. The drawback is the poor sensitivity of detectors in the MeV energy domain. As a consequence, the radioactive lines have only been detected in one core collapse supernova (SN 1987A), one Type Ia supernova (SN 2014J), and one supernova remnant (Cas A). Nevertheless these observations have provided and are providing important information about the explosion mechanisms. Unfortunately, novae are still eluding detection. These results emphasize the necessity to place as soon as possible  a new instrument in orbit with enough sensitivity to noticeably enlarge the sample of detected events.
\end{abstract}


\end{frontmatter}


\section{Introduction}

All the elements excepting some isotopes of the light elements\footnote{The isotopes of H and He are produced during the Big Bang, as predicted by \citet{alph48}, those of Be and B and $^6$Li by the interaction of cosmic rays with the interstellar medium \citep{mene71}, and $^7$Li by both scenarios with important contributions of novae and carbon-stars via the Cameron-Fowler mechanism \citep{came71}.}, are produced by stars \citep{hoyl46,burb57,came57} or by the interaction of their remnants (neutron stars, black holes...). Low and intermediate mass stars synthesize and eject through winds and planetary nebulae different amounts of elements like carbon, nitrogen and s-isotopes. Massive stars, $M \gapprox 10 M_\odot$, produce important quantities of $\alpha$-elements when their cores collapse to produce a neutron star or a black hole and a supernova outburst occurs. These simple properties allow to understand the dominance of $\alpha$-elements at low metallicities, i.e. $\alpha$/Fe ratios that are on average 2-3 times higher than the solar composition for $[{\rm Fe}/{\rm H}] \lapprox -0.5, -1 $\footnote{ $[{\rm X}/{\rm Y}]=\log_{\rm 10}[({\rm X}/{\rm Y})/({\rm X}/{\rm Y})_\odot]$}. Thermonuclear explosions in binary systems containing a white dwarf are delayed up to this point, and thereafter they inject important amounts of iron peak elements to the ISM causing a progressive reduction of $[\alpha/{\rm Fe}]$ as metallicity increases. See \citet{ilia07,dieh18,thie18,arno20} for recent overviews of nucleosynthesis in different astronomical sites.

The electromagnetic display of these explosive events, the \emph{fireworks}, are known as supernovae and novae. Its phenomenology strongly depends on the structure and composition of the  surrounding layers. For instance, if the exploding star is a Wolf-Rayet that has lost its outer envelope, hydrogen lines will be absent in the spectrum, but if it is a red supergiant they will be present (see section~\ref{sectax}). From the point of view of the source of energy, they are divided into \emph{thermonuclear} (TSN) and \emph{core collapse} (CCSN). From the nucleosynthesis point of view, the main difference is that TSN  generally proceed at larger densities  and, therefore, lower entropies, than the ejected layers of CCSN.

The synthesis of new isotopes occurs in two ways: hydrostatic and explosive. In the first case, the new isotopes are built on timescales dominated by the energy losses (photons and neutrinos) of the star, while in the second case the synthesis is governed by the dynamical time scale of the event (collapse, explosion, collision...).
Several burning stages occur during the quasi hydrostatic evolution of stars. They take place at rather well defined temperatures and they are responsible for the synthesis of nuclei placed  between hydrogen and the iron-group. To obtain heavier elements is necessary to invoke the capture of free neutrons, which can occur during hydrostatic burning (case of s-elements) or during hydrodynamic events (supernovae, collisions, mergings,...).

All stars, except brown dwarfs experience hydrogen burning. Hydrogen burning converts $^1$H into $^4$He via the pp-chains and the CNO-cycles. The most important ashes are $^4$He and the $^{14}$N that results from the processing of $^{12}$C and $^{16}$O  in the burning cycles.
Stars more massive that $\sim 0.5$~M$_\odot$ ignite helium. He-burning  is dominated by the $3\alpha$ $\rightarrow$ $^{12}$C and the $^{12}$C($\alpha$,$\gamma$)$^{16}$O reactions. During this stage, $^{14}$N is converted into $^{22}$Ne via the reaction chain $^{14}$N$(\alpha,\gamma)^{18}$F$(\beta^+)^{18}$O$(\alpha,\gamma)^{22}$Ne, being this isotope one of the main sources of free neutrons  thanks to the reaction $^{22}$Ne$(\alpha,n)^{25}$Mg. The other source of neutrons is $^{13}$C, which forms when $^{12}$C is injected into a proton rich environment as a consequence of some burning instabilities, via the reaction $^{13}$C$(\alpha,n)^{16}$O. Both sources are of capital importance to account for the abundances of s-process elements.

\begin{table}
\begin{center}
\label{t107a}
\caption{\footnotesize{Nuclear burning stages of a 15 M$_\odot$ star \cite{woos02} \citep{woos02}}}
\begin{tabular}{llllll}
\hline 
\hline  \\
Burning phase & T (GK) & $\rho$ (g/cm$^3$) & Fuel & Products & Time scale (yr) \\
\hline \\
Hydrogen & 0.035 & $5.8$ & H & He & $1.1 \times 10^7$ \\
Helium     & 0.18   &$1.4\times 10^3$ & He & C,O & $2.0 \times 10^6$ \\
Carbon     & 0.83   &$2.4\times 10 ^5$ & C  & O,Ne&$2.0 \times 10 ^3$ \\
Neon        & 1.6     &$7.2\times 10^6 $ & Ne & O,Mg&$0.7$ \\
Oxygen    & 1.9      & $6.7\times 10^6 $& O,Mg &Si,S&$ 2.6$ \\
Silicon      & 3.3     & $4.3\times 10^7 $&Si,S  & Fe,Ni  & $0.05 $ \\
\hline
\end{tabular}
\end{center}
\end{table}

The next major fusion stages  are carbon and oxygen burning, which are dominated by the $^{12}$C$+ ^{12}$C and $^{16}$O$+ ^{16}$O  reactions that mainly produce $^{20}$Ne, and $^{28}$Si and $^{32}$S, respectively. Table~\ref{t107a} displays the burning stages  of a 15 M$_\odot$ as well as the typical densities and temperatures at which they occur and their duration.
If the temperature is high enough, photons can induce the disintegration of nuclei as is the case of $^{20}$Ne, just after carbon burning, and $^{28}$Si after oxygen burning.

Stellar degenerate cores may have  different main compositions: helium, carbon plus oxygen (C/O), oxygen plus neon (O/Ne). Their behavior depends not only on the chemical composition, but also on the physical state of matter when the instability starts. Generally speaking, He-cores explode as a consequence of the extreme flammability of the fuel. C/O-cores can explode or collapse  depending on the ignition density and on the velocity of the burning front. Densities $\lapprox 5.5 \times 10^9$~g~cm$^{-3}$ and a laminar flame provide a lower limit for obtaining a collapse to a neutron star. O/Ne-cores ignite at such densities that they collapse, although depending on the velocity of the flame they can expel the outer layers and leave a bound white dwarf made of intermediate mass elements. Some models also predict the existence of hybrid structures made of a mantle of oxygen-neon surrounding a core of carbon-oxygen. Fe-cores always collapse to a neutron star or a black hole. They have been considered as one of the sites of the r-process, together with the merging of two neutron stars that are responsible of the kilonova phenomenon.   
The observational consequences of such instabilities depend on the properties of the surroundings of the degenerate cores like the mass, the dimensions and chemical composition of the envelope, the presence of circumstellar and interstellar material, and the amount and characteristics of the radioactive material injected during the explosion.

Despite the wide acceptance of this picture, important questions still remain. These questions not only have to do with the roles played by massive stars and binary systems, respectively, but also with the existing caveats that are plaguing the understanding of how stars end their life. The gamma-rays emitted by the fresh radioactive elements present in the debris provide critical information to elucidate the mechanisms that govern stellar explosions\footnote{Many details that are beyond the scope  of this review can be found in \citet{bran17, dieh18} books and in \citet{thie18}.}.

\section{ Supernova taxonomy}
\label{sectax}
Supernovae are characterized by a sudden rise of their luminosity, by a decline of different shapes and durations after maximum light that lasts several weeks, followed by an exponential decline that can last several years. The total electromagnetic output, obtained from the light curve, is $\sim 10^{49}$ erg, while the luminosity at maximum can be as high as $\sim 10^{10}$~L$_\odot$. The kinetic energy can be estimated from the expansion velocity of the ejecta, ${\rm v_{exp}} \sim 5,000 - 10,000$ km s$^{-1}$, and is of the order of $ 10^{51}$~erg\footnote{ This amount of energy is often referred as 1 Bethe.}. Such amount of energy can be obtained either from the gravitational collapse of an electron degenerate stellar core to form a neutron star \citep{zwic38}, or from the incineration of a carbon-oxygen degenerate core \citep{hoyl60}. In both cases the instability is due to the fact that relativistic electron degenerate stellar cores are weakly bound and have not a definite scale length, which means that the removal or the injection of relatively small amounts of energy can cause large variations of the radius of the structure.

The taxonomy of supernovae has been progressively developing since \citet{mink41} first recognized that at least two main types of supernova could exist, ones displaying hydrogen in their spectra and others not. Since then, several types have been added and some dismissed. 
Table~\ref {snid} displays the present supernova classification system  with some modifications from the original one \citep{silv12}. This scheme is not yet satisfactory since too many objects are still classified as peculiar but is largely used. Finally, the families of super luminous supernovae (SLSNe) and fast blue optical transients (FBOTs) have to be considered.

\begin{table}
\caption{Supernova classification}             
\label{snid}      
\centering          
\begin{tabular}{l l l l } 
\hline\hline       
  Ia  & Ib &Ic & II\\ 
\hline                    
Ia-norm     & Ib-norm & Ic-norm & IIP             \\
Ia-91T       & Ib-pec    & Ic-pec    & IIL             \\
Ia-91bg     &               & Ic-bl      & IIn             \\
Ia-csm      &                &               & IIb             \\
Ia-SCh       &                &               & II-pec       \\
Ia-Iax        &                &               &                 \\
Ia-02es     &                &               &                 \\
Ia-Ca rich  &                &               &                \\
Ia-pec       &                &               &                 \\     
\hline                  
\end{tabular}
\end{table}

Supernovae are classified according to their spectrum at maximum light and their light curve. Following Minkowski, \emph{op.cit.}, if hydrogen lines are absent, they are classified as Type I (SNI), if present as Type II (SNII). In the first case, if the spectrum exhibits a prominent Si~II absorption line at about 6100~\AA, they are called SNIa. If this line is absent but there is a prominent He~I line they are classified as SNIb. If both lines are absent they are called SNIc \citep{whee90}. Obviously, this classification only reflects the properties of the outermost layers of the exploding star.   

Type II supernovae are divided into five subclasses according to the shape of their light curve and the evolution of several strong spectral features: 1) Type IIP, that display a plateau in their light curve, 2) Type IIL that display a linear decline in  magnitude after a relatively sudden drop of the luminosity after maximum, 3) Type IIn, which have very different individual behaviors but all of them display narrow emission lines caused by the interaction of the debris with the previously ejected matter, 4) Type IIb, which display hydrogen emission lines that disappear during the nebular phase and then behave like Type Ib, and finally, 5) 87A-like or peculiar, that display long-rising light curves similar to the one exhibited by SN1987A and are associated to blue supergiants. An important characteristics of these events is that they are associated to young populations and do not occur in elliptical galaxies, where star formation stopped long time before. This property immediately suggested that they were caused by the explosion of massive, fast evolving stars as a consequence of a gravitational collapse. In particular, pre-explosion images have shown that Type IIP are associated to red supergiants, but it has not been possible to do the same with the other subtypes, except in the case of SN1987A (see \citet{smar09} for a review).

Type Ia supernovae are characterized by a sudden rise of the luminosity  up to an average maximum $\sim -19.3$ in the blue and visible bands in about two weeks \citep{ries99}, followed by a comparatively gentle decay divided in two epochs. The first one lasts $\sim 30$~days and the luminosity drops $\sim 3$~mag, while the second one slowly declines with a characteristic time  of $\sim 70$~days, suggesting the presence  of radioactive material in the debris. Infrared photometry shows that in the J, H, and K bands there is a well defined minimum at $\sim 20$~days after blue maximum and a secondary peak at $\sim 30$~days \citep{elia85} that is absent in many cases. An important property is that SNIa appear in all galaxies, including ellipticals, suggesting an association with old populations. 

One of the most striking properties of SNIa is the spectrophotometric homogeneity exhibited by a large fraction of them, i.e. the light curve  near maximum shows a relatively small dispersion, $\sigma \le 0.3^{mag}$ after normalization \citep{cado85,hamu96}, and the spectra of the different events are very similar
\citep{fili97}. These properties, together with the fact that they appear in all kind of galaxies,  immediately led to the conclusion that a thermonuclear explosion of a CO white dwarf near the Chandrasekhar's mass  in a close binary system was a plausible scenario. Unfortunately it has not been possible to identify any pre-explosion candidate up to now \citep{maoz14}, and it has only been possible to discard the nature of the companion in few events like SN2011fe. In this case the early detection of the event allowed to discard a luminous giant or a bright helium star but not a white dwarf, a main sequence or a subgiant star \citep{li11c}. In fact, the tighter constraint is provided by the delay-time distribution, i.e. the evolution of the SN rate with time in the case of an instantaneous burst of star formation, which is proportional to $t^{-1}$  \citep{maoz10}.

\begin{figure}[htb]
\center
\includegraphics[width=0.8\textwidth, clip=true, trim= 0cm  0cm 0cm 0cm]{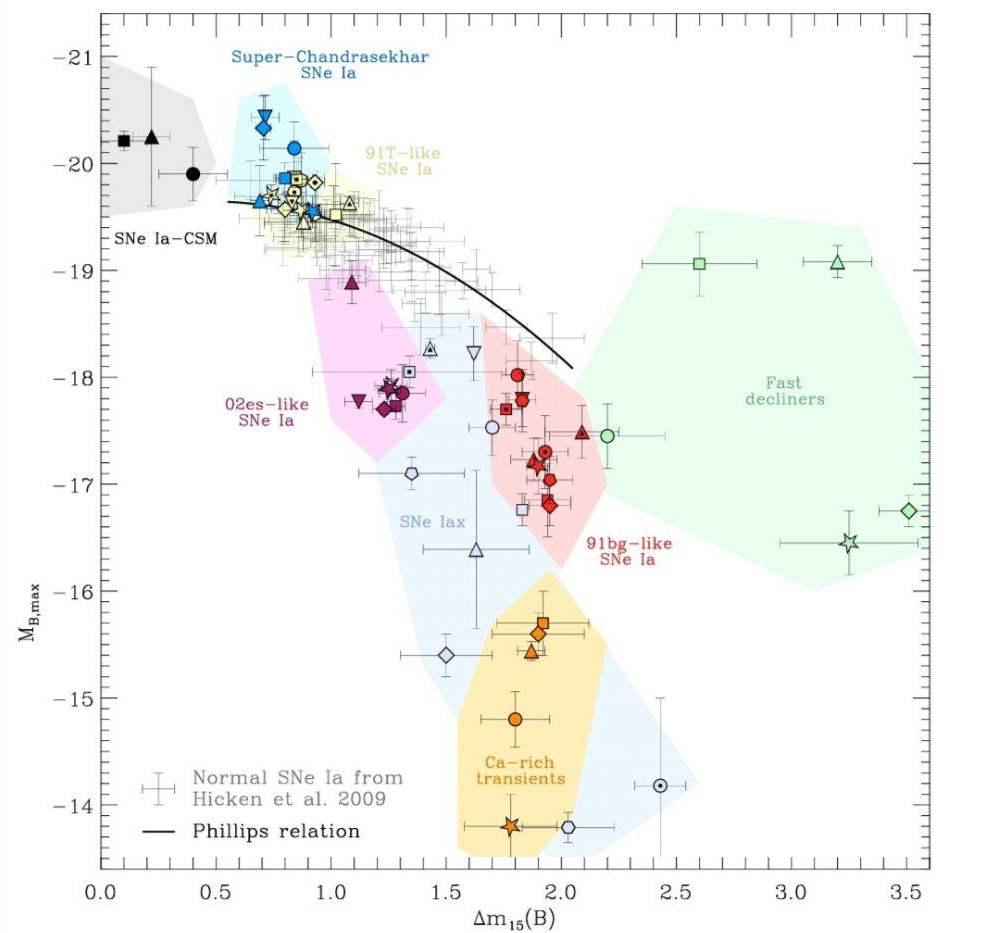}
  \caption{\footnotesize B-band peak versus the B-band decline $\Delta m_{15}(B)$ measured in SNIa. Normal supernovae are displayed in grey. Obtained from \citet{taub17}).}
\label{fsnia}
\end{figure}

For sometime it was believed that the spectrophotometric homogeneity of SNIa was a general rule with few exceptions. Consequently, these events where called \emph{normal} \citep{bran93} or \emph{Branch-normal} (SNIa-norm). Later on it was realized that the peculiarity was a quite extent property and that several subtypes do exist \citep{li11a}. Table~\ref{snid} and Figure~\ref{fsnia} display the present situation. 

In fact, even normal SNIa exhibit differences that can be grouped into a uniparametric family, the Phillips relationship \citep{phil93}. The bright extreme of this sequence is filled by the SN1991T-like and SN1999aa-like events which both show prominent Fe III features and weak Si II lines around the maximum of light. However, SN1999aa events exhibit a strong CaII absorption feature that is much weaker in SN1991T ones.
The dim extreme of the Phillips relationship is occupied by the fast declining SN1991bg-like outbursts. It has been claimed that these events constitute a separate population from the normal SNIa, because there are few 'transitional' events between them and the normal ones when using the $\Delta m_{\rm 15}$-peak luminosity relationship\footnote{$\Delta m_{15}$ is a parameter that measures the width of the supernova light curve defined as its decrease in magnitude in a 15 days interval starting at maximum light.}, but the distribution looks more continuous when other standardization methods are used.

The SNIa-CSM family events have either a normal or 91T-like spectrum that later on is dominated by the interaction with the circumstellar medium. The most relevant case is PTF11kx \citep{dild12,silv13}. The Super-Chandrasekhar SN Ia (SNIa-SCh) have also optical spectra similar to that of SNIa-norm but differ in the near infrared and show strong carbon features near the maximum light. They have a very high luminosity, slowly declining light curves, low ejecta velocities. and have a preference for low-metallicity environments \citep{taub17,inse19,khan11}. It is commonly accepted that the mass of the white dwarf involved in the explosion is larger than the Chandrasekar's mass.

Type Iax represents the most numerous group of supernovae that are off the Phillips relationship. The prototype is SN2002cx \citep{li03,jha06,fole13,jha17}. Their spectra are similar to those of normal SNIa at early times, they show important amounts of iron group elements and a light curve powered by radioactive isotopes. Near maximum they have lower photospheric velocities  than the normal ones and their luminosity is also lower ($-19 \lapprox M_V \lapprox -13$) \citep{jha17}. Interestingly enough, this category contains the only event, SNIax 2012Z, with a detected progenitor, a He-star donor \citep{mccu14}. SN2002es-like are also low luminosity, low velocity events, but their spectra are 'cooler' than those of the SNIax ones, and they  are preferently associated to old stellar populations \citep{whit15,roma18}.

The SNIa-Ca rich are characterized by a strong Ca~II emission during the nebular stage. Their luminosity is low, between that of novae and supernovae, and spectroscopically they are similar to SNIb, with prominent He features at maximum light. They occur far from star formation sites or even from the host galaxy \citep{pere11,kasl12,lyma13,lunn17}.

Finally, there is a growing number of  outliers that at present cannot be grouped into categories. They are, for instance, the twin SN2000cx and SN2013bh \citep{li01,cand03,silv13a}, the \emph{fast and faint} outbursts like SN2005et, PTF09dav, SN2010X, and SN1885A, the \emph{slow and faint} like PTF10ops, the  \emph{ fast and not so faint} like SN1939B, SN2002bj ... and so on \citep{jha19}.

This diversity has posed the question whether there are several scenarios/mechanisms, one for the Branch-normals and other for each peculiar event, or a unique one, flexible enough to account for the full range of behaviors. In any case it seems clear that all the events mentioned here were or could be caused by the explosion of a white dwarf. 

The light curves of SNIb are similar to those of SNIa but less steep and the double peak is absent. At maximum they are $\sim1.5$~mag dimmer than SNIa. The light curve of SNIc are steeper than those of SNIb and more similar, but not identical, to those of SNIa and, as SNII, they are associated to young populations. At present it is believed that they are caused by the gravitational collapse of massive stars that have experienced important mass losses. Both types  are often named stripped-envelope supernovae (SE-SNe).

 These ideas were reinforced by the observation that the mass ejected by SNIa was fairly constant and of the order of 1 M$_\odot$, while that of SNII were larger than 10 M$_\odot$ and that of SNIb/c are between both, but clearly larger than that of SNIa \citep{bran10}.

The frequency at which supernovae occur in a galaxy is fundamental to determine the nature of the progenitors and to understand the explosion mechanisms \citep{tamm70,berg91,capp03}. This quantity is hard to obtain because of the observational biases and the limited number of discoveries in controlled searches. 
These rates are presented in Supernova units (\emph{SNuX}) which are defined as
$1 SNuX=1 SN (100\, yr)^{-1}(10^{10}L_\odot^{X} )^{-1}$, where $L$ is the galaxy luminosity in solar units, and $X$ ($B$, $K$,...) the photometric band  of the survey. Since the $K$-band is sensitive to the mass of galaxies, this frequency is often expressed in terms of the mass of the galaxy obtained from an empirical relationship between the luminosity and the mass of each galaxy type, i.e. $1 SNuM=1 SN (100\, yr)^{-1}(10^{10}M_\odot )^{-1}$. The frequency, in the local Universe, $z \lapprox 0.1$, of SNII is nearly zero  in E and S0 galaxies and goes from 0.2 to 0.6 $SNuM$ in spiral and irregulars, the SE-SN have a similar behavior but within the interval 0.1-0.3 $SNuM$, and in the case of SNIa the rate is almost constant and of the order of 0.14 $SNuM$ \citep{li11a,li11b}. Interestingly enough, there is a significant correlation between the galaxy stellar mass, the specific star formation rate and the oxygen abundance. In the case of SNIa and SNII these correlations can be explained in terms of the delayed-time distributions and the ages, redshift and star formation history of the parent galaxies. Furthermore, the ratio SE-SN/SNII declines significantly in low-mass galaxies, $M\lapprox 10^{10}$~M$_\odot$, suggesting that a binary origin is dominant, while the number of normal SNIa is $\sim 30\%$ larger in them. Additionally, these low-mass galaxies have hosted all the SN1987A-like events known \citep{grau15,grau17a,grau17b}.

In the case of the Milky Way, these figures are 0.54 SNIa, 0.76 SE-SN and 2.54 SNII per century assuming that the Galaxy is a Sbc type and its stellar mass $(6.4\pm0.6)\times 10^{10}$~M$_\odot$ \citep{mcmi17}. A direct estimation based on the information provided by historical supernovae, supernova remnants, abundances of radioactive isotopes in the interstellar medium is also possible but difficult as a consequence of the incompleteness of these data. The first attempt, assuming a uniform Galaxy, was performed by \citet{berg75}. These calculations were improved by \citet{tamm94}, who included the presence of a thin and a thick galactic disks. More recently, \citet{adam13}, using the \emph{TRILEGAL} model of Galaxy \citep{gira05}, and including the presence of dust obtained a total supernova rate of $4.6^{+7.4}_{-3.0}$ per century, distributed as $3.2^{+7.3}_{-2.6}$ per century for CCSN and $1.4.^{+1.4}_{-0.8}$ per century for SNIa.  

The most extreme behavior of the supernova families is provided by the super-luminous supernovae (SLSNe), which are $\sim 10-100$ times brighter than the most luminous SNIa, and fast blue optical transients (FBOTs), which display a light curve characterized by a fast rise to the maximum followed by a rapid decay \citep{inse19}.

SLSNe are arbitrarily defined  as events that have a luminosity $ L\gapprox 10^{44}$~erg s$^{-1}$ or  $M_{AB} \lapprox -21^{\rm mag}$, at maximum and radiate a total energy of $\sim 10^{51}$ erg. At present, more than 100 events are known and this number is growing. These events are preferentially associated to low-metallicity, star-forming environments, and their frequency is estimated to be a few $\times$ 10$^{-4}$ the CCSN rate \citep{quim13,praj17}. Following the traditional classification scheme of supernovae, events that display H-lines in their spectra are classified as SLSN II and those that are deficient in hydrogen as SLSN I.

Both categories can admit a subdivision taking into account their spectrophotometric difference \citep{inse19}: i) SLSNI-F (or fast), with SN2011ke as the prototype \citep{inse13,quim18}, ii) SLSNI-S (or slow), with SN2015bn \citep{nich16,nich18}, iii) SLSNII, with SN2013hx \citep{inse18} as the prototype, and iv) SLSNIIn, with SN2006gy \citep{smit07} as the prototype, which are similar to the standard SNIIn. The observed luminosities of these events, excepting SLSNIIn  cannot be solely  accounted by the decay of $^{56}$Ni  and several possibilities have been advanced: spin-down of a rapidly rotating young magnetar, interaction of the supernova ejecta with massive C/O-rich circumstellar matter, pair instability and/or pulsating pair instability supernovae (see the references in Inserra op.c.).  

  FBOTs are characterized by a rapid raise of the light curve ($\lapprox 10$~days) and an exponential decline of $\sim 30$ days. The magnitude at maximum can go from that of a dim CCSN to those observed in SLSNe. Spectroscopically they can be represented by a hot black body T$_e$ within the range of 10,000 to 30,000 K without large features \citep{drou14,purs18}. Their origin is still a mystery and several possibilities have been advanced: i) shock-breakout or recombination of an extended envelope produced by an optically thick, low-mass circumstellar wind surrounding a CCSN, ii) a failed explosion of a blue supergiant with formation of a black hole, iii) electron capture supernova with formation of a millisecond magnetar, iv) tidal disruption by a black hole, or v) binary neutron star mergers (see the references in Inserra op.c.).

\section{The fate of massive stars} 

Massive stars, $M\gapprox 8-9$~M$_\odot$ are observed as super asymptotic giant branch stars (SAGB), blue supergiants (BSG), red supergiants (RSG), Wolf-Rayet (WR) and luminous blue variables (LBV). Their fate is a collapse to a neutron star or a black hole as a consequence of electron captures on oxygen, neon and magnesium, the ashes resulting from the carbon-burning, or by the photodisintegration of iron after central silicon burning. The nature of the bounded remnant, a neutron star or a black hole, depends on the compactness degree of the star, i.e. the central mass condensation, and on the existence of magnetic fields and fast rotation, in which case they can be at the origin of hypernovae and Gamma Ray Bursts \citep{woos15}. 

Very massive stars, those that develop a He-core with a mass $\gapprox 30$~M$_\odot$, evolve in a different way that the massive ones since, instead of developing an iron core, they have to face  an electron-positron instability as a consequence of the high temperatures and low densities they reach in their interior\footnote{At high temperatures and low densities electron positron pairs are created and, as a consequence, the equation of state softens, i.e. the adiabatic index becomes smaller than 4/3 \citep{fowl64}.}.  The corresponding mass of their ZAMS progenitors is $\gapprox 70-90$~M$_\odot$, although these values are very uncertain since they depend on rotation, metallicity, and mass losses and, usually, is better to talk in terms of the size of the helium and metallic core, or the oxygen core. 
This instability occurs after central helium burning. If this happens when the oxygen core reaches a mass $\gapprox 50$~M$_\odot$ (for low-metallicity, non-rotating stars this corresponds to a ZAMS mass $\gapprox 140$~M$_\odot$), the total amount of energy released is larger than $10^{52}$ erg and no compact remnant is left if the total mass is $\lapprox 300$~M$_\odot$. On the contrary, if stars are more massive than this value an intermediate mass black hole is produced \citep{hege03}. These events are known as pair-instability supernovae (PISN). 
 Stars in the mass range of 70-90 to 140 M$_\odot$ undergo pulsational instabilities that can end as a pair-instability supernova ( they are named pulsational PISN or PPISN) or an ordinary CCSN \citep{woos17}.  This behavior, however, strongly depends on metallicity and rotation.

\subsection{Collapse and explosion of Fe-cores}

Once the iron core of massive stars exceeds the critical mass they become gravitationally unstable and collapse. The innermost material is compressed until it reaches the nuclear density. When this happens the equation of state becomes very stiff and induces a core bounce of these inner regions provoking a shock wave that propagates outwards and 'collides' with the still infalling outer core. The energy of this shock is quickly dissipated in the dissolution of the heavy nuclei present in the infalling material and, as a consequence, the shock stalls. If nothing happened, the outcome would be a collapse to a black hole without explosion. Therefore, a revival of the shock is necessary to inject enough energy and momentum to the outer layers to blow up the parent star and to allow the birth of the neutron star\footnote{Even in the case of a successful explosion, a black hole may still form via fall-back accretion.}.

The total energy available during the collapse is $\sim 10^{53}$~ergs, but 99\% of this energy is lost via the emission of neutrinos and antineutrinos during the process of contraction and cooling of the proto-neutron star. The key point is to determine which is the exact fraction of energy deposited in the envelope and to find the mechanism responsible for the explosion. Furthermore, the revival of the shock has to occur rather soon, 1 to 1.5 seconds after bounce in order to allow the formation of a compact remnant with a mass in the range of $\sim 1.5 - 2.5$~M$_\odot$.

Several mechanisms have been proposed up to now and, probably, a combination of effects will be necessary to account for the explosion. These mechanisms include heating by neutrinos, multi-dimensional hydrodynamic instabilities in the post shock and the protoneutron star itself, protoneutron star pulsations, rotation, magnetic fields and nuclear burning, see \citet{burr13},\citet{jank16},\citet{mull16},and \citet{couc17} for recent reviews.

The \emph{neutrino heating mechanism} relies on the deposition of neutrino energy after bounce behind the stalled shock. At present, however, it seems to require the \emph{SASI mechanism} (standing-accretion shock instability) to work, except in the case of the lowest massive stars that can explode even in spherical symmetry.
The \emph{magneto rotational mechanism} is based on the amplification of the magnetic field during collapse and post-bounce times of a rapidly rotating core \citep{bisn70,lebl70}. This mechanism leads to a jet-like explosion along the rotation axis and is closely related to the long duration GRBs.
 The \emph{rotational mechanism} relies on a rapidly rotating iron core that splits into to protoneutron stars that later on merge as a consequence of the emission of gravitational waves \citep{imsh92,imsh04}.

\subsection{Collapse and explosion of O/Ne degenerate cores }
\label{onec}
Stars in the range of 8--12 M$_\odot$, the so called super-AGB stars (SAGB), develop an O/Ne core during carbon shell burning \citep{miya80}.    Those with a mass $\gapprox 10$ M$_\odot$ experience thermonuclear Ne-flashes that are not strong enough to disrupt the star, they evolve through the usual burning stages and, finally, they end their life collapsing to a neutron star after forming an Fe-core \citep{woos80,nomo84a}. The fate of stars in the mass range of 8--10 M$_\odot$ is more uncertain. Those that are able to remove the outer envelope end their life as O/Ne white dwarfs. Those that fail to do it gradually increase the mass of their core, which approaches to the Chandrasekhar's mass. The warming is gradual since the major part of the compressional work is invested in increasing the Fermi energy of electrons \citep{miya80}, in URCA neutrino emission and in thermal neutrino losses until electron captures on magnesium, $^{24}$Mg(e$^-$,$\nu_{\rm e}$)$^{24}$Na(e$^-$,$\nu_{\rm e}$)$^{24}$Ne, first, and neon, $^{20}$Ne(e$^-$,$\nu_{\rm e}$)$^{20}$F(e$^-$,$\nu_{\rm e}$)$^{20}$Ne, after, start just before the Ne-burning stage. These electron captures have a double effect, they lower the electron mole number (and with it the Chandrasekhar's mass) and they heat up the plasma to the point of triggering the ignition of oxygen. At present it is not possible to determine where the bifurcation occurs. It seems that stars from the upper mass segment, $\sim 9-10$ M$_\odot$, explode as supernovae and those from the lower mass segment end as O/Ne white dwarfs \citep{dohe17}.

The ignition of oxygen occurs at densities of $\rho_{\rm c} \sim 10^{10}$ g~cm$^{-3}$.  Therefore, the mass of the core is very near to the Chandrasekhar's limit and small differences in the input/output of energy can lead to very opposite outcomes: thermonuclear explosion or gravitational collapse \citep{nomo91, iser91, cana92,timm92}. 
Under such conditions, ignition produces the complete incineration of material to the nuclear statistical equilibrium (NSE) composition, and leads to the formation of a burning front that propagates outwards from the center of the star.  Depending on the density,  electron captures on the NSE material can remove energy and pressure so fast, as compared with the energy released by the spreading  of the burning front, that a gravitational collapse to a neutron star can occur. The associated supernova outburst is named collapsing electron-capture supernova or cECSN, to emphasize that the outburst is triggered by electron captures and the energy has a gravitational origin. On the contrary, if the density is not high enough, the flame overcomes the energy losses and a thermonuclear supernova is obtained. In this case they are represented by tECSN\footnote{Depending on the velocity of the burning front, the thermonuclear explosion can 'fail' and leave a bound degenerate remnant made of intermediate mass and iron group elements \citep{iser91}}. Notice that this scenario also applies to accreting O/Ne white dwarfs in close binary systems. In this case, however, the collapse (called accretion induced collapse or AIC) to a neutron star does not produce a supernova display as a consequence of the lack of envelope around the white dwarf.
 
The exact density at which the thermonuclear runaway occurs is crucial to understand the fate (explosion or collapse) of the O/Ne cores. It essentially depends on three factors: the degree of mixing in the region where captures occur, the relevant electron-capture rates and the detailed chemical composition.

During the process of growing in mass, the core is convectively stable thanks to the high conductivity of degenerate electrons. As soon as electron captures start on $^{24}$Mg a temperature gradient larger than the adiabatic one appears. If the Schwarzschild criterion is applied, radiative gradient larger than adiabatic gradient, i.e.

\begin{equation}
{\nabla _T} = \frac{{d\ln T}}{{d\ln P}} > {\nabla _{ad}} = {\left( {\frac{{\partial \ln T}}{{\partial \ln P}}} \right)_S} 
\end{equation}
 convection starts and mixes new material into the region where electron Fermi energies are high enough to allow electron captures proceed. As a consequence the core gradually contracts and heats up until electron captures on $^{20}$Ne  trigger the ignition of oxygen. This occurs at $\rho \sim 2 \times 10^{10}$~g~cm$^{-3}$ and, at these densities, electron captures on the NSE material ensure the collapse to a neutron star \citep{nomo79,miya80}.

Electron captures, however, not only heat up the material and steeps the temperature gradient, they also produce a positive $Y_e$-gradient, where $Y_e$ is the average electron number per nucleon. Therefore the stability analysis  has to include the influence of the chemical inhomogeneity, and the Ledoux criterion for convection has to be applied \citep{iser83,moch84}
\begin{equation}
{\nabla _L} \equiv {\nabla _{ad}} + \left[ {{{{{\left( {\frac{{\partial \ln P}}{{\partial \ln {Y_e}}}} \right)}_T}} \mathord{\left/
 {\vphantom {{{{\left( {\frac{{\partial \ln P}}{{\partial \ln {Y_e}}}} \right)}_T}} {{{\left( {\frac{{\partial \ln P}}{{\partial \ln T}}} \right)}_{{Y_e}}}}}} \right.
 \kern-\nulldelimiterspace} {{{\left( {\frac{{\partial \ln P}}{{\partial \ln T}}} \right)}_{{Y_e}}}}}} \right]{\nabla _{{Y_e}}}
\end{equation}
with
\begin{equation}
{\nabla _{{Y_e}}} \equiv -\frac{{d\ln {Y_e}}}{{d\ln P}}
\end{equation}
 and where $P$ and $T$ are the pressure and temperature, respectively.

This means that the gradient must be steeper than that corresponding to the Schwarzschild criterion to induce convection and guarantee an efficient mixing of material and heat transfer. As a consequence, the star ignites oxygen at a lower density that in the previous case, $\rho \lapprox 10^{10}$~g~cm$^{-3}$, after barely escaping ignition at the onset of electron captures on $^{24}$Mg. If convection is suppressed, ignition occurs in the interval $\rho_{\rm ign}=8.5 - 9.5\times 10^9$~g~cm$^{-3}$ \citep{miya80,moch84,miya87,cana92,guti96}.

The electron capture rates  on $^{24}$Mg and $^{20}$Ne used in the pioneering work of \citet{miya80} were obtained from the gross theory of $\beta$-decay. These rates have been noticeably improved since then thanks to the work of \citet{full80,full82,full82a,full82b,full85, taka89,oda94,mart14}.

\citet{mart14} obtained the electron capture rates for $^{20}$Ne, $^{20}$F, $^{24}$Mg, $^{24}$Na  and the $\beta$-decay rates of $^{20}$F and $^{24}$Na based on new experimental data and large-scale shell model calculations that are appropriate for the astrophysical conditions except for $^{20}$Ne that have a dominant contribution from the second-forbidden transition between ground states. The strength of this transition, which has recently been measured by \citet{kirs19}, is extremely high and increases the capture rate by several orders of magnitude.
\citet{kirs19} computed the rate of electron capture as an allowed Gamow-Teller transition using the newly obtained strength. The inclusion of such forbidden transition allows e$^-$-captures to start earlier  but, since it is $\sim 5$ orders of magnitude weaker than the allowed one it does not trigger a thermonuclear runaway but a gentle increase of temperature that allows the formation of an isothermal core of 10-60 km of radius and an off-center ignition. This is an important point as the behavior of the flame is extremely sensitive to the initial conditions.

This method of calculation has been challenged by \citet{suzu19} and \citet{zha19} who have taken into account the fact that the strength depends on the energy in forbidden transitions and have  obtained slightly different electron capture rates. An additional factor in the calculation of the ignition density has been the inclusion of the Coulomb corrections, which not only reduce the pressure but also increase the threshold for electron captures \citep{couc74} and, thus, favor the collapse.

The detailed chemical composition is also an important ingredient. Initially it was believed that $^{24}$Mg was a major constituent of the core, but detailed studies have shown that the abundance of this isotope  is much smaller than previously thought \citep{rito96,garcb97,iben97,rito99,schw17}. Electron captures on this isotope produce an increase of the temperature and a decrease of $Y_e$ in the central region and, consequently, the density at which $^{20}$Ne induces the ignition increases. However, \citet{guti05}, using the rates of \citet{oda94}  found that for a $^{24}$Mg mass fraction of x$_{24} \gapprox 0.15$ an off-center ignition occurs,  $\rho_{\rm ign}=8.3\times 10^9$~g~cm$^{-3}$ and if  x$_{24}\gapprox 0.25$ the ignition occurs at the center, $\rho_{\rm ign}=4.5\times 10^9$~g~cm$^{-3}$. In these cases a thermonuclear explosion of the star would be the most probable outcome, but these abundances are unrealistic.

Models \citep{domi93,rito96,gilp01} have also shown that carbon burning can be incomplete and O/Ne cores can contain residual amounts of carbon, x$_{12} \sim 1\%$. Some models \citep{domi93,dohe10,vent11,deni13} even predict the existence of central cores with abundances of carbon as large as 0.2 M$_\odot$. These regions are Rayleigh-Taylor unstable and quickly mix with the outer O/Ne envelope in such a way that the final abundance of carbon is reduced to values of $\sim 1\%$. In any case, the exact value is important since abundances x$_{12} \gapprox 0.01$ are enough to trigger the thermonuclear explosion of the star.

A third chemical ingredient is provided by the presence of URCA isotopes like $^{23}$Na and $^{25}$Mg \citep{gilp01}. \citet{guti05}, using the \citet{oda94} rates and including Coulomb corrections, showed that they induce a slight decrease of both temperature and $Y_e$. Because of the continuous growing of the core, URCA pairs are never in equilibrium and electron captures eventually convert all  $^{23}$Na and $^{25}$Mg into $^{23}$Ne and $^{25}$Na and this translates into a slight increase of the ignition density.
Once ignited, oxygen burning enters into a simmering phase  similar to the one that develops during the ignition and explosion of C/O degenerate cores, in such a way that the thermonuclear runaway can occur at higher densities and smaller $Y_e$. All in all, depending on the adopted physics and parameters, the runaway can start in the density range $8.9 \lapprox (\rho/10^9\,{\rm g}\,{\rm cm}^{-3})\lapprox15.8$.

The ignition density is not the unique factor that determines the final outcome. The velocity at which the flame spreads along the entire star is also critical. It is commonly accepted that the burning front propagates as a deflagration, in a laminar regime at the beginning and turbulently later on. The speed depends on the electron conductivity and on the treatment of turbulence \citep{timm92}, which demands, in fact, a full 3D treatment of the propagation process.   

The first 3D simulation (and the unique one existing at present to our knowledge) was performed by \citet{Jone16}. They considered deflagrations starting at densities $\rho/10^9\,{\rm g}\,{\rm cm}^{-3}=$~7.94, 8.91 and 19.95. In the first two cases the core experienced a thermonuclear explosion that did not completely disrupt the star leaving a bound remnant made of Fe and intermediate mass elements as predicted in \citet{iser91}, while in the third case a collapse to a neutron star of low mass seemed guaranteed. 

The existence of 'iron'-white dwarfs is a recurrent problem. There are several stars that have a radius smaller than the one predicted by the mass-radius relationship of CO and ONe white dwarfs  and can be fitted assuming that their interior contains important amounts of elements heavier than neon \citep{prov98,cata08,kepl16,beda17,joyc18}. Furthermore, \citet{radd19} have reported the existence of three new stars that, together with the prototype LP 40-365, form a class of runaway stars that are chemically peculiar that could be the inflated remnant of a tECSN or a peculiar Type Ia supernova.

 The core structure of SAGBs is very different from that of the most massive stars (those harboring an Fe-core), in the sense that they have a very steep gradient in the outermost layers and are surrounded by an extremely extended, loosely bounded H/He envelope that has a very weak inward momentum. \citet{kita06,burr07,jank08} have found successful explosions of such cores without invoking any acoustic mechanism. Probably, the most important characteristic of cECSN is the small amount of $^{56}$Ni ejected. \citet{wana09} obtained $\sim 2\times 10^{-3}$~M$_\odot$, in contrast with the  0.12 and 0.07~M$_\odot$ found in tECSN and CCSN respectively. It has been proposed that  SN 1054, the Crab Nebula, and SN2008s could have a cECSN origin.

\subsection{Pair production supernovae}
Stars  in the mass range  of 140-300 M$_\odot$ form oxygen cores with masses in excess of $\sim 50$~M$_\odot$ after helium-burning. According to stellar evolution models, these cores experience an electron-positron instability just before the oxygen ignition as a consequence of their high temperature and low density \citep{raka67,bark67,fral68}. This instability occurs off-center,  causing a violent contraction of the inner oxygen core that triggers its thermonuclear explosion. The total amount of energy released is larger than $10^{52}$ erg and no compact remnant is left \citep{hege03}. In the case of rotating stars outbursts are less energetics and eject less $^{56}$Ni and the ZAMS  mass of the progenitor can be as small as $\sim 65$~M$_\odot$ \citep{chat12,yoon12}. These events are known as pair-instability supernovae (PISN). 

Stars in the mass range of 70-90 to 140 M$_\odot$ undergo pulsational instabilities at different burning stages including O and Si-burning that induce the ejection of the outer layers without disrupting the oxygen core \citep{bark67, hege03,woos17}. These successive shell ejections can end with a full PISN outburst or a CCSN. The collision of the different shells can produce very bright events and, no matter if there is a final PISN, the total process is called pulsational pair-instability supernova (PPISN).

Stars more massive than 300 M$_\odot$, have a behavior that depends on the size of the oxygen core.   If it is smaller than $\sim 130$~M$_\odot$, the ignition of oxygen can partially reverse the collapse and a black hole of $\sim 100$~M$_\odot$ forms \citep{whal14}. If the mass of the oxygen core is larger than this critical value, oxygen burning cannot reverse the collapse and an intermediate mass black hole forms without inducing a substantial explosion \citep{hege03}.

From the nucleosynthesis point of view, the yields depend on the mass of the oxygen core and on the temperature reached during contraction \citep{hege02}. As a general trend, isotopes with even atomic number have a solar distribution while those with odd number are underproduced. The main radioactive product is $^{56}$Ni, which is produced in very large amounts. However, as a consequence of the large mass of the ejecta these explosions are not very bright.

An important problem in stellar evolution is that stars that are near the neutral stability are extremely sensitive to displacements of matter with little buoyancy since these favors mixing and the formation of 3D structures \citep{bran17}. For instance, if He is injected into the O-core a sporadic burning, followed by an expansion and cooling, can move the star from the pair-instability domain to the iron-core collapse one (Branch and Wheeler, \emph{op.cit.}).

\section{Thermonuclear explosion of electron degenerate structures in binaries} 
Low and intermediate mass stars, $ M \lapprox 8-9$~M$_\odot$, end their life as white dwarfs after expelling the outer layers during the AGB and planetary nebula phases and their destiny is to cool down forever. However, if they are members of a close binary system they can revive as a consequence of the interaction with the companion, either via mass transfer or collision. In the case of mass transfer, the radius of the white dwarf shrinks as the mass increases, as it can be easily seen from dimensional arguments. Consequently, the density increases with the mass, the degenerate electrons become relativistic, and the star is prone to explode. The outcome depends on the mass and chemical composition of the accreting white dwarf, on the nature of the accreted mass and on the accretion rate. In the case of a collision, the white dwarf explodes as a consequence of the compression and the outcome depends not only on the properties of both stars but also on the kinetic energy and impact parameter involved in the encounter. In this case, the ejected material is a mixture of pristine and thermonuclearly processed material.  

\subsection{Theoretical scenarios}

The scenarios that have been proposed to trigger the explosion can be grouped into three big  families i) single degenerate (SD), ii) double degenerate (DD), and iii) white dwarf-white dwarf collisions (WDWD). The behavior within each family of models depends on the chemical composition, mass and physical state of the accreting star as well as on the rate and chemical composition of the accreted mass. A critical ingredient is the nature of the burning front, detonation or deflagration, which depends on the state of the white dwarf.

\paragraph{Single degenerate scenario (SD)} It assumes a C/O white dwarf that accretes matter from a non-degenerate companion, that can be a normal hydrogen star (main sequence, red giant...) or a helium star and explodes when it is near the Chandrasekhar's mass  \citep{whel73, nomo82a,hach99,han04} or when enough helium has been accumulated on the top even if the mass is below the Chandrasekhar's mass \citep{woos94a,livn95,shen13}.

 The nature of the instability when the white dwarf accretes hydrogen depends on the accretion rate and the mass of the accretor. If the accretion rate is smaller than
 $ \sim 10^{-8}-10^{ -9}$ M$_\odot$~yr$^{-1}$, hydrogen accumulates on the surface of the white dwarf and becomes degenerate. When the mass of this layer reaches a critical 
value, $ \sim 10^{-4}-10^{-5}$ M$_\odot$, there is a flash,
which can be identified with a nova phenomenon, that expels almost all the accreted matter or even erodes the accretor. 
For intermediate rates, $10^{-8}- 10^{ -9}\leq \dot{M}_{\rm H} ({\rm M}_\odot/{\rm yr}) \leq 5\times 10^{-7}$, hydrogen burns steadily or through mild flashes, 
and helium accumulates on the surface of the star. If the accretion rate is high enough, this helium is converted into carbon and oxygen through weak 
flashes or steady burning and the white dwarf approaches the Chandrasekhar mass. But, if the effective accretion rate of helium is in the range  
$10^{-9}\leq \dot{M}_{\rm He} ({\rm M}_\odot{\rm yr}^{-1}) \leq 5\times 10^{-8}$, the helium layer becomes degenerate and when it reaches a critical value it experiences a thermonuclear runaway that can trigger the explosion of the white dwarf even before reaching the Chandrasekhar's limit (they are known as sub-Chandrasekhar supernovae).
If the accretion  rate is larger than $\sim 5 \times 10^{-7}$ M$_\odot$~yr$^{-1}$ a red giant-like envelope forms, a strong wind appears and the mass 
accumulates over the degenerate core.
 As in the previous case, hydrogen and helium burn peacefully and the white dwarf has the possibility to reach the Chandrasekhar mass.
Typical examples of such scenario are cataclysmic variables, classical novae, recurrent novae, symbiotic stars and supersoft X-ray sources. 

A similar behavior occurs if the companion is a helium star (AM CVn case, for instance). If the accretion rate is larger than $5\times 10^{-8}$~M$_\odot$yr$^{-1}$, helium burns peacefully and the white dwarf can approach to the Chandrasekar's mass. If it is smaller it ignites under degenerate conditions and can trigger the explosion of the star.

\paragraph{Double degenerate scenario (DD)} This class of scenarios  assumes two stars close enough that after experiencing  two episodes of common envelope evolution lead to the formation  of two close  white dwarfs. If the separation of both white dwarfs  is $\lapprox 3$~R$_\odot$, the system looses angular momentum  via the emission of gravitational waves at a rate that allows merging in less than a Hubble time. Once the secondary fills its Roche lobe, the primary starts to accrete matter \citep{iben84,webb84}.

If the secondary is massive enough, its interior is made of a mixture of carbon and oxygen and, consequently, the accreted matter has this composition. During the merging process, the secondary is destroyed after several orbital periods \citep{benz90} and a hot and thick accretion disk forms around the primary. The final outcome depends on the evolution of this disk and its interplay with the star, i.e. on the way as the angular momentum is redistributed on the system \citep{pier03a,pier03b,shen12}.  If the primary accretes spherically with a rate $\dot M \gapprox 2.7\times 10^{-6}$~M$_\odot$, carbon ignites off-center, in a similar way as in the case of SAGB stars and converts the C/O white dwarf into an O/Ne one and the final outcome can be a collapse to a neutron star or a thermonuclear explosion (see subsection~\ref{onec}).

If there is no He on the top of the primary, the impact is not able to induce a prompt ignition \citep{guer04,lore09}. During the first and more violent phase of the merger, the temperature rises quite rapidly and can trigger the ignition of carbon. However, since the degeneracy is rapidly lifted, a rapid expansion that quenches the thermonuclear runaway is ensued. Only mergers where both stars are massive enough, $M_{\rm WD} \gapprox 0.9$~M$_\odot$, can develop a detonation. It is the so called \emph{violent merger} scenario \citep{pakm10}.
 
Depending on the parameters of the system, the first newly formed C/O white dwarf can merge with the partially degenerate core of AGB companion during the second common envelope evolution. If the mass of the merger is of the order of the Chandrasekhar's mass, it will explode after sometime \citep{livi03, kash11,azna15}. This is the so called \emph{core degenerate scenario} (CD), which in some sense can be considered as a prompt merger case of a DD. The main differences are that in the canonical DD channel both stars are brought together after emitting gravitational waves for a longtime. Consequently, both white dwarfs are cool, have nearly circular orbits and, probably, they are synchronized. On the contrary, in the CD scenario the merger is triggered by the interaction with the circumbinary material and the merger occurs when the core of the AGB is still hot and the orbit of the merging objects  is very eccentric, which has a very strong influence on the duration of the coalescence process \citep{azna15}.

\paragraph{WD-WD collision scenario} In this scenario it is assumed that two white dwarfs collide and immediately ignite. These collisions can be caused by the fortuitous encounter in a dense environment like the core of a globular cluster or the central region of a galaxy \citep{ross09,rask09,lore10,rask10}\footnote{See also \url{http://sn.aubourg.net/workshop09/pdf/timmes.pdf}}, or the interaction of a binary white dwarf in a hierarchical triple or quadruple system via the Kozai-Lidov mechanism \citep{kush13,fang18,hame19}. 
\citet{azna13} performed a study of such events using a broad range of parameters covering all the main core compositions (He, C/O, O/Ne) and masses (0.4 - 1.2 M$_\odot$) of the interacting white dwarfs, as well as different impact parameters and initial velocities. There were three types of results, i)  direct collision, with one violent mass transfer episode and complete destruction of the secondary, ii) lateral collision with several mass transfer episodes during the dynamical event, and iii) formation of an eccentric binary. 
The frequency of these events is not known at present and the estimations go from those claiming they can account for a substantial fraction of all SNIa \citep{katz12} to those considering their contribution as almost negligible  \citep{toon18} passing through those advocating for a contribution $\lapprox 10\%$  \citep{fang18,hame19,hall19}.

Therefore, the question is to determine if a unique scenario is enough to account for  all the observed SNIa events or if it is necessary to invoke several ones having in common the presence of a white dwarf. The existence of many SNIa subtypes clearly favors the second option, while the $t^{-1}$ dependence of the SNIa suggests that the late events are a consequence of the merging of two white dwarfs.

The associated question is to wonder if the DD population can sustain the observed supernova rate. In this case, the main problem is that white dwarfs are dim objects and, at present, the sample is only complete up to $\sim 100$~pc, while the supernova rate is an average over all the Galaxy \citep{iben91,iser97}. In other words, is the solar neighborhood representative of all the Galaxy? When the influence of the scale height inflation induced by galactic interactions, radial migrations and other selection effects is taken into account, it turns out that, effectively, DD systems can reproduce the observed SNIa rates, provided that sub-Chandrasekhar mergings are able to produce successful explosions \citep{iser97,bade12} (see \citet{maoz14} for a detailed discussion).

\subsection{Nucleosynthesis in the different scenarios}

\begin{figure}[htb]
\center
\includegraphics[width=0.8\textwidth, clip=true, trim= 0cm  0cm 0cm 0cm]{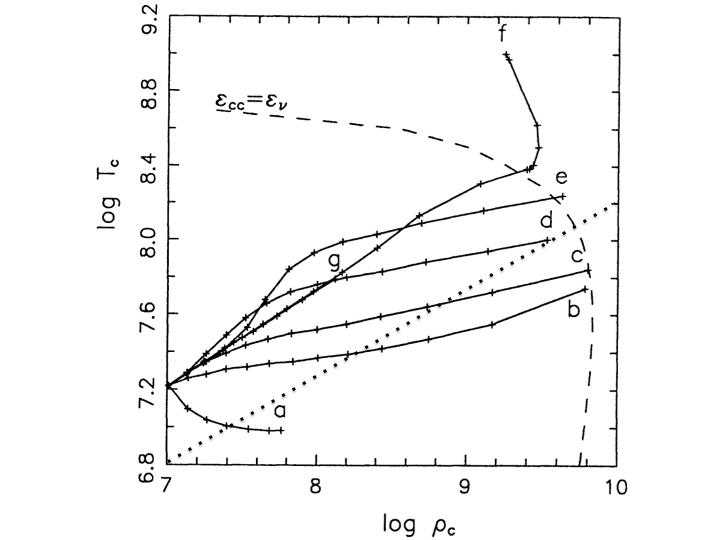}
\caption{\footnotesize{Evolution of the center of a C-O white dwarf in the $\log T$-$\log\rho$ plane for several accretion rates of pure C-O: 
a) $10^{-10}\,M_\odot/\mathrm{yr}$;
b) $5\,10^{-10}\,M_\odot/\mathrm{yr}$;
c) $10^{-9}\,M_\odot/\mathrm{yr}$;
d) $5\,10^{-9}\,M_\odot/\mathrm{yr}$;
e) $5\,10^{-8}\,M_\odot/\mathrm{yr}$;
f) $5\,10^{-7}\,M_\odot/\mathrm{yr}$;
g) $5\,10^{-6}\,M_\odot/\mathrm{yr}$. 
The dashed line represents the $^{12}$C ignition curve and the dotted line the transition from the thermonuclear to the pycnonuclear regimes. Figure obtained from  \citet{brav96}.}}
\label{fign}
\end{figure}

\begin{figure}
\center
\includegraphics[width=0.75\textwidth, clip=true, trim= 0cm  0.2cm 0cm 0.5cm]{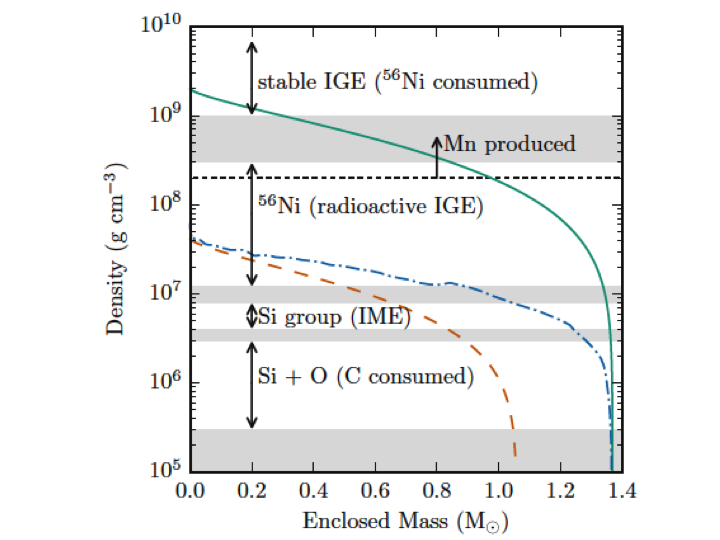}
  \caption{\footnotesize Nucleosynthesis products as a function of the density of the fuel. A WD close to the Chandrasekhar mass  (solid green) produces almost IGE only and a 1.05 M$_\odot$ (dashed orange) does not produce Mn or stable IGE because of electron capture if expansion is not allowed. If expansion is allowed, like in delayed detonations (dot-dashed blue) significant amounts of IME are produced as well as radioactive IGE. Figure from  \citet{seit17}. }
\label{f05nuc}
\end{figure}

The ignition line (Fig.~\ref{fign}), is defined as the loci where the energy release by nuclear reactions ($^{12}C + ^{12}C$ for instance) is counterbalanced by the neutrino losses. Once this line is surpassed as a consequence of the growing of the CO core, nothing can avoid the explosion. As the temperature raises, the characteristic time scale of the white dwarf evolution decreases and becomes comparable to the dynamic or sonic time, $\tau_{\rm d} \sim 0.1$~s, and a thermonuclear runaway occurs. If the CO white dwarf is incinerated with a time scale of the order of this time, the spectroscopic and photometric behavior of the ejecta is similar to that of a Type Ia supernova: H and He are absent, strong Si features are present at maximum light, and the light curve is powered by the radioactive decay of $^{56}$Ni. This is a consequence of the fact that the composition of the final products of carbon-burning strongly depends on the local density of the fuel. High densities, $\rho \gapprox 10^7$~g~cm$^{-3}$, allow the complete processing to iron group elements (IGE), although if the density is too high, $\rho \gapprox 10^9$~g~cm$^{-3}$ neutronization strongly reduces or even eliminates the presence of radioactive IGE isotopes like $^{56}$Ni. If the density is low,  $\rho \lapprox 10^7$~g~cm$^{-3}$, only Si-group and intermediate mass elements (IME) are produced (Fig.~\ref{f05nuc}).

Therefore, besides the initial density, the most important ingredient is the nature and propagation velocity of the thermonuclear burning front. Two basic mechanisms have been identified: detonations \citep{arne69} and deflagrations \citep{nomo76}. The relevancy of these two modes of propagation stems from the fact that a detonation is supersonic and matter is burned before having time to expand noticeably. This means that the structure of the white dwarf remains frozen and, since  the energetic and nucleosynthetic outcome of the combustion strongly depends on the ignition density, it is the structure of the white dwarf which determines the final results. On the contrary, in the case of a deflagration there is room for a vigorous expansion that can even quench the flame. In general, the final results depend on a complex interplay between different hydrodynamic instabilities and the energy generation.

The observed spectra of SNIa show that the innermost regions of the ejecta contain important amounts of stable IGE \cite{mazz15} thus implying an important degree of neutronization. This excess of neutrons comes partially from the presence of species like $^{22}$Ne synthesized during the previous evolution, and partially from the electron captures during the explosive phase if the densities are larger than $\sim\,10^9$~g~cm$^{-3}$.  

At densities\footnote{These densities demand carbon oxygen  white dwarfs more massive than 1.2 M$_\odot$, which can only be formed by accretion}  
$\rho\gapprox 10^9$~g~cm$^{-3}$, electron captures are rapid enough to transform NSE matter dominated by radioactive $^{56}$Ni into matter dominated by stable  $^{54}$Fe and  $^{58}$Ni in less than a dynamical time. Longer exposures can lead to the production of more neutronized species. It is important to realize that if it was possible to neglect the neutronized species present before the explosion, a comparison of their abundances with the solar ones could provide important constraints to the densities achieved during ignition. 

The most abundant nuclei containing an excess of neutrons are those of  $^{22}$Ne. This isotope is synthesized during the He-burning from the $^{14}$N left by the H-burning stage. Its abundance is approximately equal to the sum of the abundances of $^{12}$C, $^{14}$N, and $^{16}$O, but it can locally change with time due to gravitational diffusion and crystallization \citep{brav92}.

The thermonuclear runaway that triggers the explosion of the star is preceded by a phase of convective core burning that can also increase the neutron number thanks to electron captures on the products of combustion of carbon and URCA process \citep{cham08,piro08,mart16,pier17}.

\section{ Supernova remnants}
The distinction between supernova and supernova remnant (SNR) is a bit artificious since their evolution is a continuous process. Initially, the explosion debris are dominated by the physical properties  of the event, but as the time goes on, the interaction with the circumstellar medium at the beginning and later on with the interstellar medium becomes more and more important. As a rule of thumb, \citet{bran17} propose to consider supernova remnants  all supernovae older than $\sim 100$~years, but in the case of SN1987A, the optical flux started to grow around 2001 and doubled by the end of 2009 as a consequence of the energy deposited by the X--rays produced by the interaction of the ejecta with the surrounding material \citep{lars11}. Probably it is more adequate to consider that as soon as there is a shocked shell of circumstellar medium there is a supernova remnant. In that sense SN~1987A is both: the material heated by $^{44}$Ti is supernova, the hot shell seen in X-rays/optical and radio is supernova remnant.

The number of currently known galactic SNRs is of the order of 300, the majority detected through their radio emission. As they are mostly concentrated along the Galactic Plane, their optical and X-ray emission is strongly absorbed and the number of existing SNRs in the Galaxy must be certainly much larger.
Although the distances and linear sizes are not well known, it is believed that they go from few to one hundred parsecs. This means that SNRs have swept up important amounts of interstellar medium (ISM)  and consequently, excepting the young ones, they have forgotten the initial conditions of the explosion and their evolution is strongly dependent on the environment. Furthermore, the dynamics of the forward shock may be affected by the acceleration of cosmic-rays \citep{warr05}, although there is still some debate on this point as very high cosmic ray efficiencies are not certain.

The dynamical properties of historical, young, SNRs (size, shock velocities) allow one to constrain their progenitors \citep{bade07,wood17}, while their X-ray spectra provide insights into the explosion mechanism \citep{bade06,park13}. Although it is difficult to give absolute values to the ejected mass of any element in a SNR, the mass ratios of elements with close proton numbers, e.g. manganese and chromium, presumably produced in the explosion in the same mass range, allows to constrain the physical conditions experienced during their nucleosynthesis: temperature, density, electron mole number \citep{bade08,mart17}. This is particularly true for localized spots in nearby supernova remnants like Tycho's and Kepler's ones \citep{yama15,sato20}.

In general, SNIa remnants have a stratified structure and a more regular shape, while those from CCSN have an irregular shape. From the morphological point of view they are classified as \emph{shell}-type when they display a shell of shocked matter at the edge  of the remnant, like Tycho or SNR~0509-67.5 in the LMC, as \emph{plerions} when they have a filled centre dominated by the non-thermal emission powered by a pulsar wind nebula, like the Crab Nebula, and \emph{plerionic composites}, when they are essentially  pulsar wind nebulae inside a shell, in which case they may exhibit complex morphologies.

This behavior is a consequence of the fact that real supernovae are not spherical. Rotation, convective instabilities, the formation of a neutron star or the presence of a companion in the case of explosions occurring in binary systems can introduce asymmetries. Furthermore, the medium surrounding the exploding star is not necessarily uniform and can have been modified by the progenitor. For instance, it is known that Cas A originated from a highly bipolar explosion as shown by the two jets in the northeastern and southwestern directions observed in the optical \citep{fese06}, X-rays \citep{hwan04}, and through the light echo \citep{rest11}. Furthermore,  it has been possible to detect in the Cas A surroundings the influence of a fast main sequence  wind, followed by that of  a slower, clumpy red giant wind and finally that of a high-velocity wind similar to the one found in yellow supergiants \citep{weil20}.

The best studied cases are, obviously, those associated with historical supernovae such as SN 1006, the Crab Nebula, Tycho, Kepler, and Cas A\footnote{Although the supernova event has not been witnessed by Flamsteed \citep{step02}, it is generally considered as a historical SNR}. See \citet{reyn08,vink12,lope18} for a complete review.

\section{Classical Novae}

Despite their name, {\it novae} are not new stars but stars that suddenly increase their luminosity by $\gapprox 10$ magnitudes and return to their previous faint state in a few months or years. \citet{walk54}  discovered that DQ Her, a nova that exploded in 1934, was in fact an eclipsing binary and \citet{kraf64} showed that this was a common property not only of novae but also of cataclysmic variables (CV). At present it is completely accepted that nova explosions occur in white dwarfs accreting hydrogen rich matter from a  close main sequence star in a binary system  of the CV type or, less frequently, from a red giant in a symbiotic binary system. If the accretion rate is small enough, hydrogen accumulates and eventually ignites under degenerate conditions leading to mass ejection at velocities of hundreds to thousands of km s$^{-1}$ and to a large increase of luminosity that even reaches the Eddington value of $10^{37}-10^{38}$~erg~s$^{-1}$. 

The nova phenomenon is expected to be recurrent since, in contrast with SNIa, the outburst only affects the external hydrogen-rich layers, i.e. $10^{-5}-10^{-4}$~M$_\odot$. The periodicity is expected to be in the range of $10-10^5$~years, depending on the mass of the white dwarf and on the accretion rate as it represents the time necessary to accrete enough matter to reach the critical value that allows the explosion\footnote{For historical reasons, the term {\it recurrent novae} is used to describe only those events with more than one {\it recorded} nova outburst.}.

The evolution of white dwarfs experiencing the nova phenomenon is still a matter of debate since it is not clear if the mass of the white dwarf grows and approaches to the Chandrasekhar's mass or decreases after each explosion [\cite{kato15,jose19}]. Overabundances with respect to the solar values of elements like carbon, oxygen, neon and others suggest that some mixing between  the freshly accreted matter and the interior must occur which, in turn, could be an indication that the mass of the white dwarf decreases with time. However, in the case of recurrent novae, that take place in massive white dwarfs, no large overabundances are observed. In this case it is believed that the mass of the white dwarf can grow and approach to the Chandrasekhar's mass.

The light curves of classical novae are classified according to their speed class, defined either from $t_2$ or $t_3$, the time to decay by 2 or 3 visual magnitudes after maximum. Speed classes range from very fast ($t_2 <10$ days) and fast ($t_2 \sim 11-25$ days) to very slow ($t_2 \sim 115 - 250$ days) \citep{payn57}. Some examples are the fast nova N Cyg 1992, with $t_2 \sim 12 $ days, the very fast N Her 1991 with $t_2 \sim 2$ days or the slow one N  Cas 1993 with $t_2 \sim 100$ days). Empirically it has been seen that the absolute magnitude at maximum, $M_V$ is related with the decay times ($t_2$ or $t_3$ ), i.e. the brighter is the maximum the shorter is the decay time. The theoretical explanation relies on the fact that during maximum the luminosity is close to the Eddington limit and the time necessary to eject the envelope is of the order of $t_3$ \citep{livi92}. This means that $L_{\rm max}$ and $t_3$ increase and decrease, respectively, with the mass of the white dwarf allowing to establish an empirical relationship that can be used to estimate distances. The most widely accepted empirical calibration is that of \citet{dell95} and it is important to point out that there are two distinct nova populations i) disk novae, that are generally bright and fast ($M^{\rm max}_{\rm V}\approx -8$) and ii) bulge novae, that are dimmer and slower ($M^{\rm max}_{\rm V}\approx -7$). However, with the recent improvement of distances to novae provided by Gaia parallaxes, the maximum-magnitude-rate-of-decline (MMRD) relation has been proven to be not valid in several cases, and it cannot be used to determine nova distances \citep{scha18}.

Space based observations have allowed the measurement of fluxes in other energy bands than the optical. An important fact has been the discovery, by the International Ultraviolet Explorer (IUE), that the flux in the UV increases when the optical one starts to decline. The origin of such a behavior is the shift to higher energies when the receding photosphere reaches deeper and hotter regions of the expanding envelope. In the infrared there is an increase in luminosity once the UV one starts to decline. This effect is more important in the case of novae that form dust and is interpreted as a consequence of the absorption of the UV component by the dust grains and its re-emission in the infrared. All in all the bolometric luminosity of classical novae remains constant during a period of time that depends on the mass of the H-rich envelope that remains on the white dwarf after the nova eruption. Evidences of this residual H-burning were provided by the emission of soft X-rays detected with ROSAT \citep{krau96}. The bolometric luminosity obtained in this way was close or even larger than the Eddington limit suggesting that probably the radiation pressure is responsible of the nova mass ejection \citep{kato94}.

An important result has been the discovery that nova ejecta are often enriched in carbon, nitrogen and oxygen in about $\sim 1/3$ of the events in such a way that metallicities are well above the solar values.

There is a wide consensus that classical nova explosions occur in close binary systems containing a white dwarf made of carbon-oxygen (CO) or oxygen-neon (ONe) that is accreting hydrogen at a rate $\dot M \sim 10^{-9} -10^{-10}$~M$_\odot$yr$^{-1}$.  This mass accretion rates are small enough to compress the fuel under degenerate conditions in such a way that a thermonuclear runaway occurs upon ignition. Explosive hydrogen burning synthesizes $\beta^+$-unstable nuclei of short lifetime ($^{13}$N, $^{14}$O, $^{15}$O, and $^{17}$F) with lifetimes $\tau = 862,\,102,\, 176,$ and $93$~s respectively that are transported by convection to the outer layers, where they decay before having been burned\footnote{Another relevant short-lived $\beta^+$-unstable isotope is $^{18}$F, with $\tau$ =158~min. It is not crucial for the explosion itself, but for the gamma-ray emission, as explained in next section.}. The energy released in this way is responsible for the nova outburst. Mixing between the core and the accreted envelope is crucial to power the thermonuclear runaway and to explain the observed enhancement of metals in many novae. Other isotopes synthesized during the explosive hydrogen burning that are interesting as diagnostic tools are $^7$Be, $^{22}$Na and $^{26}$Al with lifetimes of $\tau$ = 7.7~d, 3.75~yr, and $10^6$~yr respectively. 

The nova rate in our Galaxy is hard to estimate because the Sun and the interstellar extinction prevent us to discover more than a fraction of the novae that explode each year. The two methods generally used to estimate the Galactic nova rate are (1) the extrapolation of Galactic nova observations to a complete sample, 
and (2) the adoption of the observation of nova rate measured in external galaxies to our own Galaxy. The first method yields to rate of 40-100 yr$^{-1}$ (e.g. \citep{alle54,shaf97}) while the second method estimate rates ranging from about 10 to 45 yr$^{-1}$ \citep{ciar90,dell94}).  The best estimate currently available of the nova rate in the Galaxy is $50^{+31}_{-23}$~yr$^{-1}$  with the first method \citep{shaf17}.

\section{Gamma-rays as a diagnostic tool}

The radiation transport in the optical and in the infrared is strongly dependent on the opacity which, in turn, depends on the detailed chemical composition and the degree of excitation and ionizations of the different chemical species. On the contrary, the radiative transfer in the gamma-ray regime is determined by the radioactive half-lives, branching ratios and relatively simple interaction processes like pair-production, Compton scattering, and photoelectric absorption. This simplicity makes gamma-rays ideal to determine the mass-velocity distribution of the explosion products.

Supernovae and novae emit low energy $\gamma$-rays because some of the nuclei they synthesize and eject are radioactive. These isotopes, which usually emit positrons or experience electron captures, decay to excited states of their daughter nuclei and de-excite to the ground state emitting $\gamma$-ray photons in the MeV domain and over a wide range of timescales.  The information they can provide depends on the lifetime of the radioisotope producing them. Short-lived isotopes provide direct information about the physical properties of the event, intermediate-lived isotopes on the remnants, and long-lived ones on the evolution of the Galaxy.  Table~\ref{isotopes} displays the most relevant isotopes produced in novae and supernovae.

\begin{table*}
\caption{Radioactive isotopes synthesized in explosive events}
\label{isotopes}
\centering
\begin{tabular}{ccccc}
\hline \hline
Isotope            & Decay chain         
              & Disintegration process               
              & Lifetime                  & Line energy (keV) \\
\hline  \hline
$^{7}$Be        & $^{7}$Be  $\rightarrow$ $^{7}$Li                
            &  $e^-$-capture
            & 77~d                          & 478 \\
$^{22}$Na      & $^{22}$Na  $\rightarrow$ $^{22}$Ne                
            &  $\beta^+$
            & 3.8~yr                       & 1275 \& 511\\
$^{26}$Al      & $^{26}$Al \  $\rightarrow$ $^{26}$Mg  
            &  82\% $\beta^+$              
            & 1.0$\times$10$^6$~yr          & 1809 \& 511\\
$^{44}$Ti       & $^{44}$Ti \  $\rightarrow$ $^{44}$Sc  $\rightarrow$ $^{44}$Ca 
 
            &  $e^-$-capture (94\% $\beta^+$)
            & 87~yr (5.7~h)              & 78, 68 (1157, 511)\\            
$^{56}$Ni       & $^{56}$Ni  $\rightarrow$  $^{56}$Co   
            &  $e^-$-capture          
            & 8.8~d                         & 158, 812, 750, 480, 270 \\
$^{56}$Co      &  $^{56}$Co   $\rightarrow$  $^{56}$Fe   
            &  80\% $e^-$-capture
            & 111~d                        & 847, 1238 \& 511 \\
$^{57}$Ni       & $^{57}$Ni  $\rightarrow$ $^{57}$Co $\rightarrow$ $^{57}$Fe 
            &  $\beta^+$ ($e^-$-capture)
            & (51~h) 390~d              & 511, 1378 (122, 136) \\
$^{60}$Fe      & $^{60}$Fe  $\rightarrow$ $^{60}$Co   $\rightarrow$ $^{60}$Ni  
            &  $\beta^-$
            & 3.8$\times$10$^6$yr (7.6~yr)    & 59 (1173, 1332) \\
\hline \hline
\end{tabular}
\end{table*}

\subsection{Theoretical models of SNIa}
The total amount and distribution of the different radioactive species  as well as the density and expansion profiles are model dependent,  leading to significant differences in the evolution of the intensity and profiles of the $\gamma$-lines they emit, thus opening the opportunity to use them as diagnostic tools able to provide a deep insight on the events. 
Several authors have examined this question and explored the predictions provided by the different model variants \citep{gehr87,ambw88,burr90,ruiz93,hoef94,kuma97,timm97,gome98,sim08,the14}. The results they obtained were self-consistent when applied to the same 1D models of supernova and assuming that the radioactive elements are buried in the inner regions of the ejecta \citep{miln04}. 

For instance, \citet{gome98}, using a code based on the methods described by \citet{colg80}, \citet{pozd83} and \citet{ambw88}, have shown that the $\gamma$--ray emission  before and around the epoch of maximum of the optical light curve  can be characterized by a spectrum dominated by the 158 and 812 keV $^{56}$Ni lines. Because of the rapid expansion, the lines are blue-shifted but their energy peak quickly evolves back to the rest wavelength as matter becomes more and more transparent. The emergent lines are broad, typically from  3\%  to 5\%, because of the Doppler effect, and the 812 keV line blends with the quickly growing 847 keV $^{56}$Co line, forming a broad feature, for which reason is advisable to put the emphasis on the 158 keV line. On the contrary the 847 and 1238 keV $^{56}$Co lines provide a reliable diagnostic tool at late times. The intensity of the $^{56}$Ni lines rises very quickly with time, after being very weak at the beginning, even in the case of Sub-Chandrasekhar's models. This fact, together with the relatively short lifetime of $^{56}$Ni, makes the observational window rather narrow (Eq.~\ref{nsigma}). As a rule of thumb it can be said that the emergence of the $^{56}$Ni lines strongly depends on the distribution of this radionuclide in the ejecta, while the late emission by $^{56}$Co provides an average over the ensemble.

\begin{table}
\caption{Mass of the white dwarf, mass of $^{56}$Ni synthesized  and kinetic energy for different models of SNIa explosion mentioned in this review.
}             
\label{tmodel}     
\centering                   
\begin{tabular}{c c c c c}       
\hline\hline                 
 Model &  M$_{\rm WD}$(M$_\odot$) & M$_{\rm Ni}$(M$_\odot$) & K(Bethe) & Author \\ 
\hline  
1D          &                 &          &         &    \\
\hline
DETO     &  1.38         & 1.16 & 1.44 &  \citet{bade03}   \\
W7         &  1.38         & 0.59 & 1.24 &  \citet{nomo84} \\
DDTc     &  1.37         &  0.74 &  1.16 & \citet{bade05}   \\
DDTe     &  1.37         &  0.51 &  1.09 & \citet{bade05}    \\
DD202c &  1.38         & 0.78 &  1.30 & \citet{hoef98} \\
SC1F      &  0.81+0.125  & 0.43 & 1.04 &  \citet{iser13}\\
SC3F      &  1.025+0.054  & 0.69 & 1.17 & \citet{iser13} \\
HED6     &  0.6+0.17 & 0.26 &  0.72 & \citet{hoef96}  \\
\hline
DDT1p4 &   1.37       & 0.65 & 1.32 &  \citet{iser16}\\
3Dbball &    1.37       &  0.65+0.05& 1.32 & \citet{iser16} \\
\hline                                   
\end{tabular}
\end{table}

Spherically symmetric models can be grouped into different categories depending on the treatment of the burning front:
\begin{itemize}
\item
Pure detonation models (DETO in Table~\ref{tmodel}) assume that carbon is ignited in the center of a C/O white dwarf near the Chandrasekhar's mass and the burning propagates supersonically in such a way that the star is completely incinerated to iron peak elements \citep{arne69}. These models are representative of the most massive models computed by \citet{fink10}.
\item
Pure deflagration models. In these models the deflagration propagates at the laminar velocity in the central regions until the Rayleigh-Taylor instability develops and wrinkles the flame surface  increasing the mass burning rate. The flame remains subsonic all the time and can be quenched by the expansion of the material. The W7 model of \citet{nomo84} is the classical one.
\item
Delayed detonation models. In these models, the flame initially propagates as a deflagration and, when it reaches a critical density $\rho_{\rm tr} =10^6 - 10^7$~g~cm$^{-3}$ makes a transition to a detonation \citep{khok91}. See the models labelled DDTc,e and DD202c of Table~\ref{tmodel}.
\item
Pulsating delayed detonation models. These models assume that the deflagration-to-detonation transition occurs after a pulsation, during the contraction phase, induced by the inefficient burning associated to a slow deflagration. The final outcome is very similar to the delayed detonation models \citep{khok91,dess14}.
\item
Sub-Chandrasekhar detonations. They assume that C/O white dwarfs with arbitrary masses  accrete helium from a companion in such a way that when the accreted mass reaches a critical value it ignites and triggers the explosion of the white dwarf \citep{woos94a}. Models SC1F and SC3F of Table~\ref{tmodel} are equivalent to models 1 and 3 of \citet{fink10}. Model HED6 corresponds to the explosion of a C/O white dwarf that accreted 0.17 M$_\odot$ of helium and exploded \citep{hoef96}. An interesting characteristic of this scenario is the production of $^{44}$Ti as a consequence of an alpha-rich freeze-out of the explosive Si-burning in the He-rich layers \citep{shen14}.
\end{itemize}

Model DDT1p4 was tailored to fit the optical light curve and the late gamma spectra of SN2014J. It is ignited at a density of $2\times 10^9$ g~cm$^{-3}$ and makes a deflagration/detonation transition at a density of $1.4\times 10^7$ g~cm$^{-3}$. The ejected mass is 1.37 M$_\odot$, the total $^{56}$Ni synthesized is 0.65 M$_\odot$ and the kinetic energy $1.32\times 10^{51}$ erg. The 3Dbball model is essentially the DDT1p4 model plus a plume similar to the one depicted on the right of Fig.~\ref{f14jbravo12}. The spectrum was computed in 3D as described in \citet{iser08}.

\subsection{Theoretical models of core collapse supernovae}

As it has been mentioned before, CCSN produce significant amounts of radioactive material, being $^{56}$Ni the dominant isotope. The total amount synthesized depends on the structure of the core of the exploding star while the observable properties of the transient also depend on the structure of the ejecta and how the radioactive  elements are distributed  in them. Therefore determining these properties is crucial for the understanding of how massive stars explode. \citet{youn06} and \citet{magk10} have performed extensive calculations of the yields of $^{44}$Ti and $^{56}$Ni, while \citet{andr20} have recently computed the yields of less abundant radioactive isotopes produced during the explosion of massive stars of ZAMS mass of 15, 20 and 25 M$_\odot$ respectively and have found that $^{43}$K, $^{47}$Ca, $^{44,47}$Sc and $^{59}$Fe are good indicators of the energetics of the explosion, while $^{48}$V and $^{51,57}$Cr are good indicators of the structure. In particular, $^{44}$Ti is produced during the $\alpha$-rich freeze out  with a yield that strongly depends on the mass cut, kinetic energy and asymmetry of the explosion \citep{naga98,magk10}. It  is produced in the innermost regions of CCSN, just at the top of the newly formed compact object.  Thus, it is an excellent diagnostic tool of the inner parts of the stellar explosions as it provides a direct probe of the supernova engine and, in particular, of how the energy is transferred to the inner ejecta.

In the case of SNII it is possible to measure the amount of $^{56}$Ni synthesized using the properties of the exponential tail of the optical light curve just after the fully recombination of hydrogen if the moment of the explosion is known and it is assumed that $\gamma$-photons are fully trapped, which seems to be the case \citep{ande14}. In the case of the stripped envelope supernovae (SE SN), the tail decreases more rapidly than expected indicating that $\gamma$-photons are not fully trapped \citep{whee15} and the mass has to be estimated via Arnett's rule\footnote{The luminosity at maximum is proportional to the mass of $^{56}$Ni, i.e.$L_{\rm max} \propto M_{\rm Ni}$.}, although there are objections \citep{dess16,khata19}. The main inconvenient of this method  comes from the conversion of the observed photometry into bolometric magnitudes caused by the uncertainties in the distance, light extinction, luminosity of the host  galaxy, and missing flux outside the observed bands.
The analysis of the values  obtained up to now suggests that the $^{56}$Ni yields synthesized in stripped envelope SN are larger than those produced in SNII \citep{ande19} and that a significant fraction of them have measured masses larger than those predicted by neutrino driven models.

The final fate of SAGBs is still uncertain as a consequence of the existing caveats in the understanding of the physical processes involved and in the computational difficulties. Therefore, the detection of a gamma-ray emission from these electron-capture supernovae would be highly valuable.

It is thought that Pair Instability Supernovae (PISN) have been common in the early Universe but are rare at present. One possible recent event is SN 2007bi, a luminous, slowly evolving supernova that exploded in a dwarf galaxy and ejected $\sim 100$~M$_\odot$ containing $\gapprox 3$~M$_\odot$ of $^{56}$Ni \citep{galy09}. The decay of this huge amount of $^{56}$Ni will occur before the maximum light as a consequence of the large mass of ejecta. Furthermore, these massive envelopes  are able to trap all the $\gamma$-rays  and the tails of the light curve follows the decay rate of $^{56}$Co as it has been seen in SN 2007bi \citep{galy09}. 
However, as mentioned before, the presence of blobs or holes in the structure cannot be discarded given the 3D nature of the evolution before the explosion and the possibility of detecting the existence of radioactive materials cannot be completely rejected.

\subsection{Theoretical models of supernova remnants}
According to the classical work of \citet{wolt72}, the evolution of a spherical SNR within a uniform ambient can be divided into several stages. At the beginning the dynamics is dominated by the ejecta that creates a shock that propagates into the interstellar medium (the forward shock). As the shock sweeps more and more mass it slows down, the ejecta collides with the swept-up material and a rarefaction wave  moving inwards forms (the reverse shock). When the reverse shock reaches the explosion center and all the material has been heated, the forward shock tends towards a Sedov-Taylor solution (at this time the forward shock front has run over an ISM mass close to that of the supernova ejecta). Depending on the ISM density, this first epoch can last several thousand years and, since the bulk of the kinetic energy is converted into thermal energy, the remnant strongly emits in the X-ray band. The total losses however are relatively small and it is possible to assume an adiabatic behavior. When the speed of the forward shock is small enough, the SNR behaves as a pressure-driven and momentum-driven bubble surrounded by a thin shell that finally dissolves into the ISM.

The most abundant radioactive isotopes produced in supernovae are $^{56}$Ni, $^{57}$Ni,  $^{55}$Fe and $^{44}$Ti together with their radioactive decay products. The first three have a short lifetime and can only be directly observed during the very first epoch of the eruption. During more advanced stages the $^{56}$Ni yield can only be measured through the shock-heated iron, the decay product of nickel, but this is not an easy task since it demands an accurate modeling of the X-ray emission of the SNR and an estimation of how much ambient iron has been swept-up by the shock. A similar problem occurs with the other non-radioactive isotopes ejected by the supernova.

Titanium-44 with its lifetime of 87 years and its associated gamma-ray lines of 67.87, 78.32 keV, produced by the dexcitation of $^{44}$Sc with 93 and 96\% of probability, and of 1157 keV by the dexcitation of $^{44}$Ca with 100\% of probability, is well taylored to study SNRs in the age range of 100 to 500 years: It is produced during the $\alpha$-rich freeze out and, consequently its yield is strongly dependent  of the physical conditions of the explosion \citep{magk10}. Theoretical models of CCSN typically predict yields in the range of $10^{-5}-10^{-4}$~M$_\odot$, while SNIa models predict a smaller production, $10^{-6}$~M$_\odot$ for a central pure deflagration to  $\sim 6\times 10^{-5}$~M$_\odot$ for a delayed detonation \citep{maed10}, but can be very large in the case of a sub-Chandrasekhar model, $\sim 10^{-3}$~M$_\odot$ \citep{fink10}.

\subsection{Theoretical models of novae}
The potential role of novae as $\gamma$-ray sources was already pointed out in the 70's \citep{clay74} and 80's \citep{audo82}.
The $\gamma$-ray emission from classical novae has its origin on the disintegration of some short- and medium-lived radioactive nuclei.  
Isotopes like $^{13}$N ($\tau$=862 s) and $^{18}$F ($\tau$=158 min) are $\beta^+$-unstable and short-lived. Other radioactive nuclei synthesized in 
novae are $^{22}$Na (again a $\beta^+$-unstable nucleus, but with a longer lifetime, $\tau$=3.75 yr), and $^{7}$Be (which experiences an electron capture giving $^{7}$Li, 
with $\tau$=77days,). As a consequence of such decays, the $\gamma$-ray emission from novae has two types of components, from the spectral 
point of view: lines (511, 478 and 1275 keV, see Table 3) and continuum (between 20-30 keV and 511 keV). From the point of view of temporal behaviour, the emission 
can be considered prompt (very early appearance and short duration) or long-lasting.

The long lasting emission originates in the decay of $^{7}$Be and $^{22}$Na, 
which emit lines at 478 and 1275 keV, respectively. The fluxes and their 
duration directly reflect the content of these isotopes in the expanding envelope and their lifetimes.
Since CO and ONe novae preferentially synthesize $^7$Be and $^{22}$Na, respectively, the long-lived
478 and 1275 keV lines can be used to identify the nature of the underlying white dwarf. The maximum of the 478 keV line is reached  one to two weeks after the explosion, 
depending on the opacity of the envelope, when white dwarf mass changes from 1.15 to 0.8 M$_\odot$, and its intensity
 is about $10^{-6}$ cm$^{-2}$s$^{-1}$ for a distance of 1 kpc. The width of the line ranges from 3 to 8 keV, for the low and high white dwarf masses, respectively. The 1275 keV line emission has a rising phase 
 that lasts 10 to 20 days for white dwarf masses from 1.25 to 1.15 M$_\odot$. Soon after maximum, the line emission declines with the same characteristic time as  $^{22}$Na, 3.75 years, and its intensity 
 directly reflects the amount of $^{22}$Na in the ejecta. The flux at maximum is $10 ^{-5}$ cm$^{-2}$s$^{-1}$ for a distance of 1 kpc and the width of the line is $\sim 20$~keV.

Short-lived $\beta^+$-radioisotopes $^{13}$N and $^{18}$F are responsible for the early or \emph{prompt} $\gamma$-ray emission. Their contribution is a line of 511 keV caused by the direct annihilation of positrons and the singlet state of positronium, plus a continuum that is related to the triplet state of positronium and the comptonization of the line \citep{gome98a,hern02}. This continuum has a sharp cut-off at energies  of 20-30 keV caused by the photoelectric absorption. The exact position of such cut-off depends on the chemical composition. The maximum of the continuum is at about 60 and 45 keV in the case of ONe and CO novae, respectively. The light curve displays two maxima, that correspond to positrons emitted by $^{13}$N and by $^{18}$F. The first one is very model dependent, as only the outer layers are contributing, and very short lived. Its detection would be very important since it could provide insight into the dynamics of the envelope. The positron annihilation feature is the strongest one, but it is very short-lived and occurs before the optical maximum, for which reason it demands well suited observation strategies.

\subsection{Past  and current missions}
\begin{figure}
\includegraphics[width=12cm, clip=true, trim= 3cm  9cm 3cm 9cm]{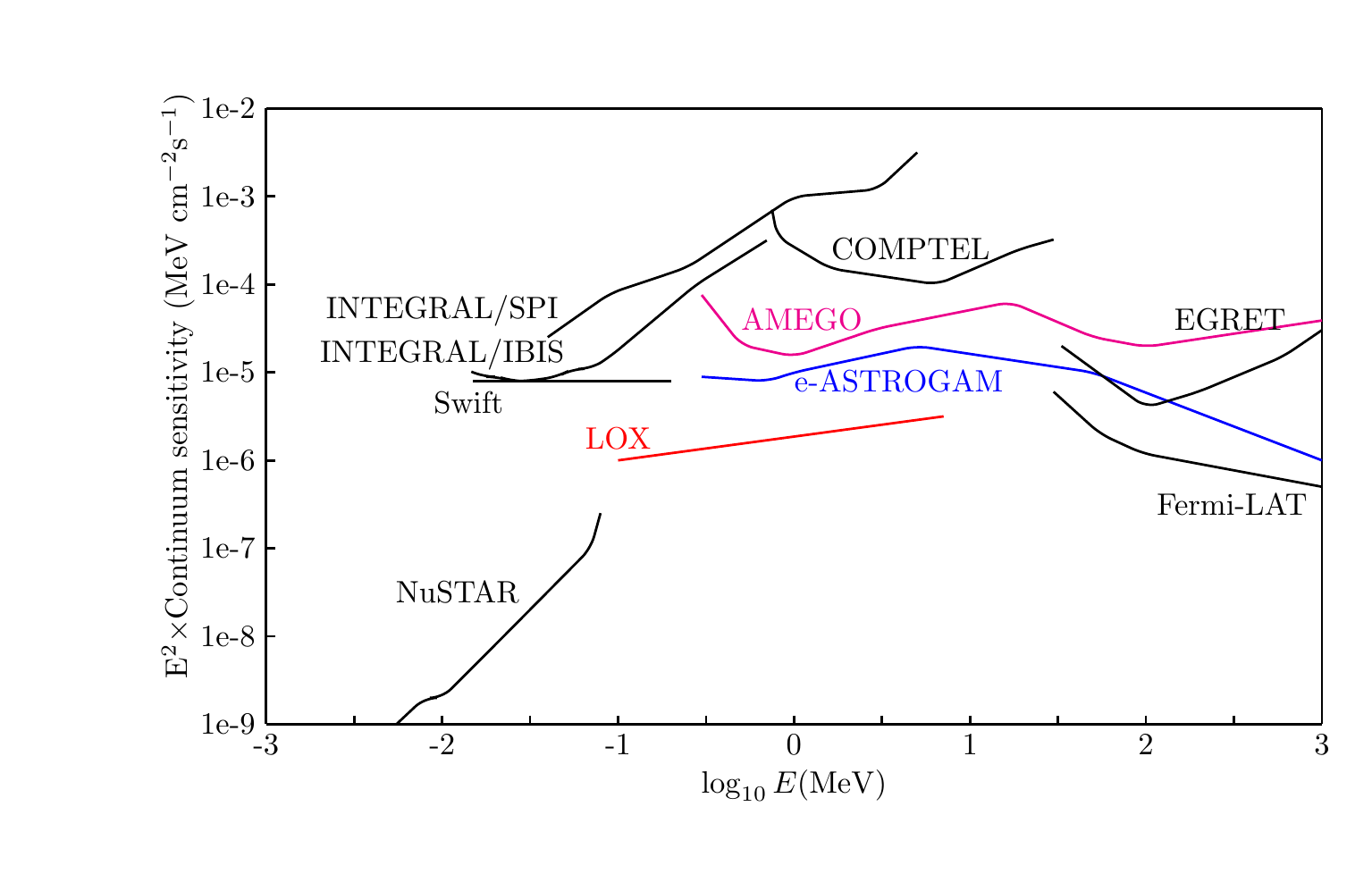}
  \caption{ 
Approximate point source continuum sensitivities of different instruments corresponding to a $3\,\sigma$ significance, $\Delta E/E=1$ and $10^6$ seconds of exposure.}
  \label{mevgap}
\end{figure}

From any point of view, gamma-ray photons  with energies of the order of MeV are excellent diagnostic tools to understand the physical processes operating behind nova and supernova outbursts. This potential is a natural consequence of their small interaction with matter which allows them to retain the information concerning their origin. 
Unfortunately, this property also makes them difficult to detect and traditionally there has been a gap of sensitivity in the MeV region (see Figure~\ref{mevgap}). In fact, the improvement of sensitivity since 1980 to now has been only a factor $\sim 10$, much smaller than those experienced in all the other domains of high-energy astrophysics.

The Solar Maximum Mission or Solar Max (SMM) was designed to investigate the high energy emission of the Sun with special emphasis to solar flares \citep{bohl80}. It was launched in 1980 and ended on 1989. The payload included the Hard Burst X-Ray  Spectrometer (HXRBS)  and the Gamma Ray Spectrometer (GRS) among other instruments. The HXRBS detector consisted of a CsI(Na)  scintillator able to work within the energy range of 30--500 keV. The GRS consisted of an array of 7 NaI(Tl) plus one CsI(Na) detectors able to work in the energy range of 0.3--9.0 MeV. It detected the emission of SN1987A \citep{matz88,leis90}. 

The \emph{Kvant 1} (or just Kvant\footnote{Kvant means 'quantum' in Russian}) module was launched on March 31, 1987 and docked to the Soviet Mir space station on April 12, 1987. There were 4 scientific experiments on its board: two high-energy instruments, HEXE and Pulsar X-1,  an X-ray imager, TTM/COMIS, and an X-ray spectrometer, GSPC. All 4 experiments pointed toward the same source at the same time. These experiments, taken together, were sometimes referred to as the international Roentgen observatory.

HEXE (High Energy X-ray Experiment, Germany) employs a phoswich of NaI/CsI. It covers the energy range 15-200 keV with a $1.6^o\times1.6^o$  field of view (FWHM). Each of the 4 identical detectors has a geometric area of $200\ \mbox{cm}^2$.
Pulsar~X-1 (Russia) consists of 4 phoswich detectors which cover the energy range 50-800 keV with a $3^o\times3^o$ field of view (FWHM). Each of the 4 identical detectors has $314\ \mbox{\rm cm}^2$ geometric area.
TTM/COMIS (COded Mask Imaging Spectrometer, Netherlands) is a wide-angle camera that uses a coded aperture mask for imaging and source localization. It covers the energy range 2-30 keV with a $7.8^o\times7.8^o$ field of view (FWHM). The geometric area is $655\ \mbox{\rm cm}^2$. It can achieve an angular resolution of $2^{\prime}$. 
GSPC (Gas Scintillation Proportional Counter, also called Sirene 2, Netherlands) is a spectrometer with enhanced energy resolution. It covers the energy range 2-100 keV with a $3^o\times3^o$ field of view (FWHM). The geometric area is $300\ \mbox{\rm cm}^2$.

The Roentgen or Mir-Kvant observatory successfully operated until fall 1989, when  operation was interrupted for a planned reconfiguration of the Mir station. Its observations were restarted again in October 1990 and continued until March 23, 2001, when the station left the orbit to burn down in the Earth's upper atmosphere. The orbital period of the Mir station was 90 min. At an inclination of $57^o$, some 20 min of each orbit were spent outside the radiation belts. The main achievement of the observatory was the discovery and study of the hard X-ray emission from Supernova 1987A. But it also contributed a lot in broad-band X-ray spectroscopy of black hole transients (X-ray novae) and X-ray imaging of the Galactic center field \citep{suny90b}.

\emph{GRIS} (Gamma-Ray Imaging Spectrometer) was an experiment on board of a balloon. It was operated from 1988 to 1995 by the NASA's Goddard Space Flight Center and was able to work in the energy range  of 20 keV to 8 MeV \citep{teeg85}. This experiment detected the emission of SN1987A \citep{teeg89}.

The Compton Gamma-Ray Observatory (CGRO) was the first gamma-ray observatory \citep{gehr93}. It was launched  in 1991 and ended in 2000. Globally, its four instruments covered the large energy range between 30 keV to 20 GeV: \emph{BATSE} (Burst and Transient Source Experiment) worked in the region of 0.2 - 1 MeV, \emph{OSSE} (Oriented Scintillation Spectrometer Experiment) in the range of 0.05 - 10 MeV, \emph{Comptel} (the Compton telescope) worked between 0.8 and 30 MeV and, finally, \emph{EGRET} (Energetic Gamma Ray Telescope) in the range of 30 MeV to 10 GeV. This mission studied SN1991T \citep{lich94,leis95} and SN1998bu \citep{geor02} and of course SN 1987A \citep{kurf92,clay92}.

\emph{INTEGRAL} is an ESA scientific mission able to operate in gamma-rays, X-rays, and visible light \citep{wink03}. It was launched on October 17, 2002 into a highly eccentric orbit with a period of about three days that spends most of this time outside the radiation belts.  One of the major objectives of this mission is the observation of the gamma-ray emission of explosive events.

The instruments on board are: i) The OMC camera,  able to operate in the visible band up to a magnitude 18 \citep{mash03}; it was used to obtain the light curve of SN 2011fe and SN2014J in the V~band, allowing an early estimate of the amount of $^{56}$Ni needed to account for the shape of the optical light curve, as well as to predict the intensity of the $^{56}$Co line at late times; ii) the X-ray monitors JEM-X, working in the 3--35 keV energy range \citep{lund03} and which were used to constrain the continuum emission of SN2014J in this band; and iii) the two main gamma instruments, SPI, a cryogenic germanium spectrometer able to operate in the energy range of 18 keV - 10 MeV  \citep{vedr03}, and IBIS/ISGRI, an imager able to operate in the energy range of 15 keV to 1 MeV \citep{lebr03,uber03}. Below 300 keV, IBIS/ISGRI is more sensitive than SPI by a factor $\sim$ 3 at $\sim$ 150 keV \citep{lebr03,roqu03}, but the sensitivity of both instruments is good enough to allow comparison of the results in the region of 40 to 200 keV.

The Neil Gehrels SWIFT observatory is a NASA mission, with international participation, specially devoted to the GRB sources. It was launched in November 2004 and comprises three instruments working in tandem: the BAT (Burst Alert Telescope) which operates in the 15-150 keV band, the XRT (Xray Telescope) that works in the band of 0.3-10 keV, and an UV/optical telescope \citep{burr05}.

NuSTAR (Nuclear Spectroscopic Telescope Array) was launched in June 2012 and is the first focusing high-energy telescope in orbit. It operates in the band of 3 to 79 keV \citep{harr13}. One of its main goals is the detection of $^{44}$Ti in supernova remnants.

\section{Results}

\subsection{Core collapse supernovae}

SN1987A was discovered on February 23rd 1987 in the Large Magellanic Cloud (51.4 kpc) and, consequently, is the closest supernova visible with naked eye since the Kepler's SN 1604. The X and $\gamma$-ray spectra were predicted by \citet{bart87}, \citet{greb87a,greb87b}, \citet{mccr87}, \citet{kuma88}, \citet{pinto88,pint88}, and \citet{xu88} and few weeks after the explosion the 847 and 1238 keV $^{56}$Co lines were detected at more than $5\sigma$ level by the Gamma Ray Spectrometer of the Solar Maximum Mission \citep{matz88,leis90}\footnote{ Observations performed in the infrared between days 255 and 576 after the explosion confirmed the presence of $^{56}$Co and $^{57}$Co in the debris \citep{vara90}.}. The analysis of data showed that to account for the observed flux it was necessary to place some amount of $^{56}$Co at very low $\gamma$-ray optical depths and that the $\gamma$-light curve was not compatible with models were the radioactive isotopes were simply buried  by an expanding uniform envelope. 
These data, together with the information provided by other wavelengths, suggested the presence of 0.07 M$_\odot$ of $^{56}$Co.

\begin{figure}
\centering 
\subfloat{\includegraphics[width = 10cm,clip=true,trim= 0cm 0cm 0cm 0cm]{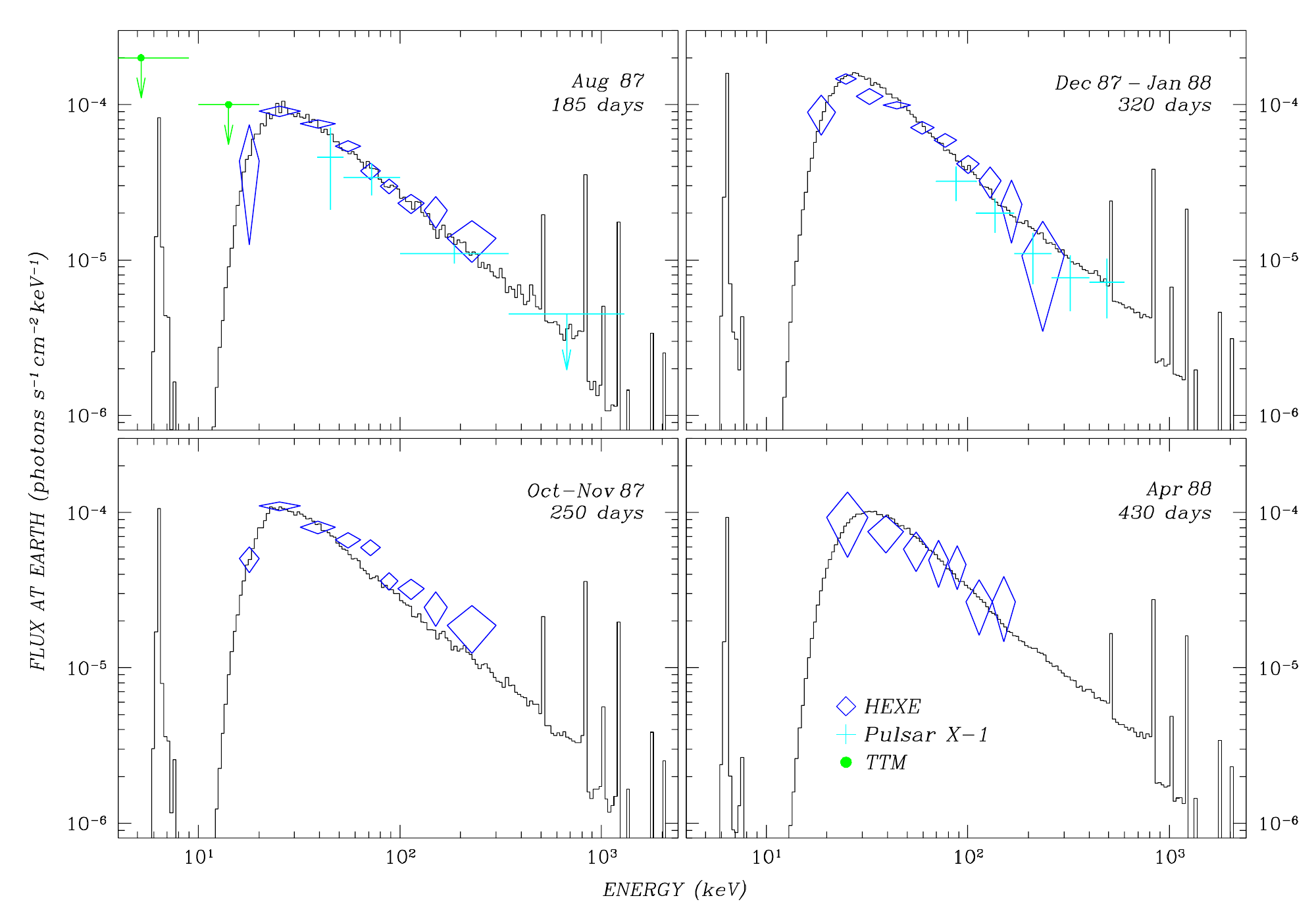}}\\ 
\subfloat{\includegraphics[width = 10cm,clip=true,trim= 0cm 7cm 0cm 7cm]{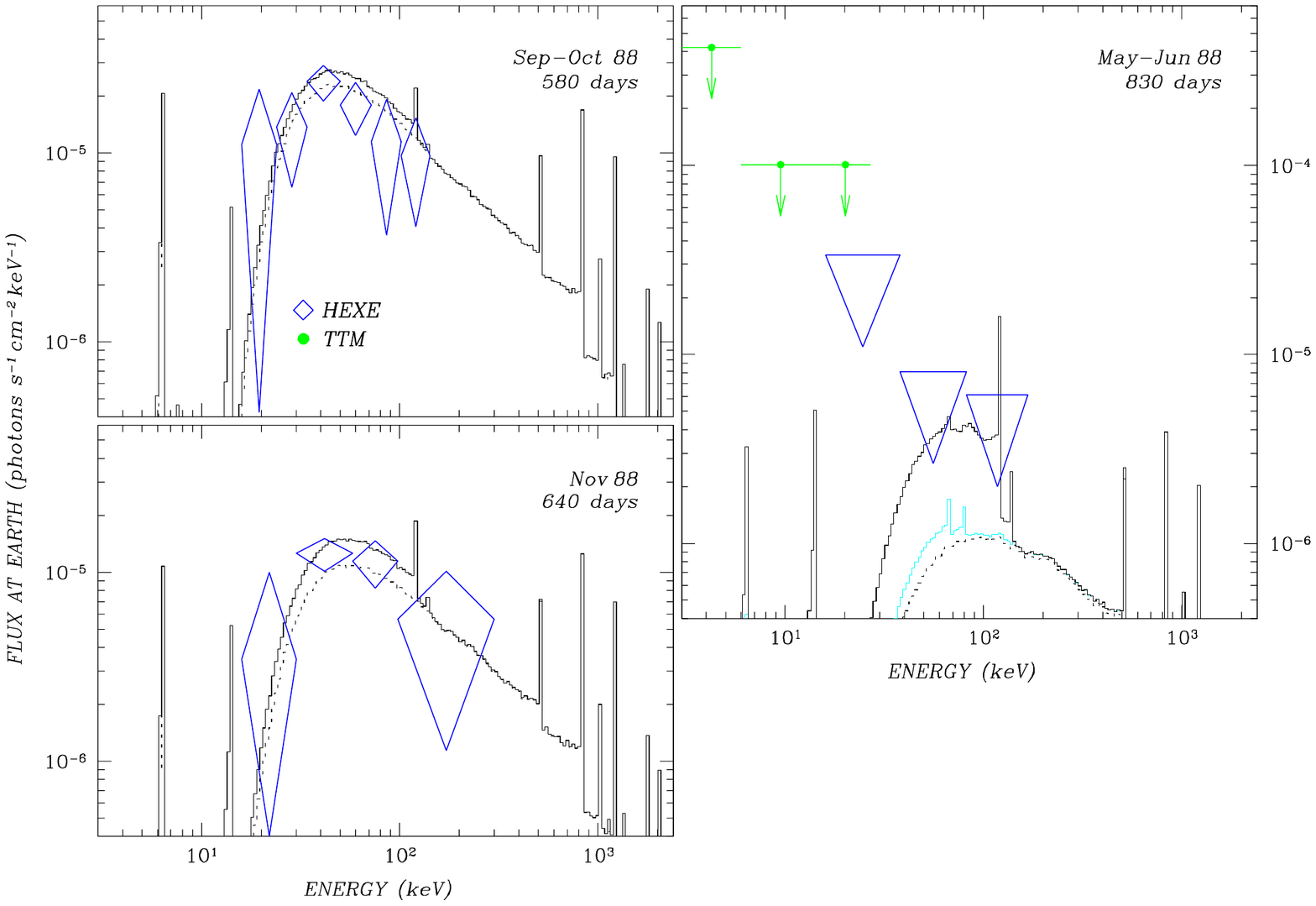}}
\caption{\footnotesize
Spectra of SN\,1987A from the data of observations with the ROENTGEN
observatory at the MIR-KVANT orbital station at 7 different epochs (days
after the explosion are indicated for each date). Taken from \citet{suny90,suny91}.}
\label{f87a}
\end{figure}

Similar conclusions were obtained from the observations performed from the MIR-KVANT observatory 'ROENTGEN'  \citep{suny87,suny90b,suny91} and \emph{Ginga} \citep{dota87}. 
Figure.~\ref{f87a} displays the observations performed by the MIR-KVANT observatory 180-830 days after the explosion. The blue diamonds show the HEXE
data, the cyan crosses - the Pulsar X-1 data, and the green crosses with central
points - the TTM upper limits (different instruments of the observatory).
The histogram shows the results of the Monte Carlo
computations of Compton down scattering of gamma-photons emitted by the
$^{56}$Co radioactive decay (for the best model of the supernova envelope
with an appropriate $^{56}$Ni radial distribution). The radiation of $^{57}$Co 
and $^{44}$Ti was also taken into account under the assumption that the 
$^{57}$Co/$^{56}$Co 
abundance ratio was  equal to 1.5 of the terrestrial $^{57}$Fe/$^{56}$Fe 
value, and the $^{44}$Ti abundance equal to $3.1\times10^{-4}\ M_{\odot}$. The 
contribution of $^{56}$Co to the total radiation spectrum is represented by a 
dotted curve, and the contribution of $^{56}$Co and $^{44}$Ti by a cyan curve. The figure clearly shows how $^{56}$Co decreases inside the envelope as a consequence of its decay and how the expansion reduces the optical depth due to electron scattering. The figure also shows how the relatively narrow $^{56}$Co $\gamma$-ray lines slowly move to lower energies where they are photoabsorbed by the heavy elements, mainly iron group. The blend of lines present in the X-ray band is produced by fluorescence of some of these photons \citep{suny90,suny91}.

Observations performed with GRIS, a Ge-spectrograph on board a balloon, allowed the detection of the 847 ($2.3\,\sigma$) and 1238 ($4.3\,\sigma$) keV lines at day 443, and 847 ($4.6\sigma$), 1238 ($3.4\sigma$) and 2599 ($1.9\sigma$) kev lines of $^{56}$Co at day 613 with a combined significance of $7.8\sigma$. These lines were Doppler broadened ($\sim 3500$~km~s$^{-1}$) and slightly reddened ($\sim 500$~km~s$^{-1}$) in contrast with the predictions of spherically symmetric mixed models. Furthermore, the intensity of the lines was consistent with an optically thin source but was $\sim 1/3$ of the one predicted by the presence of the 0.075 M$_\odot$ of $^{56}$Co deduced from the bolometric light curve \citep{tuel90}.

The OSSE instrument on board of the Compton Gamma Ray Observatory (CGRO) found evidences of the 122 keV emission of $^{57}$Co ($\sim 10^{-4}$~cm$^{-2}$s$^{-1}$), direct plus scattered component, in the period between 1600 and 1800 days after the explosion. This flux implies a $^{57}$Co/$^{56}$Co ratio of 1.5 times solar \citep{clay92,kurf92}.

\begin{figure}[h]
\center
\includegraphics[width=0.5\textwidth, clip=true, trim= 1cm  0cm 1cm 0cm]{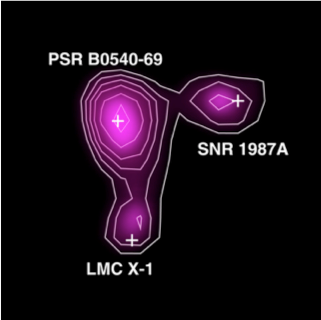}
  \caption{\footnotesize Detection of $^{44}$Ti with IBIS/ISGRI in SN 1987A \citep{greb12}.}
\label{f87ab}
\end{figure}

 \emph{INTEGRAL} invested important amounts of time to detect the $^{44}$Ti lines from SN 1987A. The first attempt was in 2003, shortly after the launch, but with an exposure of only $\sim 1.5$ Ms. The second attempt was in 2010-11 with a total exposure of $\sim 4.5$ Ms. The significance of the flux excess found by \emph{IBIS} in the region of 48-99 keV, that contains the 67 and 78 keV lines, was $\sim 4.1\,\sigma$ (Fig.~\ref{f87ab}), while that found by \emph{SPI} around the line of 1157 keV was only $1.7\,\sigma$. These measurements were translated into $^{44}$Ti mass  constraints of $M_{44}= (3.1 \pm 0.8)\times 10^{-4}$~M$_\odot$ and $M_{44} \lapprox 9\times 10^{-4}$~M$_\odot$, respectively  \citep{greb12}. A reanalysis of the SPI data using the high and low energy data has provided a slight tighter constraint, $M_{44} < 6.9\times 10^{-4}$~M$_\odot$ \citep{wein20}.

\citet{ssst14} examined the influence of the long lived radioisotopes $^{44}$Ti, $^{55}$Fe, and $^{56,57,60}$Co on the optical/IR late light curve of SN 1987A taking advantage of the fact that the V-band represents a roughly constant fraction  of the bolometric light curve during the interval 900-1900 days after the explosion and extrapolating this property to later times. They also included the heating by internal conversion and Auger electrons emitted during the decay of $^{57}$Co and Auger electrons produced by the decay of $^{57}$Fe. The masses obtained in this way for the Ni isotopes were: ${\rm M}(^{56}{\rm Ni})=(7.1\pm 0.3)\times 10^{-2}\,{\rm M}_\odot$ and ${\rm M}(^{57}{\rm Ni})=(4.1\pm 1.8)\times 10^{-3}\,{\rm M}_\odot$, with $[^{57}Ni/^{56}Ni]= 2.5\pm 1.1$, which are in agreement with the values deduced from $\gamma$-observations and with the theoretical predictions. On the contrary, the best fit obtained for titanium was ${\rm M}(^{44}{\rm Ti})=(0.55\pm 0.17)\times 10^{-4}\,{\rm M}_\odot$, much smaller than the value found by INTEGRAL. Concerning the remaining isotopes, \citet{ssst14} also obtained upper bounds for two other radioactive isotopes of cobalt: ${\rm M}(^{55}{\rm Co})< 7.2\times 10^{-3}\,{\rm M}_\odot$ and ${\rm M}(^{60}{\rm Co})< 1.7 \times 10^{-4}\,{\rm M}_\odot$.

\emph{NuSTAR} measured a flux of $(3.5\pm 0.7)\times 10^{-6}$ cm$^{-2}$s$^{-1}$ for the 67 keV line with a significance of$ \sim 8.5\,\sigma$, which corresponds to a mass of $^{44}$Ti of  $M_{44}= (1.5 \pm 0.3)\times 10^{-4}$~M$_\odot$. The profile indicated a broadening of $\lapprox 4100$~km~s$^{-1}$ and a redshift respect to the rest frame of SN 1987A of $700\pm 400$~km~s$^{-1}$, which implies an asymmetry in the radioactive ejecta probably caused by the presence of a blob or a lobe moving away from the observer. A similar behavior was found in the $^{56}$Co lines but in this case the significance was only marginal \citep{bogg15}.

 \citet{jerk20} have analysed these data using up to date 3D neutrino-heated models involving ZAMS stars of 15-20~M$_\odot$ and energies of $\sim 1.5\times10^{51}$~erg and have found that $^{56}$Ni should have a bulk asymmetry $\gapprox 400$~km~s$^{-1}$, a bulk velocity $\gapprox 1500$~km~s$^{-1}$, and a kick $\gapprox 500$~km~s$^{-1}$ for the neutron star. Furthermore, adding the UVOIR bolometric information they estimated an ejected mass of $\sim\,14$~M$_\odot$. \citet{ono20} have also performed several 3D hydrodynamical simulations of aspherical core collapse of blue supergiants putting the accent on the mixing of internal layers. The best result was obtained with a progenitor coming from a binary merger and with an asymmetric bipolar-like explosion. An interesting result is that such a model provides a prediction of the bipolar explosion axis as well as the corresponding neutron star kick velocity \citep{orla19}.

\subsection{ Thermonuclear supernovae}

There have been several attempts to detect the $\gamma$-emission of SNIa before the success of \emph{INTEGRAL} with SN2014J, but only upper limits were obtained as a consequence of the large distance of the events. These are SN1991T \citep{lich94} and SN1998bg \citep{geor02}, using the instruments on board of \emph{CGRO}, and SN2011fe \citep{iser13}, using those of \emph{INTEGRAL}.

SN 2011fe was discovered in M101 on August 24, 2011 \citep{nuge11a}. The absence of hydrogen and helium, coupled with the presence of silicon in the spectrum, clearly indicated that it was a SNIa. This supernova, placed at a distance of 6.4 Mpc, was detected one day after the explosion since it was not visible on August 23 \citep{nuge11b}. SN2011fe was, therefore, one of the brightest SNIa ever detected and this allowed to obtain very tight constraints on this supernova and its progenitor system in a variety of observational windows. Red giant and helium star companions, symbiotic systems, systems at the origin of optically thick winds or containing recurrent novae were excluded  \citep{li11c,bloo12,brow12,chom12}, leaving only either DD or a few cases of SD as possible progenitor systems. 

This distance to SN2011fe was, however, slightly larger than the maximum distance at which the \emph{INTEGRAL} gamma-ray instruments are able to detect an intrinsically luminous SNIa. Thus, SPI and IBIS/ISGRI only could provide limits to the expected emission of $^{56}$Ni. These flux upper limits amount to $7.1 \times 10^{-5}$~ph~s$^{-1}$cm$^{-2}$ for the 158~keV line and $2.3\times 10^{-4}$~ph~s$^{-1}$cm$^{-2}$ for the 812~keV line. These bounds allow the rejection at the $2\sigma$ level of explosions involving a massive white dwarf, $\sim 1$~M$_\odot$ in the sub--Chandrasekhar scenario and specifically all models that would have substantial amounts of $^{56}$Ni in the outer layers of the star causing the SN2011fe event. The optical light curve obtained with the OMC camera also suggests that SN2011fe was the outcome of the explosion of a CO white dwarf, possibly through the delayed detonation mode, although other ones are possible, that synthesized $\sim 0.55$~M$_\odot$ of $^{56}$Ni \citep{iser13}. 

SN 2014J is the nearest SNIa since SN1604 (Kepler's SN). It was discovered by \citet{foss14} on January 21st 2014 in M82 ($d = 3.5 \pm 0.3$~Mpc). The moment of the explosion was estimated to occur on January 14.72 UT 2014 or JD2456672.22 \citep{zhen14}. INTEGRAL began to observe this source on January 31st, 16.5 days after the explosion for a total observing time of 1 Ms (referred to as the \emph{early observations} in the following). Two late time observation periods (simply named \emph{late observations} thereafter), 50-100 and 130-162 days after the explosion, were also scheduled for a total exposure of 4.3 Ms. The $^{56}$Ni lines were detected by SPI  with a significance of $5\,\sigma$ during the early observations \citep{dieh14,iser14} and the $^{56}$Co lines were detected during the late observations with a significance of $5\,\sigma$ \citep{chur14a,chur14b}. 

\begin{figure}
\centering
\includegraphics[width=0.7\textwidth, clip=true, trim= 1cm  0cm 1cm 0cm]{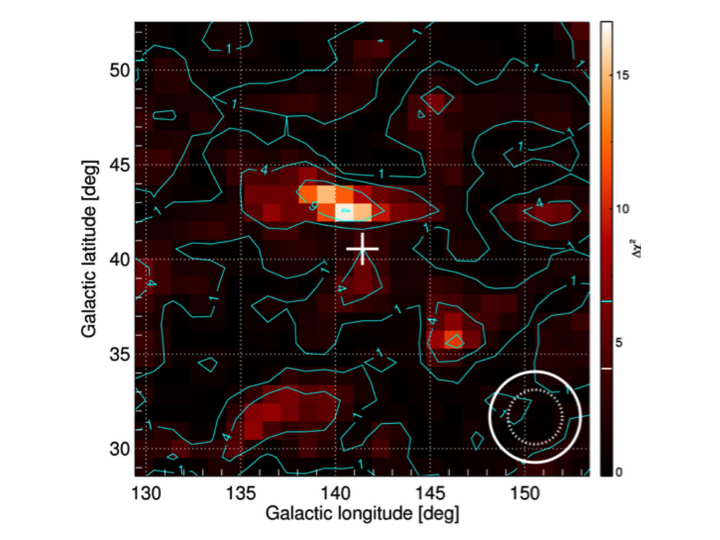}
\caption{\label{f14jmapb} \footnotesize Location of the 158 and 812 keV lines of $^{56}$Ni in the sky using the SPI data obtained during the period 16-19 days after explosion. The position of SN2014J is represented by a cross and the instrumental uncertainty by the bottom right circle. Figure from \citet{dieh14}.}
\end{figure}

\begin{figure}
\centering
\includegraphics[width=0.65\textwidth, clip=true, trim= 1cm  0cm 0cm 0.1cm]{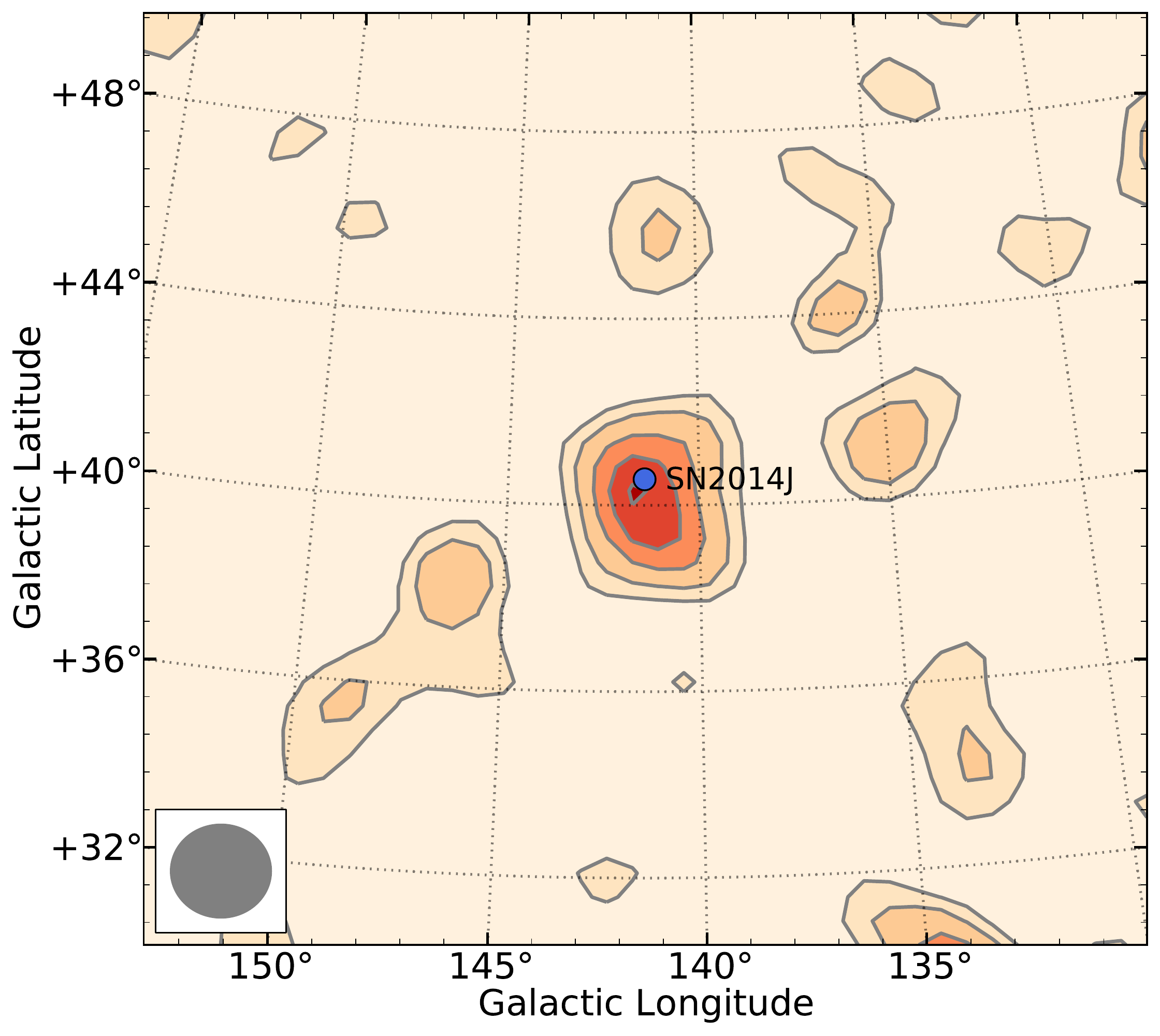}
\caption{\label{f14jmap} \footnotesize Gamma-ray signature of SN2014J data obtained with SPI  corresponding to the interval of 16-35 days after the explosion over the energy band of 145-165 keV. The maximum excess coincides with the position of the supernova ($l=140.5^o,\,b=42.5^o$), and the significance is $5\,\sigma$. Figure from \citet{iser16}.} 
\end{figure}

The early observations of \emph{INTEGRAL}  stopped 19 days after the initial trigger in order to calibrate the instruments first, and as a consequence of a giant solar flare after. 
 Figures~\ref{f14jmapb} and~\ref{f14jmap} display the signal to noise contour maps independently obtained by \citet{dieh14} and \citet{iser16} with the data provided by SPI. The analysis of these data performed by the two teams looks contradictory although both agree with the existence of an emission excess that was not present before the explosion and was clearly isolated from the surrounding sources. If real, this excess implies the presence of  radioactive material in the outer layers.

To understand the results obtained with \emph{INTEGRAL} it is necessary to take into account the following aspects. In the case of SPI, 
the influence of the temporal behavior of the line intensity and line width on the detectability of
the $\gamma$ emission can be easily seen by estimating the significance of the observation. In the limit of weak signals this significance is given by \citep{jean96}:

\begin{equation}
n_{\sigma} = \frac{{A_{eff} \int\limits_{t_i  - \Delta t}^{t_i } {\varphi \left( t \right)dt} }}{{\sqrt {bV\Delta E \Delta t} }} 
\label{nsigma}
\end{equation}
\noindent where $\Delta t$ is the observation time, $A_{eff}$ the effective area at the
corresponding energies, $\varphi$ is the flux  (cm$^{-2}$s$^{-1}$) 
in the energy band $\Delta E$,  V is the volume of the detector, and b the 
specific noise rate (cm$^{-3}$s$^{-1}$keV$^{-1}$), which, for simplicity, is assumed
to be independent of energy and time in the interval of interest.

An important point to take into account is that the number of $\gamma$-photons available is bound by the number and half lives of the radioactive isotopes. Since, in general, lines are Doppler broadened ($\Delta E/E \sim 3\%$), this relatively fixed number of photons is distributed over a relatively large energy band and, consequently, the sensitivity decreases since the total noise integrated over the energy band increases (term $b\Delta E$ of Eq.~\ref{nsigma}). Therefore, if the supernova is faint, the energy bands have to be carefully chosen to guarantee a maximum signal to noise ratio.

 If the flux grows like $ \varphi \left( t \right) = \varphi _0 e^{\alpha t} $, the significance
 reached by integrating during the time interval $\left( {t_i  - \Delta t,t_i } \right)$ is
\begin{equation} 
 n_{\sigma} = \frac{{A_{eff} \varphi \left( {t_i } \right)}}{{\sqrt {\alpha bV\Delta E} }}\frac{{1 - e^{ -
\alpha \Delta t} }}{{\sqrt {\alpha \Delta t} }} 
\label{nsigma2}
\end{equation}
 \noindent For $\alpha \Delta t <  < 1$, the significance behaves as $n_{\sigma} \propto \sqrt{\Delta t}$ and has a
maximum at $ \alpha \Delta t = 1 .26$ \citep{iser13}. Unfortunately, since the value of $\alpha$ is not known \emph{a priori}, the optimal observing time is not known in advance. Obviously, the temporal behavior of the line is not exponential and Eq.~\ref{nsigma} has to be computed using a realistic model of  light curve.  In general, models and energy ranges showing weak variations of the flux are best detected when the observation period is long whereas models with a strong variation in their flux during the first two weeks are best detected using data at those times.

Another difficulty that have to face the observations is that the flux extracted in an energy bin contains not only the photons emitted by the source at this energy, but also events produced by high energy photons from the source that do not deposit all their energy in the detector. This contribution is not negligible and depends on the injected spectrum, which is not known a priori. Therefore, to substract such a component it is necessary to convolve the response of the instrument with a set of expected theoretical spectra \citep{iser16}. 

IBIS/ISGRI covers a roughly similar energy range as \emph{SPI}, but its efficiency drops above 100 keV and becomes worse at higher energies. Although its sensitivity below 100 keV is better, it suffers the same problems of contamination by secondary photons and, since the supernova spectra are not equivalent to those of the Crab, a convolution with different theoretical spectra is necessary to obtain the correct abundances \citep{iser16}.

The behavior of the background induced by the interaction of cosmic rays and solar protons with the instrument can be distinguished thanks to the spatial and temporal modulations produced by the coded mask  and dithering. The non existence of some residual lines is not guaranteed and, in principle, since the instrumental lines are intrinsically narrow, any narrow line in the spectrum has to be considered as suspicious. This is specially true in the case of \emph{SPI} where the instrumental lines of 159 keV $^{47}$Sc and 811 keV $^{58}$Co can contaminate the 158 and 812 keV $^{56}$Ni lines.

\begin{figure}[htb]
\center
\includegraphics[width=0.75\textwidth, clip=true, trim= 0cm  0cm 0cm 0cm]{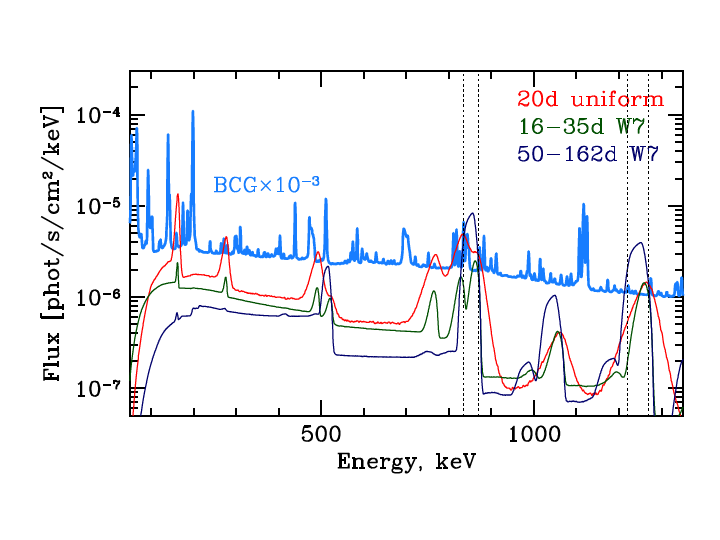}
  \caption{\footnotesize \emph{SPI} quiescent background multiplied by a factor of $10^{-3}$ (light blue) compared with the spectra predicted by a uniform (red) and a W7 (green) models near the optical maximum, and with that of W7 during the exponential tail of the light curve (blue). The dotted vertical lines represent the optimal bands to detect $^{56}$Co. Figure from \citet{chur15}.}
\label{fspin}
\end{figure}

Figure~\ref{fspin} displays the comparison between the quiescent background of SPI and the predictions obtained using the W7 model \citep{nomo84b} and a model where all the radioactive elements are uniformly distributed in the ejecta near the maximum of the optical light curve. The distance at which the source was placed in the models, 3.5 Mpc, is the same as that of SN2014J. As it can be seen, the expected signal is below 1\% of the background. Therefore, the results are sensitive to the treatment of the background. The cleanest region is around the 1238 keV $^{56}$Co line, where no strong instrumental lines are present \citep{chur15}.

\begin{figure}
\centering
\includegraphics[width=1.0\textwidth, clip=true, trim= 0cm  5cm 0cm 3.5cm]{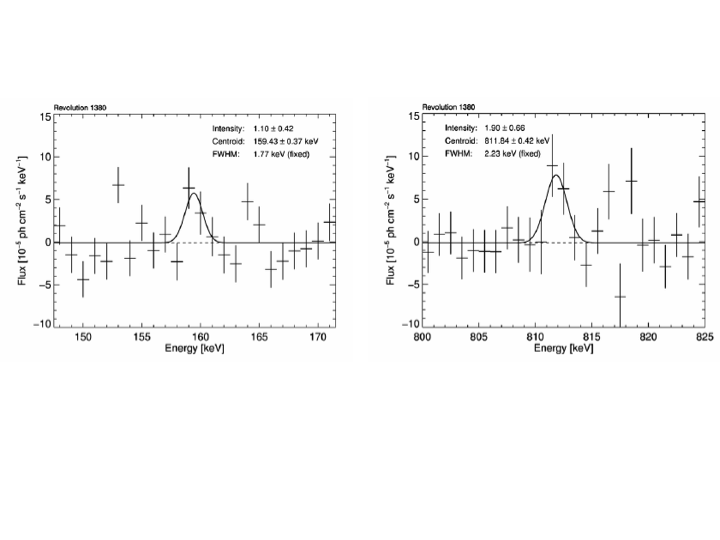}
\caption{\label{f14jscispc} \footnotesize Gamma ray spectra obtained by SPI around the 158 and 812 keV $^{56}$Ni lines during the early period.  Error bars are $1\,\sigma$. From \citet{dieh14}.} 
\end{figure}

\citet{dieh14} detected both  158 and 812 keV $^{56}$Ni lines with an intensity of $(1.10\pm 0.42) \times 10^{-4}$ and $(1.90\pm 0.66)\times 10^{-4}$ ph cm$^{-2}$s$^{-1}$ respectively. Surprisingly both lines were narrow and not significatively offset as a consequence of  the Doppler shift induced by a bulk velocity, implying that the velocity spread was $\lapprox 1500$ km/s and the bulk velocity $\lapprox 2000$ km/s (Fig.~\ref{f14jscispc}). The mass of $^{56}$Ni necessary to account for such an emission, that should be placed near the outer layers of the ejecta, would be of the order of 0.06 M$_\odot$.

\begin{figure}
\centering
\includegraphics[width=0.8\textwidth, clip=true, trim= 0cm  1cm 0cm 0.5cm]{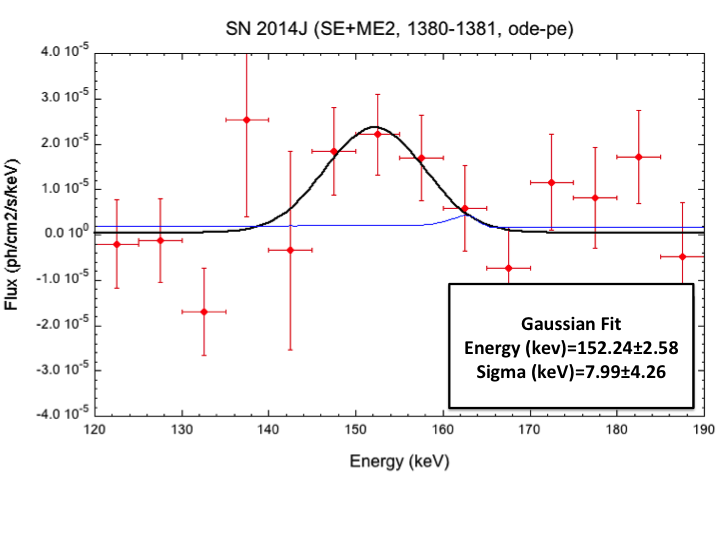}
\caption{\label{f158ni} \footnotesize Gamma ray spectrum in bins of 5 keV obtained by SPI around the 158 keV $^{56}$Ni lines during the early period \citep{iser16a}. The blue line is the predicted signal by model DDT1p4. Error bars are $1\,\sigma$. } 
\end{figure}

 \citet{iser16,iser16a} also detected an emission excess with a significance $\sim 5\,\sigma$ in the energy band of 120 - 190 keV around the 158 keV line of $^{56}$Ni during the early epoch.  Their analysis, however, showed the presence of a broad, $\sim 6$~keV, and completely unexpected red shifted feature placed at $\sim 152$ keV (Fig.~\ref{f158ni}) that could not be confused with any narrow background line because of its width and position.

\begin{figure}
\centering
\includegraphics[width=.8\textwidth, clip=true, trim= 0cm  1cm 0cm 0cm]{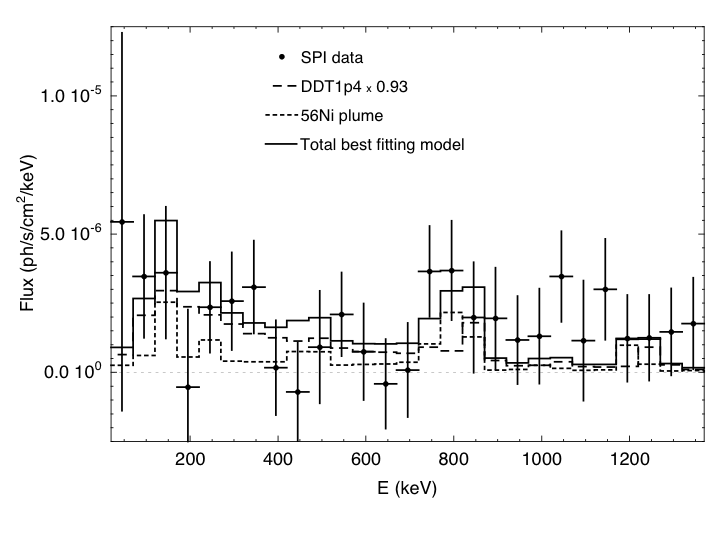}
\caption{\label{fbball8082} \footnotesize Gamma ray spectrum corresponding to the early epoch in bins of 50 keV obtained by SPI. The continuous line  represents the best fit obtained scaling the model DDT1p4, 0.605 M$_\odot$ of $^{56}$Ni, by a factor 0.93  (long dashed line), and adding a  $^{56}$Ni plume of 0.08 M$_\odot$ (short dashed line). Figure from \citep{iser16}.}
\end{figure}

Figure~\ref{fbball8082} presents, for the same epoch, the spectrum from 20 keV to 1370 keV with a binning of 50 keV where, besides the flux excess at 120-190 keV, there is another one in the 720-870 keV band. The significance of this excess is only $\sim 2.8\,\sigma$ and can be attributed to the $^{56}$Ni and $^{56}$Co decays. Unfortunately, the blending of the 812 and 847 keV lines, caused by the Doppler broadening \citep{gome98}, together with the relative weakness of the fluxes, prevented from performing a detailed spectroscopic analysis of the individual gamma-ray lines. It is also interesting to notice the presence of a feature  with a $2.6\,\sigma$ significance at  $\sim 730$ keV, the position which would correspond to the 750 keV $^{56}$Ni line  red shifted by the same amount as the 158 keV line. The gaussian fit of this feature gives  a flux of $(1.5\pm 0.7) \times 10^{-4}$ ph s$^{-1}$cm$^{-2}$ ($2.1\,\sigma$), a centroid placed at $733.4 \pm 3.8$ keV, and a FWHM of $16.9 \pm 9.0$ keV.

If it is assumed that these red shifted features are due to $^{56}$Ni, their inclusion in the analysis of the 158 keV line emission provides an additional constraint. The best Gaussian fit linking the flux, the width, and the redshift of the 158 keV line to those of the 750 keV and 812 keV lines with their corresponding branching ratio (0.50 and 0.86, respectively) gives a flux of $(1.6\pm 0.4) \times 10^{-4}$ cm$^{-2}$s$^{-1}$, centred at $155.2 ^{+1.3}_{-1.1}$ keV with a FHWM $5.2^{+3.4}_{-2.2}$ keV, implying that the radioactive material is receding from the observer with a velocity of $\sim 6000$ km/s and has a velocity dispersion of $\sim 10000$ km/s \citep{iser16,iser16a}. The associated mass of $^{56}$Ni should be $\sim 0.08$~M$_\odot$.

\begin{figure}
\center
\includegraphics[width=0.7\textwidth, clip=true, trim= 0cm  0cm 0cm 0cm]{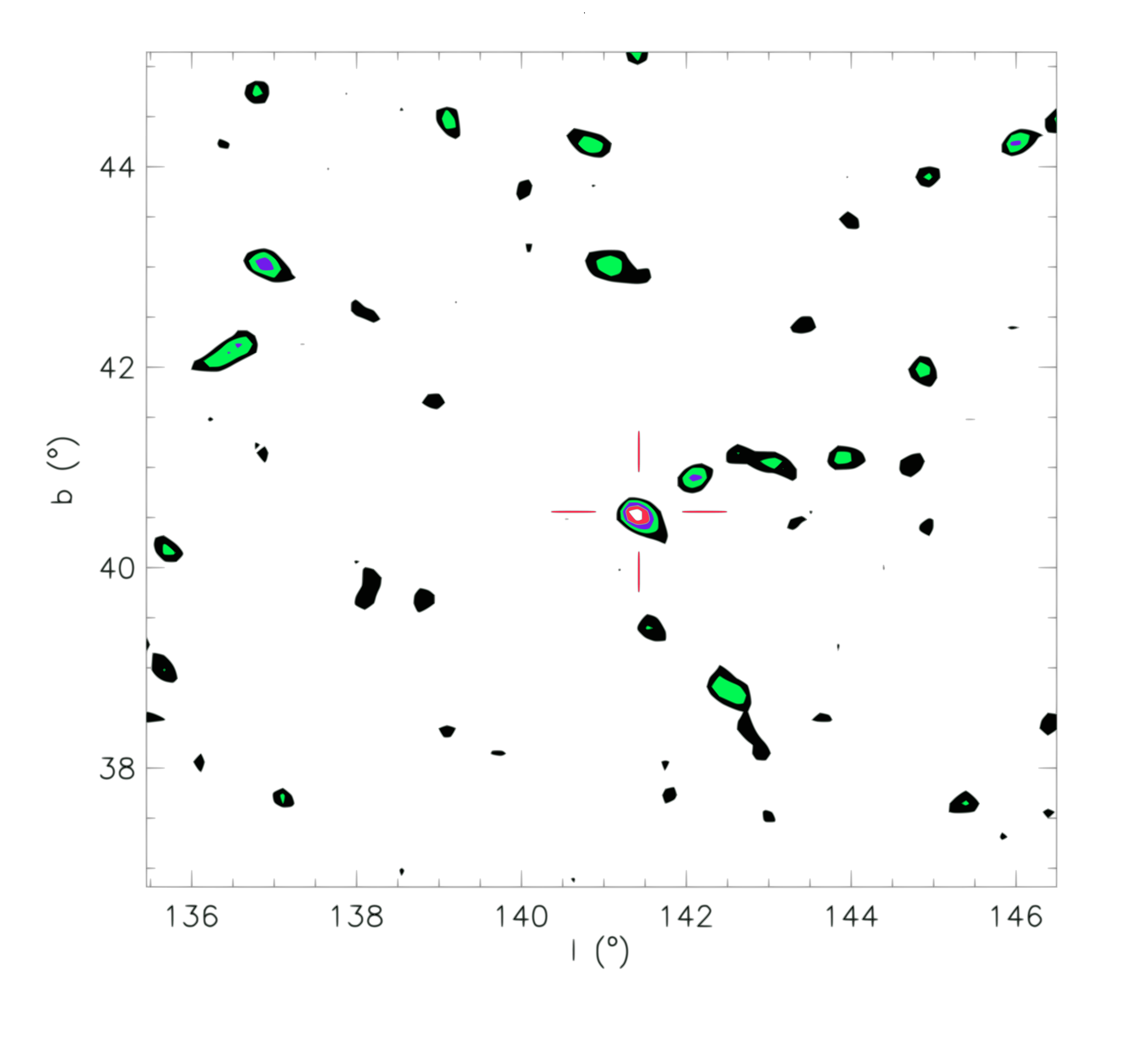}
  \caption{\footnotesize Gamma-ray signature of SN2014J obtained with the IBIS/ISGRI data. The figure displays the significance contour map for the early observations (16-35 days after the explosion) in the 67.5-189 keV energy band. Figure from \citet{iser16}.}
\label{f14jibis}
\end{figure}

\begin{figure}
\center
\includegraphics[width=0.75\textwidth, clip=true, trim= 0cm  0cm 0cm 0cm]{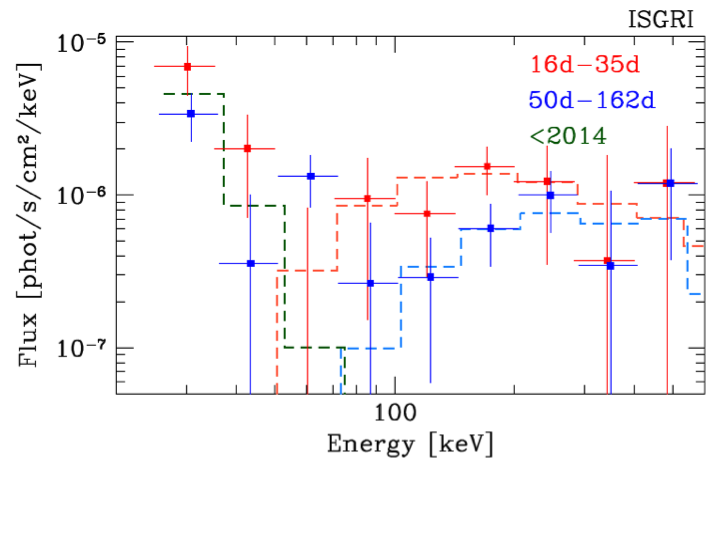}
  \caption{\footnotesize ISGRI spectra measured at the position of SN2014J during the early and late epochs. The energies corresponding to the late epoch have been multiplied by a factor 1.02 for a sake of clarity. Dashed histograms show the predictions obtained with the W7 model. The dark green line shows the spectrum of the host galaxy M82 before the explosion. Figure from \citet{chur15}.}
\label{f14jisgri}
\end{figure}

The analysis of the IBIS/ISGRI data also shows an emission excess at the position of SN2014J in the energy band 67.5-189 keV that was not present before the explosion and cannot be confused with the neighboring sources. Figure~\ref{f14jibis} displays the significance contour map for the early observation period. Contours start at the $2\,\sigma$ level and are separated by $0.5\,\sigma$ and the average flux at the position of SN2014J represents an excess of $5.4\,\sigma$. There is nothing visible in the 25-70 keV band at the position of the supernova. Figure~\ref{f14jisgri} displays the spectrum during this early epoch (red) measured with IBIS/ISGRI where the excess in the region of the 158 keV band is clearly visible as compared with the emission at late epochs. Interestingly enough, the data provided by IBIS are roughly compatible with the predictions of standard spherical models like W7 \citep{chur15}.

\begin{figure}
\center
\includegraphics[width=0.75\textwidth, clip=true, trim= 0.5cm  1cm 0.5cm 1cm]{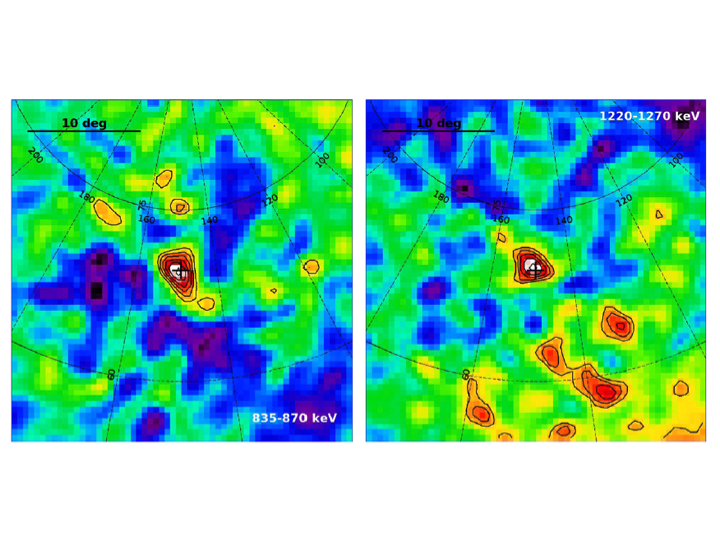}
 \caption{\footnotesize \emph{SPI} images during the late period in the narrow bands of $^{56}$Co defined in Fig.~\ref{fspin}. The cross shows the position of SN2014J and the contours are at 2,2.5,...5 $\sigma$. Figures from \citet{chur15}.}
\label{f14jmapco}
\end{figure}

The emission during the late epoch was dominated by $^{56}$Co.
Figure~\ref{f14jmapco} displays the signal to noise map obtained by \emph{SPI} during the late  observations in the 835-870 keV and 1220-1270 keV energy bands. As it can be seen from the figure, the highest signal peaks coincide well with the position of the supernova. The typical value $\Delta \chi^2 \sim 65$ suggests a $\sim 8\,\sigma$ detection.
Similar maps obtained  by ISGRI/IBIS at lower energies, 100-600 keV, show that the source was absent before the explosion. Furthermore, the images in the 25-50 keV band show that the flux due to the host galaxy M82 was similar in both epochs. All these arguments provide a robust evidence for the late gamma-ray emission from SN2014J during this period of time. 

\begin{figure}
\center
\includegraphics[width=0.75\textwidth, clip=true, trim= 0.6cm 2cm 0.5cm 1cm]{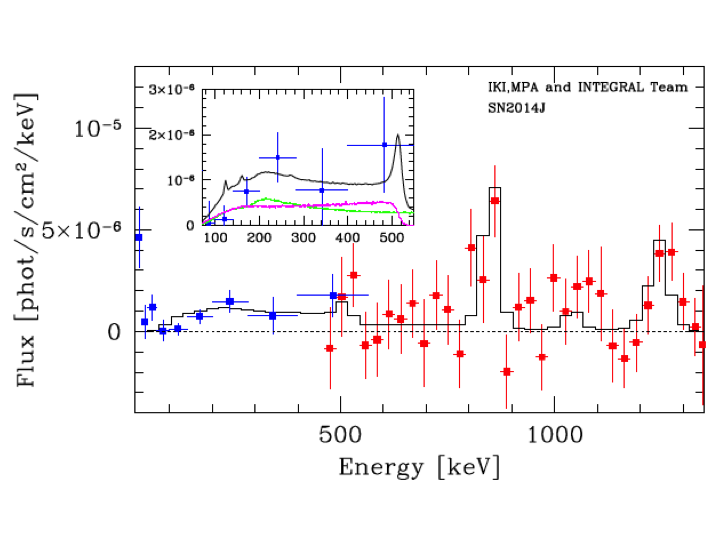}
  \caption{\footnotesize Combined  \emph{ISGRI/SPI} spectrum for the observing period of 50-100 days after maximum. Red and blue points correspond to \emph{SPI} and \emph{IBIS/ISGRI} respectively. The black curve displays a fiducial model of the supernova spectrum 75 days after the \emph{explosion}. The inlet displays the expected contributions of the three-photon positronium contribution (magenta), and the Compton down-scattered photons from the  847 and 1238 keV lines (green). Error bars are $1\,\sigma$. Figure from \citet{chur14b}.}
\label{f14jspco}
\end{figure}

Figure~\ref{f14jspco} displays the spectrum obtained from the data corresponding to 50-100 days after the explosion, when the flux due to $^{56}$Co is expected to be maximum, assuming a point source at the position of the supernova. The 847 and 1238 keV $^{56}$Co lines , with a flux of $(2.34 \pm 0.74)\times 10^{-4}$ and $(2.78 \pm 0.74)\times 10^{-4}$ photons~s$^{-1}$cm$^{-2}$, respectively, are clearly visible. These values translate into a total amount of $^{56}$Ni, $M_{\rm Ni} =(0.61 \pm 0.13)$~M$_\odot$, which agrees with the values obtained with the Arnett's rule \citep{arne82,arne85}. As expected, these lines are broadened and blue shifted because of the expansion and opacity effects of the ejecta. The blue shift corresponds to a velocity ($3,100 \pm 1,100$)~km~s$^{-1}$ and the broadening to  ($4,100 \pm 960$)~km~s$^{-1}$ \citep{chur14b,chur15}. 

Broad band light curves around the main gamma lines can provide additional information about the explosion if the signal to noise ratio is good enough. \citet{chur15} built light curves using the energy bands (100-200), (835-870) and (1200-1272) keV with a temporal resolution equal to the orbital period of \emph{INTEGRAL}, $\sim 3$~days (see their figures 3 and 4)\footnote{Concerning the problem of removing the secondary photons of the low energy band provided by IBIS/ISGRI, they found that the differences between using the Crab spectra or the predictions of theoretical models were only $\lapprox 15\%$. }.These light curves, taken at the face value, only allow to discard DETO and HED6 models and slightly favor models like W7 and DDT1p4 (Table~\ref{tmodel}).

\begin{figure}
\center
  \includegraphics[width=0.8\textwidth]{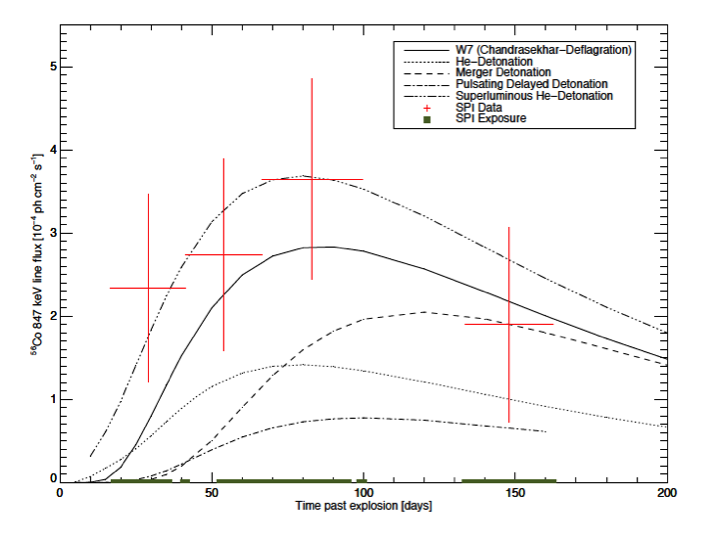}
\caption{\footnotesize Time evolution of the 847 keV line flux from SN 2014J. Models are from \citet{the14}. Figure from \citet{dieh15}}
\label{dieh15b}              
\end{figure}

\citet{dieh15} constructed a light curve dividing the entire observation period in four bins for each spectral features associated to the 847 and 1238 keV $^{56}$Co lines (see Fig.~\ref{dieh15b}). The first one covers the region of the optical maximum, the second and third ones cover the maximum of the cobalt emission  and the fourth one is representative of the period when the ejecta is completely transparent to gamma-rays.

The first low energy bin shows a broad feature centered around 827 keV that could be representative of the 847 keV line if material was receding with a velocity of $ -6920\pm 1480$~km/s, a behavior that has to be reconciled with their previous findings of  narrow unshifted 158,  and 812 keV features due to $^{56}$Ni. This broad feature is also present in the other bins but slightly blue shifted ($1600\pm 1720$~km/s) in the second and third bin and centered (-80 km/s) in the last bin.
The 1238 keV line has a centroid at 1245 keV during the last epoch but is very weak and only its intensity can be estimated in all the other epochs. Concerning broadening they adopt the values obtained with the 847 keV feature. Figure~\ref{dieh15b} displays the light curve associated to the 847 keV line. Although the spectrum during the last epoch is  featureless, the existence of blobs of radioactive material during the early epoch cannot be discarded.     

\citet{iser16} also analysed the temporal behavior of SPI and IBIS/ISGRI data during the early observations grouping the data into three bins, 16.5 - 22.2, 22.6 - 28.2, and 28.6 - 34.2 days after explosion, that were a compromise between an optimal S/N ratio and the possibility of solving the light curve in time. The Gaussian fit gives $(2.23 \pm 0.8) \times 10 ^{-4}$ ph s$^{-1}$ cm$^{-2}$ centered at $152.6 \pm 2.8$~keV and a significance of 2.8 $\sigma$ for the first bin, and only 2 $\sigma$ upper limits of  $< 1.72 \times 10 ^{-4}$ and  $<2.23 \times 10 ^{-4}$ ph s$^{-1}$ cm$^{-2}$ for the other two bins respectively. If the complete response of SPI is adopted, the flux in the line becomes $(1.59 \pm 0.57) \times 10 ^{-4}$ ph s$^{-1}$ cm$^{-2}$, centered at $154.5 \pm 0.64$ keV and a width of $ 3.7 \pm 1.5$~eV for the first bin, and $< 1.42 \times 10 ^{-4}$ and  $<1.52 \times 10 ^{-4}$ ph s$^{-1}$ cm$^{-2}$ for the remaining ones.
The data obtained by IBIS/ISGRI suggest a similar pattern. There is a 3.8 $\sigma$ emission excess during the first and the third bin, separated by a dip that is compatible with the free decay of $^{56}$Ni. This behavior is similar to that obtained with SPI but has a better significance to the point that the upturn at the end of the exposure seems real, suggesting that new radioactive layers were exposed. However, the poor S/N of the central bin prevents from drawing any solid conclusion about this possibility and an approximately constant or gently decaying behavior cannot be excluded \citep{iser16}.     

The results obtained by INTEGRAL during the early period of observations are puzzling. As it has been already mentioned, there are several spherically symmetric models in Table~\ref{tmodel} with the bulk of radioactive elements buried in the central layers of the expanding debris that, with the appropriate parameters, can reproduce the $^{56}$Co lines but not the early spectrum without adding a plume of $^{56}$Ni to the outer layers. For instance Fig.~\ref{fbball8082} compares the observed spectrum with those obtained with model DDT1p4, that was specially tailored to fit the late spectra, scaled to a central nickel  mass of 0.605 M$_\odot$ plus a plume of 0.077 M$_\odot$ in the outer layers.

The geometry and the origin of such a plume  of $^{56}$Ni  is not clear.
Optical and infrared data showed that the rising of the light curve was steeper and slightly delayed as compared with the SN2011fe data, and there are some hints about the presence of a shoulder in the early data \citep{goob14,zhen14}. The early spectra did not show the presence of carbon and oxygen absorption lines as it should be expected if there was a C/O mantle surrounding the radioactive debris \citep{goob14}. Since a single degenerate origin of SN2014J seems to be excluded by the upper limits obtained in X-rays \citep{niel14} and optical/infrared \citep{goob14,kell14} it is natural to look for an explosion triggered by the ignition of He in the outer layers of the white dwarf, i.e. a Sub-Chandrasekhar scenario or a merger involving a He-donor.

\begin{figure}
\center
\includegraphics[width=0.8\textwidth, clip=true, trim= 0cm  0cm 0cm 0cm]{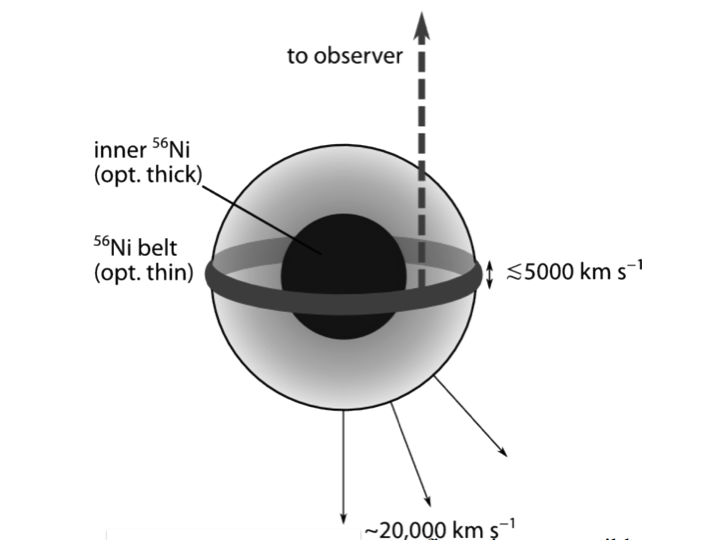}
  \caption{\footnotesize Possible geometry of the outer layers. The accreted helium forms a belt that creates another belt of $^{56}$Ni at the surface as a consequence of the explosion. The gamma-rays escape from this belt while the $^{56}$Ni in the core is still buried. Figure from   \citet{dieh14}.}
\label{f14tdiehl}
\end{figure}

If it is assumed that the early gamma-lines detected by SPI are narrow and almost centered on their laboratory value it is natural to assume that the outer radioactive material is confined into a narrow disc that must be placed almost perpendicularly to the observer as displayed in Fig.~\ref{f14tdiehl} \citep{dieh14} to satisfy the kinematic properties required by the spectra. The questions to solve in this model are the broad red shifted feature associated to the 847 keV line observed in the first bin of the light curve and the absence of a narrow spike in $^{56}$Co due to the cobalt present in the belt.

\begin{figure}
\center
\includegraphics[width=1.0\textwidth, clip=true, trim= 0cm  9cm 0cm 0cm]{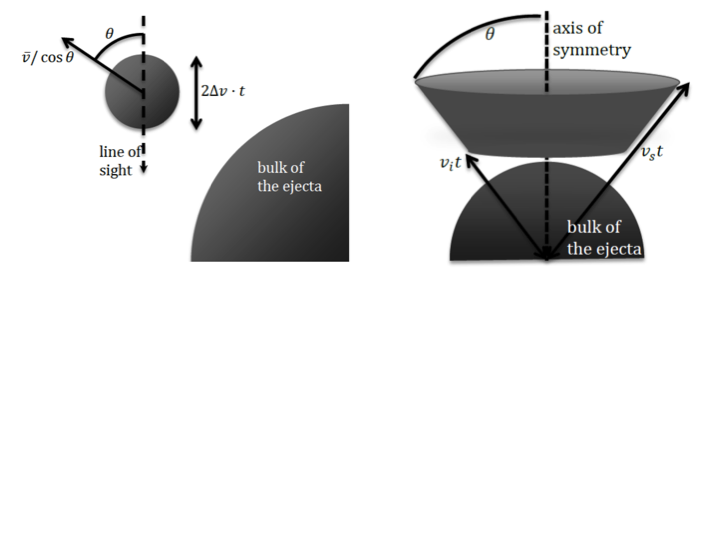}
  \caption{\footnotesize Possible geometries of the outer radioactive layers. The left panel represents a blob with mass M$\sim 0.05$~M$_\odot$ that detached from the main body during the explosion and moved with a velocity compatible with the observed redshift. The right panel represents a plume with the shape of a truncated conical ring also compatible with the observed kinematic properties. In both cases, the bulk of nickel is buried in the interior. Figure from \citet{iser16}.}
\label{f14jbravo12}
\end{figure}

If the early gamma-lines are broad and red shifted, the plume must be receding from the observer and expanding at high enough velocities to avoid interactions with the inner material at the moment of the observation. The left panel of Fig.~\ref{f14jbravo12} displays the case of a single blob of $^{56}$Ni with a mass $\sim 0.05$~M$_\odot$ expelled with a velocity of $\sim 30,000$~km/s at an angle with respect to the line of sight and a dispersion velocity compatible with the kinematic properties imposed by the gamma-ray spectra, but it is not transparent to gamma-ray photons and should be detectable in the optical. A more favourable geometry can be obtained distributing the radioactive matter in a ring with a truncated conical shape\footnote{Something like a badminton ball} (right panel of Fig.~\ref{f14jbravo12}), a mass of 0.07-0.08 M$_\odot$ and the appropriate parameters. This models is compatible with the observed properties of $^{56}$Co since, 50-100 days after the explosion, matter is transparent and the contribution of the plume is smeared over a large energy interval as a consequence of its large internal velocity dispersion \citep{iser16}.

\subsection{Gamma-ray emission from supernova remnants}
\citet{tsyg16} used the data obtained by IBIS/ISGRI during the 12 years of existence of INTEGRAL to obtain upper limits for the $^{44}$Ti emission of all the Galactic SNRs contained in the \citet{gree14} Catalogue and they only found a significant $^{44}$Ti flux in Cas A. A similar result was obtained by \citet{wein20} with the SPI data.

\emph{Cassiopeia A} (Cas A) is the remnant of a CCSN, probably a Type IIb as deduced from the similitude of its light-echo with the spectrum of SN1993J \citep{krau08}, that exploded at the year $1681\pm 19$ as inferred from the proper motions and radial velocities of the ejecta \citep{fese06}. 
The mass of the ejecta is estimated to be $\sim\, 2.4$~M$_\odot$, of which 1-2~M$_\odot$ are oxygen \citep{vink96}, the kinetic energy of the explosion $\sim\,2.3\times 10^{51}$~erg \citep{orla16}, and the mass of the progenitor in the range of 15-25~M$_\odot$ \citep{youn06}.
It is placed at a distance of 3.4 kpc, in the Galactic plane ($l= -111.74^o$; $b=-2.13^o$) and has been observed many times  with both hard X-ray and gamma-ray instruments (Beppo-SAX, CGRO, INTEGRAL, NuSTAR) thanks to the 'low' and 'high' energy emission of $^{44}$Ti.

\begin{figure}[h]
\center
\includegraphics[width=1.0\textwidth, clip=true, trim= 0cm  0cm 0cm 0cm]{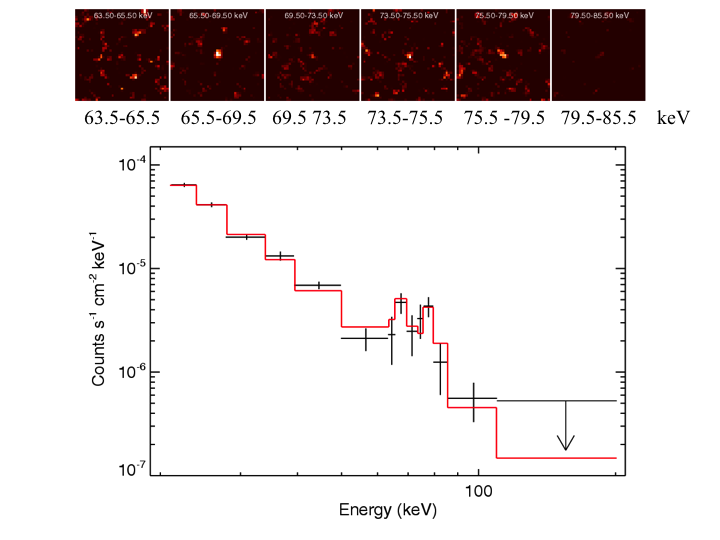}
  \caption{\footnotesize Hard X-ray spectrum of the supernova remnant Cas A obtained with IBIS/ISGRI on board of INTEGRAL. Figure from \citet{rena06}.}
\label{Ti44CasA}
\end{figure}

The high energy line was detected by \citet{iyud94} with COMPTEL-CGRO for the first time and later on by \citet{sieg15} with SPI/INTEGRAL, while the low energy lines were observed by \citet{vink01,rena06,greb12,gref14,bogg15,sieg15,tsyg16, gref17}.
NuSTAR allowed to spatially resolve the remnant and to detect the Doppler broadening of the 61.87 and 78.32 keV decay lines of $^{44}$Ti, showing that its average momentum, defined as the flux weighted average of the titanium velocities, is almost precisely opposite to the direction of movement of the compact central object, a neutron star present in the remnant. This suggests that $^{44}$Ti is tracing out the instabilities that gave the kick to the neutron star \citep{gref14,gref17}. The comparison with 3D neutrino heated models shows that these models can achieve enough bulk asymmetry to account for the observations \citep{jerk20}.

A challenging question is that, according to the results of \citet{sieg15}, the total mass of $^{44}$Ti obtained with the 1157 keV line is larger,  $(2.4\pm 0.9)\times 10^{-4}$~M$_\odot$,  than the one obtained with the 68 and 78 keV lines, $(1.5\pm 0.4)\times 10^{-4}$~M$_\odot$, even after taking into account the branching ratios and the epoch of measurements.
To solve this problem \citet{sieg15} proposed to assume the existence of a low-energy component of cosmic rays that excites the $^{44}$Ca emission (responsible of the 1157 kev line), but maintaining the induced emission of $^{44}$Sc small as a consequence of its low abundance due to its short lifetime. Another possibility has been advanced by \citet{iyud19}, to assume that the low-energy emission is preferentially absorbed by dust, in which case the true $^{44}$Ti yield would be provided by the high energy emission. It has to be noted at this point that \citet{mart09} did not detect the 1157 keV line in their analysis of the $\sim 6.6$~Ms observation of  Cas A with SPI and suggested that the line was broadened by Doppler effect with a velocity of $>500$~km~$^{-1}$ at $2\,\sigma$. \citet{wein20} have recently analysed all the available INTEGRAL/SPI data using a new background model and have obtained a mass of $^{44}$Ti of $(2.1\pm 0.6)\times 10^{-4}$~M$_\odot$ when using the 78 keV line, $(5.9\pm 1.9)\times 10^{-4}$~M$_\odot$ when using the 1157 keV line, $(2.6\pm 0.6)\times 10^{-4}$~M$_\odot$ when using both, and $(2.9\pm 0.5)\times 10^{-4}$~M$_\odot$, using the three lines but keeping the Doppler parameters of the 78 and 1157 keV lines.
Thus, it is obvious that, to solve the problem, new observations with a more sensitive instrument are necessary.

\subsection{Gamma-ray emission from novae}

The detection of nuclear $\gamma$-ray emission from novae is challenging and demands an all-sky observation capability. Up to now, no detections of such emission have been made with past or existing gamma-ray observatories. One way to overcome this difficulty is the use of the large detection area and wide-field of view provided by the anticoincidence system (ACS, made with BGO scintillators) of SPI/INTEGRAL \citep{jean99}. The idea was that the positron annihilation radiation emitted by the nova would induce an increase in the counting rate of the BGO that would depend on the spectrum and position of the nova with respect to the shield. Taking into account the predictions of the models of \cite{gome98a}, it would have been possible to detect a 0.8 M$_\odot$ CO nova up to a maximum distance of 0.7 kpc and a 1.15 M$_\odot$ ONe nova up to 7-8 kpc. However, since 1999, the predicted yields of the $\beta^+$ unstable nucleus $^{18}$F have been reduced very significantly, by more than one order of magnitude, because of  the improved (but still uncertain) nuclear reaction rates affecting its production and destruction \citep{hern99,coc2000,chafa2005,hj2006}. The detectability distances have thus been reduced by factors larger than $\sim$ 3, and 
unfortunately no novae have occurred at such short distances. 

In fact, a detailed search of the Comptonized positron annihilation emission of novae was made with the Burst Alert Telescope (BAT) on board the Swift satellite \citep{senz08}. Because of its broad field of view of $\sim$ 1/6 of the sky and 
sensitivity in the energy range (15-150 keV), Swift/BAT is well adapted to search retrospectively for the prompt $\gamma$-ray annihilation signature from
novae that have been in its field of view. The search for emission in the directions of the 24 classical novae discovered since Swift launch in 2004 until the end of 2007 did not provide detections,  
which was consistent with none of the novae being close enough. That result was also consistent with an expected detection rate of $\sim$ 2-5 novae in 10 yr, estimated with nova models 
and simulations of galactic nova observations with BAT \citep{senz08}.
 
 In 2015, the probably closest nova during the INTEGRAL mission time occurred: Nova Sgr 2015 No. 2 \citep{seac15}. This source was very bright with a maximum V-band magnitude of 4.32, and was also among the first classical novae in which the Be\,II doublet at 313\,nm was observed \citep{mola16,taji16}. From these UV measurements, it was derived that about $9 \times 10^{-9}\,\mathrm{M_{\odot}}$ of $\mathrm{^{7}Be}$ have been synthesised and ejected into the interstellar medium. This mass estimate would be on the edge of SPI's sensitivity limit. \citet{sieg18} performed a detailed study to search for the expected gamma-ray lines with SPI as well as the preceding 511\,keV flash with the SPI/ACS. Still, the 478\,keV line remained undetected, but this study showed the capabilities of SPI for individual objects, finding a lower detection limit for classical novae in $\mathrm{^{7}Be}$ of 1.2\,kpc ($3\sigma$). In addition, while there are several candidate events in the weeks before the optical maximum of V5668 Sgr, none of them could be clearly identified to be due to a possible positron annihilation flash as derived from the combined directional response of INTEGRAL's main instruments (SPI, IBIS) as well as their veto systems (SPI/ACS, IBIS-veto). Identification of individual classical novae are therefore restricted to distances below $\sim 1$\,kpc based on SPI's sensitivity, however the cumulative effect of the population of novae with a measured rate of 50--100 per year in the Milky Way \citep{shaf17}, could be readily detected.

\citet{jean2000} investigated the cumulative 1275 keV emission of $^{22}$Na produced by classical ONe novae in our Galaxy. That emission depends not only on the amount of
$^{22}$Na produced per ONe nova outburst but also on their rate and distribution which are both not well known. These authors found that the detection of that cumulative emission
is less probable than the detection of the line from a single and bright ONe nova. This was confirmed by further analyses. First, using the upper limit of the 1275 keV line flux
measured from the 6 yr of COMPTEL observations of the galactic bulge, \citet{jean2001} estimated an ejected  $^{22}$Na mass upper-limit of 2 $\times$ 10$^{-7}$ M$_{\odot}$ per
outburst from COMPTEL observation of the cumulative emission, assuming an ONe rate of 7.5 yr$^{-1}$. In the same manner and with the same ONe rate, \citet{jean2004}
estimated an upper limit of 5.7 $\times$ 10$^{-7}$ M$_{\odot}$ with 1.8 Ms of SPI observations. Those upper limits that take into account the uncertainty in the spatial
distribution of ONe novae, are less stringent than the 2$\sigma$ upper limit of 3.7 $\times$ 10$^{-8}$ M$_{\odot}$ derived from COMPTEL observations of Nova Cygni \citep{iyu95}. 
Moreover, they are far from the theoretical estimates that predict a  $^{22}$Na ejected mass of about $\sim$ 5 $\times$ 10$^{-9}$ M$_{\odot}$ \citep{jh98,jch99}.

\section{Conclusions}

As it has been seen, MeV gamma-rays provide a unique probe of the nuclear processes that occur in different explosive events like novae and supernovae since they allow the direct measurement of the radioactive decay, nuclear de-excitation and positron production associated to the freshly synthesized isotopes created during these events. However, the necessary condition for taking profit of such information is to have an instrument in orbit able to close the sensitivity gap that exists at MeV energies as depicted in Fig.~\ref{mevgap}.

The re-enter in the Earth atmosphere of \emph{INTEGRAL} will be in 2029 although the science mission will end before, depending on the status of the instruments. In view of its success, it is urgent to build a new instrument able to explore this crucial region of the electromagnetic spectrum  not covered by NuSTAR. Several proposals have been advanced up to now.

\paragraph {e-ASTROGAM} This is a proposal for an open $\gamma$-ray observatory dedicated to explore the Universe in the energy range of 0.15-100 MeV, and extending up to $\sim 3$ GeV [\citep{dean17,dean18}]. This energy range not only allows to study some of the most spectacular high energy events (Gamma Ray Bursts, blazars, pulsars) but also the $\gamma$-lines emitted as a consequence of the nuclear activity of different events as is the case of novae and supernovae.

The core science program of e-ASTROGAM addresses to: i) Processes that are characteristic of the extreme Universe (jets, outflows, explosions,\ldots) and its connection with the new astronomy windows (gravitational waves, neutrinos and ultra-high energy cosmic rays), ii) impact of high-energy particles on the evolution of the Milky Way, and iii) nucleosynthesis and chemical evolution of the Galaxy. Furthermore, thanks to its wide field of view and its polarimetric capability, it is ready for many serendipitous discoveries of unexpected transient events.

The sensitivity around the 847 keV $^{56}$Ni line is estimated to be $ 3.5 \times 10^{-6}$ cm$^{-2}$s$^{-1}$ in $10^6$ s of integration time (Fig.~\ref{mevgap}). This represents a factor $\sim 70$ of improvement with respect  to the sensitivity of INTEGRAL/SPI thus opening the possibility to detect events like SN 2014J up to a distance of 35 Mpc. This means that the number of potentially detectable SNIa is $\sim 10$ during the three years of nominal mission lifetime, a number that would be high enough to provide insight about the observed variety of SNIa events. Furthermore, the excellent sensitivity around the 511 keV annihilation line could clarify the role played by positrons in the process of energy deposition in the ejecta. This sensitivity is also enough to detect the $^{56}$Ni-$^{56}$Co signatures in core collapse and electron capture supernovae occurring in nearby galaxies thus opening the possibility to directly check the proposed models of explosion.

The sensitivity for the 1157 keV $^{44}$Ti is $ 3.6 \times 10^{-6}$ cm$^{-2}$s$^{-1}$ in $10^6$ s of integration time. This represents a gain of a factor 27 with respect to INTEGRAL capabilities and should allow one to detect this line emission from a large fraction of young supernova remnants in the Milky Way and to clarify  the  discrepant measurements in the SN 1987A.

\paragraph{AMEGO}
The All-sky Medium Energy Gamma-ray Observatory (AMEGO) is a mission that will cover the energy range comprised between 200 keV and 20 GeV with a continuum sensitivity in the region of MeV one order of magnitude better than previous missions (Fig.\ref{mevgap}). 
It will be placed in a low-inclination, low-Earth orbit, with a $\sim \, 90$~min period and, thanks to its large field of view will be able to scan the entire sky every three hours, making it an excellent instrument to follow the early emissions of novae and supernovae \citep{mcen19}.

\paragraph{LOX}
  The Lunar Occultation eXplorer (LOX) is a lunar orbiting mission that uses the occultations by the disk of the Moon to systematically explore transient phenomena, SNIa mainly, in the region of 0.1-10 MeV, see \citet{mill19} for a full description. The instruments consists of a spectrometer that is pointing to the nadir, i.e. the center of the Moon, in such a way that cosmic gamma-ray sources rise and set along the lunar limb. The spectrometer (BAGEL: Big Array of individual Gamma-ray Energy Logging)  is a large array of individual spectrometers that have been inherited from multiple planetary and astrophysics missions with a total area of $\sim 1.5$~m$^2$. The individual spectrometers have no imaging capacity and the localization of the sources  is achieved thanks to the modulations introduced by the occultations. Figure~\ref{mevgap} displays the expected sensitivity. One of the ingredients to achieve such a sensitivity is the relatively weak, slowly changing and well calibrated background of the Moon. Notice that its high sensitivity around the 158~keV~$^{56}$Ni line makes this mission specially valuable to study the early stages of supernovae

COSI (Compton Spectrometer and Imager) is a high-energy resolution Compton telescope with an  improved sensitivity for the low energy gamma-ray lines emitted by supernovae and novae. It  has been developed with Ge detectors and was launched with a superpressure balloon at stratospheric altitudes in 2016 \citep{kier17}. COSI is proposed for a SMEX mission \citep{toms19}; its wide field of view is adapted to enhance the exposure on gamma-ray sources and to detect transients, in the 0.2-5 MeV band. 

One possibility to noticeably increase the sensitivity to MeV gamma-rays is offered by the so called Electron-Tracking Compton Camera that is able to provide all the parameters of Compton scattering by measuring the 3D recoil tracks, as well as the energy loss rate which help to reduce the background. All these improvements offer the possibility to reach sensitivities of $10^{-12}$~erg~cm$^{-2}$~s$^{-1}$ at 1 MeV for expositions of 1 Ms \citep{tani15,tani17}.

An additional way to increase the sensitivity without increasing the noise consists in the introduction of a lens. In the case of $\gamma$-rays this can be achieved using Bragg diffraction with crystals made of high-Z elements like germanium arranged circularly around an optical axis forming concentric rings. The main limitation of this method is that a given ring focuses only a very narrow energy band on a detector placed at a fixed distance in such a way that it is necessary to use multiple ring or a moving detector to explore a wide band of energies \citep{barr17}. This means that the objectives have to be very focused to be successful. This concept has been demonstrated with the balloon-borne instrument \emph{CLAIRE} \citep{ball05}.

\section*{Acknowledgements}
This work  has been supported  by MINECO grants  AYA2015-63588-P and PGC2018-095317-B-C21 (EB), ESP2017-82674-R (MH, JI), 
by EU FEDER funds and by grants 2014SGR1458 and CERCA Programe of  the Generalitat de
Catalunya (MH, JI), 
by the German Research Society, DFG-Forschungsstipedium SI 2502/1-1 (TS), by project no. 18-12-00522 of the Russian Science Foundation (SG).


\bibliography{sn}

\begin{thebibliography}{357}
\expandafter\ifx\csname natexlab\endcsname\relax\def\natexlab#1{#1}\fi
\providecommand{\url}[1]{\texttt{#1}}
\providecommand{\href}[2]{#2}
\providecommand{\path}[1]{#1}
\providecommand{\DOIprefix}{doi:}
\providecommand{\ArXivprefix}{arXiv:}
\providecommand{\URLprefix}{URL: }
\providecommand{\Pubmedprefix}{pmid:}
\providecommand{\doi}[1]{\href{http://dx.doi.org/#1}{\path{#1}}}
\providecommand{\Pubmed}[1]{\href{pmid:#1}{\path{#1}}}
\providecommand{\bibinfo}[2]{#2}
\ifx\xfnm\relax \def\xfnm[#1]{\unskip,\space#1}\fi
\bibitem[{{Adams} et~al.(2013){Adams}, {Kochanek}, {Beacom}, {Vagins} and
  {Stanek}}]{adam13}
\bibinfo{author}{{Adams}, S.M.}, \bibinfo{author}{{Kochanek}, C.S.},
  \bibinfo{author}{{Beacom}, J.F.}, \bibinfo{author}{{Vagins}, M.R.},
  \bibinfo{author}{{Stanek}, K.Z.}, \bibinfo{year}{2013}.
\newblock \bibinfo{title}{{Observing the Next Galactic Supernova}}.
\newblock \bibinfo{journal}{Astrophys. J.} \bibinfo{volume}{778},
  \bibinfo{pages}{164}.
\newblock \DOIprefix\doi{10.1088/0004-637X/778/2/164}.
\bibitem[{{Allen}(1954)}]{alle54}
\bibinfo{author}{{Allen}, C.W.}, \bibinfo{year}{1954}.
\newblock \bibinfo{title}{{Whole-sky statistics of celestial objects}}.
\newblock \bibinfo{journal}{Mon. Not. R. Astron. Soc.} \bibinfo{volume}{114},
  \bibinfo{pages}{387}.
\newblock \DOIprefix\doi{10.1093/mnras/114.4.387}.
\bibitem[{{Alpher} et~al.(1948){Alpher}, {Bethe} and {Gamow}}]{alph48}
\bibinfo{author}{{Alpher}, R.A.}, \bibinfo{author}{{Bethe}, H.},
  \bibinfo{author}{{Gamow}, G.}, \bibinfo{year}{1948}.
\newblock \bibinfo{title}{{The Origin of Chemical Elements}}.
\newblock \bibinfo{journal}{Physical Review} \bibinfo{volume}{73},
  \bibinfo{pages}{803--804}.
\newblock \DOIprefix\doi{10.1103/PhysRev.73.803}.
\bibitem[{{Ambwani} and {Sutherland}(1988)}]{ambw88}
\bibinfo{author}{{Ambwani}, K.}, \bibinfo{author}{{Sutherland}, P.},
  \bibinfo{year}{1988}.
\newblock \bibinfo{title}{{Gamma-ray spectra and energy deposition for type IA
  supernovae}}.
\newblock \bibinfo{journal}{Astrophys. J.} \bibinfo{volume}{325},
  \bibinfo{pages}{820--827}.
\newblock \DOIprefix\doi{10.1086/166052}.
\bibitem[{{Anderson}(2019)}]{ande19}
\bibinfo{author}{{Anderson}, J.P.}, \bibinfo{year}{2019}.
\newblock \bibinfo{title}{{A meta-analysis of core-collapse supernova $^{56}$Ni
  masses}}.
\newblock \bibinfo{journal}{A\&A} \bibinfo{volume}{628}, \bibinfo{pages}{A7}.
\newblock \DOIprefix\doi{10.1051/0004-6361/201935027}.
\bibitem[{{Anderson} et~al.(2014){Anderson}, {Dessart} and
  {Gutierrez}}]{ande14}
\bibinfo{author}{{Anderson}, J.P.}, \bibinfo{author}{{Dessart}, L.},
  \bibinfo{author}{{Gutierrez}, C.P.e.a.}, \bibinfo{year}{2014}.
\newblock \bibinfo{title}{{Analysis of blueshifted emission peaks in Type II
  supernovae}}.
\newblock \bibinfo{journal}{Mon. Not. R. Astron. Soc.} \bibinfo{volume}{441},
  \bibinfo{pages}{671--680}.
\newblock \DOIprefix\doi{10.1093/mnras/stu610}.
\bibitem[{{Andrews} et~al.(2020){Andrews}, {Fryer}, {Even}, {Jones} and
  {Pignatari}}]{andr20}
\bibinfo{author}{{Andrews}, S.}, \bibinfo{author}{{Fryer}, C.},
  \bibinfo{author}{{Even}, W.}, \bibinfo{author}{{Jones}, S.},
  \bibinfo{author}{{Pignatari}, M.}, \bibinfo{year}{2020}.
\newblock \bibinfo{title}{{The Nucleosynthetic Yields of Core-collapse
  Supernovae: Prospects for the Next Generation of Gamma-Ray Astronomy}}.
\newblock \bibinfo{journal}{Astrophys. J.} \bibinfo{volume}{890},
  \bibinfo{pages}{35}.
\newblock \DOIprefix\doi{10.3847/1538-4357/ab64f8},
  \href{http://arxiv.org/abs/1912.10542}{\tt arXiv:1912.10542}.
\bibitem[{{Arnett}(1969)}]{arne69}
\bibinfo{author}{{Arnett}, W.D.}, \bibinfo{year}{1969}.
\newblock \bibinfo{title}{{A Possible Model of Supernovae: Detonation of
  $^{12}$C}}.
\newblock \bibinfo{journal}{Astrophys. Space Sci.} \bibinfo{volume}{5},
  \bibinfo{pages}{180--212}.
\newblock \DOIprefix\doi{10.1007/BF00650291}.
\bibitem[{{Arnett}(1982)}]{arne82}
\bibinfo{author}{{Arnett}, W.D.}, \bibinfo{year}{1982}.
\newblock \bibinfo{title}{{Type I supernovae. I - Analytic solutions for the
  early part of the light curve}}.
\newblock \bibinfo{journal}{Astrophys. J.} \bibinfo{volume}{253},
  \bibinfo{pages}{785--797}.
\newblock \DOIprefix\doi{10.1086/159681}.
\bibitem[{{Arnett} et~al.(1985){Arnett}, {Branch} and {Wheeler}}]{arne85}
\bibinfo{author}{{Arnett}, W.D.}, \bibinfo{author}{{Branch}, D.},
  \bibinfo{author}{{Wheeler}, J.C.}, \bibinfo{year}{1985}.
\newblock \bibinfo{title}{{Hubble's constant and exploding carbon-oxygen white
  dwarf models for Type I supernovae}}.
\newblock \bibinfo{journal}{Nature} \bibinfo{volume}{314},
  \bibinfo{pages}{337--338}.
\newblock \DOIprefix\doi{10.1038/314337a0}.
\bibitem[{{Arnould} and {Goriely}(2020)}]{arno20}
\bibinfo{author}{{Arnould}, M.}, \bibinfo{author}{{Goriely}, S.},
  \bibinfo{year}{2020}.
\newblock \bibinfo{title}{{Astronuclear Physics: A tale of the atomic nuclei in
  the skies}}.
\newblock \bibinfo{journal}{Progress in Particle and Nuclear Physics}
  \bibinfo{volume}{112}, \bibinfo{pages}{103766}.
\newblock \DOIprefix\doi{10.1016/j.ppnp.2020.103766}.
\bibitem[{{Audouze} and {Reeves}(1982)}]{audo82}
\bibinfo{author}{{Audouze}, J.}, \bibinfo{author}{{Reeves}, H.},
  \bibinfo{year}{1982}.
\newblock \bibinfo{title}{{The Origin of the Light Elements}}, in:
  \bibinfo{editor}{{Barnes}, C.A.}, \bibinfo{editor}{{Clayton}, D.D.},
  \bibinfo{editor}{{Schramm}, D.N.} (Eds.), \bibinfo{booktitle}{Essays in
  Nuclear Astrophysics}, p. \bibinfo{pages}{355}.
\bibitem[{{Aznar-Sigu{\'a}n} et~al.(2013){Aznar-Sigu{\'a}n},
  {Garc{\'{\i}}a-Berro}, {Lor{\'e}n-Aguilar}, {Jos{\'e}} and {Isern}}]{azna13}
\bibinfo{author}{{Aznar-Sigu{\'a}n}, G.},
  \bibinfo{author}{{Garc{\'{\i}}a-Berro}, E.},
  \bibinfo{author}{{Lor{\'e}n-Aguilar}, P.}, \bibinfo{author}{{Jos{\'e}}, J.},
  \bibinfo{author}{{Isern}, J.}, \bibinfo{year}{2013}.
\newblock \bibinfo{title}{{Detonations in white dwarf dynamical interactions}}.
\newblock \bibinfo{journal}{Mon. Not. R. Astron. Soc.} \bibinfo{volume}{434},
  \bibinfo{pages}{2539--2555}.
\newblock \DOIprefix\doi{10.1093/mnras/stt1198}.
\bibitem[{{Aznar-Sigu{\'a}n} et~al.(2015){Aznar-Sigu{\'a}n},
  {Garc{\'\i}a-Berro}, {Lor{\'e}n-Aguilar}, {Soker} and {Kashi}}]{azna15}
\bibinfo{author}{{Aznar-Sigu{\'a}n}, G.}, \bibinfo{author}{{Garc{\'\i}a-Berro},
  E.}, \bibinfo{author}{{Lor{\'e}n-Aguilar}, P.}, \bibinfo{author}{{Soker},
  N.}, \bibinfo{author}{{Kashi}, A.}, \bibinfo{year}{2015}.
\newblock \bibinfo{title}{{Smoothed particle hydrodynamics simulations of the
  core-degenerate scenario for Type Ia supernovae}}.
\newblock \bibinfo{journal}{Mon. Not. R. Astron. Soc.} \bibinfo{volume}{450},
  \bibinfo{pages}{2948--2962}.
\newblock \DOIprefix\doi{10.1093/mnras/stv824}.
\bibitem[{{Badenes} et~al.(2005){Badenes}, {Borkowski} and {Bravo}}]{bade05}
\bibinfo{author}{{Badenes}, C.}, \bibinfo{author}{{Borkowski}, K.J.},
  \bibinfo{author}{{Bravo}, E.}, \bibinfo{year}{2005}.
\newblock \bibinfo{title}{{Thermal X-Ray Emission from Shocked Ejecta in Type
  Ia Supernova Remnants. II. Parameters Affecting the Spectrum}}.
\newblock \bibinfo{journal}{Astrophys. J.} \bibinfo{volume}{624},
  \bibinfo{pages}{198--212}.
\newblock \DOIprefix\doi{10.1086/428829}.
\bibitem[{{Badenes} et~al.(2006){Badenes}, {Borkowski}, {Hughes}, {Hwang} and
  {Bravo}}]{bade06}
\bibinfo{author}{{Badenes}, C.}, \bibinfo{author}{{Borkowski}, K.J.},
  \bibinfo{author}{{Hughes}, J.P.}, \bibinfo{author}{{Hwang}, U.},
  \bibinfo{author}{{Bravo}, E.}, \bibinfo{year}{2006}.
\newblock \bibinfo{title}{{Constraints on the Physics of Type Ia Supernovae
  from the X-Ray Spectrum of the Tycho Supernova Remnant}}.
\newblock \bibinfo{journal}{Astrophys. J.} \bibinfo{volume}{645},
  \bibinfo{pages}{1373--1391}.
\newblock \DOIprefix\doi{10.1086/504399}.
\bibitem[{{Badenes} et~al.(2003){Badenes}, {Bravo}, {Borkowski} and
  {Dom{\'\i}nguez}}]{bade03}
\bibinfo{author}{{Badenes}, C.}, \bibinfo{author}{{Bravo}, E.},
  \bibinfo{author}{{Borkowski}, K.J.}, \bibinfo{author}{{Dom{\'\i}nguez}, I.},
  \bibinfo{year}{2003}.
\newblock \bibinfo{title}{{Thermal X-Ray Emission from Shocked Ejecta in Type
  Ia Supernova Remnants: Prospects for Explosion Mechanism Identification}}.
\newblock \bibinfo{journal}{Astrophys. J.} \bibinfo{volume}{593},
  \bibinfo{pages}{358--369}.
\newblock \DOIprefix\doi{10.1086/376448}.
\bibitem[{{Badenes} et~al.(2008){Badenes}, {Bravo} and {Hughes}}]{bade08}
\bibinfo{author}{{Badenes}, C.}, \bibinfo{author}{{Bravo}, E.},
  \bibinfo{author}{{Hughes}, J.P.}, \bibinfo{year}{2008}.
\newblock \bibinfo{title}{{The End of Amnesia: A New Method for Measuring the
  Metallicity of Type Ia Supernova Progenitors Using Manganese Lines in
  Supernova Remnants}}.
\newblock \bibinfo{journal}{Astrophys. J. Lett.} \bibinfo{volume}{680},
  \bibinfo{pages}{L33}.
\newblock \DOIprefix\doi{10.1086/589832},
  \href{http://arxiv.org/abs/0805.3344}{\tt arXiv:0805.3344}.
\bibitem[{{Badenes} et~al.(2007){Badenes}, {Hughes}, {Bravo} and
  {Langer}}]{bade07}
\bibinfo{author}{{Badenes}, C.}, \bibinfo{author}{{Hughes}, J.P.},
  \bibinfo{author}{{Bravo}, E.}, \bibinfo{author}{{Langer}, N.},
  \bibinfo{year}{2007}.
\newblock \bibinfo{title}{{Are the Models for Type Ia Supernova Progenitors
  Consistent with the Properties of Supernova Remnants?}}
\newblock \bibinfo{journal}{Astrophys. J.} \bibinfo{volume}{662},
  \bibinfo{pages}{472--486}.
\newblock \DOIprefix\doi{10.1086/518022}.
\bibitem[{{Badenes} and {Maoz}(2012)}]{bade12}
\bibinfo{author}{{Badenes}, C.}, \bibinfo{author}{{Maoz}, D.},
  \bibinfo{year}{2012}.
\newblock \bibinfo{title}{{The Merger Rate of Binary White Dwarfs in the
  Galactic Disk}}.
\newblock \bibinfo{journal}{Astrophys. J. Lett.} \bibinfo{volume}{749},
  \bibinfo{pages}{L11}.
\newblock \DOIprefix\doi{10.1088/2041-8205/749/1/L11}.
\bibitem[{{Barkat} et~al.(1967){Barkat}, {Rakvy} and {Sack}}]{bark67}
\bibinfo{author}{{Barkat}, Z.}, \bibinfo{author}{{Rakvy}, G.},
  \bibinfo{author}{{Sack}, N.}, \bibinfo{year}{1967}.
\newblock \bibinfo{title}{{Dynamics of Supernova Explosion resulting from Pair
  Formation}}.
\newblock \bibinfo{journal}{Phys. Rev. Lett} \bibinfo{volume}{18},
  \bibinfo{pages}{379}.
\bibitem[{{Barri{\`e}re} et~al.(2017){Barri{\`e}re}, {von Ballmoos} and
  {Natalucci}}]{barr17}
\bibinfo{author}{{Barri{\`e}re}, N.}, \bibinfo{author}{{von Ballmoos}, P.},
  \bibinfo{author}{{Natalucci}, L.e.a.}, \bibinfo{year}{2017}.
\newblock \bibinfo{title}{{Laue lens: the challenge of focusing gamma rays}},
  in: \bibinfo{booktitle}{SPIE}, p. \bibinfo{pages}{1056603}.
\newblock \DOIprefix\doi{10.1117/12.2308289}.
\bibitem[{{Bartunov} et~al.(1987){Bartunov}, {Blinnikov}, {Levakhina} and
  {Nadyiozhin}}]{bart87}
\bibinfo{author}{{Bartunov}, O.S.}, \bibinfo{author}{{Blinnikov}, S.I.},
  \bibinfo{author}{{Levakhina}, L.V.}, \bibinfo{author}{{Nadyiozhin}, D.K.},
  \bibinfo{year}{1987}.
\newblock \bibinfo{title}{{The gamma radiation expected from the supernova
  1987A in the Large Magellanic Cloud}}.
\newblock \bibinfo{journal}{Soviet Astronomy Letters} \bibinfo{volume}{13},
  \bibinfo{pages}{313}.
\bibitem[{{B{\'e}dard} et~al.(2017){B{\'e}dard}, {Bergeron} and
  {Fontaine}}]{beda17}
\bibinfo{author}{{B{\'e}dard}, A.}, \bibinfo{author}{{Bergeron}, P.},
  \bibinfo{author}{{Fontaine}, G.}, \bibinfo{year}{2017}.
\newblock \bibinfo{title}{{Measurements of Physical Parameters of White Dwarfs:
  A Test of the Mass-Radius Relation}}.
\newblock \bibinfo{journal}{Astrophys. J.} \bibinfo{volume}{848},
  \bibinfo{pages}{11}.
\newblock \DOIprefix\doi{10.3847/1538-4357/aa8bb6}.
\bibitem[{{Benz} et~al.(1990){Benz}, {Cameron}, {Press} and {Bowers}}]{benz90}
\bibinfo{author}{{Benz}, W.}, \bibinfo{author}{{Cameron}, A.G.W.},
  \bibinfo{author}{{Press}, W.H.}, \bibinfo{author}{{Bowers}, R.L.},
  \bibinfo{year}{1990}.
\newblock \bibinfo{title}{{Dynamic mass exchange in doubly degenerate binaries.
  I - 0.9 and 1.2 solar mass stars}}.
\newblock \bibinfo{journal}{Astrophys. J.} \bibinfo{volume}{348},
  \bibinfo{pages}{647--667}.
\newblock \DOIprefix\doi{10.1086/168273}.
\bibitem[{{Bisnovatyi-Kogan}(1970)}]{bisn70}
\bibinfo{author}{{Bisnovatyi-Kogan}, G.S.}, \bibinfo{year}{1970}.
\newblock \bibinfo{title}{{The Explosion of a Rotating Star As a Supernova
  Mechanism.}}
\newblock \bibinfo{journal}{Astron. Zh+} \bibinfo{volume}{47},
  \bibinfo{pages}{813}.
\bibitem[{{Bloom} et~al.(2012){Bloom}, {Kasen} and {Shen}}]{bloo12}
\bibinfo{author}{{Bloom}, J.S.}, \bibinfo{author}{{Kasen}, D.},
  \bibinfo{author}{{Shen}, K.J.e.a.}, \bibinfo{year}{2012}.
\newblock \bibinfo{title}{{A Compact Degenerate Primary-star Progenitor of
  SN2011fe}}.
\newblock \bibinfo{journal}{Astrophys. J. Lett.} \bibinfo{volume}{744},
  \bibinfo{pages}{L17}.
\newblock \DOIprefix\doi{10.1088/2041-8205/744/2/L17}.
\bibitem[{{Boggs} et~al.(2015){Boggs}, {Harrison} and {Miyasaka}}]{bogg15}
\bibinfo{author}{{Boggs}, S.E.}, \bibinfo{author}{{Harrison}, F.A.},
  \bibinfo{author}{{Miyasaka}, H.e.a.}, \bibinfo{year}{2015}.
\newblock \bibinfo{title}{{$^{44}$Ti gamma-ray emission lines from SN1987A
  reveal an asymmetric explosion}}.
\newblock \bibinfo{journal}{Science} \bibinfo{volume}{348},
  \bibinfo{pages}{670--671}.
\bibitem[{{Bohlin} et~al.(1980){Bohlin}, {Frost}, {Burr}, {Guha} and
  {Withbroe}}]{bohl80}
\bibinfo{author}{{Bohlin}, J.D.}, \bibinfo{author}{{Frost}, K.J.},
  \bibinfo{author}{{Burr}, P.T.}, \bibinfo{author}{{Guha}, A.K.},
  \bibinfo{author}{{Withbroe}, G.L.}, \bibinfo{year}{1980}.
\newblock \bibinfo{title}{{Solar Maximum Mission}}.
\newblock \bibinfo{journal}{Solar Physics} \bibinfo{volume}{65},
  \bibinfo{pages}{5--14}.
\newblock \DOIprefix\doi{10.1007/BF00151380}.
\bibitem[{{Branch}(2010)}]{bran10}
\bibinfo{author}{{Branch}, D.}, \bibinfo{year}{2010}.
\newblock \bibinfo{title}{{Supernovae: New explosions of old stars?}}
\newblock \bibinfo{journal}{Nature} \bibinfo{volume}{465},
  \bibinfo{pages}{303--304}.
\newblock \DOIprefix\doi{10.1038/465303a}.
\bibitem[{{Branch} et~al.(1993){Branch}, {Fisher} and {Nugent}}]{bran93}
\bibinfo{author}{{Branch}, D.}, \bibinfo{author}{{Fisher}, A.},
  \bibinfo{author}{{Nugent}, P.}, \bibinfo{year}{1993}.
\newblock \bibinfo{title}{{On the relative frequencies of spectroscopically
  normal and peculiar type IA supernovae}}.
\newblock \bibinfo{journal}{Astron. J.} \bibinfo{volume}{106},
  \bibinfo{pages}{2383--2391}.
\newblock \DOIprefix\doi{10.1086/116810}.
\bibitem[{{Branch} and {Wheeler}(2017)}]{bran17}
\bibinfo{author}{{Branch}, D.}, \bibinfo{author}{{Wheeler}, J.C.},
  \bibinfo{year}{2017}.
\newblock \bibinfo{title}{{Supernova Explosions}}.
\newblock \DOIprefix\doi{10.1007/978-3-662-55054-0}.
\bibitem[{{Bravo} et~al.(1992){Bravo}, {Isern}, {Canal} and {Labay}}]{brav92}
\bibinfo{author}{{Bravo}, E.}, \bibinfo{author}{{Isern}, J.},
  \bibinfo{author}{{Canal}, R.}, \bibinfo{author}{{Labay}, J.},
  \bibinfo{year}{1992}.
\newblock \bibinfo{title}{{On the contribution of Ne-22 to the synthesis of
  Fe-54 and Ni-58 in thermonuclear supernovae}}.
\newblock \bibinfo{journal}{A\&A} \bibinfo{volume}{257},
  \bibinfo{pages}{534--538}.
\bibitem[{{Bravo} et~al.(1996){Bravo}, {Tornambe}, {Dominguez} and
  {Isern}}]{brav96}
\bibinfo{author}{{Bravo}, E.}, \bibinfo{author}{{Tornambe}, A.},
  \bibinfo{author}{{Dominguez}, I.}, \bibinfo{author}{{Isern}, J.},
  \bibinfo{year}{1996}.
\newblock \bibinfo{title}{{Clues to Type IA SN progenitors from degenerate
  carbon ignition models.}}
\newblock \bibinfo{journal}{A\&A} \bibinfo{volume}{306},
  \bibinfo{pages}{811--+}.
\bibitem[{{Brown} et~al.(2012){Brown}, {Dawson} and {de Pasquale}}]{brow12}
\bibinfo{author}{{Brown}, P.J.}, \bibinfo{author}{{Dawson}, K.S.},
  \bibinfo{author}{{de Pasquale}, M.e.a.}, \bibinfo{year}{2012}.
\newblock \bibinfo{title}{{A Swift Look at SN 2011fe: The Earliest Ultraviolet
  Observations of a Type Ia Supernova}}.
\newblock \bibinfo{journal}{Astrophys. J.} \bibinfo{volume}{753},
  \bibinfo{pages}{22}.
\newblock \DOIprefix\doi{10.1088/0004-637X/753/1/22}.
\bibitem[{{Burbidge} et~al.(1957){Burbidge}, {Burbidge}, {Fowler} and
  {Hoyle}}]{burb57}
\bibinfo{author}{{Burbidge}, E.M.}, \bibinfo{author}{{Burbidge}, G.R.},
  \bibinfo{author}{{Fowler}, W.A.}, \bibinfo{author}{{Hoyle}, F.},
  \bibinfo{year}{1957}.
\newblock \bibinfo{title}{{Synthesis of the Elements in Stars}}.
\newblock \bibinfo{journal}{Reviews of Modern Physics} \bibinfo{volume}{29},
  \bibinfo{pages}{547--650}.
\newblock \DOIprefix\doi{10.1103/RevModPhys.29.547}.
\bibitem[{{Burrows}(2013)}]{burr13}
\bibinfo{author}{{Burrows}, A.}, \bibinfo{year}{2013}.
\newblock \bibinfo{title}{{Colloquium: Perspectives on core-collapse supernova
  theory}}.
\newblock \bibinfo{journal}{Reviews of Modern Physics} \bibinfo{volume}{85},
  \bibinfo{pages}{245--261}.
\newblock \DOIprefix\doi{10.1103/RevModPhys.85.245},
  \href{http://arxiv.org/abs/1210.4921}{\tt arXiv:1210.4921}.
\bibitem[{{Burrows} et~al.(2007){Burrows}, {Dessart}, {Ott} and
  {Livne}}]{burr07}
\bibinfo{author}{{Burrows}, A.}, \bibinfo{author}{{Dessart}, L.},
  \bibinfo{author}{{Ott}, C.D.}, \bibinfo{author}{{Livne}, E.},
  \bibinfo{year}{2007}.
\newblock \bibinfo{title}{{Multi-dimensional explorations in supernova
  theory}}.
\newblock \bibinfo{journal}{Physics Repports} \bibinfo{volume}{442},
  \bibinfo{pages}{23--37}.
\newblock \DOIprefix\doi{10.1016/j.physrep.2007.02.001}.
\bibitem[{{Burrows} and {The}(1990)}]{burr90}
\bibinfo{author}{{Burrows}, A.}, \bibinfo{author}{{The}, L.},
  \bibinfo{year}{1990}.
\newblock \bibinfo{title}{{X- and gamma-ray signatures of type Ia supernovae}}.
\newblock \bibinfo{journal}{Astrophys. J.} \bibinfo{volume}{360},
  \bibinfo{pages}{626--638}.
\newblock \DOIprefix\doi{10.1086/169150}.
\bibitem[{{Burrows} et~al.(2005){Burrows}, {Hill} and {Nousek}}]{burr05}
\bibinfo{author}{{Burrows}, D.N.}, \bibinfo{author}{{Hill}, J.E.},
  \bibinfo{author}{{Nousek}, J.A.e.a.}, \bibinfo{year}{2005}.
\newblock \bibinfo{title}{{The Swift X-Ray Telescope}}.
\newblock \bibinfo{journal}{Space Sci. Rev.} \bibinfo{volume}{120},
  \bibinfo{pages}{165--195}.
\newblock \DOIprefix\doi{10.1007/s11214-005-5097-2}.
\bibitem[{{Cadonau} et~al.(1985){Cadonau}, {Tammann} and {Sandage}}]{cado85}
\bibinfo{author}{{Cadonau}, R.}, \bibinfo{author}{{Tammann}, G.A.},
  \bibinfo{author}{{Sandage}, A.}, \bibinfo{year}{1985}.
\newblock \bibinfo{title}{{Type I supernovae as standard candles}}, in:
  \bibinfo{editor}{{N.~Bartel}} (Ed.), \bibinfo{booktitle}{Supernovae as
  Distance Indicators}, pp. \bibinfo{pages}{151--165}.
\newblock \DOIprefix\doi{10.1007/3-540-15206-7_56}.
\bibitem[{{Cameron}(1957)}]{came57}
\bibinfo{author}{{Cameron}, A.G.W.}, \bibinfo{year}{1957}.
\newblock \bibinfo{title}{{Nuclear Reactions in Stars and Nucleogenesis}}.
\newblock \bibinfo{journal}{Pub. Astro. Soc. Pacific} \bibinfo{volume}{69},
  \bibinfo{pages}{201}.
\newblock \DOIprefix\doi{10.1086/127051}.
\bibitem[{{Cameron} and {Fowler}(1971)}]{came71}
\bibinfo{author}{{Cameron}, A.G.W.}, \bibinfo{author}{{Fowler}, W.A.},
  \bibinfo{year}{1971}.
\newblock \bibinfo{title}{{Lithium and the s-PROCESS in Red-Giant Stars}}.
\newblock \bibinfo{journal}{Astrophys. J.} \bibinfo{volume}{164},
  \bibinfo{pages}{111}.
\newblock \DOIprefix\doi{10.1086/150821}.
\bibitem[{{Canal} et~al.(1992){Canal}, {Isern} and {Labay}}]{cana92}
\bibinfo{author}{{Canal}, R.}, \bibinfo{author}{{Isern}, J.},
  \bibinfo{author}{{Labay}, J.}, \bibinfo{year}{1992}.
\newblock \bibinfo{title}{{The Quasi-static Evolution of ONeMg Cores: Explosive
  Ignition Densities and the Collapse/Explosion Alternative}}.
\newblock \bibinfo{journal}{Astrophys. J. Lett.} \bibinfo{volume}{398},
  \bibinfo{pages}{L49}.
\bibitem[{{Candia} et~al.(2003){Candia}, {Krisciunas} and {Suntzeff}}]{cand03}
\bibinfo{author}{{Candia}, P.}, \bibinfo{author}{{Krisciunas}, K.},
  \bibinfo{author}{{Suntzeff}, N.B.a.}, \bibinfo{year}{2003}.
\newblock \bibinfo{title}{{Optical and Infrared Photometry of the Unusual Type
  Ia Supernova 2000cx}}.
\newblock \bibinfo{journal}{Publ. Astron. Soc. Pac.} \bibinfo{volume}{115},
  \bibinfo{pages}{277--294}.
\newblock \DOIprefix\doi{10.1086/368229}.
\bibitem[{{Cappellaro}(2003)}]{capp03}
\bibinfo{author}{{Cappellaro}, E.}, \bibinfo{year}{2003}.
\newblock \bibinfo{title}{{Supernova Rates}}. volume \bibinfo{volume}{598}.
\newblock pp. \bibinfo{pages}{37--46}.
\newblock \DOIprefix\doi{10.1007/3-540-45863-8_4}.
\bibitem[{{Catal{\'a}n} et~al.(2008){Catal{\'a}n}, {Ribas}, {Isern} and
  {Garc{\'\i}a-Berro}}]{cata08}
\bibinfo{author}{{Catal{\'a}n}, S.}, \bibinfo{author}{{Ribas}, I.},
  \bibinfo{author}{{Isern}, J.}, \bibinfo{author}{{Garc{\'\i}a-Berro}, E.},
  \bibinfo{year}{2008}.
\newblock \bibinfo{title}{{WD0433+270: an old Hyades stream member or an
  Fe-core white dwarf?}}
\newblock \bibinfo{journal}{A\&A} \bibinfo{volume}{477},
  \bibinfo{pages}{901--906}.
\newblock \DOIprefix\doi{10.1051/0004-6361:20078230}.
\bibitem[{{Chafa} et~al.(2005){Chafa}, {Tatischeff}, {Aguer}, {Barhoumi},
  {Coc}, {Garrido}, {Hernanz}, {Jos{\'e}}, {Kiener}, {Lefebvre-Schuhl},
  {Ouichaoui}, {de S{\'e}r{\'e}ville} and {Thibaud}}]{chafa2005}
\bibinfo{author}{{Chafa}, A.}, \bibinfo{author}{{Tatischeff}, V.},
  \bibinfo{author}{{Aguer}, P.}, \bibinfo{author}{{Barhoumi}, S.},
  \bibinfo{author}{{Coc}, A.}, \bibinfo{author}{{Garrido}, F.},
  \bibinfo{author}{{Hernanz}, M.}, \bibinfo{author}{{Jos{\'e}}, J.},
  \bibinfo{author}{{Kiener}, J.}, \bibinfo{author}{{Lefebvre-Schuhl}, A.},
  \bibinfo{author}{{Ouichaoui}, S.}, \bibinfo{author}{{de S{\'e}r{\'e}ville},
  N.}, \bibinfo{author}{{Thibaud}, J.P.}, \bibinfo{year}{2005}.
\newblock \bibinfo{title}{{Hydrogen Burning of $^{17}$O in Classical Novae}}.
\newblock \bibinfo{journal}{Phys. Rev. Lett} \bibinfo{volume}{95},
  \bibinfo{pages}{031101}.
\newblock \DOIprefix\doi{10.1103/PhysRevLett.95.031101}.
\bibitem[{{Chamulak} et~al.(2008){Chamulak}, {Brown}, {Timmes} and
  {Dupczak}}]{cham08}
\bibinfo{author}{{Chamulak}, D.A.}, \bibinfo{author}{{Brown}, E.F.},
  \bibinfo{author}{{Timmes}, F.X.}, \bibinfo{author}{{Dupczak}, K.},
  \bibinfo{year}{2008}.
\newblock \bibinfo{title}{{The Reduction of the Electron Abundance during the
  Pre-explosion Simmering in White Dwarf Supernovae}}.
\newblock \bibinfo{journal}{Astrophys. J.} \bibinfo{volume}{677},
  \bibinfo{pages}{160--168}.
\newblock \DOIprefix\doi{10.1086/528944}.
\bibitem[{{Chatzopoulos} and {Wheeler}(2012)}]{chat12}
\bibinfo{author}{{Chatzopoulos}, E.}, \bibinfo{author}{{Wheeler}, J.C.},
  \bibinfo{year}{2012}.
\newblock \bibinfo{title}{{Effects of Rotation on the Minimum Mass of
  Primordial Progenitors of Pair-instability Supernovae}}.
\newblock \bibinfo{journal}{Astrophys. J.} \bibinfo{volume}{748},
  \bibinfo{pages}{42}.
\newblock \DOIprefix\doi{10.1088/0004-637X/748/1/42}.
\bibitem[{{Chomiuk} et~al.(2012){Chomiuk}, {Soderberg} and {Moe}}]{chom12}
\bibinfo{author}{{Chomiuk}, L.}, \bibinfo{author}{{Soderberg}, A.M.},
  \bibinfo{author}{{Moe}, M.e.a.}, \bibinfo{year}{2012}.
\newblock \bibinfo{title}{{EVLA Observations Constrain the Environment and
  Progenitor System of Type Ia Supernova 2011fe}}.
\newblock \bibinfo{journal}{Astrophys. J.} \bibinfo{volume}{750},
  \bibinfo{pages}{164}.
\newblock \DOIprefix\doi{10.1088/0004-637X/750/2/164}.
\bibitem[{{Churazov} et~al.(2014a){Churazov}, {Sunyaev}, {Grebenev} and
  et~al.}]{chur14a}
\bibinfo{author}{{Churazov}, E.}, \bibinfo{author}{{Sunyaev}, R.},
  \bibinfo{author}{{Grebenev}, S.}, \bibinfo{author}{et~al.},
  \bibinfo{year}{2014}a.
\newblock \bibinfo{title}{{Detection of the 847 keV gamma-ray line of
  radio-active Co56 from the Type Ia Supernova SN2014J in M82 with INTEGRAL.}}
\newblock \bibinfo{journal}{The Astronomer's Telegram} \bibinfo{volume}{5992},
  \bibinfo{pages}{1}.
\bibitem[{{Churazov} et~al.(2014b){Churazov}, {Sunyaev}, {Isern} and
  et~al.}]{chur14b}
\bibinfo{author}{{Churazov}, E.}, \bibinfo{author}{{Sunyaev}, R.},
  \bibinfo{author}{{Isern}, J.}, \bibinfo{author}{et~al.},
  \bibinfo{year}{2014}b.
\newblock \bibinfo{title}{{Cobalt-56 {\ensuremath{\gamma}}-ray emission lines
  from the type Ia supernova 2014J}}.
\newblock \bibinfo{journal}{Nature} \bibinfo{volume}{512},
  \bibinfo{pages}{406--408}.
\newblock \DOIprefix\doi{10.1038/nature13672}.
\bibitem[{{Churazov} et~al.(2015){Churazov}, {Sunyaev}, {Isern} and
  et~al.}]{chur15}
\bibinfo{author}{{Churazov}, E.}, \bibinfo{author}{{Sunyaev}, R.},
  \bibinfo{author}{{Isern}, J.}, \bibinfo{author}{et~al.},
  \bibinfo{year}{2015}.
\newblock \bibinfo{title}{{Gamma-rays from Type Ia Supernova SN2014J}}.
\newblock \bibinfo{journal}{Astrophys. J.} \bibinfo{volume}{812},
  \bibinfo{pages}{62}.
\newblock \DOIprefix\doi{10.1088/0004-637X/812/1/62}.
\bibitem[{{Ciardullo} et~al.(1990){Ciardullo}, {Ford}, {Williams}, {Tamblyn}
  and {Jacoby}}]{ciar90}
\bibinfo{author}{{Ciardullo}, R.}, \bibinfo{author}{{Ford}, H.C.},
  \bibinfo{author}{{Williams}, R.E.}, \bibinfo{author}{{Tamblyn}, P.},
  \bibinfo{author}{{Jacoby}, G.H.}, \bibinfo{year}{1990}.
\newblock \bibinfo{title}{{The Nova Rate in the Elliptical Component of NGC
  5128}}.
\newblock \bibinfo{journal}{Astron. J.} \bibinfo{volume}{99},
  \bibinfo{pages}{1079}.
\newblock \DOIprefix\doi{10.1086/115397}.
\bibitem[{{Clayton} and {Hoyle}(1974)}]{clay74}
\bibinfo{author}{{Clayton}, D.D.}, \bibinfo{author}{{Hoyle}, F.},
  \bibinfo{year}{1974}.
\newblock \bibinfo{title}{{Gamma-Ray Lines from Novae}}.
\newblock \bibinfo{journal}{Astrophys. J. Lett.} \bibinfo{volume}{187},
  \bibinfo{pages}{L101}.
\newblock \DOIprefix\doi{10.1086/181406}.
\bibitem[{{Clayton} et~al.(1992){Clayton}, {Leising}, {The}, {Johnson} and
  {Kurfess}}]{clay92}
\bibinfo{author}{{Clayton}, D.D.}, \bibinfo{author}{{Leising}, M.D.},
  \bibinfo{author}{{The}, L.S.}, \bibinfo{author}{{Johnson}, W.N.},
  \bibinfo{author}{{Kurfess}, J.D.}, \bibinfo{year}{1992}.
\newblock \bibinfo{title}{{The 57 CO Abundance in SN 1987A}}.
\newblock \bibinfo{journal}{Astrophys. J. Lett.} \bibinfo{volume}{399},
  \bibinfo{pages}{L141}.
\newblock \DOIprefix\doi{10.1086/186627}.
\bibitem[{{Coc} et~al.(2000){Coc}, {Hernanz}, {Jos{\'e}} and
  {Thibaud}}]{coc2000}
\bibinfo{author}{{Coc}, A.}, \bibinfo{author}{{Hernanz}, M.},
  \bibinfo{author}{{Jos{\'e}}, J.}, \bibinfo{author}{{Thibaud}, J.P.},
  \bibinfo{year}{2000}.
\newblock \bibinfo{title}{{Influence of new reaction rates on $^{18}$F
  production in novae}}.
\newblock \bibinfo{journal}{A\&A} \bibinfo{volume}{357},
  \bibinfo{pages}{561--571}.
\bibitem[{{Colgate} et~al.(1980){Colgate}, {Petschek} and {Kriese}}]{colg80}
\bibinfo{author}{{Colgate}, S.A.}, \bibinfo{author}{{Petschek}, A.G.},
  \bibinfo{author}{{Kriese}, J.T.}, \bibinfo{year}{1980}.
\newblock \bibinfo{title}{{The luminosity of type I supernovae}}.
\newblock \bibinfo{journal}{Astrophys. J. Lett.} \bibinfo{volume}{237},
  \bibinfo{pages}{L81--L85}.
\newblock \DOIprefix\doi{10.1086/183239}.
\bibitem[{{Couch} and {Loumos}(1974)}]{couc74}
\bibinfo{author}{{Couch}, R.G.}, \bibinfo{author}{{Loumos}, G.L.},
  \bibinfo{year}{1974}.
\newblock \bibinfo{title}{{The Urca process in dense stellar interiors.}}
\newblock \bibinfo{journal}{Astrophys. J.} \bibinfo{volume}{194},
  \bibinfo{pages}{385--392}.
\newblock \DOIprefix\doi{10.1086/153255}.
\bibitem[{{Couch}(2017)}]{couc17}
\bibinfo{author}{{Couch}, S.M.}, \bibinfo{year}{2017}.
\newblock \bibinfo{title}{{The mechanism(s) of core-collapse supernovae}}.
\newblock \bibinfo{journal}{Philosophical Transactions of the Royal Society of
  London Series A} \bibinfo{volume}{375}, \bibinfo{pages}{20160271}.
\newblock \DOIprefix\doi{10.1098/rsta.2016.0271}.
\bibitem[{{de Angelis} et~al.(2018){de Angelis}, {Tatischeff}, {Grenier},
  {McEnery}, {Mallamaci}, {Tavani}, {Oberlack}, {Hanlon}, {Walter}, {Argan} and
  et~al.}]{dean18}
\bibinfo{author}{{de Angelis}, A.}, \bibinfo{author}{{Tatischeff}, V.},
  \bibinfo{author}{{Grenier}, I.A.}, \bibinfo{author}{{McEnery}, J.},
  \bibinfo{author}{{Mallamaci}, M.}, \bibinfo{author}{{Tavani}, M.},
  \bibinfo{author}{{Oberlack}, U.}, \bibinfo{author}{{Hanlon}, L.},
  \bibinfo{author}{{Walter}, R.}, \bibinfo{author}{{Argan}, A.},
  \bibinfo{author}{et~al.}, \bibinfo{year}{2018}.
\newblock \bibinfo{title}{{Science with e-ASTROGAM. A space mission for MeV-GeV
  gamma-ray astrophysics}}.
\newblock \bibinfo{journal}{Journal of High Energy Astrophysics}
  \bibinfo{volume}{19}, \bibinfo{pages}{1--106}.
\bibitem[{{De Angelis} et~al.(2017){De Angelis}, {Tatischeff} and
  {Tavani}}]{dean17}
\bibinfo{author}{{De Angelis}, A.}, \bibinfo{author}{{Tatischeff}, V.},
  \bibinfo{author}{{Tavani}, M.e.a.}, \bibinfo{year}{2017}.
\newblock \bibinfo{title}{{The e-ASTROGAM mission. Exploring the extreme
  Universe with gamma rays in the MeV - GeV range}}.
\newblock \bibinfo{journal}{Experimental Astronomy} \bibinfo{volume}{44},
  \bibinfo{pages}{25--82}.
\newblock \DOIprefix\doi{10.1007/s10686-017-9533-6}.
\bibitem[{{della Valle} and {Livio}(1994)}]{dell94}
\bibinfo{author}{{della Valle}, M.}, \bibinfo{author}{{Livio}, M.},
  \bibinfo{year}{1994}.
\newblock \bibinfo{title}{{On the nova rate in the Galaxy}}.
\newblock \bibinfo{journal}{A\&A} \bibinfo{volume}{286},
  \bibinfo{pages}{786--788}.
\bibitem[{{della Valle} and {Livio}(1995)}]{dell95}
\bibinfo{author}{{della Valle}, M.}, \bibinfo{author}{{Livio}, M.},
  \bibinfo{year}{1995}.
\newblock \bibinfo{title}{{The Calibration of Novae as Distance Indicators}}.
\newblock \bibinfo{journal}{Astrophys. J.} \bibinfo{volume}{452},
  \bibinfo{pages}{704}.
\newblock \DOIprefix\doi{10.1086/176342}.
\bibitem[{{Denissenkov} et~al.(2013){Denissenkov}, {Herwig}, {Truran} and
  {Paxton}}]{deni13}
\bibinfo{author}{{Denissenkov}, P.A.}, \bibinfo{author}{{Herwig}, F.},
  \bibinfo{author}{{Truran}, J.W.}, \bibinfo{author}{{Paxton}, B.},
  \bibinfo{year}{2013}.
\newblock \bibinfo{title}{{The C-flame Quenching by Convective Boundary Mixing
  in Super-AGB Stars and the Formation of Hybrid C/O/Ne White Dwarfs and SN
  Progenitors}}.
\newblock \bibinfo{journal}{Astrophys. J.} \bibinfo{volume}{772},
  \bibinfo{pages}{37}.
\newblock \DOIprefix\doi{10.1088/0004-637X/772/1/37}.
\bibitem[{{Dessart} et~al.(2014){Dessart}, {Blondin}, {Hillier} and
  {Khokhlov}}]{dess14}
\bibinfo{author}{{Dessart}, L.}, \bibinfo{author}{{Blondin}, S.},
  \bibinfo{author}{{Hillier}, D.J.}, \bibinfo{author}{{Khokhlov}, A.},
  \bibinfo{year}{2014}.
\newblock \bibinfo{title}{{Constraints on the explosion mechanism and
  progenitors of Type Ia supernovae}}.
\newblock \bibinfo{journal}{Mon. Not. R. Astron. Soc.} \bibinfo{volume}{441},
  \bibinfo{pages}{532--550}.
\newblock \DOIprefix\doi{10.1093/mnras/stu598},
  \href{http://arxiv.org/abs/1310.7747}{\tt arXiv:1310.7747}.
\bibitem[{{Dessart} et~al.(2016){Dessart}, {Hillier}, {Woosley}, {Livne},
  {Waldman}, {Yoon} and {Langer}}]{dess16}
\bibinfo{author}{{Dessart}, L.}, \bibinfo{author}{{Hillier}, D.J.},
  \bibinfo{author}{{Woosley}, S.}, \bibinfo{author}{{Livne}, E.},
  \bibinfo{author}{{Waldman}, R.}, \bibinfo{author}{{Yoon}, S.C.},
  \bibinfo{author}{{Langer}, N.}, \bibinfo{year}{2016}.
\newblock \bibinfo{title}{{Inferring supernova IIb/Ib/Ic ejecta properties from
  light curves and spectra: correlations from radiative-transfer models}}.
\newblock \bibinfo{journal}{Mon. Not. R. Astron. Soc.} \bibinfo{volume}{458},
  \bibinfo{pages}{1618--1635}.
\bibitem[{{Diehl} et~al.(2018){Diehl}, {Hartmann} and {Prantzos}}]{dieh18}
\bibinfo{author}{{Diehl}, R.}, \bibinfo{author}{{Hartmann}, D.H.},
  \bibinfo{author}{{Prantzos}, N.}, \bibinfo{year}{2018}.
\newblock \bibinfo{title}{{Astrophysics with Radioactive Isotopes}}. volume
  \bibinfo{volume}{453}.
\newblock \DOIprefix\doi{10.1007/978-3-319-91929-4_11}.
\bibitem[{{Diehl} et~al.(2014){Diehl}, {Siegert} and {Hillebrandt}}]{dieh14}
\bibinfo{author}{{Diehl}, R.}, \bibinfo{author}{{Siegert}, T.},
  \bibinfo{author}{{Hillebrandt}, W.e.a.}, \bibinfo{year}{2014}.
\newblock \bibinfo{title}{{Early $^{56}$Ni decay gamma rays from SN2014J
  suggest an unusual explosion}}.
\newblock \bibinfo{journal}{Science} \bibinfo{volume}{345},
  \bibinfo{pages}{1162--1165}.
\newblock \DOIprefix\doi{10.1126/science.1254738}.
\bibitem[{{Diehl} et~al.(2015){Diehl}, {Siegert} and {Hillebrandt}}]{dieh15}
\bibinfo{author}{{Diehl}, R.}, \bibinfo{author}{{Siegert}, T.},
  \bibinfo{author}{{Hillebrandt}, W.e.a.}, \bibinfo{year}{2015}.
\newblock \bibinfo{title}{{SN2014J gamma rays from the $^{56}$Ni decay chain}}.
\newblock \bibinfo{journal}{A\&A} \bibinfo{volume}{574}, \bibinfo{pages}{A72}.
\newblock \DOIprefix\doi{10.1051/0004-6361/201424991}.
\bibitem[{{Dilday} et~al.(2012){Dilday}, {Howell} and {Cenko}}]{dild12}
\bibinfo{author}{{Dilday}, B.}, \bibinfo{author}{{Howell}, D.A.},
  \bibinfo{author}{{Cenko}, S.B.e.a.}, \bibinfo{year}{2012}.
\newblock \bibinfo{title}{{PTF 11kx: A Type Ia Supernova with a Symbiotic Nova
  Progenitor}}.
\newblock \bibinfo{journal}{Science} \bibinfo{volume}{337},
  \bibinfo{pages}{942--}.
\newblock \DOIprefix\doi{10.1126/science.1219164}.
\bibitem[{{Doherty} et~al.(2017){Doherty}, {Gil-Pons}, {Siess} and
  {Lattanzio}}]{dohe17}
\bibinfo{author}{{Doherty}, C.L.}, \bibinfo{author}{{Gil-Pons}, P.},
  \bibinfo{author}{{Siess}, L.}, \bibinfo{author}{{Lattanzio}, J.C.},
  \bibinfo{year}{2017}.
\newblock \bibinfo{title}{{Super-AGB Stars and their Role as Electron Capture
  Supernova Progenitors}}.
\newblock \bibinfo{journal}{Publ. Astron. Soc. Aust.} \bibinfo{volume}{34},
  \bibinfo{pages}{e056}.
\bibitem[{{Doherty} et~al.(2010){Doherty}, {Siess}, {Lattanzio} and
  {Gil-Pons}}]{dohe10}
\bibinfo{author}{{Doherty}, C.L.}, \bibinfo{author}{{Siess}, L.},
  \bibinfo{author}{{Lattanzio}, J.C.}, \bibinfo{author}{{Gil-Pons}, P.},
  \bibinfo{year}{2010}.
\newblock \bibinfo{title}{{Super asymptotic giant branch stars. I - Evolution
  code comparison}}.
\newblock \bibinfo{journal}{Mon. Not. R. Astron. Soc.} \bibinfo{volume}{401},
  \bibinfo{pages}{1453--1464}.
\newblock \DOIprefix\doi{10.1111/j.1365-2966.2009.15772.x}.
\bibitem[{{Dominguez} et~al.(1993){Dominguez}, {Tornambe} and {Isern}}]{domi93}
\bibinfo{author}{{Dominguez}, I.}, \bibinfo{author}{{Tornambe}, A.},
  \bibinfo{author}{{Isern}, J.}, \bibinfo{year}{1993}.
\newblock \bibinfo{title}{{On the Formation of O-Ne White Dwarfs in Metal-rich
  Close Binary Systems}}.
\newblock \bibinfo{journal}{Astrophys. J.} \bibinfo{volume}{419},
  \bibinfo{pages}{268}.
\newblock \DOIprefix\doi{10.1086/173480}.
\bibitem[{{Dotani} et~al.(1987){Dotani}, {Hayashida}, {Inoue}, {Itoh},
  {Koyama}, {Makino}, {Mitsuda}, {Murakami}, {Oda}, {Ogawara}, {Takano},
  {Tanaka}, {Yoshida}, {Makishima}, {Ohashi}, {Kawai}, {Matsuoka}, {Hoshi},
  {Hayakawa}, {Kii}, {Kunieda}, {Nagase}, {Tawara}, {Hatsukade}, {Kitamoto},
  {Miyamoto}, {Tsunemi}, {Yamashita}, {Nakagawa}, {Yamauchi}, {Turner},
  {Pounds}, {Thomas}, {Stewart}, {Cruise}, {Patchett} and {Reading}}]{dota87}
\bibinfo{author}{{Dotani}, T.}, \bibinfo{author}{{Hayashida}, K.},
  \bibinfo{author}{{Inoue}, H.}, \bibinfo{author}{{Itoh}, M.},
  \bibinfo{author}{{Koyama}, K.}, \bibinfo{author}{{Makino}, F.},
  \bibinfo{author}{{Mitsuda}, K.}, \bibinfo{author}{{Murakami}, T.},
  \bibinfo{author}{{Oda}, M.}, \bibinfo{author}{{Ogawara}, Y.},
  \bibinfo{author}{{Takano}, S.}, \bibinfo{author}{{Tanaka}, Y.},
  \bibinfo{author}{{Yoshida}, A.}, \bibinfo{author}{{Makishima}, K.},
  \bibinfo{author}{{Ohashi}, T.}, \bibinfo{author}{{Kawai}, N.},
  \bibinfo{author}{{Matsuoka}, M.}, \bibinfo{author}{{Hoshi}, R.},
  \bibinfo{author}{{Hayakawa}, S.}, \bibinfo{author}{{Kii}, T.},
  \bibinfo{author}{{Kunieda}, H.}, \bibinfo{author}{{Nagase}, F.},
  \bibinfo{author}{{Tawara}, Y.}, \bibinfo{author}{{Hatsukade}, I.},
  \bibinfo{author}{{Kitamoto}, S.}, \bibinfo{author}{{Miyamoto}, S.},
  \bibinfo{author}{{Tsunemi}, H.}, \bibinfo{author}{{Yamashita}, K.},
  \bibinfo{author}{{Nakagawa}, M.}, \bibinfo{author}{{Yamauchi}, M.},
  \bibinfo{author}{{Turner}, M.J.L.}, \bibinfo{author}{{Pounds}, K.A.},
  \bibinfo{author}{{Thomas}, H.D.}, \bibinfo{author}{{Stewart}, G.C.},
  \bibinfo{author}{{Cruise}, A.M.}, \bibinfo{author}{{Patchett}, B.E.},
  \bibinfo{author}{{Reading}, D.H.}, \bibinfo{year}{1987}.
\newblock \bibinfo{title}{{Discovery of an unusual hard X-ray source in the
  region of supernova 1987A}}.
\newblock \bibinfo{journal}{Nature} \bibinfo{volume}{330},
  \bibinfo{pages}{230--231}.
\newblock \DOIprefix\doi{10.1038/330230a0}.
\bibitem[{{Drout} et~al.(2014){Drout}, {Chornock} and {Soderberg}}]{drou14}
\bibinfo{author}{{Drout}, M.R.}, \bibinfo{author}{{Chornock}, R.},
  \bibinfo{author}{{Soderberg}, A.M.e.a.}, \bibinfo{year}{2014}.
\newblock \bibinfo{title}{{Rapidly Evolving and Luminous Transients from
  Pan-STARRS1}}.
\newblock \bibinfo{journal}{Astrophys. J.} \bibinfo{volume}{794},
  \bibinfo{pages}{23}.
\bibitem[{{Elias} et~al.(1985){Elias}, {Matthews}, {Neugebauer} and
  {Persson}}]{elia85}
\bibinfo{author}{{Elias}, J.H.}, \bibinfo{author}{{Matthews}, K.},
  \bibinfo{author}{{Neugebauer}, G.}, \bibinfo{author}{{Persson}, S.E.},
  \bibinfo{year}{1985}.
\newblock \bibinfo{title}{{Type I supernovae in the infrared and their use as
  distance indicators}}.
\newblock \bibinfo{journal}{Astrophys. J.} \bibinfo{volume}{296},
  \bibinfo{pages}{379--389}.
\newblock \DOIprefix\doi{10.1086/163456}.
\bibitem[{{Fang} et~al.(2018){Fang}, {Thompson} and {Hirata}}]{fang18}
\bibinfo{author}{{Fang}, X.}, \bibinfo{author}{{Thompson}, T.A.},
  \bibinfo{author}{{Hirata}, C.M.}, \bibinfo{year}{2018}.
\newblock \bibinfo{title}{{Dynamics of quadruple systems composed of two
  binaries: stars, white dwarfs, and implications for Ia supernovae}}.
\newblock \bibinfo{journal}{Mon. Not. R. Astron. Soc.} \bibinfo{volume}{476},
  \bibinfo{pages}{4234--4262}.
\newblock \DOIprefix\doi{10.1093/mnras/sty472}.
\bibitem[{{Fesen} et~al.(2006){Fesen}, {Hammell} and {Morse}}]{fese06}
\bibinfo{author}{{Fesen}, R.A.}, \bibinfo{author}{{Hammell}, M.C.},
  \bibinfo{author}{{Morse}, J.e.a.}, \bibinfo{year}{2006}.
\newblock \bibinfo{title}{{The Expansion Asymmetry and Age of the Cassiopeia A
  Supernova Remnant}}.
\newblock \bibinfo{journal}{Astrophys. J.} \bibinfo{volume}{645},
  \bibinfo{pages}{283--292}.
\newblock \DOIprefix\doi{10.1086/504254}.
\bibitem[{{Filippenko}(1997)}]{fili97}
\bibinfo{author}{{Filippenko}, A.V.}, \bibinfo{year}{1997}.
\newblock \bibinfo{title}{{Optical Spectra of Supernovae}}.
\newblock \bibinfo{journal}{Annu. Rev. Astron. Astrophys.}
  \bibinfo{volume}{35}, \bibinfo{pages}{309--355}.
\newblock \DOIprefix\doi{10.1146/annurev.astro.35.1.309}.
\bibitem[{{Fink} et~al.(2010){Fink}, {R{\"o}pke}, {Hillebrandt}, {Seitenzahl},
  {Sim} and {Kromer}}]{fink10}
\bibinfo{author}{{Fink}, M.}, \bibinfo{author}{{R{\"o}pke}, F.K.},
  \bibinfo{author}{{Hillebrandt}, W.}, \bibinfo{author}{{Seitenzahl}, I.R.},
  \bibinfo{author}{{Sim}, S.A.}, \bibinfo{author}{{Kromer}, M.},
  \bibinfo{year}{2010}.
\newblock \bibinfo{title}{{Double-detonation sub-Chandrasekhar supernovae: can
  minimum helium shell masses detonate the core?}}
\newblock \bibinfo{journal}{A\&A} \bibinfo{volume}{514}, \bibinfo{pages}{53}.
\newblock \DOIprefix\doi{10.1051/0004-6361/200913892}.
\bibitem[{{Foley} et~al.(2013){Foley}, {Challis} and {Chornock}}]{fole13}
\bibinfo{author}{{Foley}, R.J.}, \bibinfo{author}{{Challis}, P.J.},
  \bibinfo{author}{{Chornock}, R.e.a.}, \bibinfo{year}{2013}.
\newblock \bibinfo{title}{{Type Iax Supernovae: A New Class of Stellar
  Explosion}}.
\newblock \bibinfo{journal}{Astrophys. J.} \bibinfo{volume}{767},
  \bibinfo{pages}{57}.
\newblock \DOIprefix\doi{10.1088/0004-637X/767/1/57}.
\bibitem[{{Fossey} et~al.(2014){Fossey}, {Cooke}, {Pollack}, {Wilde} and
  {Wright}}]{foss14}
\bibinfo{author}{{Fossey}, S.J.}, \bibinfo{author}{{Cooke}, B.},
  \bibinfo{author}{{Pollack}, G.}, \bibinfo{author}{{Wilde}, M.},
  \bibinfo{author}{{Wright}, T.}, \bibinfo{year}{2014}.
\newblock \bibinfo{title}{{Supernova 2014J in M82 = Psn J09554214+6940260}}.
\newblock \bibinfo{journal}{Central Bureau Electronic Telegrams}
  \bibinfo{volume}{3792}, \bibinfo{pages}{1}.
\bibitem[{{Fowler} and {Hoyle}(1964)}]{fowl64}
\bibinfo{author}{{Fowler}, W.A.}, \bibinfo{author}{{Hoyle}, F.},
  \bibinfo{year}{1964}.
\newblock \bibinfo{title}{{Neutrino Processes and Pair Formation in Massive
  Stars and Supernovae.}}
\newblock \bibinfo{journal}{Astrophys. J. Suppl. S} \bibinfo{volume}{9},
  \bibinfo{pages}{201}.
\newblock \DOIprefix\doi{10.1086/190103}.
\bibitem[{{Fraley}(1968)}]{fral68}
\bibinfo{author}{{Fraley}, G.S.}, \bibinfo{year}{1968}.
\newblock \bibinfo{title}{{Supernovae Explosions Induced by Pair-Production
  Instability}}.
\newblock \bibinfo{journal}{Astrophys. Space Sci.} \bibinfo{volume}{2},
  \bibinfo{pages}{96--114}.
\newblock \DOIprefix\doi{10.1007/BF00651498}.
\bibitem[{{Fuller}(1982)}]{full82}
\bibinfo{author}{{Fuller}, G.M.}, \bibinfo{year}{1982}.
\newblock \bibinfo{title}{{Neutron shell blocking of electron capture during
  gravitational collapse}}.
\newblock \bibinfo{journal}{Astrophys. J.} \bibinfo{volume}{252},
  \bibinfo{pages}{741--764}.
\newblock \DOIprefix\doi{10.1086/159598}.
\bibitem[{{Fuller} et~al.(1980){Fuller}, {Fowler} and {Newman}}]{full80}
\bibinfo{author}{{Fuller}, G.M.}, \bibinfo{author}{{Fowler}, W.A.},
  \bibinfo{author}{{Newman}, M.J.}, \bibinfo{year}{1980}.
\newblock \bibinfo{title}{{Stellar weak-interaction rates for sd-shell nuclei.
  I - Nuclear matrix element systematics with application to Al-26 and selected
  nuclei of importance to the supernova problem}}.
\newblock \bibinfo{journal}{Astrophys. J. Suppl. S} \bibinfo{volume}{42},
  \bibinfo{pages}{447--473}.
\newblock \DOIprefix\doi{10.1086/190657}.
\bibitem[{{Fuller} et~al.(1982a){Fuller}, {Fowler} and {Newman}}]{full82a}
\bibinfo{author}{{Fuller}, G.M.}, \bibinfo{author}{{Fowler}, W.A.},
  \bibinfo{author}{{Newman}, M.J.}, \bibinfo{year}{1982}a.
\newblock \bibinfo{title}{{Stellar weak interaction rates for intermediate-mass
  nuclei. II - A = 21 to A = 60}}.
\newblock \bibinfo{journal}{Astrophys. J.} \bibinfo{volume}{252},
  \bibinfo{pages}{715--740}.
\newblock \DOIprefix\doi{10.1086/159597}.
\bibitem[{{Fuller} et~al.(1982b){Fuller}, {Fowler} and {Newman}}]{full82b}
\bibinfo{author}{{Fuller}, G.M.}, \bibinfo{author}{{Fowler}, W.A.},
  \bibinfo{author}{{Newman}, M.J.}, \bibinfo{year}{1982}b.
\newblock \bibinfo{title}{{Stellar weak interaction rates for intermediate mass
  nuclei. III - Rate tables for the free nucleons and nuclei with A = 21 to A =
  60}}.
\newblock \bibinfo{journal}{Astrophys. J. Suppl. S} \bibinfo{volume}{48},
  \bibinfo{pages}{279--319}.
\newblock \DOIprefix\doi{10.1086/190779}.
\bibitem[{{Fuller} et~al.(1985){Fuller}, {Fowler} and {Newman}}]{full85}
\bibinfo{author}{{Fuller}, G.M.}, \bibinfo{author}{{Fowler}, W.A.},
  \bibinfo{author}{{Newman}, M.J.}, \bibinfo{year}{1985}.
\newblock \bibinfo{title}{{Stellar weak interaction rates for intermediate-mass
  nuclei. IV - Interpolation procedures for rapidly varying lepton capture
  rates using effective log (ft)-values}}.
\newblock \bibinfo{journal}{Astrophys. J.} \bibinfo{volume}{293},
  \bibinfo{pages}{1--16}.
\newblock \DOIprefix\doi{10.1086/163208}.
\bibitem[{{Gal-Yam} et~al.(2009){Gal-Yam}, {Mazzali}, {Ofek}, {Nugent},
  {Kulkarni}, {Kasliwal}, {Quimby}, {Filippenko}, {Cenko} and
  {Chornock}}]{galy09}
\bibinfo{author}{{Gal-Yam}, A.}, \bibinfo{author}{{Mazzali}, P.},
  \bibinfo{author}{{Ofek}, E.O.}, \bibinfo{author}{{Nugent}, P.E.},
  \bibinfo{author}{{Kulkarni}, S.R.}, \bibinfo{author}{{Kasliwal}, M.M.},
  \bibinfo{author}{{Quimby}, R.M.}, \bibinfo{author}{{Filippenko}, A.V.},
  \bibinfo{author}{{Cenko}, S.B.}, \bibinfo{author}{{Chornock}, R.},
  \bibinfo{year}{2009}.
\newblock \bibinfo{title}{{Supernova 2007bi as a pair-instability explosion}}.
\newblock \bibinfo{journal}{Nature} \bibinfo{volume}{462},
  \bibinfo{pages}{624--627}.
\bibitem[{{Garc{\'\i}a-Berro} et~al.(1997){Garc{\'\i}a-Berro}, {Ritossa} and
  {Iben}}]{garcb97}
\bibinfo{author}{{Garc{\'\i}a-Berro}, E.}, \bibinfo{author}{{Ritossa}, C.},
  \bibinfo{author}{{Iben}, Icko, J.}, \bibinfo{year}{1997}.
\newblock \bibinfo{title}{{On the Evolution of Stars that Form
  Electron-Degenerate Cores Processed by Carbon Burning. III. The Inward
  Propagation of a Carbon-Burning Flame and Other Properties of a 9 M$_{\odot}$
  Model Star}}.
\newblock \bibinfo{journal}{Astrophys. J.} \bibinfo{volume}{485},
  \bibinfo{pages}{765--784}.
\newblock \DOIprefix\doi{10.1086/304444}.
\bibitem[{{Gehrels} et~al.(1993){Gehrels}, {Chipman} and {Kniffen}}]{gehr93}
\bibinfo{author}{{Gehrels}, N.}, \bibinfo{author}{{Chipman}, E.},
  \bibinfo{author}{{Kniffen}, D.A.}, \bibinfo{year}{1993}.
\newblock \bibinfo{title}{{The Compton Gamma Ray Observatory.}}
\newblock \bibinfo{journal}{A\&AS} \bibinfo{volume}{97},
  \bibinfo{pages}{5--12}.
\bibitem[{{Gehrels} et~al.(1987){Gehrels}, {Leventhal} and
  {MacCallum}}]{gehr87}
\bibinfo{author}{{Gehrels}, N.}, \bibinfo{author}{{Leventhal}, M.},
  \bibinfo{author}{{MacCallum}, C.J.}, \bibinfo{year}{1987}.
\newblock \bibinfo{title}{{Prospects for gamma-ray line observations of
  individual supernovae}}.
\newblock \bibinfo{journal}{Astrophys. J.} \bibinfo{volume}{322},
  \bibinfo{pages}{215--233}.
\newblock \DOIprefix\doi{10.1086/165717}.
\bibitem[{{Georgii} et~al.(2002){Georgii}, {Pl{\"u}schke} and {Diehl}}]{geor02}
\bibinfo{author}{{Georgii}, R.}, \bibinfo{author}{{Pl{\"u}schke}, S.},
  \bibinfo{author}{{Diehl}, R.e.a.}, \bibinfo{year}{2002}.
\newblock \bibinfo{title}{{COMPTEL upper limits for the $^{56}$Co gamma -ray
  emission from SN1998bu}}.
\newblock \bibinfo{journal}{A\&A} \bibinfo{volume}{394},
  \bibinfo{pages}{517--523}.
\newblock \DOIprefix\doi{10.1051/0004-6361:20021133}.
\bibitem[{{Gil-Pons} and {Garc{\'\i}a-Berro}(2001)}]{gilp01}
\bibinfo{author}{{Gil-Pons}, P.}, \bibinfo{author}{{Garc{\'\i}a-Berro}, E.},
  \bibinfo{year}{2001}.
\newblock \bibinfo{title}{{On the formation of oxygen-neon white dwarfs in
  close binary systems}}.
\newblock \bibinfo{journal}{A\&A} \bibinfo{volume}{375},
  \bibinfo{pages}{87--99}.
\newblock \DOIprefix\doi{10.1051/0004-6361:20010828}.
\bibitem[{{Girardi} et~al.(2005){Girardi}, {Groenewegen}, {Hatziminaoglou} and
  {da Costa}}]{gira05}
\bibinfo{author}{{Girardi}, L.}, \bibinfo{author}{{Groenewegen}, M.A.T.},
  \bibinfo{author}{{Hatziminaoglou}, E.}, \bibinfo{author}{{da Costa}, L.},
  \bibinfo{year}{2005}.
\newblock \bibinfo{title}{{Star counts in the Galaxy. Simulating from very deep
  to very shallow photometric surveys with the TRILEGAL code}}.
\newblock \bibinfo{journal}{A\&A} \bibinfo{volume}{436},
  \bibinfo{pages}{895--915}.
\newblock \DOIprefix\doi{10.1051/0004-6361:20042352}.
\bibitem[{{Gomez-Gomar} et~al.(1998){Gomez-Gomar}, {Hernanz}, {Jose} and
  {Isern}}]{gome98a}
\bibinfo{author}{{Gomez-Gomar}, J.}, \bibinfo{author}{{Hernanz}, M.},
  \bibinfo{author}{{Jose}, J.}, \bibinfo{author}{{Isern}, J.},
  \bibinfo{year}{1998}.
\newblock \bibinfo{title}{{Gamma-ray emission from individual classical
  novae}}.
\newblock \bibinfo{journal}{Mon. Not. R. Astron. Soc.} \bibinfo{volume}{296},
  \bibinfo{pages}{913--920}.
\newblock \DOIprefix\doi{10.1046/j.1365-8711.1998.01421.x}.
\bibitem[{{G{\'o}mez-Gomar} et~al.(1998){G{\'o}mez-Gomar}, {Isern} and
  {Jean}}]{gome98}
\bibinfo{author}{{G{\'o}mez-Gomar}, J.}, \bibinfo{author}{{Isern}, J.},
  \bibinfo{author}{{Jean}, P.}, \bibinfo{year}{1998}.
\newblock \bibinfo{title}{{Prospects for Type IA supernova explosion mechanism
  identification with gamma rays}}.
\newblock \bibinfo{journal}{Mon. Not. R. Astron. Soc.} \bibinfo{volume}{295},
  \bibinfo{pages}{913--920}.
\newblock \DOIprefix\doi{10.1046/j.1365-8711.1998.29511115.x}.
\bibitem[{{Goobar} et~al.(2014){Goobar}, {Johansson}, {Amanullah} and
  et~al.}]{goob14}
\bibinfo{author}{{Goobar}, A.}, \bibinfo{author}{{Johansson}, J.},
  \bibinfo{author}{{Amanullah}, R.}, \bibinfo{author}{et~al.},
  \bibinfo{year}{2014}.
\newblock \bibinfo{title}{{The Rise of SN 2014J in the Nearby Galaxy M82}}.
\newblock \bibinfo{journal}{Astrophys. J. Lett.} \bibinfo{volume}{784},
  \bibinfo{pages}{L12}.
\newblock \DOIprefix\doi{10.1088/2041-8205/784/1/L12}.
\bibitem[{{Graur} et~al.(2017a){Graur}, {Bianco}, {Huang}, {Modjaz},
  {Shivvers}, {Filippenko}, {Li} and {Eldridge}}]{grau17a}
\bibinfo{author}{{Graur}, O.}, \bibinfo{author}{{Bianco}, F.B.},
  \bibinfo{author}{{Huang}, S.}, \bibinfo{author}{{Modjaz}, M.},
  \bibinfo{author}{{Shivvers}, I.}, \bibinfo{author}{{Filippenko}, A.V.},
  \bibinfo{author}{{Li}, W.}, \bibinfo{author}{{Eldridge}, J.J.},
  \bibinfo{year}{2017}a.
\newblock \bibinfo{title}{{LOSS Revisited. I. Unraveling Correlations Between
  Supernova Rates and Galaxy Properties, as Measured in a Reanalysis of the
  Lick Observatory Supernova Search}}.
\newblock \bibinfo{journal}{Astrophys. J.} \bibinfo{volume}{837},
  \bibinfo{pages}{120}.
\newblock \DOIprefix\doi{10.3847/1538-4357/aa5eb8}.
\bibitem[{{Graur} et~al.(2015){Graur}, {Bianco} and {Modjaz}}]{grau15}
\bibinfo{author}{{Graur}, O.}, \bibinfo{author}{{Bianco}, F.B.},
  \bibinfo{author}{{Modjaz}, M.}, \bibinfo{year}{2015}.
\newblock \bibinfo{title}{{A unified explanation for the supernova rate-galaxy
  mass dependence based on supernovae detected in Sloan galaxy spectra}}.
\newblock \bibinfo{journal}{Mon. Not. R. Astron. Soc.} \bibinfo{volume}{450},
  \bibinfo{pages}{905--925}.
\newblock \DOIprefix\doi{10.1093/mnras/stv713}.
\bibitem[{{Graur} et~al.(2017b){Graur}, {Bianco}, {Modjaz}, {Shivvers},
  {Filippenko}, {Li} and {Smith}}]{grau17b}
\bibinfo{author}{{Graur}, O.}, \bibinfo{author}{{Bianco}, F.B.},
  \bibinfo{author}{{Modjaz}, M.}, \bibinfo{author}{{Shivvers}, I.},
  \bibinfo{author}{{Filippenko}, A.V.}, \bibinfo{author}{{Li}, W.},
  \bibinfo{author}{{Smith}, N.}, \bibinfo{year}{2017}b.
\newblock \bibinfo{title}{{LOSS Revisited. II. The Relative Rates of Different
  Types of Supernovae Vary between Low- and High-mass Galaxies}}.
\newblock \bibinfo{journal}{Astrophys. J.} \bibinfo{volume}{837},
  \bibinfo{pages}{121}.
\newblock \DOIprefix\doi{10.3847/1538-4357/aa5eb7}.
\bibitem[{{Grebenev} et~al.(2012){Grebenev}, {Lutovinov}, {Tsygankov} and
  {Winkler}}]{greb12}
\bibinfo{author}{{Grebenev}, S.A.}, \bibinfo{author}{{Lutovinov}, A.A.},
  \bibinfo{author}{{Tsygankov}, S.S.}, \bibinfo{author}{{Winkler}, C.},
  \bibinfo{year}{2012}.
\newblock \bibinfo{title}{{Hard-X-ray emission lines from the decay of
  $^{44}$Ti in the remnant of supernova 1987A}}.
\newblock \bibinfo{journal}{Nature} \bibinfo{volume}{490},
  \bibinfo{pages}{373--375}.
\bibitem[{{Grebenev} and {Sunyaev}(1987a)}]{greb87b}
\bibinfo{author}{{Grebenev}, S.A.}, \bibinfo{author}{{Sunyaev}, R.A.},
  \bibinfo{year}{1987}a.
\newblock \bibinfo{title}{{The Expected X-Ray Emission from Supernova 1987A -
  Analytic Considerations}}.
\newblock \bibinfo{journal}{Soviet Astronomy Letters} \bibinfo{volume}{13},
  \bibinfo{pages}{438}.
\bibitem[{{Grebenev} and {Sunyaev}(1987b)}]{greb87a}
\bibinfo{author}{{Grebenev}, S.A.}, \bibinfo{author}{{Sunyaev}, R.A.},
  \bibinfo{year}{1987}b.
\newblock \bibinfo{title}{{The Expected X-Ray Emission from Supernova 1987A -
  Monte-Carlo Calculations}}.
\newblock \bibinfo{journal}{Soviet Astronomy Letters} \bibinfo{volume}{13},
  \bibinfo{pages}{397}.
\bibitem[{{Green}(2014)}]{gree14}
\bibinfo{author}{{Green}, D.A.}, \bibinfo{year}{2014}.
\newblock \bibinfo{title}{{A catalogue of 294 Galactic supernova remnants}}.
\newblock \bibinfo{journal}{Bulletin of the Astronomical Society of India}
  \bibinfo{volume}{42}, \bibinfo{pages}{47--58}.
\newblock \href{http://arxiv.org/abs/1409.0637}{\tt arXiv:1409.0637}.
\bibitem[{{Grefenstette} et~al.(2017){Grefenstette}, {Fryer} and
  {Harrison}}]{gref17}
\bibinfo{author}{{Grefenstette}, B.W.}, \bibinfo{author}{{Fryer}, C.L.},
  \bibinfo{author}{{Harrison}, F.A.e.a.}, \bibinfo{year}{2017}.
\newblock \bibinfo{title}{{The Distribution of Radioactive $^{44}$Ti in
  Cassiopeia A}}.
\newblock \bibinfo{journal}{Astrophys. J.} \bibinfo{volume}{834},
  \bibinfo{pages}{19}.
\newblock \DOIprefix\doi{10.3847/1538-4357/834/1/19}.
\bibitem[{{Grefenstette} et~al.(2014){Grefenstette}, {Harrison} and
  {Boggs}}]{gref14}
\bibinfo{author}{{Grefenstette}, B.W.}, \bibinfo{author}{{Harrison}, F.A.},
  \bibinfo{author}{{Boggs}, S.E.e.a.}, \bibinfo{year}{2014}.
\newblock \bibinfo{title}{{Asymmetries in core-collapse supernovae from maps of
  radioactive $^{44}$Ti in Cassiopeia A}}.
\newblock \bibinfo{journal}{Nature} \bibinfo{volume}{506},
  \bibinfo{pages}{339--342}.
\newblock \DOIprefix\doi{10.1038/nature12997}.
\bibitem[{{Guerrero} et~al.(2004){Guerrero}, {Garc{\'{\i}}a-Berro} and
  {Isern}}]{guer04}
\bibinfo{author}{{Guerrero}, J.}, \bibinfo{author}{{Garc{\'{\i}}a-Berro}, E.},
  \bibinfo{author}{{Isern}, J.}, \bibinfo{year}{2004}.
\newblock \bibinfo{title}{{Smoothed Particle Hydrodynamics simulations of
  merging white dwarfs}}.
\newblock \bibinfo{journal}{A\&A} \bibinfo{volume}{413},
  \bibinfo{pages}{257--272}.
\newblock \DOIprefix\doi{10.1051/0004-6361:20031504}.
\bibitem[{{Guti{\'e}rrez} et~al.(2005){Guti{\'e}rrez}, {Canal} and
  {Garc{\'\i}a-Berro}}]{guti05}
\bibinfo{author}{{Guti{\'e}rrez}, J.}, \bibinfo{author}{{Canal}, R.},
  \bibinfo{author}{{Garc{\'\i}a-Berro}, E.}, \bibinfo{year}{2005}.
\newblock \bibinfo{title}{{The gravitational collapse of ONe
  electron-degenerate cores and white dwarfs: The role of $^{24}$Mg and
  $^{12}$C revisited}}.
\newblock \bibinfo{journal}{A\&A} \bibinfo{volume}{435},
  \bibinfo{pages}{231--237}.
\bibitem[{{Gutierrez} et~al.(1996){Gutierrez}, {Garcia-Berro} and
  {Iben}}]{guti96}
\bibinfo{author}{{Gutierrez}, J.}, \bibinfo{author}{{Garcia-Berro}, E.},
  \bibinfo{author}{{Iben}, Jr., I.e.a.}, \bibinfo{year}{1996}.
\newblock \bibinfo{title}{{The Final Evolution of ONeMg Electron-Degenerate
  Cores}}.
\newblock \bibinfo{journal}{Astrophys. J.} \bibinfo{volume}{459}.
\bibitem[{{Hachisu} et~al.(1999){Hachisu}, {Kato} and {Nomoto}}]{hach99}
\bibinfo{author}{{Hachisu}, I.}, \bibinfo{author}{{Kato}, M.},
  \bibinfo{author}{{Nomoto}, K.}, \bibinfo{year}{1999}.
\newblock \bibinfo{title}{{A Wide Symbiotic Channel to Type IA Supernovae}}.
\newblock \bibinfo{journal}{Astrophys. J.} \bibinfo{volume}{522},
  \bibinfo{pages}{487--503}.
\newblock \DOIprefix\doi{10.1086/307608}.
\bibitem[{{Hallakoun} and {Maoz}(2019)}]{hall19}
\bibinfo{author}{{Hallakoun}, N.}, \bibinfo{author}{{Maoz}, D.},
  \bibinfo{year}{2019}.
\newblock \bibinfo{title}{{Limits on a population of collisional-triples as
  progenitors of Type-Ia supernovae}}.
\newblock \bibinfo{journal}{Mon. Not. R. Astron. Soc.} \bibinfo{volume}{490},
  \bibinfo{pages}{657}.
\bibitem[{{Hamers} and {Thompson}(2019)}]{hame19}
\bibinfo{author}{{Hamers}, A.S.}, \bibinfo{author}{{Thompson}, T.A.},
  \bibinfo{year}{2019}.
\newblock \bibinfo{title}{{The Impact of White Dwarf Natal Kicks and Stellar
  Flybys on the Rates of Type Ia Supernovae in Triple-star Systems}}.
\newblock \bibinfo{journal}{Astrophys. J.} \bibinfo{volume}{882},
  \bibinfo{pages}{24}.
\newblock \DOIprefix\doi{10.3847/1538-4357/ab321f},
  \href{http://arxiv.org/abs/1904.12881}{\tt arXiv:1904.12881}.
\bibitem[{{Hamuy} et~al.(1996){Hamuy}, {Phillips} and {Suntzeff}}]{hamu96}
\bibinfo{author}{{Hamuy}, M.}, \bibinfo{author}{{Phillips}, M.M.},
  \bibinfo{author}{{Suntzeff}, N.B.e.a.}, \bibinfo{year}{1996}.
\newblock \bibinfo{title}{{BVRI Light Curves for 29 Type IA Supernovae}}.
\newblock \bibinfo{journal}{Astron. J.} \bibinfo{volume}{112},
  \bibinfo{pages}{2408}.
\newblock \DOIprefix\doi{10.1086/118192}.
\bibitem[{{Han} and {Podsiadlowski}(2004)}]{han04}
\bibinfo{author}{{Han}, Z.}, \bibinfo{author}{{Podsiadlowski}, P.},
  \bibinfo{year}{2004}.
\newblock \bibinfo{title}{{The single-degenerate channel for the progenitors of
  Type Ia supernovae}}.
\newblock \bibinfo{journal}{Mon. Not. R. Astron. Soc.} \bibinfo{volume}{350},
  \bibinfo{pages}{1301--1309}.
\newblock \DOIprefix\doi{10.1111/j.1365-2966.2004.07713.x}.
\bibitem[{{Harrison} et~al.(2013){Harrison}, {Craig} and
  {Christensen}}]{harr13}
\bibinfo{author}{{Harrison}, F.A.}, \bibinfo{author}{{Craig}, W.W.},
  \bibinfo{author}{{Christensen}, F.E.e.a.}, \bibinfo{year}{2013}.
\newblock \bibinfo{title}{{The Nuclear Spectroscopic Telescope Array (NuSTAR)
  High-energy X-Ray Mission}}.
\newblock \bibinfo{journal}{Astrophys. J.} \bibinfo{volume}{770},
  \bibinfo{pages}{103}.
\newblock \DOIprefix\doi{10.1088/0004-637X/770/2/103}.
\bibitem[{{Heger} et~al.(2003){Heger}, {Fryer}, {Woosley}, {Langer} and
  {Hartmann}}]{hege03}
\bibinfo{author}{{Heger}, A.}, \bibinfo{author}{{Fryer}, C.L.},
  \bibinfo{author}{{Woosley}, S.E.}, \bibinfo{author}{{Langer}, N.},
  \bibinfo{author}{{Hartmann}, D.H.}, \bibinfo{year}{2003}.
\newblock \bibinfo{title}{{How Massive Single Stars End Their Life}}.
\newblock \bibinfo{journal}{Astrophys. J.} \bibinfo{volume}{591},
  \bibinfo{pages}{288--300}.
\newblock \DOIprefix\doi{10.1086/375341}.
\bibitem[{{Heger} and {Woosley}(2002)}]{hege02}
\bibinfo{author}{{Heger}, A.}, \bibinfo{author}{{Woosley}, S.E.},
  \bibinfo{year}{2002}.
\newblock \bibinfo{title}{{The Nucleosynthetic Signature of Population III}}.
\newblock \bibinfo{journal}{Astrophys. J.} \bibinfo{volume}{567},
  \bibinfo{pages}{532--543}.
\newblock \DOIprefix\doi{10.1086/338487}.
\bibitem[{{Hernanz} et~al.(2002){Hernanz}, {G{\'o}mez-Gomar} and
  {Jos{\'e}}}]{hern02}
\bibinfo{author}{{Hernanz}, M.}, \bibinfo{author}{{G{\'o}mez-Gomar}, J.},
  \bibinfo{author}{{Jos{\'e}}, J.}, \bibinfo{year}{2002}.
\newblock \bibinfo{title}{{The prompt gamma-ray emission of novae}}.
\newblock \bibinfo{journal}{New Astro. Rev.} \bibinfo{volume}{46},
  \bibinfo{pages}{559--563}.
\newblock \DOIprefix\doi{10.1016/S1387-6473(02)00201-4}.
\bibitem[{{Hernanz} and {Jos{\'e}}(2006)}]{hj2006}
\bibinfo{author}{{Hernanz}, M.}, \bibinfo{author}{{Jos{\'e}}, J.},
  \bibinfo{year}{2006}.
\newblock \bibinfo{title}{{Radioactivities from novae}}.
\newblock \bibinfo{journal}{New Astron. Rev.} \bibinfo{volume}{50},
  \bibinfo{pages}{504--508}.
\newblock \DOIprefix\doi{10.1016/j.newar.2006.06.012}.
\bibitem[{{Hernanz} et~al.(1999){Hernanz}, {Jos{\'e}}, {Coc}, {G{\'o}mez-Gomar}
  and {Isern}}]{hern99}
\bibinfo{author}{{Hernanz}, M.}, \bibinfo{author}{{Jos{\'e}}, J.},
  \bibinfo{author}{{Coc}, A.}, \bibinfo{author}{{G{\'o}mez-Gomar}, J.},
  \bibinfo{author}{{Isern}, J.}, \bibinfo{year}{1999}.
\newblock \bibinfo{title}{{Gamma-Ray Emission from Novae Related to Positron
  Annihilation: Constraints on its Observability Posed by New Experimental
  Nuclear Data}}.
\newblock \bibinfo{journal}{Astrophys. J. Lett.} \bibinfo{volume}{526},
  \bibinfo{pages}{L97--L100}.
\newblock \DOIprefix\doi{10.1086/312372}.
\bibitem[{{Hoeflich} and {Khokhlov}(1996)}]{hoef96}
\bibinfo{author}{{Hoeflich}, P.}, \bibinfo{author}{{Khokhlov}, A.},
  \bibinfo{year}{1996}.
\newblock \bibinfo{title}{{Explosion Models for Type IA Supernovae: A
  Comparison with Observed Light Curves, Distances, H 0, and Q 0}}.
\newblock \bibinfo{journal}{Astrophys. J.} \bibinfo{volume}{457},
  \bibinfo{pages}{500--+}.
\bibitem[{{Hoeflich} et~al.(1994){Hoeflich}, {Khokhlov} and {Mueller}}]{hoef94}
\bibinfo{author}{{Hoeflich}, P.}, \bibinfo{author}{{Khokhlov}, A.},
  \bibinfo{author}{{Mueller}, E.}, \bibinfo{year}{1994}.
\newblock \bibinfo{title}{{Gamma-ray light curves and spectra of models for
  Type IA supernovae}}.
\newblock \bibinfo{journal}{Astrophys. J. Suppl. S} \bibinfo{volume}{92},
  \bibinfo{pages}{501--504}.
\newblock \DOIprefix\doi{10.1086/192004}.
\bibitem[{{H{\"o}flich} et~al.(1998){H{\"o}flich}, {Wheeler} and
  {Thielemann}}]{hoef98}
\bibinfo{author}{{H{\"o}flich}, P.}, \bibinfo{author}{{Wheeler}, J.C.},
  \bibinfo{author}{{Thielemann}, F.K.}, \bibinfo{year}{1998}.
\newblock \bibinfo{title}{{Type Ia Supernovae: Influence of the Initial
  Composition on the Nucleosynthesis, Light Curves, and Spectra and
  Consequences for the Determination of {\ensuremath{\Omega}}$_{M}$ and
  {\ensuremath{\Lambda}}}}.
\newblock \bibinfo{journal}{Astrophys. J.} \bibinfo{volume}{495},
  \bibinfo{pages}{617--629}.
\newblock \DOIprefix\doi{10.1086/305327}.
\bibitem[{{Hoyle}(1946)}]{hoyl46}
\bibinfo{author}{{Hoyle}, F.}, \bibinfo{year}{1946}.
\newblock \bibinfo{title}{{The synthesis of the elements from hydrogen}}.
\newblock \bibinfo{journal}{MNRAS} \bibinfo{volume}{106}, \bibinfo{pages}{343}.
\bibitem[{{Hoyle} and {Fowler}(1960)}]{hoyl60}
\bibinfo{author}{{Hoyle}, F.}, \bibinfo{author}{{Fowler}, W.A.},
  \bibinfo{year}{1960}.
\newblock \bibinfo{title}{{Nucleosynthesis in Supernovae.}}
\newblock \bibinfo{journal}{Astrophys. J.} \bibinfo{volume}{132},
  \bibinfo{pages}{565}.
\newblock \DOIprefix\doi{10.1086/146963}.
\bibitem[{{Hwang} et~al.(2004){Hwang}, {Laming}, {Badenes}, {Berendse},
  {Blondin}, {Cioffi}, {DeLaney}, {Dewey}, {Fesen}, {Flanagan}, {Fryer},
  {Ghavamian}, {Hughes}, {Morse}, {Plucinsky}, {Petre}, {Pohl}, {Rudnick},
  {Sankrit}, {Slane}, {Smith}, {Vink} and {Warren}}]{hwan04}
\bibinfo{author}{{Hwang}, U.}, \bibinfo{author}{{Laming}, J.M.},
  \bibinfo{author}{{Badenes}, C.}, \bibinfo{author}{{Berendse}, F.},
  \bibinfo{author}{{Blondin}, J.}, \bibinfo{author}{{Cioffi}, D.},
  \bibinfo{author}{{DeLaney}, T.}, \bibinfo{author}{{Dewey}, D.},
  \bibinfo{author}{{Fesen}, R.}, \bibinfo{author}{{Flanagan}, K.A.},
  \bibinfo{author}{{Fryer}, C.L.}, \bibinfo{author}{{Ghavamian}, P.},
  \bibinfo{author}{{Hughes}, J.P.}, \bibinfo{author}{{Morse}, J.A.},
  \bibinfo{author}{{Plucinsky}, P.P.}, \bibinfo{author}{{Petre}, R.},
  \bibinfo{author}{{Pohl}, M.}, \bibinfo{author}{{Rudnick}, L.},
  \bibinfo{author}{{Sankrit}, R.}, \bibinfo{author}{{Slane}, P.O.},
  \bibinfo{author}{{Smith}, R.K.}, \bibinfo{author}{{Vink}, J.},
  \bibinfo{author}{{Warren}, J.S.}, \bibinfo{year}{2004}.
\newblock \bibinfo{title}{{A Million Second Chandra View of Cassiopeia A}}.
\newblock \bibinfo{journal}{Astrophys. J. Lett.} \bibinfo{volume}{615},
  \bibinfo{pages}{L117--L120}.
\newblock \DOIprefix\doi{10.1086/426186}.
\bibitem[{{Iben} and {Tutukov}(1984)}]{iben84}
\bibinfo{author}{{Iben}, Jr., I.}, \bibinfo{author}{{Tutukov}, A.V.},
  \bibinfo{year}{1984}.
\newblock \bibinfo{title}{{Supernovae of type I as end products of the
  evolution of binaries with components of moderate initial mass (M not greater
  than about 9 solar masses)}}.
\newblock \bibinfo{journal}{Astrophys. J. Suppl. S} \bibinfo{volume}{54},
  \bibinfo{pages}{335--372}.
\newblock \DOIprefix\doi{10.1086/190932}.
\bibitem[{{Iben} and {Tutukov}(1991)}]{iben91}
\bibinfo{author}{{Iben}, I., J.}, \bibinfo{author}{{Tutukov}, A.V.},
  \bibinfo{year}{1991}.
\newblock \bibinfo{title}{{Close Binaries with Evolved Components}}, in:
  \bibinfo{editor}{{Lambert}, D.L.} (Ed.), \bibinfo{booktitle}{Frontiers of
  Stellar Evolution}, p. \bibinfo{pages}{403}.
\bibitem[{{Iben} et~al.(1997){Iben}, {Ritossa} and
  {Garc{\'\i}a-Berro}}]{iben97}
\bibinfo{author}{{Iben}, Icko, J.}, \bibinfo{author}{{Ritossa}, C.},
  \bibinfo{author}{{Garc{\'\i}a-Berro}, E.}, \bibinfo{year}{1997}.
\newblock \bibinfo{title}{{On the Evolution of Stars that Form
  Electron-degenerate Cores Processed by Carbon Burning. IV. Outward Mixing
  During the Second Dredge-up Phase and Other Properties of a 10.5 M$_{\odot}$
  Model Star}}.
\newblock \bibinfo{journal}{Astrophys. J.} \bibinfo{volume}{489},
  \bibinfo{pages}{772--790}.
\newblock \DOIprefix\doi{10.1086/304822}.
\bibitem[{{Iliadis}(2007)}]{ilia07}
\bibinfo{author}{{Iliadis}, C.}, \bibinfo{year}{2007}.
\newblock \bibinfo{title}{{Nuclear Physics of Stars}}.
\newblock \DOIprefix\doi{10.1002/9783527692668}.
\bibitem[{{Imshennik}(1992)}]{imsh92}
\bibinfo{author}{{Imshennik}, V.S.}, \bibinfo{year}{1992}.
\newblock \bibinfo{title}{{Scenario for a supernova explosion in the
  gravitational collapse of a massive stellar core}}.
\newblock \bibinfo{journal}{Soviet Astronomy Letters} \bibinfo{volume}{18},
  \bibinfo{pages}{194}.
\bibitem[{{Imshennik} and {Ryazhskaya}(2004)}]{imsh04}
\bibinfo{author}{{Imshennik}, V.S.}, \bibinfo{author}{{Ryazhskaya}, O.G.},
  \bibinfo{year}{2004}.
\newblock \bibinfo{title}{{A Rotating Collapsar and Possible Interpretation of
  the LSD Neutrino Signal from SN 1987 A}}.
\newblock \bibinfo{journal}{Astronomy Letters} \bibinfo{volume}{30},
  \bibinfo{pages}{14--31}.
\newblock \DOIprefix\doi{10.1134/1.1647473},
  \href{http://arxiv.org/abs/astro-ph/0401613}{\tt arXiv:astro-ph/0401613}.
\bibitem[{{Inserra}(2019)}]{inse19}
\bibinfo{author}{{Inserra}, C.}, \bibinfo{year}{2019}.
\newblock \bibinfo{title}{{Observational properties of extreme supernovae}}.
\newblock \bibinfo{journal}{Nature Astronomy} \bibinfo{volume}{3},
  \bibinfo{pages}{697--705}.
\bibitem[{{Inserra} et~al.(2018){Inserra}, {Smartt} and {Gall}}]{inse18}
\bibinfo{author}{{Inserra}, C.}, \bibinfo{author}{{Smartt}, S.J.},
  \bibinfo{author}{{Gall}, E.E.E.e.a.}, \bibinfo{year}{2018}.
\newblock \bibinfo{title}{{On the nature of hydrogen-rich superluminous
  supernovae}}.
\newblock \bibinfo{journal}{MNRAS} \bibinfo{volume}{475},
  \bibinfo{pages}{1046--1072}.
\newblock \DOIprefix\doi{10.1093/mnras/stx3179}.
\bibitem[{{Inserra} et~al.(2013){Inserra}, {Smartt} and {Jerkstrand}}]{inse13}
\bibinfo{author}{{Inserra}, C.}, \bibinfo{author}{{Smartt}, S.J.},
  \bibinfo{author}{{Jerkstrand}, A.e.a.}, \bibinfo{year}{2013}.
\newblock \bibinfo{title}{{Super-luminous Type Ic Supernovae: Catching a
  Magnetar by the Tail}}.
\newblock \bibinfo{journal}{APJ} \bibinfo{volume}{770}, \bibinfo{pages}{128}.
\newblock \DOIprefix\doi{10.1088/0004-637X/770/2/128}.
\bibitem[{{Isern} et~al.(2008){Isern}, {Bravo} and {Hirschmann}}]{iser08}
\bibinfo{author}{{Isern}, J.}, \bibinfo{author}{{Bravo}, E.},
  \bibinfo{author}{{Hirschmann}, A.}, \bibinfo{year}{2008}.
\newblock \bibinfo{title}{{Detection and interpretation of {$\gamma$}-ray
  emission from SNIa}}.
\newblock \bibinfo{journal}{New Astron. Rev.} \bibinfo{volume}{52},
  \bibinfo{pages}{377--380}.
\newblock \DOIprefix\doi{10.1016/j.newar.2008.06.021}.
\bibitem[{{Isern} et~al.(2016a){Isern}, {Bravo}, {Jean} and
  {Knodlseder}}]{iser16a}
\bibinfo{author}{{Isern}, J.}, \bibinfo{author}{{Bravo}, E.},
  \bibinfo{author}{{Jean}, P.}, \bibinfo{author}{{Knodlseder}, J.},
  \bibinfo{year}{2016}a.
\newblock \bibinfo{title}{{Insights on the physics of SNIa obtained from their
  gamma-ray emission}}, in: \bibinfo{booktitle}{Proceedings of the 11th
  INTEGRAL Conference Gamma-Ray Astrophysics in Multi-Wavelength Perspective.
  10-14 October 2016 Amsterdam}, p.~\bibinfo{pages}{54}.
\bibitem[{{Isern} et~al.(1991){Isern}, {Canal} and {Labay}}]{iser91}
\bibinfo{author}{{Isern}, J.}, \bibinfo{author}{{Canal}, R.},
  \bibinfo{author}{{Labay}, J.}, \bibinfo{year}{1991}.
\newblock \bibinfo{title}{{The Outcome of Explosive Ignition of ONeMg Cores:
  Supernovae, Neutron Stars, or ``Iron'' White Dwarfs?}}
\newblock \bibinfo{journal}{Astrophys. J. Lett.} \bibinfo{volume}{372},
  \bibinfo{pages}{L83}.
\newblock \DOIprefix\doi{10.1086/186029}.
\bibitem[{{Isern} et~al.(1997){Isern}, {Hernanz}, {Salaris}, {Bravo},
  {Garcia-Berro} and {Tornambe}}]{iser97}
\bibinfo{author}{{Isern}, J.}, \bibinfo{author}{{Hernanz}, M.},
  \bibinfo{author}{{Salaris}, M.}, \bibinfo{author}{{Bravo}, E.},
  \bibinfo{author}{{Garcia-Berro}, E.}, \bibinfo{author}{{Tornambe}, A.},
  \bibinfo{year}{1997}.
\newblock \bibinfo{title}{{The double degenerate population in the solar
  neighborhood}}, in: \bibinfo{editor}{{P.~Ruiz-Lapuente, R.~Canal, \&
  J.~Isern}} (Ed.), \bibinfo{booktitle}{NATO ASIC Proc. 486: Thermonuclear
  Supernovae}, pp. \bibinfo{pages}{127--+}.
\bibitem[{{Isern} et~al.(2016b){Isern}, {Jean}, {Bravo} and et~al.}]{iser16}
\bibinfo{author}{{Isern}, J.}, \bibinfo{author}{{Jean}, P.},
  \bibinfo{author}{{Bravo}, E.}, \bibinfo{author}{et~al.},
  \bibinfo{year}{2016}b.
\newblock \bibinfo{title}{{Gamma-ray emission from SN2014J near maximum optical
  light}}.
\newblock \bibinfo{journal}{A\&A} \bibinfo{volume}{588}, \bibinfo{pages}{A67}.
\bibitem[{{Isern} et~al.(2013){Isern}, {Jean}, {Bravo}, {Diehl},
  {Kn{\"o}dlseder}, {Domingo}, {Hirschmann}, {Hoeflich}, {Lebrun} and
  {Renaud}}]{iser13}
\bibinfo{author}{{Isern}, J.}, \bibinfo{author}{{Jean}, P.},
  \bibinfo{author}{{Bravo}, E.}, \bibinfo{author}{{Diehl}, R.},
  \bibinfo{author}{{Kn{\"o}dlseder}, J.}, \bibinfo{author}{{Domingo}, A.},
  \bibinfo{author}{{Hirschmann}, A.}, \bibinfo{author}{{Hoeflich}, P.},
  \bibinfo{author}{{Lebrun}, F.}, \bibinfo{author}{{Renaud}, M.},
  \bibinfo{year}{2013}.
\newblock \bibinfo{title}{{Observation of SN2011fe with INTEGRAL. I.
  Pre-maximum phase}}.
\newblock \bibinfo{journal}{A\&A} \bibinfo{volume}{552}, \bibinfo{pages}{A97}.
\bibitem[{{Isern} et~al.(2014){Isern}, {Knoedlseder} and {Jean}}]{iser14}
\bibinfo{author}{{Isern}, J.}, \bibinfo{author}{{Knoedlseder}, J.},
  \bibinfo{author}{{Jean}, P.e.a.}, \bibinfo{year}{2014}.
\newblock \bibinfo{title}{{Early gamma--ray emission from SN2014J during the
  optical maximum as obtained by INTEGRAL}}.
\newblock \bibinfo{journal}{The Astronomer's Telegram} \bibinfo{volume}{6099},
  \bibinfo{pages}{1}.
\bibitem[{{Isern} et~al.(1983){Isern}, {Labay}, {Hernanz} and {Canal}}]{iser83}
\bibinfo{author}{{Isern}, J.}, \bibinfo{author}{{Labay}, J.},
  \bibinfo{author}{{Hernanz}, M.}, \bibinfo{author}{{Canal}, R.},
  \bibinfo{year}{1983}.
\newblock \bibinfo{title}{{Collapse and explosion of white dwarfs. I -
  Precollapse evolution}}.
\newblock \bibinfo{journal}{Astrophys. J.} \bibinfo{volume}{273},
  \bibinfo{pages}{320--329}.
\bibitem[{{Iyudin} et~al.(1995){Iyudin}, {Bennett}, {Bloemen}, {Diehl},
  {Hermsen}, {Lichti}, {Morris}, {Ryan}, {Schoenfelder}, {Steinle}, {Strong},
  {Varendorff} and {Winkler}}]{iyu95}
\bibinfo{author}{{Iyudin}, A.F.}, \bibinfo{author}{{Bennett}, K.},
  \bibinfo{author}{{Bloemen}, H.}, \bibinfo{author}{{Diehl}, R.},
  \bibinfo{author}{{Hermsen}, W.}, \bibinfo{author}{{Lichti}, G.G.},
  \bibinfo{author}{{Morris}, D.}, \bibinfo{author}{{Ryan}, J.},
  \bibinfo{author}{{Schoenfelder}, V.}, \bibinfo{author}{{Steinle}, H.},
  \bibinfo{author}{{Strong}, A.}, \bibinfo{author}{{Varendorff}, M.},
  \bibinfo{author}{{Winkler}, C.}, \bibinfo{year}{1995}.
\newblock \bibinfo{title}{{COMPTEL search for $^{22}$Na line emission from
  recent novae.}}
\newblock \bibinfo{journal}{A\&A} \bibinfo{volume}{300}, \bibinfo{pages}{422}.
\bibitem[{{Iyudin} et~al.(1994){Iyudin}, {Diehl} and {Bloemen}}]{iyud94}
\bibinfo{author}{{Iyudin}, A.F.}, \bibinfo{author}{{Diehl}, R.},
  \bibinfo{author}{{Bloemen}, H.a.}, \bibinfo{year}{1994}.
\newblock \bibinfo{title}{{COMPTEL observations of Ti-44 gamma-ray line
  emission from CAS A}}.
\newblock \bibinfo{journal}{A\&A} \bibinfo{volume}{284},
  \bibinfo{pages}{L1--L4}.
\bibitem[{{Iyudin} et~al.(2019){Iyudin}, {M{\"u}ller} and
  {Obergaulinger}}]{iyud19}
\bibinfo{author}{{Iyudin}, A.F.}, \bibinfo{author}{{M{\"u}ller}, E.},
  \bibinfo{author}{{Obergaulinger}, M.}, \bibinfo{year}{2019}.
\newblock \bibinfo{title}{{Titanium hidden in dust}}.
\newblock \bibinfo{journal}{Mon. Not. R. Astron. Soc.} \bibinfo{volume}{485},
  \bibinfo{pages}{3288--3295}.
\newblock \DOIprefix\doi{10.1093/mnras/stz419}.
\bibitem[{{Janka} et~al.(2016){Janka}, {Melson} and {Summa}}]{jank16}
\bibinfo{author}{{Janka}, H.T.}, \bibinfo{author}{{Melson}, T.},
  \bibinfo{author}{{Summa}, A.}, \bibinfo{year}{2016}.
\newblock \bibinfo{title}{{Physics of Core-Collapse Supernovae in Three
  Dimensions: A Sneak Preview}}.
\newblock \bibinfo{journal}{Annual Review of Nuclear and Particle Science}
  \bibinfo{volume}{66}, \bibinfo{pages}{341--375}.
\newblock \DOIprefix\doi{10.1146/annurev-nucl-102115-044747}.
\bibitem[{{Janka} et~al.(2008){Janka}, {M{\"u}ller}, {Kitaura} and
  {Buras}}]{jank08}
\bibinfo{author}{{Janka}, H.T.}, \bibinfo{author}{{M{\"u}ller}, B.},
  \bibinfo{author}{{Kitaura}, F.S.}, \bibinfo{author}{{Buras}, R.},
  \bibinfo{year}{2008}.
\newblock \bibinfo{title}{{Dynamics of shock propagation and nucleosynthesis
  conditions in O-Ne-Mg core supernovae}}.
\newblock \bibinfo{journal}{A\&A} \bibinfo{volume}{485},
  \bibinfo{pages}{199--208}.
\newblock \DOIprefix\doi{10.1051/0004-6361:20079334}.
\bibitem[{Jean(1996)}]{jean96}
\bibinfo{author}{Jean, P.}, \bibinfo{year}{1996}.
\newblock \bibinfo{title}{Etudes des performances et modelisation de
  spectrometres gamma pour l'Astrophysique Nucleaire}.
\newblock Ph.D. thesis. Universite Paul Sabatier Toulouse.
\bibitem[{{Jean} et~al.(1999){Jean}, {G{\'o}mez-Gomar}, {Hernanz}, {Jos{\'e}},
  {Isern}, {Vedrenne}, {Mandrou}, {Sch{\"o}nfelder}, {Lichti} and
  {Georgii}}]{jean99}
\bibinfo{author}{{Jean}, P.}, \bibinfo{author}{{G{\'o}mez-Gomar}, J.},
  \bibinfo{author}{{Hernanz}, M.}, \bibinfo{author}{{Jos{\'e}}, J.},
  \bibinfo{author}{{Isern}, J.}, \bibinfo{author}{{Vedrenne}, G.},
  \bibinfo{author}{{Mandrou}, P.}, \bibinfo{author}{{Sch{\"o}nfelder}, V.},
  \bibinfo{author}{{Lichti}, G.G.}, \bibinfo{author}{{Georgii}, R.},
  \bibinfo{year}{1999}.
\newblock \bibinfo{title}{{Possibility of the Detection of Classical Novae with
  the Shield of the Integral Spectrometer SPI}}.
\newblock \bibinfo{journal}{Astrophysical Letters and Communications}
  \bibinfo{volume}{38}, \bibinfo{pages}{421}.
\bibitem[{{Jean} et~al.(2000){Jean}, {Hernanz}, {G{\'o}mez-Gomar} and
  {Jos{\'e}}}]{jean2000}
\bibinfo{author}{{Jean}, P.}, \bibinfo{author}{{Hernanz}, M.},
  \bibinfo{author}{{G{\'o}mez-Gomar}, J.}, \bibinfo{author}{{Jos{\'e}}, J.},
  \bibinfo{year}{2000}.
\newblock \bibinfo{title}{{Galactic 1.275-MeV emission from ONe novae and its
  detectability by INTEGRAL/SPI}}.
\newblock \bibinfo{journal}{Mon. Not. R. Astron. Soc.} \bibinfo{volume}{319},
  \bibinfo{pages}{350--364}.
\bibitem[{{Jean} et~al.(2004){Jean}, {Kn{\"o}dlseder}, {Hernanz}, {Cisana},
  {Valsesia}, {von Kienlin}, {Leleux}, {Strong}, {Winkler} and
  {Wunderer}}]{jean2004}
\bibinfo{author}{{Jean}, P.}, \bibinfo{author}{{Kn{\"o}dlseder}, J.},
  \bibinfo{author}{{Hernanz}, M.}, \bibinfo{author}{{Cisana}, E.},
  \bibinfo{author}{{Valsesia}, M.}, \bibinfo{author}{{von Kienlin}, A.},
  \bibinfo{author}{{Leleux}, P.}, \bibinfo{author}{{Strong}, A.},
  \bibinfo{author}{{Winkler}, C.}, \bibinfo{author}{{Wunderer}, C.},
  \bibinfo{year}{2004}.
\newblock \bibinfo{title}{{Search for Galactic 1275 keV Line Emission with
  SPI/INTEGRAL}}, in: \bibinfo{editor}{{Schoenfelder}, V.},
  \bibinfo{editor}{{Lichti}, G.}, \bibinfo{editor}{{Winkler}, C.} (Eds.),
  \bibinfo{booktitle}{5th INTEGRAL Workshop on the INTEGRAL Universe}, p.
  \bibinfo{pages}{119}.
\bibitem[{{Jean} et~al.(2001){Jean}, {Kn{\"o}dlseder}, {von Ballmoos},
  {G{\'o}mez-Gomar}, {Hernanz} and {Jos{\'e}}}]{jean2001}
\bibinfo{author}{{Jean}, P.}, \bibinfo{author}{{Kn{\"o}dlseder}, J.},
  \bibinfo{author}{{von Ballmoos}, P.}, \bibinfo{author}{{G{\'o}mez-Gomar},
  J.}, \bibinfo{author}{{Hernanz}, M.}, \bibinfo{author}{{Jos{\'e}}, J.},
  \bibinfo{year}{2001}.
\newblock \bibinfo{title}{{Upper limits of the $^{22}$Na yield from O-Ne
  nova}}, in: \bibinfo{editor}{{Gimenez}, A.}, \bibinfo{editor}{{Reglero}, V.},
  \bibinfo{editor}{{Winkler}, C.} (Eds.), \bibinfo{booktitle}{Exploring the
  Gamma-Ray Universe}, pp. \bibinfo{pages}{73--77}.
\bibitem[{{Jerkstrand} et~al.(2020){Jerkstrand}, {Wongwathanarat}, {Janka},
  {Gabler}, {Alp}, {Diehl}, {Maeda}, {Larsson}, {Fransson}, {Menon} and
  {Heger}}]{jerk20}
\bibinfo{author}{{Jerkstrand}, A.}, \bibinfo{author}{{Wongwathanarat}, A.},
  \bibinfo{author}{{Janka}, H.T.}, \bibinfo{author}{{Gabler}, M.},
  \bibinfo{author}{{Alp}, D.}, \bibinfo{author}{{Diehl}, R.},
  \bibinfo{author}{{Maeda}, K.}, \bibinfo{author}{{Larsson}, J.},
  \bibinfo{author}{{Fransson}, C.}, \bibinfo{author}{{Menon}, A.},
  \bibinfo{author}{{Heger}, A.}, \bibinfo{year}{2020}.
\newblock \bibinfo{title}{{Properties of gamma-ray decay lines in 3D
  core-collapse supernova models, with application to SN 1987A and Cas A}}.
\newblock \bibinfo{journal}{Mon. Not. R. Astron. Soc.} \bibinfo{volume}{494},
  \bibinfo{pages}{2471--2497}.
\newblock \DOIprefix\doi{10.1093/mnras/staa736},
  \href{http://arxiv.org/abs/2003.05156}{\tt arXiv:2003.05156}.
\bibitem[{{Jha} et~al.(2006){Jha}, {Branch} and {Chornock}}]{jha06}
\bibinfo{author}{{Jha}, S.}, \bibinfo{author}{{Branch}, D.},
  \bibinfo{author}{{Chornock}, R.e.a.}, \bibinfo{year}{2006}.
\newblock \bibinfo{title}{{Late-Time Spectroscopy of SN 2002cx: The Prototype
  of a New Subclass of Type Ia Supernovae}}.
\newblock \bibinfo{journal}{Astron. J.} \bibinfo{volume}{132},
  \bibinfo{pages}{189--196}.
\newblock \DOIprefix\doi{10.1086/504599}.
\bibitem[{{Jha}(2017)}]{jha17}
\bibinfo{author}{{Jha}, S.W.}, \bibinfo{year}{2017}.
\newblock \bibinfo{title}{{Type Iax Supernovae}}.
\newblock p. \bibinfo{pages}{375}.
\newblock \DOIprefix\doi{10.1007/978-3-319-21846-5_42}.
\bibitem[{{Jha} et~al.(2019){Jha}, {Maguire} and {Sullivan}}]{jha19}
\bibinfo{author}{{Jha}, S.W.}, \bibinfo{author}{{Maguire}, K.},
  \bibinfo{author}{{Sullivan}, M.}, \bibinfo{year}{2019}.
\newblock \bibinfo{title}{{Observational properties of thermonuclear
  supernovae}}.
\newblock \bibinfo{journal}{Nature Astronomy} \bibinfo{volume}{3},
  \bibinfo{pages}{706--716}.
\newblock \DOIprefix\doi{10.1038/s41550-019-0858-0}.
\bibitem[{{Jones} et~al.(2016){Jones}, {R{\"o}pke} and {Pakmor}}]{Jone16}
\bibinfo{author}{{Jones}, S.}, \bibinfo{author}{{R{\"o}pke}, F.K.},
  \bibinfo{author}{{Pakmor}, R.e.a.}, \bibinfo{year}{2016}.
\newblock \bibinfo{title}{{Do electron-capture supernovae make neutron stars?.
  First multidimensional hydrodynamic simulations of the oxygen deflagration}}.
\newblock \bibinfo{journal}{A\&A} \bibinfo{volume}{593}, \bibinfo{pages}{A72}.
\newblock \DOIprefix\doi{10.1051/0004-6361/201628321}.
\bibitem[{{Jos{\'e}} et~al.(1999){Jos{\'e}}, {Coc} and {Hernanz}}]{jch99}
\bibinfo{author}{{Jos{\'e}}, J.}, \bibinfo{author}{{Coc}, A.},
  \bibinfo{author}{{Hernanz}, M.}, \bibinfo{year}{1999}.
\newblock \bibinfo{title}{{Nuclear Uncertainties in the NeNa-MgAl Cycles and
  Production of $^{22}$Na and $^{26}$Al during Nova Outbursts}}.
\newblock \bibinfo{journal}{Astrophys. J.} \bibinfo{volume}{520},
  \bibinfo{pages}{347--360}.
\newblock \DOIprefix\doi{10.1086/307445}.
\bibitem[{{Jos{\'e}} and {Hernanz}(1998)}]{jh98}
\bibinfo{author}{{Jos{\'e}}, J.}, \bibinfo{author}{{Hernanz}, M.},
  \bibinfo{year}{1998}.
\newblock \bibinfo{title}{{Nucleosynthesis in Classical Novae: CO versus ONe
  White Dwarfs}}.
\newblock \bibinfo{journal}{Astrophys. J.} \bibinfo{volume}{494},
  \bibinfo{pages}{680--690}.
\newblock \DOIprefix\doi{10.1086/305244}.
\bibitem[{{Jose} et~al.(2019){Jose}, {Shore} and {Casanova}}]{jose19}
\bibinfo{author}{{Jose}, J.}, \bibinfo{author}{{Shore}, S.N.},
  \bibinfo{author}{{Casanova}, J.}, \bibinfo{year}{2019}.
\newblock \bibinfo{title}{{123-321 Models of Classical Novae}}.
\newblock \bibinfo{journal}{arXiv e-prints} ,
  \bibinfo{pages}{arXiv:1912.08443}.
\bibitem[{{Joyce} et~al.(2018){Joyce}, {Barstow}, {Casewell}, {Burleigh},
  {Holberg} and {Bond}}]{joyc18}
\bibinfo{author}{{Joyce}, S.R.G.}, \bibinfo{author}{{Barstow}, M.A.},
  \bibinfo{author}{{Casewell}, S.L.}, \bibinfo{author}{{Burleigh}, M.R.},
  \bibinfo{author}{{Holberg}, J.B.}, \bibinfo{author}{{Bond}, H.E.},
  \bibinfo{year}{2018}.
\newblock \bibinfo{title}{{Testing the white dwarf mass-radius relation and
  comparing optical and far-UV spectroscopic results with Gaia DR2, HST, and
  FUSE}}.
\newblock \bibinfo{journal}{Mon. Not. R. Astron. Soc.} \bibinfo{volume}{479},
  \bibinfo{pages}{1612--1626}.
\newblock \DOIprefix\doi{10.1093/mnras/sty1425}.
\bibitem[{{Kashi} and {Soker}(2011)}]{kash11}
\bibinfo{author}{{Kashi}, A.}, \bibinfo{author}{{Soker}, N.},
  \bibinfo{year}{2011}.
\newblock \bibinfo{title}{{A circumbinary disc in the final stages of common
  envelope and the core-degenerate scenario for Type Ia supernovae}}.
\newblock \bibinfo{journal}{Mon. Not. R. Astron. Soc.} \bibinfo{volume}{417},
  \bibinfo{pages}{1466--1479}.
\newblock \DOIprefix\doi{10.1111/j.1365-2966.2011.19361.x}.
\bibitem[{{Kasliwal} et~al.(2012){Kasliwal}, {Kulkarni} and {Gal-Yam}}]{kasl12}
\bibinfo{author}{{Kasliwal}, M.M.}, \bibinfo{author}{{Kulkarni}, S.R.},
  \bibinfo{author}{{Gal-Yam}, A.e.a.}, \bibinfo{year}{2012}.
\newblock \bibinfo{title}{{Calcium-rich Gap Transients in the Remote Outskirts
  of Galaxies}}.
\newblock \bibinfo{journal}{Astrophys. J.} \bibinfo{volume}{755},
  \bibinfo{pages}{161}.
\newblock \DOIprefix\doi{10.1088/0004-637X/755/2/161}.
\bibitem[{{Kato} and {Hachisu}(1994)}]{kato94}
\bibinfo{author}{{Kato}, M.}, \bibinfo{author}{{Hachisu}, I.},
  \bibinfo{year}{1994}.
\newblock \bibinfo{title}{{Optically Thick Winds in Nova Outbursts}}.
\newblock \bibinfo{journal}{Astrophys. J.} \bibinfo{volume}{437},
  \bibinfo{pages}{802}.
\newblock \DOIprefix\doi{10.1086/175041}.
\bibitem[{{Kato} and {Hachisu}(2015)}]{kato15}
\bibinfo{author}{{Kato}, M.}, \bibinfo{author}{{Hachisu}, I.},
  \bibinfo{year}{2015}.
\newblock \bibinfo{title}{{Theory of Classical Novae}}, in:
  \bibinfo{booktitle}{The Golden Age of Cataclysmic Variables and Related
  Objects - III (Golden2015)}, p.~\bibinfo{pages}{52}.
\bibitem[{{Katz} and {Dong}(2012)}]{katz12}
\bibinfo{author}{{Katz}, B.}, \bibinfo{author}{{Dong}, S.},
  \bibinfo{year}{2012}.
\newblock \bibinfo{title}{{The rate of WD-WD head-on collisions may be as high
  as the SNe Ia rate}}.
\newblock \bibinfo{journal}{arXiv e-prints} ,
  \bibinfo{pages}{arXiv:1211.4584}\href{http://arxiv.org/abs/1211.4584}{\tt
  arXiv:1211.4584}.
\bibitem[{{Kelly} et~al.(2014){Kelly}, {Fox} and {Filippenko}}]{kell14}
\bibinfo{author}{{Kelly}, P.L.}, \bibinfo{author}{{Fox}, O.D.},
  \bibinfo{author}{{Filippenko}, A.V.e.a.}, \bibinfo{year}{2014}.
\newblock \bibinfo{title}{{Constraints on the Progenitor System of the Type Ia
  Supernova 2014J from Pre-explosion Hubble Space Telescope Imaging}}.
\newblock \bibinfo{journal}{Astrophys. J.} \bibinfo{volume}{790},
  \bibinfo{pages}{3}.
\newblock \DOIprefix\doi{10.1088/0004-637X/790/1/3}.
\bibitem[{{Kepler} et~al.(2016){Kepler}, {Pelisoli}, {Koester} and
  {Ourique}}]{kepl16}
\bibinfo{author}{{Kepler}, S.O.}, \bibinfo{author}{{Pelisoli}, I.},
  \bibinfo{author}{{Koester}, D.}, \bibinfo{author}{{Ourique}, G.e.a.},
  \bibinfo{year}{2016}.
\newblock \bibinfo{title}{{New white dwarf and subdwarf stars in the Sloan
  Digital Sky Survey Data Release 12}}.
\newblock \bibinfo{journal}{Mon. Not. R. Astron. Soc.} \bibinfo{volume}{455},
  \bibinfo{pages}{3413--3423}.
\newblock \DOIprefix\doi{10.1093/mnras/stv2526}.
\bibitem[{{Khan} et~al.(2011){Khan}, {Stanek}, {Stoll} and {Prieto}}]{khan11}
\bibinfo{author}{{Khan}, R.}, \bibinfo{author}{{Stanek}, K.Z.},
  \bibinfo{author}{{Stoll}, R.}, \bibinfo{author}{{Prieto}, J.L.},
  \bibinfo{year}{2011}.
\newblock \bibinfo{title}{{Super-Chandrasekhar SNe Ia Strongly Prefer
  Metal-poor Environments}}.
\newblock \bibinfo{journal}{Astrophys. J. Lett.} \bibinfo{volume}{737},
  \bibinfo{pages}{L24}.
\newblock \DOIprefix\doi{10.1088/2041-8205/737/1/L24}.
\bibitem[{{Khatami} and {Kasen}(2019)}]{khata19}
\bibinfo{author}{{Khatami}, D.K.}, \bibinfo{author}{{Kasen}, D.N.},
  \bibinfo{year}{2019}.
\newblock \bibinfo{title}{{Physics of Luminous Transient Light Curves: A New
  Relation between Peak Time and Luminosity}}.
\newblock \bibinfo{journal}{Astrophys. J.} \bibinfo{volume}{878},
  \bibinfo{pages}{56}.
\newblock \DOIprefix\doi{10.3847/1538-4357/ab1f09}.
\bibitem[{{Khokhlov}(1991)}]{khok91}
\bibinfo{author}{{Khokhlov}, A.M.}, \bibinfo{year}{1991}.
\newblock \bibinfo{title}{{Delayed detonation model for type IA supernovae}}.
\newblock \bibinfo{journal}{A\&A} \bibinfo{volume}{245},
  \bibinfo{pages}{114--128}.
\bibitem[{{Kierans} et~al.(2017){Kierans}, {Boggs}, {Chiu}, {Lowell},
  {Sleator}, {Tomsick}, {Zoglauer}, {Amman}, {Chang}, {Tseng}, {Yang}, {Lin},
  {Jean} and {von Ballmoos}}]{kier17}
\bibinfo{author}{{Kierans}, C.A.}, \bibinfo{author}{{Boggs}, S.E.},
  \bibinfo{author}{{Chiu}, J.L.}, \bibinfo{author}{{Lowell}, A.},
  \bibinfo{author}{{Sleator}, C.}, \bibinfo{author}{{Tomsick}, J.A.},
  \bibinfo{author}{{Zoglauer}, A.}, \bibinfo{author}{{Amman}, M.},
  \bibinfo{author}{{Chang}, H.K.}, \bibinfo{author}{{Tseng}, C.H.},
  \bibinfo{author}{{Yang}, C.Y.}, \bibinfo{author}{{Lin}, C.H.},
  \bibinfo{author}{{Jean}, P.}, \bibinfo{author}{{von Ballmoos}, P.},
  \bibinfo{year}{2017}.
\newblock \bibinfo{title}{{The 2016 Super Pressure Balloon flight of the
  Compton Spectrometer and Imager}}.
\newblock \bibinfo{journal}{arXiv e-prints} ,
  \bibinfo{pages}{arXiv:1701.05558}\href{http://arxiv.org/abs/1701.05558}{\tt
  arXiv:1701.05558}.
\bibitem[{{Kirsebom} et~al.(2019){Kirsebom}, {Jones}, {Str{\"o}mberg},
  {Mart{\'\i}nez-Pinedo}, {Langanke}, {R{\"o}pke}, {Brown}, {Eronen}, {Fynbo},
  {Hukkanen}, {Idini}, {Jokinen}, {Kankainen}, {Kostensalo}, {Moore},
  {M{\"o}ller}, {Ohlmann}, {Penttil{\"a}}, {Riisager}, {Rinta-Antila},
  {Srivastava}, {Suhonen}, {Trzaska} and {{\'n}yst{\"o}}}]{kirs19}
\bibinfo{author}{{Kirsebom}, O.S.}, \bibinfo{author}{{Jones}, S.},
  \bibinfo{author}{{Str{\"o}mberg}, D.F.},
  \bibinfo{author}{{Mart{\'\i}nez-Pinedo}, G.}, \bibinfo{author}{{Langanke},
  K.}, \bibinfo{author}{{R{\"o}pke}, F.K.}, \bibinfo{author}{{Brown}, B.A.},
  \bibinfo{author}{{Eronen}, T.}, \bibinfo{author}{{Fynbo}, H.O.U.},
  \bibinfo{author}{{Hukkanen}, M.}, \bibinfo{author}{{Idini}, A.},
  \bibinfo{author}{{Jokinen}, A.}, \bibinfo{author}{{Kankainen}, A.},
  \bibinfo{author}{{Kostensalo}, J.}, \bibinfo{author}{{Moore}, I.},
  \bibinfo{author}{{M{\"o}ller}, H.}, \bibinfo{author}{{Ohlmann}, S.T.},
  \bibinfo{author}{{Penttil{\"a}}, H.}, \bibinfo{author}{{Riisager}, K.},
  \bibinfo{author}{{Rinta-Antila}, S.}, \bibinfo{author}{{Srivastava}, P.C.},
  \bibinfo{author}{{Suhonen}, J.}, \bibinfo{author}{{Trzaska}, W.H.},
  \bibinfo{author}{{{\'n}yst{\"o}}, J.}, \bibinfo{year}{2019}.
\newblock \bibinfo{title}{{Discovery of an Exceptionally Strong
  {\ensuremath{\beta}} -Decay Transition of $^{20}$F and Implications for the
  Fate of Intermediate-Mass Stars}}.
\newblock \bibinfo{journal}{Phys. Rev. Lett} \bibinfo{volume}{123},
  \bibinfo{pages}{262701}.
\newblock \DOIprefix\doi{10.1103/PhysRevLett.123.262701},
  \href{http://arxiv.org/abs/1905.09407}{\tt arXiv:1905.09407}.
\bibitem[{{Kitaura} et~al.(2006){Kitaura}, {Janka} and {Hillebrandt}}]{kita06}
\bibinfo{author}{{Kitaura}, F.S.}, \bibinfo{author}{{Janka}, H.T.},
  \bibinfo{author}{{Hillebrandt}, W.}, \bibinfo{year}{2006}.
\newblock \bibinfo{title}{{Explosions of O-Ne-Mg cores, the Crab supernova, and
  subluminous type II-P supernovae}}.
\newblock \bibinfo{journal}{A\&A} \bibinfo{volume}{450},
  \bibinfo{pages}{345--350}.
\bibitem[{{Kraft}(1964)}]{kraf64}
\bibinfo{author}{{Kraft}, R.P.}, \bibinfo{year}{1964}.
\newblock \bibinfo{title}{{Binary Stars among Cataclysmic Variables. III. Ten
  Old Novae.}}
\newblock \bibinfo{journal}{Astrophys. J.} \bibinfo{volume}{139},
  \bibinfo{pages}{457}.
\newblock \DOIprefix\doi{10.1086/147776}.
\bibitem[{{Krause} et~al.(2008){Krause}, {Birkmann} and {Usuda}}]{krau08}
\bibinfo{author}{{Krause}, O.}, \bibinfo{author}{{Birkmann}, S.M.},
  \bibinfo{author}{{Usuda}, T.a.}, \bibinfo{year}{2008}.
\newblock \bibinfo{title}{{The Cassiopeia A Supernova Was of Type IIb}}.
\newblock \bibinfo{journal}{Science} \bibinfo{volume}{320},
  \bibinfo{pages}{1195}.
\newblock \DOIprefix\doi{10.1126/science.1155788}.
\bibitem[{{Krautter} et~al.(1996){Krautter}, {Oegelman} and
  {Starrfield}}]{krau96}
\bibinfo{author}{{Krautter}, J.}, \bibinfo{author}{{Oegelman}, H.},
  \bibinfo{author}{{Starrfield}, S.e.a.}, \bibinfo{year}{1996}.
\newblock \bibinfo{title}{{ROSAT X-Ray Observations of Nova V1974 Cygni: The
  Rise and Fall of the Brightest Supersoft X-Ray Source}}.
\newblock \bibinfo{journal}{Astrophys. J.} \bibinfo{volume}{456},
  \bibinfo{pages}{788}.
\newblock \DOIprefix\doi{10.1086/176697}.
\bibitem[{{Kumagai} et~al.(1988){Kumagai}, {Itoh}, {Shigeyama}, {Nomoto} and
  {Nishimura}}]{kuma88}
\bibinfo{author}{{Kumagai}, S.}, \bibinfo{author}{{Itoh}, M.},
  \bibinfo{author}{{Shigeyama}, T.}, \bibinfo{author}{{Nomoto}, K.},
  \bibinfo{author}{{Nishimura}, J.}, \bibinfo{year}{1988}.
\newblock \bibinfo{title}{{Hard X-rays and gamma-rays from SN 1987A and mixing
  of the supernovaejecta.}}
\newblock \bibinfo{journal}{A\&A} \bibinfo{volume}{197},
  \bibinfo{pages}{L7--L10}.
\bibitem[{{Kumagai} and {Nomoto}(1997)}]{kuma97}
\bibinfo{author}{{Kumagai}, S.}, \bibinfo{author}{{Nomoto}, K.},
  \bibinfo{year}{1997}.
\newblock \bibinfo{title}{{Gamma-rays and X-rays from Type Ia supernovae}}, in:
  \bibinfo{editor}{{P.~Ruiz-Lapuente, R.~Canal, \& J.~Isern}} (Ed.),
  \bibinfo{booktitle}{NATO ASIC Proc. 486: Thermonuclear Supernovae}, pp.
  \bibinfo{pages}{515--+}.
\bibitem[{{Kurfess} et~al.(1992){Kurfess}, {Johnson} and {Kinzer}}]{kurf92}
\bibinfo{author}{{Kurfess}, J.D.}, \bibinfo{author}{{Johnson}, W.N.},
  \bibinfo{author}{{Kinzer}, R.L.e.a.}, \bibinfo{year}{1992}.
\newblock \bibinfo{title}{{Oriented Scintillation Spectrometer Experiment
  Observations of 57Co in SN 1987A}}.
\newblock \bibinfo{journal}{Astrophys. J. Lett.} \bibinfo{volume}{399},
  \bibinfo{pages}{L137}.
\bibitem[{{Kushnir} et~al.(2013){Kushnir}, {Katz}, {Dong}, {Livne} and
  {Fern{\'a}ndez}}]{kush13}
\bibinfo{author}{{Kushnir}, D.}, \bibinfo{author}{{Katz}, B.},
  \bibinfo{author}{{Dong}, S.}, \bibinfo{author}{{Livne}, E.},
  \bibinfo{author}{{Fern{\'a}ndez}, R.}, \bibinfo{year}{2013}.
\newblock \bibinfo{title}{{Head-on Collisions of White Dwarfs in Triple Systems
  Could Explain Type Ia Supernovae}}.
\newblock \bibinfo{journal}{Astrophys. J. Lett.} \bibinfo{volume}{778},
  \bibinfo{pages}{L37}.
\newblock \DOIprefix\doi{10.1088/2041-8205/778/2/L37}.
\bibitem[{{Larsson} et~al.(2011){Larsson}, {Fransson} and
  {{\"O}stlin}}]{lars11}
\bibinfo{author}{{Larsson}, J.}, \bibinfo{author}{{Fransson}, C.},
  \bibinfo{author}{{{\"O}stlin}, G.e.a.}, \bibinfo{year}{2011}.
\newblock \bibinfo{title}{{X-ray illumination of the ejecta of supernova
  1987A}}.
\newblock \bibinfo{journal}{Nature} \bibinfo{volume}{474},
  \bibinfo{pages}{484--486}.
\newblock \DOIprefix\doi{10.1038/nature10090}.
\bibitem[{{LeBlanc} and {Wilson}(1970)}]{lebl70}
\bibinfo{author}{{LeBlanc}, J.M.}, \bibinfo{author}{{Wilson}, J.R.},
  \bibinfo{year}{1970}.
\newblock \bibinfo{title}{{A Numerical Example of the Collapse of a Rotating
  Magnetized Star}}.
\newblock \bibinfo{journal}{APJ} \bibinfo{volume}{161}, \bibinfo{pages}{541}.
\newblock \DOIprefix\doi{10.1086/150558}.
\bibitem[{{Lebrun} et~al.(2003){Lebrun}, {Leray} and {Lavocat}}]{lebr03}
\bibinfo{author}{{Lebrun}, F.}, \bibinfo{author}{{Leray}, J.P.},
  \bibinfo{author}{{Lavocat}, P.e.a.}, \bibinfo{year}{2003}.
\newblock \bibinfo{title}{{ISGRI: The INTEGRAL Soft Gamma-Ray Imager}}.
\newblock \bibinfo{journal}{A\&A} \bibinfo{volume}{411},
  \bibinfo{pages}{L141--L148}.
\newblock \DOIprefix\doi{10.1051/0004-6361:20031367}.
\bibitem[{{Leising} et~al.(1995){Leising}, {Johnson}, {Kurfess}, {Clayton},
  {Grabelsky}, {Jung}, {Kinzer}, {Purcell}, {Strickman}, {The} and
  {Ulmer}}]{leis95}
\bibinfo{author}{{Leising}, M.D.}, \bibinfo{author}{{Johnson}, W.N.},
  \bibinfo{author}{{Kurfess}, J.D.}, \bibinfo{author}{{Clayton}, D.D.},
  \bibinfo{author}{{Grabelsky}, D.A.}, \bibinfo{author}{{Jung}, G.V.},
  \bibinfo{author}{{Kinzer}, R.L.}, \bibinfo{author}{{Purcell}, W.R.},
  \bibinfo{author}{{Strickman}, M.S.}, \bibinfo{author}{{The}, L.S.},
  \bibinfo{author}{{Ulmer}, M.P.}, \bibinfo{year}{1995}.
\newblock \bibinfo{title}{{Compton Gamma Ray Observatory OSSE Observations of
  SN 1991T}}.
\newblock \bibinfo{journal}{Astrophys. J.} \bibinfo{volume}{450},
  \bibinfo{pages}{805}.
\newblock \DOIprefix\doi{10.1086/176185}.
\bibitem[{{Leising} and {Share}(1990)}]{leis90}
\bibinfo{author}{{Leising}, M.D.}, \bibinfo{author}{{Share}, G.H.},
  \bibinfo{year}{1990}.
\newblock \bibinfo{title}{{The Gamma-Ray Light Curves of SN 1987A}}.
\newblock \bibinfo{journal}{Astrophys. J.} \bibinfo{volume}{357},
  \bibinfo{pages}{638}.
\bibitem[{{Li} et~al.(2011a){Li}, {Bloom} and {Podsiadlowski}}]{li11c}
\bibinfo{author}{{Li}, W.}, \bibinfo{author}{{Bloom}, J.S.},
  \bibinfo{author}{{Podsiadlowski}, P.e.a.}, \bibinfo{year}{2011}a.
\newblock \bibinfo{title}{{Exclusion of a luminous red giant as a companion
  star to the progenitor of supernova SN 2011fe}}.
\newblock \bibinfo{journal}{Nature} \bibinfo{volume}{480},
  \bibinfo{pages}{348--350}.
\newblock \DOIprefix\doi{10.1038/nature10646}.
\bibitem[{{Li} et~al.(2011b){Li}, {Chornock}, {Leaman}, {Filippenko},
  {Poznanski}, {Wang}, {Ganeshalingam} and {Mannucci}}]{li11b}
\bibinfo{author}{{Li}, W.}, \bibinfo{author}{{Chornock}, R.},
  \bibinfo{author}{{Leaman}, J.}, \bibinfo{author}{{Filippenko}, A.V.},
  \bibinfo{author}{{Poznanski}, D.}, \bibinfo{author}{{Wang}, X.},
  \bibinfo{author}{{Ganeshalingam}, M.}, \bibinfo{author}{{Mannucci}, F.},
  \bibinfo{year}{2011}b.
\newblock \bibinfo{title}{{Nearby supernova rates from the Lick Observatory
  Supernova Search - III. The rate-size relation, and the rates as a function
  of galaxy Hubble type and colour}}.
\newblock \bibinfo{journal}{Mon. Not. R. Astron. Soc.} \bibinfo{volume}{412},
  \bibinfo{pages}{1473--1507}.
\newblock \DOIprefix\doi{10.1111/j.1365-2966.2011.18162.x}.
\bibitem[{{Li} et~al.(2003){Li}, {Filippenko} and {Chornock}}]{li03}
\bibinfo{author}{{Li}, W.}, \bibinfo{author}{{Filippenko}, A.V.},
  \bibinfo{author}{{Chornock}, R.e.a.}, \bibinfo{year}{2003}.
\newblock \bibinfo{title}{{SN 2002cx: The Most Peculiar Known Type Ia
  Supernova}}.
\newblock \bibinfo{journal}{Publ. Astron. Soc. Pac.} \bibinfo{volume}{115},
  \bibinfo{pages}{453--473}.
\newblock \DOIprefix\doi{10.1086/374200}.
\bibitem[{{Li} et~al.(2001){Li}, {Filippenko} and {Gates}}]{li01}
\bibinfo{author}{{Li}, W.}, \bibinfo{author}{{Filippenko}, A.V.},
  \bibinfo{author}{{Gates}, E.e.a.}, \bibinfo{year}{2001}.
\newblock \bibinfo{title}{{The Unique Type Ia Supernova 2000cx in NGC 524}}.
\newblock \bibinfo{journal}{pasp} \bibinfo{volume}{113},
  \bibinfo{pages}{1178--1204}.
\newblock \DOIprefix\doi{10.1086/323355}.
\bibitem[{{Li} et~al.(2011c){Li}, {Leaman}, {Chornock} and
  {Filippenko}}]{li11a}
\bibinfo{author}{{Li}, W.}, \bibinfo{author}{{Leaman}, J.},
  \bibinfo{author}{{Chornock}, R.}, \bibinfo{author}{{Filippenko}, e.a.},
  \bibinfo{year}{2011}c.
\newblock \bibinfo{title}{{Nearby supernova rates from the Lick Observatory
  Supernova Search - II. The observed luminosity functions and fractions of
  supernovae in a complete sample}}.
\newblock \bibinfo{journal}{Mon. Not. R. Astron. Soc.} \bibinfo{volume}{412},
  \bibinfo{pages}{1441--1472}.
\newblock \DOIprefix\doi{10.1111/j.1365-2966.2011.18160.x}.
\bibitem[{{Lichti} et~al.(1994){Lichti}, {Bennett}, {den Herder}, {Diehl},
  {Morris}, {Ryan}, {Schoenfelder}, {Steinle}, {Strong} and {Winkler}}]{lich94}
\bibinfo{author}{{Lichti}, G.G.}, \bibinfo{author}{{Bennett}, K.},
  \bibinfo{author}{{den Herder}, J.W.}, \bibinfo{author}{{Diehl}, R.},
  \bibinfo{author}{{Morris}, D.}, \bibinfo{author}{{Ryan}, J.},
  \bibinfo{author}{{Schoenfelder}, V.}, \bibinfo{author}{{Steinle}, H.},
  \bibinfo{author}{{Strong}, A.W.}, \bibinfo{author}{{Winkler}, C.},
  \bibinfo{year}{1994}.
\newblock \bibinfo{title}{{COMPTEL upper limits on gamma-ray line emission from
  Supernova 1991T.}}
\newblock \bibinfo{journal}{A\&A} \bibinfo{volume}{292}, \bibinfo{pages}{569}.
\bibitem[{{Livio}(1992)}]{livi92}
\bibinfo{author}{{Livio}, M.}, \bibinfo{year}{1992}.
\newblock \bibinfo{title}{{Classical Novae and the Extragalactic Distance
  Scale}}.
\newblock \bibinfo{journal}{Astrophys. J.} \bibinfo{volume}{393},
  \bibinfo{pages}{516}.
\newblock \DOIprefix\doi{10.1086/171524}.
\bibitem[{{Livio} and {Riess}(2003)}]{livi03}
\bibinfo{author}{{Livio}, M.}, \bibinfo{author}{{Riess}, A.G.},
  \bibinfo{year}{2003}.
\newblock \bibinfo{title}{{Have the Elusive Progenitors of Type Ia Supernovae
  Been Discovered?}}
\newblock \bibinfo{journal}{Astrophys. J. Lett.} \bibinfo{volume}{594},
  \bibinfo{pages}{L93--L94}.
\newblock \DOIprefix\doi{10.1086/378765}.
\bibitem[{{Livne} and {Arnett}(1995)}]{livn95}
\bibinfo{author}{{Livne}, E.}, \bibinfo{author}{{Arnett}, D.},
  \bibinfo{year}{1995}.
\newblock \bibinfo{title}{{Explosions of Sub--Chandrasekhar Mass White Dwarfs
  in Two Dimensions}}.
\newblock \bibinfo{journal}{Astrophys. J.} \bibinfo{volume}{452},
  \bibinfo{pages}{62}.
\newblock \DOIprefix\doi{10.1086/176279}.
\bibitem[{{Lopez} and {Fesen}(2018)}]{lope18}
\bibinfo{author}{{Lopez}, L.A.}, \bibinfo{author}{{Fesen}, R.A.},
  \bibinfo{year}{2018}.
\newblock \bibinfo{title}{{The Morphologies and Kinematics of Supernova
  Remnants}}.
\newblock \bibinfo{journal}{Space Sci. Rev.} \bibinfo{volume}{214},
  \bibinfo{pages}{44}.
\newblock \DOIprefix\doi{10.1007/s11214-018-0481-x},
  \href{http://arxiv.org/abs/1804.00024}{\tt arXiv:1804.00024}.
\bibitem[{{Lor{\'e}n-Aguilar} et~al.(2009){Lor{\'e}n-Aguilar}, {Isern} and
  {Garc{\'{\i}}a-Berro}}]{lore09}
\bibinfo{author}{{Lor{\'e}n-Aguilar}, P.}, \bibinfo{author}{{Isern}, J.},
  \bibinfo{author}{{Garc{\'{\i}}a-Berro}, E.}, \bibinfo{year}{2009}.
\newblock \bibinfo{title}{{High-resolution smoothed particle hydrodynamics
  simulations of the merger of binary white dwarfs}}.
\newblock \bibinfo{journal}{A\&A} \bibinfo{volume}{500},
  \bibinfo{pages}{1193--1205}.
\newblock \DOIprefix\doi{10.1051/0004-6361/200811060}.
\bibitem[{{Lor{\'e}n-Aguilar} et~al.(2010){Lor{\'e}n-Aguilar}, {Isern} and
  {Garc{\'\i}a-Berro}}]{lore10}
\bibinfo{author}{{Lor{\'e}n-Aguilar}, P.}, \bibinfo{author}{{Isern}, J.},
  \bibinfo{author}{{Garc{\'\i}a-Berro}, E.}, \bibinfo{year}{2010}.
\newblock \bibinfo{title}{{Smoothed particle hydrodynamics simulations of white
  dwarf collisions and close encounters}}.
\newblock \bibinfo{journal}{Mon. Not. R. Astron. Soc.} \bibinfo{volume}{406},
  \bibinfo{pages}{2749--2763}.
\newblock \DOIprefix\doi{10.1111/j.1365-2966.2010.16878.x}.
\bibitem[{{Lund} et~al.(2003){Lund}, {Budtz-J{\o}rgensen} and
  {Westergaard}}]{lund03}
\bibinfo{author}{{Lund}, N.}, \bibinfo{author}{{Budtz-J{\o}rgensen}, C.},
  \bibinfo{author}{{Westergaard}, N.J.e.a.}, \bibinfo{year}{2003}.
\newblock \bibinfo{title}{{JEM-X: The X-ray monitor aboard INTEGRAL}}.
\newblock \bibinfo{journal}{A\&A} \bibinfo{volume}{411},
  \bibinfo{pages}{L231--L238}.
\newblock \DOIprefix\doi{10.1051/0004-6361:20031358}.
\bibitem[{{Lunnan} et~al.(2017){Lunnan}, {Kasliwal} and {Cao}}]{lunn17}
\bibinfo{author}{{Lunnan}, R.}, \bibinfo{author}{{Kasliwal}, M.M.},
  \bibinfo{author}{{Cao}, Y.e.a.}, \bibinfo{year}{2017}.
\newblock \bibinfo{title}{{Two New Calcium-rich Gap Transients in Group and
  Cluster Environments}}.
\newblock \bibinfo{journal}{Astrophys. J.} \bibinfo{volume}{836},
  \bibinfo{pages}{60}.
\newblock \DOIprefix\doi{10.3847/1538-4357/836/1/60}.
\bibitem[{{Lyman} et~al.(2013){Lyman}, {James} and {Perets}}]{lyma13}
\bibinfo{author}{{Lyman}, J.D.}, \bibinfo{author}{{James}, P.A.},
  \bibinfo{author}{{Perets}, H.B.e.a.}, \bibinfo{year}{2013}.
\newblock \bibinfo{title}{{Environment-derived constraints on the progenitors
  of low-luminosity Type I supernovae}}.
\newblock \bibinfo{journal}{Mon. Not. R. Astron. Soc.} \bibinfo{volume}{434},
  \bibinfo{pages}{527--541}.
\newblock \DOIprefix\doi{10.1093/mnras/stt1038}.
\bibitem[{{Maeda} et~al.(2010){Maeda}, {R{\"o}pke}, {Fink}, {Hillebrandt},
  {Travaglio} and {Thielemann}}]{maed10}
\bibinfo{author}{{Maeda}, K.}, \bibinfo{author}{{R{\"o}pke}, F.K.},
  \bibinfo{author}{{Fink}, M.}, \bibinfo{author}{{Hillebrandt}, W.},
  \bibinfo{author}{{Travaglio}, C.}, \bibinfo{author}{{Thielemann}, F.K.},
  \bibinfo{year}{2010}.
\newblock \bibinfo{title}{{Nucleosynthesis in Two-Dimensional Delayed
  Detonation Models of Type Ia Supernova Explosions}}.
\newblock \bibinfo{journal}{Astrophys. J.} \bibinfo{volume}{712},
  \bibinfo{pages}{624--638}.
\newblock \DOIprefix\doi{10.1088/0004-637X/712/1/624},
  \href{http://arxiv.org/abs/1002.2153}{\tt arXiv:1002.2153}.
\bibitem[{{Magkotsios} et~al.(2010){Magkotsios}, {Timmes}, {Hungerford},
  {Fryer}, {Young} and {Wiescher}}]{magk10}
\bibinfo{author}{{Magkotsios}, G.}, \bibinfo{author}{{Timmes}, F.X.},
  \bibinfo{author}{{Hungerford}, A.L.}, \bibinfo{author}{{Fryer}, C.L.},
  \bibinfo{author}{{Young}, P.A.}, \bibinfo{author}{{Wiescher}, M.},
  \bibinfo{year}{2010}.
\newblock \bibinfo{title}{{Trends in $^{44}$Ti and $^{56}$Ni from Core-collapse
  Supernovae}}.
\newblock \bibinfo{journal}{Astrophys. J. Suppl. S} \bibinfo{volume}{191},
  \bibinfo{pages}{66--95}.
\newblock \DOIprefix\doi{10.1088/0067-0049/191/1/66},
  \href{http://arxiv.org/abs/1009.3175}{\tt arXiv:1009.3175}.
\bibitem[{{Maoz} et~al.(2014){Maoz}, {Mannucci} and {Nelemans}}]{maoz14}
\bibinfo{author}{{Maoz}, D.}, \bibinfo{author}{{Mannucci}, F.},
  \bibinfo{author}{{Nelemans}, G.}, \bibinfo{year}{2014}.
\newblock \bibinfo{title}{{Observational Clues to the Progenitors of Type Ia
  Supernovae}}.
\newblock \bibinfo{journal}{Annu. Rev. Astron. Astrophys.}
  \bibinfo{volume}{52}, \bibinfo{pages}{107--170}.
\newblock \DOIprefix\doi{10.1146/annurev-astro-082812-141031}.
\bibitem[{{Maoz} et~al.(2010){Maoz}, {Sharon} and {Gal-Yam}}]{maoz10}
\bibinfo{author}{{Maoz}, D.}, \bibinfo{author}{{Sharon}, K.},
  \bibinfo{author}{{Gal-Yam}, A.}, \bibinfo{year}{2010}.
\newblock \bibinfo{title}{{The Supernova Delay Time Distribution in Galaxy
  Clusters and Implications for Type-Ia Progenitors and Metal Enrichment}}.
\newblock \bibinfo{journal}{Astrophys. J.} \bibinfo{volume}{722},
  \bibinfo{pages}{1879--1894}.
\newblock \DOIprefix\doi{10.1088/0004-637X/722/2/1879}.
\bibitem[{{Martin} et~al.(2009){Martin}, {Kn{\"o}dlseder}, {Vink},
  {Decourchelle} and {Renaud}}]{mart09}
\bibinfo{author}{{Martin}, P.}, \bibinfo{author}{{Kn{\"o}dlseder}, J.},
  \bibinfo{author}{{Vink}, J.}, \bibinfo{author}{{Decourchelle}, A.},
  \bibinfo{author}{{Renaud}, M.}, \bibinfo{year}{2009}.
\newblock \bibinfo{title}{{Constraints on the kinematics of the $^{44}$Ti
  ejecta of Cassiopeia A from INTEGRAL/SPI}}.
\newblock \bibinfo{journal}{A\&A} \bibinfo{volume}{502},
  \bibinfo{pages}{131--137}.
\newblock \DOIprefix\doi{10.1051/0004-6361/200809735}.
\bibitem[{{Mart{\'\i}nez-Pinedo} et~al.(2014){Mart{\'\i}nez-Pinedo}, {Lam},
  {Langanke}, {Zegers} and {Sullivan}}]{mart14}
\bibinfo{author}{{Mart{\'\i}nez-Pinedo}, G.}, \bibinfo{author}{{Lam}, Y.H.},
  \bibinfo{author}{{Langanke}, K.}, \bibinfo{author}{{Zegers}, R.G.T.},
  \bibinfo{author}{{Sullivan}, C.}, \bibinfo{year}{2014}.
\newblock \bibinfo{title}{{Astrophysical weak-interaction rates for selected A
  =20 and A =24 nuclei}}.
\newblock \bibinfo{journal}{Phys. Rev. C} \bibinfo{volume}{89}.
\bibitem[{{Mart{\'\i}nez-Rodr{\'\i}guez}
  et~al.(2017){Mart{\'\i}nez-Rodr{\'\i}guez}, {Badenes}, {Yamaguchi}, {Bravo},
  {Timmes}, {Miles}, {Townsley}, {Piro}, {Mori}, {Andrews} and {Park}}]{mart17}
\bibinfo{author}{{Mart{\'\i}nez-Rodr{\'\i}guez}, H.},
  \bibinfo{author}{{Badenes}, C.}, \bibinfo{author}{{Yamaguchi}, H.},
  \bibinfo{author}{{Bravo}, E.}, \bibinfo{author}{{Timmes}, F.X.},
  \bibinfo{author}{{Miles}, B.J.}, \bibinfo{author}{{Townsley}, D.M.},
  \bibinfo{author}{{Piro}, A.L.}, \bibinfo{author}{{Mori}, H.},
  \bibinfo{author}{{Andrews}, B.}, \bibinfo{author}{{Park}, S.},
  \bibinfo{year}{2017}.
\newblock \bibinfo{title}{{Observational Evidence for High Neutronization in
  Supernova Remnants: Implications for Type Ia Supernova Progenitors}}.
\newblock \bibinfo{journal}{Astrophys. J.} \bibinfo{volume}{843},
  \bibinfo{pages}{35}.
\newblock \DOIprefix\doi{10.3847/1538-4357/aa72f8},
  \href{http://arxiv.org/abs/1701.07073}{\tt arXiv:1701.07073}.
\bibitem[{{Mart{\'\i}nez-Rodr{\'\i}guez}
  et~al.(2016){Mart{\'\i}nez-Rodr{\'\i}guez}, {Piro}, {Schwab} and
  {Badenes}}]{mart16}
\bibinfo{author}{{Mart{\'\i}nez-Rodr{\'\i}guez}, H.}, \bibinfo{author}{{Piro},
  A.L.}, \bibinfo{author}{{Schwab}, J.}, \bibinfo{author}{{Badenes}, C.},
  \bibinfo{year}{2016}.
\newblock \bibinfo{title}{{Neutronization During Carbon Simmering In Type Ia
  Supernova Progenitors}}.
\newblock \bibinfo{journal}{Astrophys. J.} \bibinfo{volume}{825},
  \bibinfo{pages}{57}.
\newblock \DOIprefix\doi{10.3847/0004-637X/825/1/57}.
\bibitem[{{Mas-Hesse} et~al.(2003){Mas-Hesse}, {Gim{\'e}nez} and
  {Culhane}}]{mash03}
\bibinfo{author}{{Mas-Hesse}, J.M.}, \bibinfo{author}{{Gim{\'e}nez}, A.},
  \bibinfo{author}{{Culhane}, J.L.e.a.}, \bibinfo{year}{2003}.
\newblock \bibinfo{title}{{OMC: An Optical Monitoring Camera for INTEGRAL.
  Instrument description and performance}}.
\newblock \bibinfo{journal}{A\&A} \bibinfo{volume}{411},
  \bibinfo{pages}{L261--L268}.
\newblock \DOIprefix\doi{10.1051/0004-6361:20031418}.
\bibitem[{{Matz} et~al.(1988){Matz}, {Share}, {Leising}, {Chupp} and
  {Vestrand}}]{matz88}
\bibinfo{author}{{Matz}, S.M.}, \bibinfo{author}{{Share}, G.H.},
  \bibinfo{author}{{Leising}, M.D.}, \bibinfo{author}{{Chupp}, E.L.},
  \bibinfo{author}{{Vestrand}, W.T.}, \bibinfo{year}{1988}.
\newblock \bibinfo{title}{{Gamma-ray line emission from SN1987A}}.
\newblock \bibinfo{journal}{Nature} \bibinfo{volume}{331},
  \bibinfo{pages}{416--418}.
\bibitem[{{Mazzali} et~al.(2015){Mazzali}, {Sullivan} and
  {Filippenko}}]{mazz15}
\bibinfo{author}{{Mazzali}, P.A.}, \bibinfo{author}{{Sullivan}, M.},
  \bibinfo{author}{{Filippenko}, A.V.e.a.}, \bibinfo{year}{2015}.
\newblock \bibinfo{title}{{Nebular spectra and abundance tomography of the Type
  Ia supernova SN 2011fe: a normal SN Ia with a stable Fe core}}.
\newblock \bibinfo{journal}{Mon. Not. R. Astron. Soc.} \bibinfo{volume}{450},
  \bibinfo{pages}{2631--2643}.
\newblock \DOIprefix\doi{10.1093/mnras/stv761}.
\bibitem[{{McCray} et~al.(1987){McCray}, {Shull} and {Sutherland}}]{mccr87}
\bibinfo{author}{{McCray}, R.}, \bibinfo{author}{{Shull}, J.M.},
  \bibinfo{author}{{Sutherland}, P.}, \bibinfo{year}{1987}.
\newblock \bibinfo{title}{{Inside Supernova 1987A}}.
\newblock \bibinfo{journal}{Astrophys. J. Lett.} \bibinfo{volume}{317},
  \bibinfo{pages}{L73}.
\newblock \DOIprefix\doi{10.1086/184915}.
\bibitem[{{McCully} et~al.(2014){McCully}, {Jha} and {Foley}}]{mccu14}
\bibinfo{author}{{McCully}, C.}, \bibinfo{author}{{Jha}, S.W.},
  \bibinfo{author}{{Foley}, R.J.e.a.}, \bibinfo{year}{2014}.
\newblock \bibinfo{title}{{A luminous, blue progenitor system for the type Iax
  supernova 2012Z}}.
\newblock \bibinfo{journal}{Nature} \bibinfo{volume}{512},
  \bibinfo{pages}{54--56}.
\newblock \DOIprefix\doi{10.1038/nature13615}.
\bibitem[{{McEnery} et~al.(2019){McEnery}, {van der Horst}, {Dominguez},
  {Moiseev}, {Marcowith}, {Harding}, {Lien}, {Giuliani}, {Inglis}, {Ansoldi},
  {Stamerra}, {Manousakis}, {Strong}, {Bambi}, {Patricelli}, {Baring},
  {Barrio}, {Bastieri}, {Fields}, {Beacom}, {Beckmann}, {Bednarek}, {Rani},
  {Boggs}, {Bolotnikov}, {Cenko}, {Buckley}, {Grefenstette}, {Hui}, {Pittori},
  {Prescod-Weinstein}, {Shrader}, {Gouiffes}, {Kierans}, {Wilson-Hodge},
  {D'Ammando}, {Castro}, {Kocveski}, {Gasparrini}, {Thompson}, {Williams}, {De
  Angelis}, {Bernard}, {Digel}, {Morcuende}, {Charles}, {Bissaldi}, {Hays},
  {Ferrara}, {Bozzo}, {Grove}, {Wulf}, {Bottacini}, {Caroli}, {Kislat},
  {Oikonomou}, {Giordano}, {Longo}, {Fryer}, {Fukazawa}, {Georganopoulos}, {De
  Nolfo}, {Vianello}, {Kanbach}, {Younes}, {Blumer}, {Hartmann}, {Hernanz},
  {Takahashi}, {Li}, {Agudo}, {Moskalenko}, {Stumke}, {Grenier}, {Smith},
  {Rodi}, {Perkins}, {Gelfand}, {Holder}, {Knodlseder}, {Kopp}, {Lenain},
  {{\'A}lvarez}, {Metcalfe}, {Krizmanic}, {Stephen}, {Hewitt}, {Mitchell},
  {Harding}, {Tomsick}, {Racusin}, {Finke}, {Kargaltsev}, {Klimenko},
  {Krawczynski}, {Smith}, {Kubo}, {Di Venere}, {Marcotulli}, {Lommler},
  {Parker}, {Baldini}, {Foffano}, {Zampieri}, {Tibaldo}, {Petropoulou},
  {Ajello}, {Meyer}, {L{\'o}pez}, {McConnell}, {Boettcher}, {Cardillo},
  {Martinez}, {Kerr}, {Mazziotta}, {McEnery}, {Di Mauro}, {Wood}, {Meyer},
  {Briggs}, {De Becker}, {Lovellette}, {Doro}, {Sanchez-Conde}, {Moss},
  {Mizuno}, {Rib{\'o}}, {Nakazawa}, {Neilson}, {Auricchio}, {Omodei},
  {Oberlack}, {Ohno}, {Orland o}, {Otte}, {Coppi}, {Bloser}, {Zhang},
  {Laurent}, {Pohl}, {Prand ini}, {Shawhan}, {Caputo}, {Campana}, {Rando},
  {Woolf}, {Johnson}, {Mignani}, {Walter}, {Ojha}, {da Silva}, {Dietrich},
  {Funk}, {Zane}, {Anton}, {Buson}, {Cutini}, {Saz Parkinson}, {Schirato},
  {Griffin}, {Kaufmann}, {Stawarz}, {Ciprini}, {Del Sordo}, {Jones}, {Guiriec},
  {Tajima}, {Cheung}, {The}, {Venters}, {Porter}, {Linden}, {Barres}, {Paliya},
  {Bozhilov}, {Vestrand}, {Tatischeff}, {Chen}, {Wang}, {Tanaka}, {Uhm},
  {Zhang}, {Zimmer}, {Zoglauer} and {Wadiasingh}}]{mcen19}
\bibinfo{author}{{McEnery}, J.}, \bibinfo{author}{{van der Horst}, A.},
  \bibinfo{author}{{Dominguez}, A.}, \bibinfo{author}{{Moiseev}, A.},
  \bibinfo{author}{{Marcowith}, A.r.}, \bibinfo{author}{{Harding}, A.},
  \bibinfo{author}{{Lien}, A.}, \bibinfo{author}{{Giuliani}, A.},
  \bibinfo{author}{{Inglis}, A.}, \bibinfo{author}{{Ansoldi}, S.},
  \bibinfo{author}{{Stamerra}, A.}, \bibinfo{author}{{Manousakis}, A.},
  \bibinfo{author}{{Strong}, A.}, \bibinfo{author}{{Bambi}, C.},
  \bibinfo{author}{{Patricelli}, B.}, \bibinfo{author}{{Baring}, M.},
  \bibinfo{author}{{Barrio}, J.A.}, \bibinfo{author}{{Bastieri}, D.},
  \bibinfo{author}{{Fields}, B.}, \bibinfo{author}{{Beacom}, J.},
  \bibinfo{author}{{Beckmann}, V.}, \bibinfo{author}{{Bednarek}, W.},
  \bibinfo{author}{{Rani}, B.}, \bibinfo{author}{{Boggs}, S.},
  \bibinfo{author}{{Bolotnikov}, A.}, \bibinfo{author}{{Cenko}, S.B.},
  \bibinfo{author}{{Buckley}, J.}, \bibinfo{author}{{Grefenstette}, B.},
  \bibinfo{author}{{Hui}, M.}, \bibinfo{author}{{Pittori}, C.},
  \bibinfo{author}{{Prescod-Weinstein}, C.}, \bibinfo{author}{{Shrader}, C.},
  \bibinfo{author}{{Gouiffes}, C.}, \bibinfo{author}{{Kierans}, C.},
  \bibinfo{author}{{Wilson-Hodge}, C.}, \bibinfo{author}{{D'Ammando}, F.},
  \bibinfo{author}{{Castro}, D.}, \bibinfo{author}{{Kocveski}, D.},
  \bibinfo{author}{{Gasparrini}, D.}, \bibinfo{author}{{Thompson}, D.},
  \bibinfo{author}{{Williams}, D.}, \bibinfo{author}{{De Angelis}, A.},
  \bibinfo{author}{{Bernard}, D.}, \bibinfo{author}{{Digel}, S.},
  \bibinfo{author}{{Morcuende}, D.}, \bibinfo{author}{{Charles}, E.},
  \bibinfo{author}{{Bissaldi}, E.}, \bibinfo{author}{{Hays}, E.},
  \bibinfo{author}{{Ferrara}, E.}, \bibinfo{author}{{Bozzo}, E.},
  \bibinfo{author}{{Grove}, E.}, \bibinfo{author}{{Wulf}, E.},
  \bibinfo{author}{{Bottacini}, E.}, \bibinfo{author}{{Caroli}, E.},
  \bibinfo{author}{{Kislat}, F.}, \bibinfo{author}{{Oikonomou}, F.},
  \bibinfo{author}{{Giordano}, F.}, \bibinfo{author}{{Longo}, F.},
  \bibinfo{author}{{Fryer}, C.}, \bibinfo{author}{{Fukazawa}, Y.},
  \bibinfo{author}{{Georganopoulos}, M.}, \bibinfo{author}{{De Nolfo}, G.},
  \bibinfo{author}{{Vianello}, G.}, \bibinfo{author}{{Kanbach}, G.},
  \bibinfo{author}{{Younes}, G.}, \bibinfo{author}{{Blumer}, H.},
  \bibinfo{author}{{Hartmann}, D.}, \bibinfo{author}{{Hernanz}, M.},
  \bibinfo{author}{{Takahashi}, H.}, \bibinfo{author}{{Li}, H.},
  \bibinfo{author}{{Agudo}, I.}, \bibinfo{author}{{Moskalenko}, I.},
  \bibinfo{author}{{Stumke}, I.}, \bibinfo{author}{{Grenier}, I.},
  \bibinfo{author}{{Smith}, J.}, \bibinfo{author}{{Rodi}, J.},
  \bibinfo{author}{{Perkins}, J.}, \bibinfo{author}{{Gelfand}, J.},
  \bibinfo{author}{{Holder}, J.}, \bibinfo{author}{{Knodlseder}, J.},
  \bibinfo{author}{{Kopp}, J.}, \bibinfo{author}{{Lenain}, J.P.},
  \bibinfo{author}{{{\'A}lvarez}, J.M.}, \bibinfo{author}{{Metcalfe}, J.},
  \bibinfo{author}{{Krizmanic}, J.}, \bibinfo{author}{{Stephen}, J.B.},
  \bibinfo{author}{{Hewitt}, J.}, \bibinfo{author}{{Mitchell}, J.},
  \bibinfo{author}{{Harding}, P.}, \bibinfo{author}{{Tomsick}, J.},
  \bibinfo{author}{{Racusin}, J.}, \bibinfo{author}{{Finke}, J.},
  \bibinfo{author}{{Kargaltsev}, O.}, \bibinfo{author}{{Klimenko}, A.V.},
  \bibinfo{author}{{Krawczynski}, H.}, \bibinfo{author}{{Smith}, K.},
  \bibinfo{author}{{Kubo}, H.}, \bibinfo{author}{{Di Venere}, L.},
  \bibinfo{author}{{Marcotulli}, L.}, \bibinfo{author}{{Lommler}, J.},
  \bibinfo{author}{{Parker}, L.}, \bibinfo{author}{{Baldini}, L.},
  \bibinfo{author}{{Foffano}, L.}, \bibinfo{author}{{Zampieri}, L.},
  \bibinfo{author}{{Tibaldo}, L.}, \bibinfo{author}{{Petropoulou}, M.},
  \bibinfo{author}{{Ajello}, M.}, \bibinfo{author}{{Meyer}, M.},
  \bibinfo{author}{{L{\'o}pez}, M.}, \bibinfo{author}{{McConnell}, M.},
  \bibinfo{author}{{Boettcher}, M.}, \bibinfo{author}{{Cardillo}, M.},
  \bibinfo{author}{{Martinez}, M.}, \bibinfo{author}{{Kerr}, M.},
  \bibinfo{author}{{Mazziotta}, M.N.}, \bibinfo{author}{{McEnery}, J.},
  \bibinfo{author}{{Di Mauro}, M.}, \bibinfo{author}{{Wood}, M.},
  \bibinfo{author}{{Meyer}, E.}, \bibinfo{author}{{Briggs}, M.},
  \bibinfo{author}{{De Becker}, M.}, \bibinfo{author}{{Lovellette}, M.},
  \bibinfo{author}{{Doro}, M.}, \bibinfo{author}{{Sanchez-Conde}, M.A.},
  \bibinfo{author}{{Moss}, M.}, \bibinfo{author}{{Mizuno}, T.},
  \bibinfo{author}{{Rib{\'o}}, M.}, \bibinfo{author}{{Nakazawa}, K.},
  \bibinfo{author}{{Neilson}, N.K.}, \bibinfo{author}{{Auricchio}, N.},
  \bibinfo{author}{{Omodei}, N.}, \bibinfo{author}{{Oberlack}, U.},
  \bibinfo{author}{{Ohno}, M.}, \bibinfo{author}{{Orland o}, E.},
  \bibinfo{author}{{Otte}, N.}, \bibinfo{author}{{Coppi}, P.},
  \bibinfo{author}{{Bloser}, P.}, \bibinfo{author}{{Zhang}, H.},
  \bibinfo{author}{{Laurent}, P.}, \bibinfo{author}{{Pohl}, M.},
  \bibinfo{author}{{Prand ini}, E.}, \bibinfo{author}{{Shawhan}, P.},
  \bibinfo{author}{{Caputo}, R.}, \bibinfo{author}{{Campana}, R.},
  \bibinfo{author}{{Rando}, R.}, \bibinfo{author}{{Woolf}, R.},
  \bibinfo{author}{{Johnson}, R.}, \bibinfo{author}{{Mignani}, R.},
  \bibinfo{author}{{Walter}, R.}, \bibinfo{author}{{Ojha}, R.},
  \bibinfo{author}{{da Silva}, R.C.}, \bibinfo{author}{{Dietrich}, S.},
  \bibinfo{author}{{Funk}, S.}, \bibinfo{author}{{Zane}, S.},
  \bibinfo{author}{{Anton}, S.}, \bibinfo{author}{{Buson}, S.},
  \bibinfo{author}{{Cutini}, S.}, \bibinfo{author}{{Saz Parkinson}, P.},
  \bibinfo{author}{{Schirato}, R.}, \bibinfo{author}{{Griffin}, S.},
  \bibinfo{author}{{Kaufmann}, S.}, \bibinfo{author}{{Stawarz}, L.},
  \bibinfo{author}{{Ciprini}, S.}, \bibinfo{author}{{Del Sordo}, S.},
  \bibinfo{author}{{Jones}, S.}, \bibinfo{author}{{Guiriec}, S.},
  \bibinfo{author}{{Tajima}, H.}, \bibinfo{author}{{Cheung}, T.},
  \bibinfo{author}{{The}, L.S.}, \bibinfo{author}{{Venters}, T.},
  \bibinfo{author}{{Porter}, T.}, \bibinfo{author}{{Linden}, T.},
  \bibinfo{author}{{Barres}, U.}, \bibinfo{author}{{Paliya}, V.S.},
  \bibinfo{author}{{Bozhilov}, V.}, \bibinfo{author}{{Vestrand}, T.},
  \bibinfo{author}{{Tatischeff}, V.}, \bibinfo{author}{{Chen}, W.},
  \bibinfo{author}{{Wang}, X.}, \bibinfo{author}{{Tanaka}, Y.},
  \bibinfo{author}{{Uhm}, L.}, \bibinfo{author}{{Zhang}, B.},
  \bibinfo{author}{{Zimmer}, S.}, \bibinfo{author}{{Zoglauer}, A.},
  \bibinfo{author}{{Wadiasingh}, Z.}, \bibinfo{year}{2019}.
\newblock \bibinfo{title}{{All-sky Medium Energy Gamma-ray Observatory:
  Exploring the Extreme Multimessenger Universe}}, in:
  \bibinfo{booktitle}{Bulletin of the American Astronomical Society}, p.
  \bibinfo{pages}{245}.
\newblock \href{http://arxiv.org/abs/1907.07558}{\tt arXiv:1907.07558}.
\bibitem[{{McMillan}(2017)}]{mcmi17}
\bibinfo{author}{{McMillan}, P.J.}, \bibinfo{year}{2017}.
\newblock \bibinfo{title}{{The mass distribution and gravitational potential of
  the Milky Way}}.
\newblock \bibinfo{journal}{Mon. Not. R. Astron. Soc.} \bibinfo{volume}{465},
  \bibinfo{pages}{76--94}.
\newblock \DOIprefix\doi{10.1093/mnras/stw2759}.
\bibitem[{{Meneguzzi} et~al.(1971){Meneguzzi}, {Audouze} and {Reeves}}]{mene71}
\bibinfo{author}{{Meneguzzi}, M.}, \bibinfo{author}{{Audouze}, J.},
  \bibinfo{author}{{Reeves}, H.}, \bibinfo{year}{1971}.
\newblock \bibinfo{title}{{The production of the elements Li, Be, B by galactic
  cosmic rays in space and its relation with stellar observations.}}
\newblock \bibinfo{journal}{A\&A} \bibinfo{volume}{15}, \bibinfo{pages}{337}.
\bibitem[{{Miller} et~al.(2019){Miller}, {Ajello}, {Beacom}, {Bloser},
  {Burrows}, {Fryer}, {Goldsten}, {Hartmann}, {Hoeflich}, {Hungerford},
  {Lawrence}, {Leising}, {Milne}, {Peplowski}, {Shirazi}, {Sukhbold}, {The},
  {Yokley} and {Young}}]{mill19}
\bibinfo{author}{{Miller}, R.}, \bibinfo{author}{{Ajello}, M.},
  \bibinfo{author}{{Beacom}, J.F.}, \bibinfo{author}{{Bloser}, P.F.},
  \bibinfo{author}{{Burrows}, A.}, \bibinfo{author}{{Fryer}, C.L.},
  \bibinfo{author}{{Goldsten}, J.O.}, \bibinfo{author}{{Hartmann}, D.H.},
  \bibinfo{author}{{Hoeflich}, P.}, \bibinfo{author}{{Hungerford}, A.},
  \bibinfo{author}{{Lawrence}, D.J.}, \bibinfo{author}{{Leising}, M.D.},
  \bibinfo{author}{{Milne}, P.}, \bibinfo{author}{{Peplowski}, P.N.},
  \bibinfo{author}{{Shirazi}, F.}, \bibinfo{author}{{Sukhbold}, T.},
  \bibinfo{author}{{The}, L.S.}, \bibinfo{author}{{Yokley}, Z.},
  \bibinfo{author}{{Young}, C.A.}, \bibinfo{year}{2019}.
\newblock \bibinfo{title}{{Ex Luna Scientia: The Lunar Occultation eXplorer
  (LOX)}}, in: \bibinfo{booktitle}{Bulletin of the American Astronomical
  Society}, p. \bibinfo{pages}{123}.
\newblock \href{http://arxiv.org/abs/1907.07005}{\tt arXiv:1907.07005}.
\bibitem[{{Milne} et~al.(2004){Milne}, {Hungerford} and {Fryer}}]{miln04}
\bibinfo{author}{{Milne}, P.A.}, \bibinfo{author}{{Hungerford}, A.L.},
  \bibinfo{author}{{Fryer}, C.L.a.}, \bibinfo{year}{2004}.
\newblock \bibinfo{title}{{Unified One-Dimensional Simulations of Gamma-Ray
  Line Emission from Type Ia Supernovae}}.
\newblock \bibinfo{journal}{Astrophys. J.} \bibinfo{volume}{613},
  \bibinfo{pages}{1101--1119}.
\newblock \DOIprefix\doi{10.1086/423235}.
\bibitem[{{Minkowski}(1941)}]{mink41}
\bibinfo{author}{{Minkowski}, R.}, \bibinfo{year}{1941}.
\newblock \bibinfo{title}{{Spectra of Supernovae}}.
\newblock \bibinfo{journal}{Pub. Astro. Soc. Pacific} \bibinfo{volume}{53},
  \bibinfo{pages}{224--225}.
\newblock \DOIprefix\doi{10.1086/125315}.
\bibitem[{{Miyaji} and {Nomoto}(1987)}]{miya87}
\bibinfo{author}{{Miyaji}, S.}, \bibinfo{author}{{Nomoto}, K.},
  \bibinfo{year}{1987}.
\newblock \bibinfo{title}{{On the Collapse of 8--10 M$_{odot}$ Stars Due to
  Electron Capture}}.
\newblock \bibinfo{journal}{Astrophys. J.} \bibinfo{volume}{318},
  \bibinfo{pages}{307}.
\bibitem[{{Miyaji} et~al.(1980){Miyaji}, {Nomoto}, {Yokoi} and
  {Sugimoto}}]{miya80}
\bibinfo{author}{{Miyaji}, S.}, \bibinfo{author}{{Nomoto}, K.},
  \bibinfo{author}{{Yokoi}, K.}, \bibinfo{author}{{Sugimoto}, D.},
  \bibinfo{year}{1980}.
\newblock \bibinfo{title}{{Supernova triggered by electron captures.}}
\newblock \bibinfo{journal}{Publ. Astron. Soc. Jpn.} \bibinfo{volume}{32},
  \bibinfo{pages}{303--329}.
\bibitem[{{Mochkovitch}(1984)}]{moch84}
\bibinfo{author}{{Mochkovitch}, R.}, \bibinfo{year}{1984}.
\newblock \bibinfo{title}{{Final Evolution of 8-10 M$_\odot$ Stars}}, in:
  \bibinfo{editor}{{Bancel}, D.}, \bibinfo{editor}{{Signore}, M.} (Eds.),
  \bibinfo{booktitle}{NATO Advanced Science Institutes (ASI) Series C}, p.
  \bibinfo{pages}{125}.
\bibitem[{{Molaro} et~al.(2016){Molaro}, {Izzo}, {Mason}, {Bonifacio} and
  {Della Valle}}]{mola16}
\bibinfo{author}{{Molaro}, P.}, \bibinfo{author}{{Izzo}, L.},
  \bibinfo{author}{{Mason}, E.}, \bibinfo{author}{{Bonifacio}, P.},
  \bibinfo{author}{{Della Valle}, M.}, \bibinfo{year}{2016}.
\newblock \bibinfo{title}{{Highly enriched $^{7}$Be in the ejecta of Nova
  Sagittarii 2015 No. 2 (V5668 Sgr) and the Galactic $^{7}$Li origin}}.
\newblock \bibinfo{journal}{Mon. Not. R. Astron. Soc.} \bibinfo{volume}{463},
  \bibinfo{pages}{L117--L121}.
\newblock \DOIprefix\doi{10.1093/mnrasl/slw169}.
\bibitem[{{M{\"u}ller}(2016)}]{mull16}
\bibinfo{author}{{M{\"u}ller}, B.}, \bibinfo{year}{2016}.
\newblock \bibinfo{title}{{The Status of Multi-Dimensional Core-Collapse
  Supernova Models}}.
\newblock \bibinfo{journal}{PASA} \bibinfo{volume}{33}, \bibinfo{pages}{e048}.
\newblock \DOIprefix\doi{10.1017/pasa.2016.40},
  \href{http://arxiv.org/abs/1608.03274}{\tt arXiv:1608.03274}.
\bibitem[{{Nagataki} et~al.(1998){Nagataki}, {Hashimoto}, {Sato}, {Yamada} and
  {Mochizuki}}]{naga98}
\bibinfo{author}{{Nagataki}, S.}, \bibinfo{author}{{Hashimoto}, M.a.},
  \bibinfo{author}{{Sato}, K.}, \bibinfo{author}{{Yamada}, S.},
  \bibinfo{author}{{Mochizuki}, Y.S.}, \bibinfo{year}{1998}.
\newblock \bibinfo{title}{{The High Ratio of $^{44}$Ti/$^{56}$Ni in Cassiopeia
  A and the Axisymmetric Collapse-driven Supernova Explosion}}.
\newblock \bibinfo{journal}{Astrophys. J. Lett.} \bibinfo{volume}{492},
  \bibinfo{pages}{L45--L48}.
\newblock \DOIprefix\doi{10.1086/311089},
  \href{http://arxiv.org/abs/astro-ph/9807015}{\tt arXiv:astro-ph/9807015}.
\bibitem[{{Nicholl} et~al.(2016){Nicholl}, {Berger}, {Smartt} and
  et~al.}]{nich16}
\bibinfo{author}{{Nicholl}, M.}, \bibinfo{author}{{Berger}, E.},
  \bibinfo{author}{{Smartt}, S.J.}, \bibinfo{author}{et~al.},
  \bibinfo{year}{2016}.
\newblock \bibinfo{title}{{SN 2015BN: A Detailed Multi-wavelength View of a
  Nearby Superluminous Supernova}}.
\newblock \bibinfo{journal}{Astrophys. J.} \bibinfo{volume}{826},
  \bibinfo{pages}{39}.
\newblock \DOIprefix\doi{10.3847/0004-637X/826/1/39}.
\bibitem[{{Nicholl} et~al.(2018){Nicholl}, {Blanchard} and {Berger}}]{nich18}
\bibinfo{author}{{Nicholl}, M.}, \bibinfo{author}{{Blanchard}, P.K.},
  \bibinfo{author}{{Berger}, Edo, e.a.}, \bibinfo{year}{2018}.
\newblock \bibinfo{title}{{One Thousand Days of SN2015bn: HST Imaging Shows a
  Light Curve Flattening Consistent with Magnetar Predictions}}.
\newblock \bibinfo{journal}{Astrophys. J. Lett.} \bibinfo{volume}{866},
  \bibinfo{pages}{L24}.
\newblock \DOIprefix\doi{10.3847/2041-8213/aae70d}.
\bibitem[{{Nielsen} et~al.(2014){Nielsen}, {Gilfanov}, {Bogd{\'a}n}, {Woods}
  and {Nelemans}}]{niel14}
\bibinfo{author}{{Nielsen}, M.T.B.}, \bibinfo{author}{{Gilfanov}, M.},
  \bibinfo{author}{{Bogd{\'a}n}, {\'A}.}, \bibinfo{author}{{Woods}, T.E.},
  \bibinfo{author}{{Nelemans}, G.}, \bibinfo{year}{2014}.
\newblock \bibinfo{title}{{Upper limits on the luminosity of the progenitor of
  Type Ia supernova SN 2014J}}.
\newblock \bibinfo{journal}{Mon. Not. R. Astron. Soc.} \bibinfo{volume}{442},
  \bibinfo{pages}{3400--3406}.
\newblock \DOIprefix\doi{10.1093/mnras/stu913}.
\bibitem[{{Nomoto}(1982)}]{nomo82a}
\bibinfo{author}{{Nomoto}, K.}, \bibinfo{year}{1982}.
\newblock \bibinfo{title}{{Accreting white dwarf models for type I supernovae.
  I - Presupernova evolution and triggering mechanisms}}.
\newblock \bibinfo{journal}{Astrophys. J.} \bibinfo{volume}{253},
  \bibinfo{pages}{798--810}.
\newblock \DOIprefix\doi{10.1086/159682}.
\bibitem[{{Nomoto}(1984)}]{nomo84a}
\bibinfo{author}{{Nomoto}, K.}, \bibinfo{year}{1984}.
\newblock \bibinfo{title}{{Evolution of 8-10 solar mass stars toward electron
  capture supernovae. I - Formation of electron-degenerate O + NE + MG cores.}}
\newblock \bibinfo{journal}{Astrophys. J.} \bibinfo{volume}{277},
  \bibinfo{pages}{791--805}.
\bibitem[{{Nomoto} and {Kondo}(1991)}]{nomo91}
\bibinfo{author}{{Nomoto}, K.}, \bibinfo{author}{{Kondo}, Y.},
  \bibinfo{year}{1991}.
\newblock \bibinfo{title}{{Conditions for accretion-induced collapse of white
  dwarfs}}.
\newblock \bibinfo{journal}{Astrophys. J. Lett.} \bibinfo{volume}{367},
  \bibinfo{pages}{L19--L22}.
\newblock \DOIprefix\doi{10.1086/185922}.
\bibitem[{{Nomoto} et~al.(1979){Nomoto}, {Miyaji}, {Sugimoto} and
  {Yokoi}}]{nomo79}
\bibinfo{author}{{Nomoto}, K.}, \bibinfo{author}{{Miyaji}, S.},
  \bibinfo{author}{{Sugimoto}, D.}, \bibinfo{author}{{Yokoi}, K.},
  \bibinfo{year}{1979}.
\newblock \bibinfo{title}{{Collapse of Accreting White Dwarf to Form a Neutron
  Star}}, in: \bibinfo{editor}{{van Horn}, H.M.}, \bibinfo{editor}{{Weidemann},
  V.}, \bibinfo{editor}{{Savedoff}, M.P.} (Eds.), \bibinfo{booktitle}{IAU
  Colloq. 53: White Dwarfs and Variable Degenerate Stars},
  p.~\bibinfo{pages}{56}.
\bibitem[{{Nomoto} et~al.(1976){Nomoto}, {Sugimoto} and {Neo}}]{nomo76}
\bibinfo{author}{{Nomoto}, K.}, \bibinfo{author}{{Sugimoto}, D.},
  \bibinfo{author}{{Neo}, S.}, \bibinfo{year}{1976}.
\newblock \bibinfo{title}{{Carbon deflagration supernova, an alternative to
  carbon detonation}}.
\newblock \bibinfo{journal}{Astrop. Space Sci.} \bibinfo{volume}{39},
  \bibinfo{pages}{L37--L42}.
\newblock \DOIprefix\doi{10.1007/BF00648354}.
\bibitem[{{Nomoto} et~al.(1984a){Nomoto}, {Thielemann} and {Yokoi}}]{nomo84}
\bibinfo{author}{{Nomoto}, K.}, \bibinfo{author}{{Thielemann}, F.},
  \bibinfo{author}{{Yokoi}, K.}, \bibinfo{year}{1984}a.
\newblock \bibinfo{title}{{Accreting white dwarf models of Type I supernovae.
  III - Carbon deflagration supernovae}}.
\newblock \bibinfo{journal}{Astrophys. J.} \bibinfo{volume}{286},
  \bibinfo{pages}{644--658}.
\newblock \DOIprefix\doi{10.1086/162639}.
\bibitem[{{Nomoto} et~al.(1984b){Nomoto}, {Thielemann} and {Yokoi}}]{nomo84b}
\bibinfo{author}{{Nomoto}, K.}, \bibinfo{author}{{Thielemann}, F.},
  \bibinfo{author}{{Yokoi}, K.}, \bibinfo{year}{1984}b.
\newblock \bibinfo{title}{{Accreting white dwarf models of Type I supernovae.
  III - Carbon deflagration supernovae}}.
\newblock \bibinfo{journal}{Astrophys. J.} \bibinfo{volume}{286},
  \bibinfo{pages}{644--658}.
\newblock \DOIprefix\doi{10.1086/162639}.
\bibitem[{{Nugent} et~al.(2011a){Nugent}, {Sullivan} and {Bersier}}]{nuge11a}
\bibinfo{author}{{Nugent}, P.}, \bibinfo{author}{{Sullivan}, M.},
  \bibinfo{author}{{Bersier}, D.e.a.}, \bibinfo{year}{2011}a.
\newblock \bibinfo{title}{{Young Type Ia Supernova PTF11kly in M101}}.
\newblock \bibinfo{journal}{The Astronomer's Telegram} \bibinfo{volume}{3581},
  \bibinfo{pages}{1--+}.
\bibitem[{{Nugent} et~al.(2011b){Nugent}, {Sullivan} and {Cenko}}]{nuge11b}
\bibinfo{author}{{Nugent}, P.E.}, \bibinfo{author}{{Sullivan}, M.},
  \bibinfo{author}{{Cenko}, S.B.e.a.}, \bibinfo{year}{2011}b.
\newblock \bibinfo{title}{{Supernova SN 2011fe from an exploding carbon-oxygen
  white dwarf star}}.
\newblock \bibinfo{journal}{Nature} \bibinfo{volume}{480},
  \bibinfo{pages}{344--347}.
\newblock \DOIprefix\doi{10.1038/nature10644}.
\bibitem[{{Oda} et~al.(1994){Oda}, {Hino}, {Muto} and et~al.}]{oda94}
\bibinfo{author}{{Oda}, T.}, \bibinfo{author}{{Hino}, M.},
  \bibinfo{author}{{Muto}, K.}, \bibinfo{author}{et~al.}, \bibinfo{year}{1994}.
\newblock \bibinfo{title}{{Rate Tables for the Weak Processes of sd-Shell
  Nuclei in Stellar Matter}}.
\newblock \bibinfo{journal}{Atomic Data and Nuclear Data Tables}
  \bibinfo{volume}{56}, \bibinfo{pages}{231--403}.
\newblock \DOIprefix\doi{10.1006/adnd.1994.1007}.
\bibitem[{{Ono} et~al.(2020){Ono}, {Nagataki}, {Ferrand}, {Takahashi}, {Umeda},
  {Yoshida}, {Orland o} and {Miceli}}]{ono20}
\bibinfo{author}{{Ono}, M.}, \bibinfo{author}{{Nagataki}, S.},
  \bibinfo{author}{{Ferrand}, G.}, \bibinfo{author}{{Takahashi}, K.},
  \bibinfo{author}{{Umeda}, H.}, \bibinfo{author}{{Yoshida}, T.},
  \bibinfo{author}{{Orland o}, S.}, \bibinfo{author}{{Miceli}, M.},
  \bibinfo{year}{2020}.
\newblock \bibinfo{title}{{Matter Mixing in Aspherical Core-collapse
  Supernovae: Three-dimensional Simulations with Single-star and Binary Merger
  Progenitor Models for SN 1987A}}.
\newblock \bibinfo{journal}{Astrophys. J.} \bibinfo{volume}{888},
  \bibinfo{pages}{111}.
\newblock \DOIprefix\doi{10.3847/1538-4357/ab5dba}.
\bibitem[{{Orlando} et~al.(2016){Orlando}, {Miceli}, {Pumo} and
  {Bocchino}}]{orla16}
\bibinfo{author}{{Orlando}, S.}, \bibinfo{author}{{Miceli}, M.},
  \bibinfo{author}{{Pumo}, M.L.}, \bibinfo{author}{{Bocchino}, F.},
  \bibinfo{year}{2016}.
\newblock \bibinfo{title}{{Modeling SNR Cassiopeia A from the Supernova
  Explosion to its Current Age: The Role of Post-explosion Anisotropies of
  Ejecta}}.
\newblock \bibinfo{journal}{Astrophys. J.} \bibinfo{volume}{822},
  \bibinfo{pages}{22}.
\newblock \DOIprefix\doi{10.3847/0004-637X/822/1/22}.
\bibitem[{{Orlando} et~al.(2020){Orlando}, {Ono}, {Nagataki}, {Miceli},
  {Umeda}, {Ferrand}, {Bocchino}, {Petruk}, {Peres}, {Takahashi} and
  {Yoshida}}]{orla19}
\bibinfo{author}{{Orlando}, S.}, \bibinfo{author}{{Ono}, M.},
  \bibinfo{author}{{Nagataki}, S.}, \bibinfo{author}{{Miceli}, M.},
  \bibinfo{author}{{Umeda}, H.}, \bibinfo{author}{{Ferrand}, G.},
  \bibinfo{author}{{Bocchino}, F.}, \bibinfo{author}{{Petruk}, O.},
  \bibinfo{author}{{Peres}, G.}, \bibinfo{author}{{Takahashi}, K.},
  \bibinfo{author}{{Yoshida}, T.}, \bibinfo{year}{2020}.
\newblock \bibinfo{title}{{Hydrodynamic simulations unravel the
  progenitor-supernova-remnant connection in SN 1987A}}.
\newblock \bibinfo{journal}{Astronom. Astrophys.} \bibinfo{volume}{636},
  \bibinfo{pages}{A22}.
\newblock \DOIprefix\doi{10.1051/0004-6361/201936718},
  \href{http://arxiv.org/abs/1912.03070}{\tt arXiv:1912.03070}.
\bibitem[{{Pakmor} et~al.(2010){Pakmor}, {Kromer}, {R{\"o}pke}, {Sim}, {Ruiter}
  and {Hillebrandt}}]{pakm10}
\bibinfo{author}{{Pakmor}, R.}, \bibinfo{author}{{Kromer}, M.},
  \bibinfo{author}{{R{\"o}pke}, F.K.}, \bibinfo{author}{{Sim}, S.A.},
  \bibinfo{author}{{Ruiter}, A.J.}, \bibinfo{author}{{Hillebrandt}, W.},
  \bibinfo{year}{2010}.
\newblock \bibinfo{title}{{Sub-luminous type Ia supernovae from the mergers of
  equal-mass white dwarfs with mass
  \raisebox{-0.5ex}\textasciitilde0.9M$_\odot$}}.
\newblock \bibinfo{journal}{Nature} \bibinfo{volume}{463},
  \bibinfo{pages}{61--64}.
\bibitem[{{Park} et~al.(2013){Park}, {Badenes}, {Mori}, {Kaida}, {Bravo},
  {Schenck}, {Eriksen}, {Hughes}, {Slane}, {Burrows} and {Lee}}]{park13}
\bibinfo{author}{{Park}, S.}, \bibinfo{author}{{Badenes}, C.},
  \bibinfo{author}{{Mori}, K.}, \bibinfo{author}{{Kaida}, R.},
  \bibinfo{author}{{Bravo}, E.}, \bibinfo{author}{{Schenck}, A.},
  \bibinfo{author}{{Eriksen}, K.A.}, \bibinfo{author}{{Hughes}, J.P.},
  \bibinfo{author}{{Slane}, P.O.}, \bibinfo{author}{{Burrows}, D.N.},
  \bibinfo{author}{{Lee}, J.J.}, \bibinfo{year}{2013}.
\newblock \bibinfo{title}{{A Super-solar Metallicity for the Progenitor of
  Kepler's Supernova}}.
\newblock \bibinfo{journal}{Astrophys. J. Lett.} \bibinfo{volume}{767},
  \bibinfo{pages}{L10}.
\newblock \DOIprefix\doi{10.1088/2041-8205/767/1/L10},
  \href{http://arxiv.org/abs/1302.5435}{\tt arXiv:1302.5435}.
\bibitem[{{Payne-Gaposchkin}(1957)}]{payn57}
\bibinfo{author}{{Payne-Gaposchkin}, C.}, \bibinfo{year}{1957}.
\newblock \bibinfo{title}{{Spectrophotometric Study of Stellar Rotation: an
  Analysis of {\ensuremath{\beta}} Cassiopeiae}}.
\newblock \bibinfo{journal}{Publ. Astron. Soc. Pac.} \bibinfo{volume}{69},
  \bibinfo{pages}{46}.
\newblock \DOIprefix\doi{10.1086/127011}.
\bibitem[{{Perets} et~al.(2011){Perets}, {Gal-yam} and {Crockett}}]{pere11}
\bibinfo{author}{{Perets}, H.B.}, \bibinfo{author}{{Gal-yam}, A.},
  \bibinfo{author}{{Crockett}, R.M.e.a.}, \bibinfo{year}{2011}.
\newblock \bibinfo{title}{{The Old Environment of the Faint Calcium-rich
  Supernova SN 2005cz}}.
\newblock \bibinfo{journal}{Astrophys. J. Lett.} \bibinfo{volume}{728},
  \bibinfo{pages}{L36}.
\newblock \DOIprefix\doi{10.1088/2041-8205/728/2/L36}.
\bibitem[{{Phillips}(1993)}]{phil93}
\bibinfo{author}{{Phillips}, M.M.}, \bibinfo{year}{1993}.
\newblock \bibinfo{title}{{The Absolute Magnitudes of Type IA Supernovae}}.
\newblock \bibinfo{journal}{Astrophys. J. Lett.} \bibinfo{volume}{413},
  \bibinfo{pages}{L105}.
\newblock \DOIprefix\doi{10.1086/186970}.
\bibitem[{{Piersanti} et~al.(2017){Piersanti}, {Bravo}, {Cristallo},
  {Dom{\'\i}nguez}, {Straniero}, {Tornamb{\'e}} and
  {Mart{\'\i}nez-Pinedo}}]{pier17}
\bibinfo{author}{{Piersanti}, L.}, \bibinfo{author}{{Bravo}, E.},
  \bibinfo{author}{{Cristallo}, S.}, \bibinfo{author}{{Dom{\'\i}nguez}, I.},
  \bibinfo{author}{{Straniero}, O.}, \bibinfo{author}{{Tornamb{\'e}}, A.},
  \bibinfo{author}{{Mart{\'\i}nez-Pinedo}, G.}, \bibinfo{year}{2017}.
\newblock \bibinfo{title}{{Type Ia Supernovae Keep Memory of their Progenitor
  Metallicity}}.
\newblock \bibinfo{journal}{Astrophys. J. Lett.} \bibinfo{volume}{836},
  \bibinfo{pages}{L9}.
\newblock \DOIprefix\doi{10.3847/2041-8213/aa5c7e}.
\bibitem[{{Piersanti} et~al.(2003a){Piersanti}, {Gagliardi}, {Iben} and
  {Tornamb{\'e}}}]{pier03b}
\bibinfo{author}{{Piersanti}, L.}, \bibinfo{author}{{Gagliardi}, S.},
  \bibinfo{author}{{Iben}, Icko, J.}, \bibinfo{author}{{Tornamb{\'e}}, A.},
  \bibinfo{year}{2003}a.
\newblock \bibinfo{title}{{Carbon-Oxygen White Dwarf Accreting CO-Rich Matter.
  II. Self-Regulating Accretion Process up to the Explosive Stage}}.
\newblock \bibinfo{journal}{Astrophys. J.} \bibinfo{volume}{598},
  \bibinfo{pages}{1229--1238}.
\newblock \DOIprefix\doi{10.1086/378952}.
\bibitem[{{Piersanti} et~al.(2003b){Piersanti}, {Gagliardi}, {Iben} and
  {Tornamb{\'e}}}]{pier03a}
\bibinfo{author}{{Piersanti}, L.}, \bibinfo{author}{{Gagliardi}, S.},
  \bibinfo{author}{{Iben}, Icko, J.}, \bibinfo{author}{{Tornamb{\'e}}, A.},
  \bibinfo{year}{2003}b.
\newblock \bibinfo{title}{{Carbon-Oxygen White Dwarfs Accreting CO-rich Matter.
  I. A Comparison between Rotating and Nonrotating Models}}.
\newblock \bibinfo{journal}{Astrophys. J.} \bibinfo{volume}{583},
  \bibinfo{pages}{885--901}.
\newblock \DOIprefix\doi{10.1086/345444}.
\bibitem[{{Pinto} and {Woosley}(1988a)}]{pinto88}
\bibinfo{author}{{Pinto}, P.A.}, \bibinfo{author}{{Woosley}, S.E.},
  \bibinfo{year}{1988}a.
\newblock \bibinfo{title}{{The theory of gamma-ray emergence in supernova
  1987A}}.
\newblock \bibinfo{journal}{Nature} \bibinfo{volume}{333},
  \bibinfo{pages}{534--537}.
\newblock \DOIprefix\doi{10.1038/333534a0}.
\bibitem[{{Pinto} and {Woosley}(1988b)}]{pint88}
\bibinfo{author}{{Pinto}, P.A.}, \bibinfo{author}{{Woosley}, S.E.},
  \bibinfo{year}{1988}b.
\newblock \bibinfo{title}{{X-Ray and Gamma-Ray Emission from Supernova 1987A}}.
\newblock \bibinfo{journal}{Astrophys. J.} \bibinfo{volume}{329},
  \bibinfo{pages}{820}.
\newblock \DOIprefix\doi{10.1086/166426}.
\bibitem[{{Piro} and {Bildsten}(2008)}]{piro08}
\bibinfo{author}{{Piro}, A.L.}, \bibinfo{author}{{Bildsten}, L.},
  \bibinfo{year}{2008}.
\newblock \bibinfo{title}{{Neutronization during Type Ia Supernova Simmering}}.
\newblock \bibinfo{journal}{Astrophys. J.} \bibinfo{volume}{673},
  \bibinfo{pages}{1009--1013}.
\newblock \DOIprefix\doi{10.1086/524189}.
\bibitem[{{Pozdnyakov} et~al.(1983){Pozdnyakov}, {Sobol} and
  {Syunyaev}}]{pozd83}
\bibinfo{author}{{Pozdnyakov}, L.A.}, \bibinfo{author}{{Sobol}, I.M.},
  \bibinfo{author}{{Syunyaev}, R.A.}, \bibinfo{year}{1983}.
\newblock \bibinfo{title}{{Comptonization and the shaping of X-ray source
  spectra - Monte Carlo calculations}}.
\newblock \bibinfo{journal}{Astrophysics and Space Physics Reviews}
  \bibinfo{volume}{2}, \bibinfo{pages}{189--331}.
\bibitem[{{Prajs} et~al.(2017){Prajs}, {Sullivan} and {Smith}}]{praj17}
\bibinfo{author}{{Prajs}, S.}, \bibinfo{author}{{Sullivan}, M.},
  \bibinfo{author}{{Smith}, M.e.a.}, \bibinfo{year}{2017}.
\newblock \bibinfo{title}{{The volumetric rate of superluminous supernovae at
  $z\sim$1}}.
\newblock \bibinfo{journal}{Mon. Not. R. Astron. Soc.} \bibinfo{volume}{464},
  \bibinfo{pages}{3568--3579}.
\newblock \DOIprefix\doi{10.1093/mnras/stw1942}.
\bibitem[{{Provencal} et~al.(1998){Provencal}, {Shipman}, {H{\o}g} and
  {Thejll}}]{prov98}
\bibinfo{author}{{Provencal}, J.L.}, \bibinfo{author}{{Shipman}, H.L.},
  \bibinfo{author}{{H{\o}g}, E.}, \bibinfo{author}{{Thejll}, P.},
  \bibinfo{year}{1998}.
\newblock \bibinfo{title}{{Testing the White Dwarf Mass-Radius Relation with
  HIPPARCOS}}.
\newblock \bibinfo{journal}{Astrophys. J.} \bibinfo{volume}{494},
  \bibinfo{pages}{759--767}.
\newblock \DOIprefix\doi{10.1086/305238}.
\bibitem[{{Pursiainen} et~al.(2018){Pursiainen}, {Childress} and
  {Smith}}]{purs18}
\bibinfo{author}{{Pursiainen}, M.}, \bibinfo{author}{{Childress}, M.},
  \bibinfo{author}{{Smith}, e.a.}, \bibinfo{year}{2018}.
\newblock \bibinfo{title}{{Rapidly evolving transients in the Dark Energy
  Survey}}.
\newblock \bibinfo{journal}{Mon. Not. R. Astron. Soc.} \bibinfo{volume}{481},
  \bibinfo{pages}{894--917}.
\bibitem[{{Quimby} et~al.(2018){Quimby}, {De Cia} and {Gal-Yam}}]{quim18}
\bibinfo{author}{{Quimby}, R.M.}, \bibinfo{author}{{De Cia}, A.},
  \bibinfo{author}{{Gal-Yam}, A.e.a.}, \bibinfo{year}{2018}.
\newblock \bibinfo{title}{{Spectra of Hydrogen-poor Superluminous Supernovae
  from the Palomar Transient Factory}}.
\newblock \bibinfo{journal}{APJ} \bibinfo{volume}{855}, \bibinfo{pages}{2}.
\newblock \DOIprefix\doi{10.3847/1538-4357/aaac2f}.
\bibitem[{{Quimby} et~al.(2013){Quimby}, {Yuan}, {Akerlof} and
  {Wheeler}}]{quim13}
\bibinfo{author}{{Quimby}, R.M.}, \bibinfo{author}{{Yuan}, F.},
  \bibinfo{author}{{Akerlof}, C.}, \bibinfo{author}{{Wheeler}, J.C.},
  \bibinfo{year}{2013}.
\newblock \bibinfo{title}{{Rates of superluminous supernovae at $z\sim$0.2}}.
\newblock \bibinfo{journal}{Mon. Not. R. Astron. Soc.} \bibinfo{volume}{431},
  \bibinfo{pages}{912--922}.
\newblock \DOIprefix\doi{10.1093/mnras/stt213}.
\bibitem[{{Raddi} et~al.(2019){Raddi}, {Hollands} and {Koester}}]{radd19}
\bibinfo{author}{{Raddi}, R.}, \bibinfo{author}{{Hollands}, M.A.},
  \bibinfo{author}{{Koester}, D.e.a.}, \bibinfo{year}{2019}.
\newblock \bibinfo{title}{{Partly burnt runaway stellar remnants from peculiar
  thermonuclear supernovae}}.
\newblock \bibinfo{journal}{Mon. Not. R. Astron. Soc.} \bibinfo{volume}{489},
  \bibinfo{pages}{1489--1508}.
\newblock \DOIprefix\doi{10.1093/mnras/stz1618}.
\bibitem[{{Rakavy} and {Shaviv}(1967)}]{raka67}
\bibinfo{author}{{Rakavy}, G.}, \bibinfo{author}{{Shaviv}, G.},
  \bibinfo{year}{1967}.
\newblock \bibinfo{title}{{Instabilities in Highly Evolved Stellar Models}}.
\newblock \bibinfo{journal}{Astrophys. J.} \bibinfo{volume}{148},
  \bibinfo{pages}{803}.
\newblock \DOIprefix\doi{10.1086/149204}.
\bibitem[{{Raskin} et~al.(2010){Raskin}, {Scannapieco}, {Rockefeller}, {Fryer},
  {Diehl} and {Timmes}}]{rask10}
\bibinfo{author}{{Raskin}, C.}, \bibinfo{author}{{Scannapieco}, E.},
  \bibinfo{author}{{Rockefeller}, G.}, \bibinfo{author}{{Fryer}, C.},
  \bibinfo{author}{{Diehl}, S.}, \bibinfo{author}{{Timmes}, F.X.},
  \bibinfo{year}{2010}.
\newblock \bibinfo{title}{{$^{56}$Ni Production in Double-degenerate White
  Dwarf Collisions}}.
\newblock \bibinfo{journal}{Astrophys. J.} \bibinfo{volume}{724},
  \bibinfo{pages}{111--125}.
\newblock \DOIprefix\doi{10.1088/0004-637X/724/1/111}.
\bibitem[{{Raskin} et~al.(2009){Raskin}, {Timmes}, {Scannapieco}, {Diehl} and
  {Fryer}}]{rask09}
\bibinfo{author}{{Raskin}, C.}, \bibinfo{author}{{Timmes}, F.X.},
  \bibinfo{author}{{Scannapieco}, E.}, \bibinfo{author}{{Diehl}, S.},
  \bibinfo{author}{{Fryer}, C.}, \bibinfo{year}{2009}.
\newblock \bibinfo{title}{{On Type Ia supernovae from the collisions of two
  white dwarfs}}.
\newblock \bibinfo{journal}{Mon. Not. R. Astron. Soc.} \bibinfo{volume}{399},
  \bibinfo{pages}{L156--L159}.
\newblock \DOIprefix\doi{10.1111/j.1745-3933.2009.00743.x}.
\bibitem[{{Renaud} et~al.(2006){Renaud}, {Vink} and {Decourchelle}}]{rena06}
\bibinfo{author}{{Renaud}, M.}, \bibinfo{author}{{Vink}, J.},
  \bibinfo{author}{{Decourchelle}, A.e.a.}, \bibinfo{year}{2006}.
\newblock \bibinfo{title}{{The Signature of $^{44}$Ti in Cassiopeia A Revealed
  by IBIS/ISGRI on INTEGRAL}}.
\newblock \bibinfo{journal}{Astrophys. J. Lett.} \bibinfo{volume}{647},
  \bibinfo{pages}{L41--L44}.
\newblock \DOIprefix\doi{10.1086/507300}.
\bibitem[{{Rest} et~al.(2011){Rest}, {Foley}, {Sinnott}, {Welch}, {Badenes},
  {Filippenko}, {Bergmann}, {Bhatti}, {Blondin}, {Challis}, {Damke}, {Finley},
  {Huber}, {Kasen}, {Kirshner}, {Matheson}, {Mazzali}, {Minniti}, {Nakajima},
  {Narayan}, {Olsen}, {Sauer}, {Smith} and {Suntzeff}}]{rest11}
\bibinfo{author}{{Rest}, A.}, \bibinfo{author}{{Foley}, R.J.},
  \bibinfo{author}{{Sinnott}, B.}, \bibinfo{author}{{Welch}, D.L.},
  \bibinfo{author}{{Badenes}, C.}, \bibinfo{author}{{Filippenko}, A.V.},
  \bibinfo{author}{{Bergmann}, M.}, \bibinfo{author}{{Bhatti}, W.A.},
  \bibinfo{author}{{Blondin}, S.}, \bibinfo{author}{{Challis}, P.},
  \bibinfo{author}{{Damke}, G.}, \bibinfo{author}{{Finley}, H.},
  \bibinfo{author}{{Huber}, M.E.}, \bibinfo{author}{{Kasen}, D.},
  \bibinfo{author}{{Kirshner}, R.P.}, \bibinfo{author}{{Matheson}, T.},
  \bibinfo{author}{{Mazzali}, P.}, \bibinfo{author}{{Minniti}, D.},
  \bibinfo{author}{{Nakajima}, R.}, \bibinfo{author}{{Narayan}, G.},
  \bibinfo{author}{{Olsen}, K.}, \bibinfo{author}{{Sauer}, D.},
  \bibinfo{author}{{Smith}, R.C.}, \bibinfo{author}{{Suntzeff}, N.B.},
  \bibinfo{year}{2011}.
\newblock \bibinfo{title}{{Direct Confirmation of the Asymmetry of the Cas A
  Supernova with Light Echoes}}.
\newblock \bibinfo{journal}{Astrophys. J.} \bibinfo{volume}{732},
  \bibinfo{pages}{3}.
\newblock \DOIprefix\doi{10.1088/0004-637X/732/1/3},
  \href{http://arxiv.org/abs/1003.5660}{\tt arXiv:1003.5660}.
\bibitem[{{Reynolds}(2008)}]{reyn08}
\bibinfo{author}{{Reynolds}, S.P.}, \bibinfo{year}{2008}.
\newblock \bibinfo{title}{{Supernova remnants at high energy.}}
\newblock \bibinfo{journal}{Annu. Rev. Astron. Astrophys.}
  \bibinfo{volume}{46}, \bibinfo{pages}{89--126}.
\newblock \DOIprefix\doi{10.1146/annurev.astro.46.060407.145237}.
\bibitem[{{Riess} et~al.(1999){Riess}, {Filippenko}, {Li}, {Treffers},
  {Schmidt}, {Qiu}, {Hu}, {Armstrong}, {Faranda}, {Thouvenot} and
  {Buil}}]{ries99}
\bibinfo{author}{{Riess}, A.G.}, \bibinfo{author}{{Filippenko}, A.V.},
  \bibinfo{author}{{Li}, W.}, \bibinfo{author}{{Treffers}, R.R.},
  \bibinfo{author}{{Schmidt}, B.P.}, \bibinfo{author}{{Qiu}, Y.},
  \bibinfo{author}{{Hu}, J.}, \bibinfo{author}{{Armstrong}, M.},
  \bibinfo{author}{{Faranda}, C.}, \bibinfo{author}{{Thouvenot}, E.},
  \bibinfo{author}{{Buil}, C.}, \bibinfo{year}{1999}.
\newblock \bibinfo{title}{{The Rise Time of Nearby Type IA Supernovae}}.
\newblock \bibinfo{journal}{Astron. J.} \bibinfo{volume}{118},
  \bibinfo{pages}{2675--2688}.
\newblock \DOIprefix\doi{10.1086/301143}.
\bibitem[{{Ritossa} et~al.(1996){Ritossa}, {Garcia-Berro} and {Iben}}]{rito96}
\bibinfo{author}{{Ritossa}, C.}, \bibinfo{author}{{Garcia-Berro}, E.},
  \bibinfo{author}{{Iben}, Icko, J.}, \bibinfo{year}{1996}.
\newblock \bibinfo{title}{{On the Evolution of Stars That Form
  Electron-degenerate Cores Processed by Carbon Burning. II. Isotope Abundances
  and Thermal Pulses in a 10 M$_{odot}$ Model with an ONe Core and Applications
  to Long-Period Variables, Classical Novae, and Accretion-induced Collapse}}.
\newblock \bibinfo{journal}{Astrophys. J.} \bibinfo{volume}{460},
  \bibinfo{pages}{489}.
\newblock \DOIprefix\doi{10.1086/176987}.
\bibitem[{{Ritossa} et~al.(1999){Ritossa}, {Garc{\'\i}a-Berro} and
  {Iben}}]{rito99}
\bibinfo{author}{{Ritossa}, C.}, \bibinfo{author}{{Garc{\'\i}a-Berro}, E.},
  \bibinfo{author}{{Iben}, Icko, J.}, \bibinfo{year}{1999}.
\newblock \bibinfo{title}{{On the Evolution of Stars that Form
  Electron-degenerate Cores Processed by Carbon Burning. V. Shell Convection
  Sustained by Helium Burning, Transient Neon Burning, Dredge-out, Urca
  Cooling, and Other Properties of an 11 M$_{\odot}$ Population I Model Star}}.
\newblock \bibinfo{journal}{Astrophys. J.} \bibinfo{volume}{515},
  \bibinfo{pages}{381--397}.
\newblock \DOIprefix\doi{10.1086/307017}.
\bibitem[{{Roman} et~al.(2018){Roman}, {Hardin} and {Betoule}}]{roma18}
\bibinfo{author}{{Roman}, M.}, \bibinfo{author}{{Hardin}, D.},
  \bibinfo{author}{{Betoule}, M.e.a.}, \bibinfo{year}{2018}.
\newblock \bibinfo{title}{{Dependence of Type Ia supernova luminosities on
  their local environment}}.
\newblock \bibinfo{journal}{A\&A} \bibinfo{volume}{615}, \bibinfo{pages}{A68}.
\newblock \DOIprefix\doi{10.1051/0004-6361/201731425}.
\bibitem[{{Roques} et~al.(2003){Roques}, {Schanne} and {von Kienlin}}]{roqu03}
\bibinfo{author}{{Roques}, J.P.}, \bibinfo{author}{{Schanne}, S.},
  \bibinfo{author}{{von Kienlin}, A.e.a.}, \bibinfo{year}{2003}.
\newblock \bibinfo{title}{{SPI/INTEGRAL in-flight performance}}.
\newblock \bibinfo{journal}{A\&A} \bibinfo{volume}{411},
  \bibinfo{pages}{L91--L100}.
\newblock \DOIprefix\doi{10.1051/0004-6361:20031501}.
\bibitem[{{Rosswog} et~al.(2009){Rosswog}, {Ramirez-Ruiz} and {Hix}}]{ross09}
\bibinfo{author}{{Rosswog}, S.}, \bibinfo{author}{{Ramirez-Ruiz}, E.},
  \bibinfo{author}{{Hix}, W.R.}, \bibinfo{year}{2009}.
\newblock \bibinfo{title}{{Tidal Disruption and Ignition of White Dwarfs by
  Moderately Massive Black Holes}}.
\newblock \bibinfo{journal}{Astrophys. J.} \bibinfo{volume}{695},
  \bibinfo{pages}{404--419}.
\newblock \DOIprefix\doi{10.1088/0004-637X/695/1/404}.
\bibitem[{{Ruiz-Lapuente} et~al.(1993){Ruiz-Lapuente}, {Lichti}, {Lehoucq},
  {Canal} and {Casse}}]{ruiz93}
\bibinfo{author}{{Ruiz-Lapuente}, P.}, \bibinfo{author}{{Lichti}, G.G.},
  \bibinfo{author}{{Lehoucq}, R.}, \bibinfo{author}{{Canal}, R.},
  \bibinfo{author}{{Casse}, M.}, \bibinfo{year}{1993}.
\newblock \bibinfo{title}{{Gamma-Ray Escape in Type IA Supernovae: The 847
  keV--m B Diagram}}.
\newblock \bibinfo{journal}{Astrophys. J.} \bibinfo{volume}{417},
  \bibinfo{pages}{547}.
\newblock \DOIprefix\doi{10.1086/173333}.
\bibitem[{{Sato} et~al.(2020){Sato}, {Bravo}, {Badenes}, {P. Hughes},
  {Williams} and {Yamaguchi}}]{sato20}
\bibinfo{author}{{Sato}, T.}, \bibinfo{author}{{Bravo}, E.},
  \bibinfo{author}{{Badenes}, C.}, \bibinfo{author}{{P. Hughes}, J.},
  \bibinfo{author}{{Williams}, B.J.}, \bibinfo{author}{{Yamaguchi}, H.},
  \bibinfo{year}{2020}.
\newblock \bibinfo{title}{{A Nucleosynthetic Origin for the Southwestern
  Fe-rich Structure in Kepler's Supernova Remnant}}.
\newblock \bibinfo{journal}{Astrophys. J.} \bibinfo{volume}{890},
  \bibinfo{pages}{104}.
\newblock \DOIprefix\doi{10.3847/1538-4357/ab6aa2},
  \href{http://arxiv.org/abs/2001.02662}{\tt arXiv:2001.02662}.
\bibitem[{{Schaefer}(2018)}]{scha18}
\bibinfo{author}{{Schaefer}, B.E.}, \bibinfo{year}{2018}.
\newblock \bibinfo{title}{{The distances to Novae as seen by Gaia}}.
\newblock \bibinfo{journal}{Mon. Not. R. Astron. Soc.} \bibinfo{volume}{481},
  \bibinfo{pages}{3033--3051}.
\newblock \DOIprefix\doi{10.1093/mnras/sty2388}.
\bibitem[{{Schwab} et~al.(2017){Schwab}, {Bildsten} and {Quataert}}]{schw17}
\bibinfo{author}{{Schwab}, J.}, \bibinfo{author}{{Bildsten}, L.},
  \bibinfo{author}{{Quataert}, E.}, \bibinfo{year}{2017}.
\newblock \bibinfo{title}{{The importance of Urca-process cooling in accreting
  ONe white dwarfs}}.
\newblock \bibinfo{journal}{Mon. Not. R. Astron. Soc.} ,
  \bibinfo{pages}{3390--3406}\DOIprefix\doi{10.1093/mnras/stx2169}.
\bibitem[{{Seach} et~al.(2015){Seach}, {Guido}, {Howes}, {Powles}, {Kazarovets}
  and {Samus}}]{seac15}
\bibinfo{author}{{Seach}, J.}, \bibinfo{author}{{Guido}, E.},
  \bibinfo{author}{{Howes}, N.}, \bibinfo{author}{{Powles}, J.},
  \bibinfo{author}{{Kazarovets}, E.}, \bibinfo{author}{{Samus}, N.},
  \bibinfo{year}{2015}.
\newblock \bibinfo{title}{{V5668 Sagittarii = N Sgr 2015 No. 2 = Pnv
  J18365700-2855420}}.
\newblock \bibinfo{journal}{International Astronomical Union Circular}
  \bibinfo{volume}{9275}, \bibinfo{pages}{1}.
\bibitem[{{Seitenzahl} et~al.(2014){Seitenzahl}, {Timmes} and
  {Magkotsios}}]{ssst14}
\bibinfo{author}{{Seitenzahl}, I.R.}, \bibinfo{author}{{Timmes}, F.X.},
  \bibinfo{author}{{Magkotsios}, G.}, \bibinfo{year}{2014}.
\newblock \bibinfo{title}{{The Light Curve of SN 1987A Revisited: Constraining
  Production Masses of Radioactive Nuclides}}.
\newblock \bibinfo{journal}{Astrophys. J.} \bibinfo{volume}{792},
  \bibinfo{pages}{10}.
\newblock \DOIprefix\doi{10.1088/0004-637X/792/1/10}.
\bibitem[{{Seitenzahl} and {Townsley}(2017)}]{seit17}
\bibinfo{author}{{Seitenzahl}, I.R.}, \bibinfo{author}{{Townsley}, D.M.},
  \bibinfo{year}{2017}.
\newblock \bibinfo{title}{{Nucleosynthesis in Thermonuclear Supernovae}}, in:
  \bibinfo{booktitle}{Handbook of Supernovae, ISBN 978-3-319-21845-8. Springer
  International Publishing AG, 2017, p. 1955}, p. \bibinfo{pages}{1955}.
\newblock \DOIprefix\doi{10.1007/978-3-319-21846-5_87}.
\bibitem[{{Senziani} et~al.(2008){Senziani}, {Skinner}, {Jean} and
  {Hernanz}}]{senz08}
\bibinfo{author}{{Senziani}, F.}, \bibinfo{author}{{Skinner}, G.},
  \bibinfo{author}{{Jean}, P.}, \bibinfo{author}{{Hernanz}, M.},
  \bibinfo{year}{2008}.
\newblock \bibinfo{title}{{Results from SWIFT/BAT Searches for Prompt Gamma-ray
  Emission from RS Ophiuchi and other Novae}}. volume \bibinfo{volume}{401} of
  \textit{\bibinfo{series}{Astronomical Society of the Pacific Conference
  Series}}.
\newblock p. \bibinfo{pages}{323}.
\bibitem[{{Shafter}(1997)}]{shaf97}
\bibinfo{author}{{Shafter}, A.W.}, \bibinfo{year}{1997}.
\newblock \bibinfo{title}{{On the Nova Rate in the Galaxy}}.
\newblock \bibinfo{journal}{Astrophys. J.} \bibinfo{volume}{487},
  \bibinfo{pages}{226--236}.
\newblock \DOIprefix\doi{10.1086/304609}.
\bibitem[{{Shafter}(2017)}]{shaf17}
\bibinfo{author}{{Shafter}, A.W.}, \bibinfo{year}{2017}.
\newblock \bibinfo{title}{{The Galactic Nova Rate Revisited}}.
\newblock \bibinfo{journal}{Astrophys. J.} \bibinfo{volume}{834},
  \bibinfo{pages}{196}.
\newblock \DOIprefix\doi{10.3847/1538-4357/834/2/196}.
\bibitem[{{Shen} and {Bildsten}(2014)}]{shen14}
\bibinfo{author}{{Shen}, K.J.}, \bibinfo{author}{{Bildsten}, L.},
  \bibinfo{year}{2014}.
\newblock \bibinfo{title}{{The Ignition of Carbon Detonations via Converging
  Shock Waves in White Dwarfs}}.
\newblock \bibinfo{journal}{Astrophys. J.} \bibinfo{volume}{785},
  \bibinfo{pages}{61}.
\newblock \DOIprefix\doi{10.1088/0004-637X/785/1/61}.
\bibitem[{{Shen} et~al.(2012){Shen}, {Bildsten}, {Kasen} and
  {Quataert}}]{shen12}
\bibinfo{author}{{Shen}, K.J.}, \bibinfo{author}{{Bildsten}, L.},
  \bibinfo{author}{{Kasen}, D.}, \bibinfo{author}{{Quataert}, E.},
  \bibinfo{year}{2012}.
\newblock \bibinfo{title}{{The Long-term Evolution of Double White Dwarf
  Mergers}}.
\newblock \bibinfo{journal}{Astrophys. J.} \bibinfo{volume}{748}.
\newblock \DOIprefix\doi{10.1088/0004-637X/748/1/35}.
\bibitem[{{Shen} et~al.(2013){Shen}, {Guillochon} and {Foley}}]{shen13}
\bibinfo{author}{{Shen}, K.J.}, \bibinfo{author}{{Guillochon}, J.},
  \bibinfo{author}{{Foley}, R.J.}, \bibinfo{year}{2013}.
\newblock \bibinfo{title}{{Circumstellar Absorption in Double Detonation Type
  Ia Supernovae}}.
\newblock \bibinfo{journal}{Astrophys. J. Lett.} \bibinfo{volume}{770},
  \bibinfo{pages}{L35}.
\newblock \DOIprefix\doi{10.1088/2041-8205/770/2/L35}.
\bibitem[{{Siegert} et~al.(2018){Siegert}, {Coc} and {Delgado}}]{sieg18}
\bibinfo{author}{{Siegert}, T.}, \bibinfo{author}{{Coc}, A.},
  \bibinfo{author}{{Delgado}, L.e.a.}, \bibinfo{year}{2018}.
\newblock \bibinfo{title}{{Gamma-ray observations of Nova Sgr 2015 No. 2 with
  INTEGRAL}}.
\newblock \bibinfo{journal}{A\&A} \bibinfo{volume}{615}, \bibinfo{pages}{A107}.
\newblock \DOIprefix\doi{10.1051/0004-6361/201732514}.
\bibitem[{{Siegert} et~al.(2015){Siegert}, {Diehl}, {Krause} and
  {Greiner}}]{sieg15}
\bibinfo{author}{{Siegert}, T.}, \bibinfo{author}{{Diehl}, R.},
  \bibinfo{author}{{Krause}, M.G.H.}, \bibinfo{author}{{Greiner}, J.},
  \bibinfo{year}{2015}.
\newblock \bibinfo{title}{{Revisiting INTEGRAL/SPI observations of $^{44}$Ti
  from Cassiopeia A}}.
\newblock \bibinfo{journal}{A\&A} \bibinfo{volume}{579}, \bibinfo{pages}{A124}.
\newblock \DOIprefix\doi{10.1051/0004-6361/201525877}.
\bibitem[{{Silverman} et~al.(2012){Silverman}, {Foley} and
  {Filippenko}}]{silv12}
\bibinfo{author}{{Silverman}, J.M.}, \bibinfo{author}{{Foley}, R.J.},
  \bibinfo{author}{{Filippenko}, A.~V., e.a.}, \bibinfo{year}{2012}.
\newblock \bibinfo{title}{{Berkeley Supernova Ia Program - I. Observations,
  data reduction and spectroscopic sample of 582 low-redshift Type Ia
  supernovae}}.
\newblock \bibinfo{journal}{Mon. Not. R. Astron. Soc.} \bibinfo{volume}{425},
  \bibinfo{pages}{1789--1818}.
\newblock \DOIprefix\doi{10.1111/j.1365-2966.2012.21270.x}.
\bibitem[{{Silverman} et~al.(2013a){Silverman}, {Nugent} and
  {Gal-Yam}}]{silv13}
\bibinfo{author}{{Silverman}, J.M.}, \bibinfo{author}{{Nugent}, P.E.},
  \bibinfo{author}{{Gal-Yam}, e.a.}, \bibinfo{year}{2013}a.
\newblock \bibinfo{title}{{Late-time Spectral Observations of the Strongly
  Interacting Type Ia Supernova PTF11kx}}.
\newblock \bibinfo{journal}{Astrophys. J.} \bibinfo{volume}{772},
  \bibinfo{pages}{125}.
\newblock \DOIprefix\doi{10.1088/0004-637X/772/2/125}.
\bibitem[{{Silverman} et~al.(2013b){Silverman}, {Vinko} and
  {Kasliwal}}]{silv13a}
\bibinfo{author}{{Silverman}, J.M.}, \bibinfo{author}{{Vinko}, J.},
  \bibinfo{author}{{Kasliwal}, M.M.e.a.}, \bibinfo{year}{2013}b.
\newblock \bibinfo{title}{{SN 2000cx and SN 2013bh: extremely rare, nearly twin
  Type Ia supernovae}}.
\newblock \bibinfo{journal}{Mon. Not. R. Astron. Soc.} \bibinfo{volume}{436},
  \bibinfo{pages}{1225--1237}.
\newblock \DOIprefix\doi{10.1093/mnras/stt1647}.
\bibitem[{{Sim} and {Mazzali}(2008)}]{sim08}
\bibinfo{author}{{Sim}, S.A.}, \bibinfo{author}{{Mazzali}, P.A.},
  \bibinfo{year}{2008}.
\newblock \bibinfo{title}{{On the {$\gamma$}-ray emission of Type Ia
  supernovae}}.
\newblock \bibinfo{journal}{Mon. Not. R. Astron. Soc.} \bibinfo{volume}{385},
  \bibinfo{pages}{1681--1690}.
\newblock \DOIprefix\doi{10.1111/j.1365-2966.2008.12600.x}.
\bibitem[{{Smartt}(2009)}]{smar09}
\bibinfo{author}{{Smartt}, S.J.}, \bibinfo{year}{2009}.
\newblock \bibinfo{title}{{Progenitors of Core-Collapse Supernovae}}.
\newblock \bibinfo{journal}{Annu. Rev. Astron. Astrophys.}
  \bibinfo{volume}{47}, \bibinfo{pages}{63--106}.
\newblock \DOIprefix\doi{10.1146/annurev-astro-082708-101737}.
\bibitem[{{Smith} and {McCray}(2007)}]{smit07}
\bibinfo{author}{{Smith}, N.}, \bibinfo{author}{{McCray}, R.},
  \bibinfo{year}{2007}.
\newblock \bibinfo{title}{{Shell-shocked Diffusion Model for the Light Curve of
  SN 2006gy}}.
\newblock \bibinfo{journal}{Astrophys. J. Lett.} \bibinfo{volume}{671},
  \bibinfo{pages}{L17--L20}.
\newblock \DOIprefix\doi{10.1086/524681}.
\bibitem[{{Stephenson} and {Green}(2002)}]{step02}
\bibinfo{author}{{Stephenson}, F.R.}, \bibinfo{author}{{Green}, D.A.},
  \bibinfo{year}{2002}.
\newblock \bibinfo{title}{{Historical supernovae and their remnants}}.
\newblock \bibinfo{journal}{Historical supernovae and their remnants}
  \bibinfo{volume}{5}.
\bibitem[{{Sunyaev} et~al.(1991){Sunyaev}, {Grebenev}, {Kaniovsky}, {Efremov},
  {Kuznetsov}, {Pavlinsky}, {Yamburenko}, {Englhauser}, {Doebereiner},
  {Pietsch}, {Reppin}, {Truemper}, {Kendziorra}, {Maisack}, {Mony} and
  {Staubert}}]{suny91}
\bibinfo{author}{{Sunyaev}, R.}, \bibinfo{author}{{Grebenev}, S.},
  \bibinfo{author}{{Kaniovsky}, A.}, \bibinfo{author}{{Efremov}, V.},
  \bibinfo{author}{{Kuznetsov}, A.}, \bibinfo{author}{{Pavlinsky}, M.},
  \bibinfo{author}{{Yamburenko}, N.}, \bibinfo{author}{{Englhauser}, J.},
  \bibinfo{author}{{Doebereiner}, S.}, \bibinfo{author}{{Pietsch}, W.},
  \bibinfo{author}{{Reppin}, C.}, \bibinfo{author}{{Truemper}, J.},
  \bibinfo{author}{{Kendziorra}, E.}, \bibinfo{author}{{Maisack}, M.},
  \bibinfo{author}{{Mony}, B.}, \bibinfo{author}{{Staubert}, R.},
  \bibinfo{year}{1991}.
\newblock \bibinfo{title}{{Hard x-rays from supernova 1987A: results of
  Mir-Kvant and Granat in 1987-1990 and expectations}}, in:
  \bibinfo{editor}{{Durouchoux}, P.}, \bibinfo{editor}{{Prantzos}, N.} (Eds.),
  \bibinfo{booktitle}{Gamma-Ray Line Astrophysics}, pp.
  \bibinfo{pages}{211--217}.
\newblock \DOIprefix\doi{10.1063/1.40938}.
\bibitem[{{Sunyaev} et~al.(1987){Sunyaev}, {Kaniovsky} and {Efremov}}]{suny87}
\bibinfo{author}{{Sunyaev}, R.}, \bibinfo{author}{{Kaniovsky}, A.},
  \bibinfo{author}{{Efremov}, V., e.a.}, \bibinfo{year}{1987}.
\newblock \bibinfo{title}{{Discovery of hard X-ray emission from supernova
  1987A.}}
\newblock \bibinfo{journal}{Nature} \bibinfo{volume}{330},
  \bibinfo{pages}{227--229}.
\newblock \DOIprefix\doi{10.1038/330227a0}.
\bibitem[{{Sunyaev} et~al.(1990a){Sunyaev}, {Churazov}, {Efremov}, {Gilfanov},
  {Grebenev}, {Kaniovsky}, {Stepanov}, {Yunin}, {Kuznetsov}, {Melioransky},
  {Yamburenko}, {Kiselev}, {Lapshov}, {Pappe}, {Boyarsky}, {Loznikov},
  {Prudkoglyad}, {Terekhov}, {Reppin}, {Pietsch}, {Englhauser}, {Tr{\"u}mper},
  {Voges}, {Kendziorra}, {Bezler}, {Staubert}, {Brinkman}, {Heise}, {in't
  Zand}, {Jager}, {Skinner}, {Al-Emam}, {Patterson}, {Willmore}, {Smith} and
  {Parmar}}]{suny90b}
\bibinfo{author}{{Sunyaev}, R.A.}, \bibinfo{author}{{Churazov}, E.},
  \bibinfo{author}{{Efremov}, V.}, \bibinfo{author}{{Gilfanov}, M.},
  \bibinfo{author}{{Grebenev}, S.}, \bibinfo{author}{{Kaniovsky}, A.},
  \bibinfo{author}{{Stepanov}, D.}, \bibinfo{author}{{Yunin}, S.},
  \bibinfo{author}{{Kuznetsov}, A.}, \bibinfo{author}{{Melioransky}, A.},
  \bibinfo{author}{{Yamburenko}, N.}, \bibinfo{author}{{Kiselev}, S.},
  \bibinfo{author}{{Lapshov}, I.}, \bibinfo{author}{{Pappe}, N.},
  \bibinfo{author}{{Boyarsky}, M.}, \bibinfo{author}{{Loznikov}, V.},
  \bibinfo{author}{{Prudkoglyad}, A.}, \bibinfo{author}{{Terekhov}, O.},
  \bibinfo{author}{{Reppin}, C.}, \bibinfo{author}{{Pietsch}, W.},
  \bibinfo{author}{{Englhauser}, J.}, \bibinfo{author}{{Tr{\"u}mper}, J.},
  \bibinfo{author}{{Voges}, W.}, \bibinfo{author}{{Kendziorra}, E.},
  \bibinfo{author}{{Bezler}, M.}, \bibinfo{author}{{Staubert}, R.},
  \bibinfo{author}{{Brinkman}, A.C.}, \bibinfo{author}{{Heise}, J.},
  \bibinfo{author}{{in't Zand}, J.J.M.}, \bibinfo{author}{{Jager}, R.},
  \bibinfo{author}{{Skinner}, G.K.}, \bibinfo{author}{{Al-Emam}, O.},
  \bibinfo{author}{{Patterson}, T.G.}, \bibinfo{author}{{Willmore}, A.P.},
  \bibinfo{author}{{Smith}, A.}, \bibinfo{author}{{Parmar}, A.},
  \bibinfo{year}{1990}a.
\newblock \bibinfo{title}{{Highlights from the KVANT mission}}.
\newblock \bibinfo{journal}{Advances in Space Research} \bibinfo{volume}{10},
  \bibinfo{pages}{41--46}.
\bibitem[{{Sunyaev} et~al.(1990b){Sunyaev}, {Kaniovsky}, {Efremov}, {Grebenev},
  {Kuznetsov}, {Englhauser}, {Doebereiner}, {Pietsch}, {Reppin}, {Truemper},
  {Kendziorra}, {Maisack}, {Mony} and {Staubert}}]{suny90}
\bibinfo{author}{{Sunyaev}, R.A.}, \bibinfo{author}{{Kaniovsky}, A.S.},
  \bibinfo{author}{{Efremov}, V.V.}, \bibinfo{author}{{Grebenev}, S.A.},
  \bibinfo{author}{{Kuznetsov}, A.V.}, \bibinfo{author}{{Englhauser}, J.},
  \bibinfo{author}{{Doebereiner}, S.}, \bibinfo{author}{{Pietsch}, W.},
  \bibinfo{author}{{Reppin}, C.}, \bibinfo{author}{{Truemper}, J.},
  \bibinfo{author}{{Kendziorra}, E.}, \bibinfo{author}{{Maisack}, M.},
  \bibinfo{author}{{Mony}, B.}, \bibinfo{author}{{Staubert}, R.},
  \bibinfo{year}{1990}b.
\newblock \bibinfo{title}{{Hard X-ray radiation from supernova 1987A. The
  results of the ``Roentgen'' Observatory observations onboard the ``Kvant''
  module in 1987-1989.}}
\newblock \bibinfo{journal}{Sov. Astron. Lett.} \bibinfo{volume}{16},
  \bibinfo{pages}{171--176}.
\bibitem[{{Suzuki} et~al.(2019){Suzuki}, {Zha}, {Leung} and {Nomoto}}]{suzu19}
\bibinfo{author}{{Suzuki}, T.}, \bibinfo{author}{{Zha}, S.},
  \bibinfo{author}{{Leung}, S.C.}, \bibinfo{author}{{Nomoto}, K.},
  \bibinfo{year}{2019}.
\newblock \bibinfo{title}{{Electron-capture Rates in $^{20}$Ne for a Forbidden
  Transition to the Ground State of $^{20}$F Relevant to the Final Evolution of
  High-density O-Ne-Mg Cores}}.
\newblock \bibinfo{journal}{Astrophys. J.} \bibinfo{volume}{881},
  \bibinfo{pages}{64}.
\newblock \DOIprefix\doi{10.3847/1538-4357/ab2b93}.
\bibitem[{{Tajitsu} et~al.(2016){Tajitsu}, {Sadakane} and {Naito}}]{taji16}
\bibinfo{author}{{Tajitsu}, A.}, \bibinfo{author}{{Sadakane}, K.},
  \bibinfo{author}{{Naito}, H.e.a.}, \bibinfo{year}{2016}.
\newblock \bibinfo{title}{{The $^{7}$Be II Resonance Lines in Two Classical
  Novae V5668 Sgr and V2944 Oph}}.
\newblock \bibinfo{journal}{Astrophys. J.} \bibinfo{volume}{818},
  \bibinfo{pages}{191}.
\newblock \DOIprefix\doi{10.3847/0004-637X/818/2/191}.
\bibitem[{{Takahara} et~al.(1989){Takahara}, {Hino} and {Oda}}]{taka89}
\bibinfo{author}{{Takahara}, M.}, \bibinfo{author}{{Hino}, M.},
  \bibinfo{author}{{Oda}, T.e.a.}, \bibinfo{year}{1989}.
\newblock \bibinfo{title}{{Microscopic calculation of the rates of electron
  captures which induce the collapse of O+Ne+Mg cores}}.
\newblock \bibinfo{journal}{Nuc. Phys. A} \bibinfo{volume}{504},
  \bibinfo{pages}{167--192}.
\newblock \DOIprefix\doi{10.1016/0375-9474(89)90288-1}.
\bibitem[{{Tammann}(1970)}]{tamm70}
\bibinfo{author}{{Tammann}, G.A.}, \bibinfo{year}{1970}.
\newblock \bibinfo{title}{{On the Frequency of Supernovae as a Function of the
  Integral Properties of Inter mediate and Late Type Spiral Galaxies}}.
\newblock \bibinfo{journal}{A\&A} \bibinfo{volume}{8}, \bibinfo{pages}{458}.
\bibitem[{{Tammann} et~al.(1994){Tammann}, {Loeffler} and {Schroeder}}]{tamm94}
\bibinfo{author}{{Tammann}, G.A.}, \bibinfo{author}{{Loeffler}, W.},
  \bibinfo{author}{{Schroeder}, A.}, \bibinfo{year}{1994}.
\newblock \bibinfo{title}{{The Galactic Supernova Rate}}.
\newblock \bibinfo{journal}{Astrophys. J. Suppl. S} \bibinfo{volume}{92},
  \bibinfo{pages}{487}.
\newblock \DOIprefix\doi{10.1086/192002}.
\bibitem[{{Tanimori} et~al.(2015){Tanimori}, {Kubo} and {Takada}}]{tani15}
\bibinfo{author}{{Tanimori}, T.}, \bibinfo{author}{{Kubo}, H.},
  \bibinfo{author}{{Takada}, A.e.a.}, \bibinfo{year}{2015}.
\newblock \bibinfo{title}{{An Electron-Tracking Compton Telescope for a Survey
  of the Deep Universe by MeV Gamma-Rays}}.
\newblock \bibinfo{journal}{Astrophys. J.} \bibinfo{volume}{810},
  \bibinfo{pages}{28}.
\newblock \DOIprefix\doi{10.1088/0004-637X/810/1/28}.
\bibitem[{{Tanimori} et~al.(2017){Tanimori}, {Mizumura} and {Takada}}]{tani17}
\bibinfo{author}{{Tanimori}, T.}, \bibinfo{author}{{Mizumura}, Y.},
  \bibinfo{author}{{Takada}, A.e.a.}, \bibinfo{year}{2017}.
\newblock \bibinfo{title}{{Establishment of Imaging Spectroscopy of Nuclear
  Gamma-Rays based on Geometrical Optics}}.
\newblock \bibinfo{journal}{Nature Scientific Reports} \bibinfo{volume}{7},
  \bibinfo{pages}{41511}.
\newblock \DOIprefix\doi{10.1038/srep41511}.
\bibitem[{{Taubenberger}(2017)}]{taub17}
\bibinfo{author}{{Taubenberger}, S.}, \bibinfo{year}{2017}.
\newblock \bibinfo{title}{{The Extremes of Thermonuclear Supernovae}}, in:
  \bibinfo{booktitle}{Handbook of Supernovae, ISBN 978-3-319-21845-8. Springer
  International Publishing AG, 2017, p. 317}, p. \bibinfo{pages}{317}.
\newblock \DOIprefix\doi{10.1007/978-3-319-21846-5_37}.
\bibitem[{{Teegarden} et~al.(1989){Teegarden}, {Barthelmy}, {Gehrels},
  {Tueller} and {Leventhal}}]{teeg89}
\bibinfo{author}{{Teegarden}, B.J.}, \bibinfo{author}{{Barthelmy}, S.D.},
  \bibinfo{author}{{Gehrels}, N.}, \bibinfo{author}{{Tueller}, J.},
  \bibinfo{author}{{Leventhal}, M.}, \bibinfo{year}{1989}.
\newblock \bibinfo{title}{{Resolution of the 1,238-keV
  {\ensuremath{\gamma}}-ray line from supernova 1987A}}.
\newblock \bibinfo{journal}{Nature} \bibinfo{volume}{339},
  \bibinfo{pages}{122--123}.
\newblock \DOIprefix\doi{10.1038/339122a0}.
\bibitem[{{Teegarden} et~al.(1985){Teegarden}, {Cline} and {Gehrels}}]{teeg85}
\bibinfo{author}{{Teegarden}, B.J.}, \bibinfo{author}{{Cline}, T.L.},
  \bibinfo{author}{{Gehrels}, N.e.a.}, \bibinfo{year}{1985}.
\newblock \bibinfo{title}{{The Gamma-Ray Imaging Spectrometer (GRIS): A new
  balloon-borne experiment for gamma-ray line astronomy}}, in:
  \bibinfo{booktitle}{19th International Cosmic Ray Conference (ICRC19), Volume
  3}, pp. \bibinfo{pages}{307--310}.
\bibitem[{{The} and {Burrows}(2014)}]{the14}
\bibinfo{author}{{The}, L.S.}, \bibinfo{author}{{Burrows}, A.},
  \bibinfo{year}{2014}.
\newblock \bibinfo{title}{{Expectations for the Hard X-Ray Continuum and
  Gamma-Ray Line Fluxes from the Type Ia Supernova SN 2014J in M82}}.
\newblock \bibinfo{journal}{Astrophys. J.} \bibinfo{volume}{786},
  \bibinfo{pages}{141}.
\newblock \DOIprefix\doi{10.1088/0004-637X/786/2/141}.
\bibitem[{{Thielemann} et~al.(2018){Thielemann}, {Isern}, {Perego} and {von
  Ballmoos}}]{thie18}
\bibinfo{author}{{Thielemann}, F.K.}, \bibinfo{author}{{Isern}, J.},
  \bibinfo{author}{{Perego}, A.}, \bibinfo{author}{{von Ballmoos}, P.},
  \bibinfo{year}{2018}.
\newblock \bibinfo{title}{{Nucleosynthesis in Supernovae}}.
\newblock \bibinfo{journal}{Space Sci. Rev.} \bibinfo{volume}{214},
  \bibinfo{pages}{62}.
\newblock \DOIprefix\doi{10.1007/s11214-018-0494-5}.
\bibitem[{{Timmes} and {Woosley}(1992)}]{timm92}
\bibinfo{author}{{Timmes}, F.X.}, \bibinfo{author}{{Woosley}, S.E.},
  \bibinfo{year}{1992}.
\newblock \bibinfo{title}{{The conductive propagation of nuclear flames. I -
  Degenerate C + O and O + NE + MG white dwarfs}}.
\newblock \bibinfo{journal}{Astrophys. J.} \bibinfo{volume}{396},
  \bibinfo{pages}{649--667}.
\bibitem[{{Timmes} and {Woosley}(1997)}]{timm97}
\bibinfo{author}{{Timmes}, F.X.}, \bibinfo{author}{{Woosley}, S.E.},
  \bibinfo{year}{1997}.
\newblock \bibinfo{title}{{Gamma-Ray Line Signals from Supernovae within 100
  MPC}}.
\newblock \bibinfo{journal}{Astrophys. J.} \bibinfo{volume}{489},
  \bibinfo{pages}{160}.
\newblock \DOIprefix\doi{10.1086/304768}.
\bibitem[{{Tomsick} et~al.(2019){Tomsick}, {Zoglauer}, {Sleator}, {Lazar},
  {Beechert}, {Boggs}, {Roberts}, {Siegert}, {Lowell}, {Wulf}, {Grove},
  {Phlips}, {Brand t}, {Smale}, {Kierans}, {Burns}, {Hartmann}, {Leising},
  {Ajello}, {Fryer}, {Amman}, {Chang}, {Jean} and {von Ballmoos}}]{toms19}
\bibinfo{author}{{Tomsick}, J.}, \bibinfo{author}{{Zoglauer}, A.},
  \bibinfo{author}{{Sleator}, C.}, \bibinfo{author}{{Lazar}, H.},
  \bibinfo{author}{{Beechert}, J.}, \bibinfo{author}{{Boggs}, S.},
  \bibinfo{author}{{Roberts}, J.}, \bibinfo{author}{{Siegert}, T.},
  \bibinfo{author}{{Lowell}, A.}, \bibinfo{author}{{Wulf}, E.},
  \bibinfo{author}{{Grove}, E.}, \bibinfo{author}{{Phlips}, B.},
  \bibinfo{author}{{Brand t}, T.}, \bibinfo{author}{{Smale}, A.},
  \bibinfo{author}{{Kierans}, C.}, \bibinfo{author}{{Burns}, E.},
  \bibinfo{author}{{Hartmann}, D.}, \bibinfo{author}{{Leising}, M.},
  \bibinfo{author}{{Ajello}, M.}, \bibinfo{author}{{Fryer}, C.},
  \bibinfo{author}{{Amman}, M.}, \bibinfo{author}{{Chang}, H.K.},
  \bibinfo{author}{{Jean}, P.}, \bibinfo{author}{{von Ballmoos}, P.},
  \bibinfo{year}{2019}.
\newblock \bibinfo{title}{{The Compton Spectrometer and Imager}}, in:
  \bibinfo{booktitle}{Bull. Aust. Acoust. Soc}, p.~\bibinfo{pages}{98}.
\newblock \href{http://arxiv.org/abs/1908.04334}{\tt arXiv:1908.04334}.
\bibitem[{{Toonen} et~al.(2018){Toonen}, {Perets} and {Hamers}}]{toon18}
\bibinfo{author}{{Toonen}, S.}, \bibinfo{author}{{Perets}, H.B.},
  \bibinfo{author}{{Hamers}, A.S.}, \bibinfo{year}{2018}.
\newblock \bibinfo{title}{{Rate of WD-WD head-on collisions in isolated triples
  is too low to explain standard type Ia supernovae}}.
\newblock \bibinfo{journal}{A\&A} \bibinfo{volume}{610}, \bibinfo{pages}{A22}.
\newblock \DOIprefix\doi{10.1051/0004-6361/201731874}.
\bibitem[{{Tsygankov} et~al.(2016){Tsygankov}, {Krivonos} and
  {Lutovinov}}]{tsyg16}
\bibinfo{author}{{Tsygankov}, S.S.}, \bibinfo{author}{{Krivonos}, R.A.},
  \bibinfo{author}{{Lutovinov}, A.A.a.}, \bibinfo{year}{2016}.
\newblock \bibinfo{title}{{Galactic survey of $^{44}$Ti sources with the IBIS
  telescope onboard INTEGRAL}}.
\newblock \bibinfo{journal}{Mon. Not. R. Astron. Soc.} \bibinfo{volume}{458},
  \bibinfo{pages}{3411--3419}.
\newblock \DOIprefix\doi{10.1093/mnras/stw549}.
\bibitem[{{Tueller} et~al.(1990){Tueller}, {Barthelmy}, {Gehrels}, {Teegarden},
  {Leventhal} and {MacCallum}}]{tuel90}
\bibinfo{author}{{Tueller}, J.}, \bibinfo{author}{{Barthelmy}, S.},
  \bibinfo{author}{{Gehrels}, N.}, \bibinfo{author}{{Teegarden}, B.J.},
  \bibinfo{author}{{Leventhal}, M.}, \bibinfo{author}{{MacCallum}, C.J.},
  \bibinfo{year}{1990}.
\newblock \bibinfo{title}{{Observations of gamma-ray line profiles from SN
  1987A}}.
\newblock \bibinfo{journal}{Astrophys. J. Lett.} \bibinfo{volume}{351},
  \bibinfo{pages}{L41--L44}.
\bibitem[{{Ubertini} et~al.(2003){Ubertini}, {Lebrun} and {Di Cocco}}]{uber03}
\bibinfo{author}{{Ubertini}, P.}, \bibinfo{author}{{Lebrun}, F.},
  \bibinfo{author}{{Di Cocco}, G.e.a.}, \bibinfo{year}{2003}.
\newblock \bibinfo{title}{{IBIS: The Imager on-board INTEGRAL}}.
\newblock \bibinfo{journal}{A\&A} \bibinfo{volume}{411},
  \bibinfo{pages}{L131--L139}.
\newblock \DOIprefix\doi{10.1051/0004-6361:20031224}.
\bibitem[{{van den Bergh}(1975)}]{berg75}
\bibinfo{author}{{van den Bergh}, S.}, \bibinfo{year}{1975}.
\newblock \bibinfo{title}{{The Next Galactic Supernova}}.
\newblock \bibinfo{journal}{Astrophys. Space Sci.} \bibinfo{volume}{38},
  \bibinfo{pages}{447}.
\bibitem[{{van den Bergh} and {Tammann}(1991)}]{berg91}
\bibinfo{author}{{van den Bergh}, S.}, \bibinfo{author}{{Tammann}, G.A.},
  \bibinfo{year}{1991}.
\newblock \bibinfo{title}{{Galactic and extragalactic supernova rates.}}
\newblock \bibinfo{journal}{Annu. Rev. Astron. Astrophys.}
  \bibinfo{volume}{29}, \bibinfo{pages}{363--407}.
\newblock \DOIprefix\doi{10.1146/annurev.aa.29.090191.002051}.
\bibitem[{{Varani} et~al.(1990){Varani}, {Meikle}, {Spyromilio} and
  {Allen}}]{vara90}
\bibinfo{author}{{Varani}, G.F.}, \bibinfo{author}{{Meikle}, W.P.S.},
  \bibinfo{author}{{Spyromilio}, J.}, \bibinfo{author}{{Allen}, D.A.},
  \bibinfo{year}{1990}.
\newblock \bibinfo{title}{{Direct observation of radioactive cobalt decay in
  supernova 1987A.}}
\newblock \bibinfo{journal}{Mon. Not. R. Astron. Soc.} \bibinfo{volume}{245},
  \bibinfo{pages}{570}.
\bibitem[{{Vedrenne} et~al.(2003){Vedrenne}, {Roques} and
  {Sch{\"o}nfelder}}]{vedr03}
\bibinfo{author}{{Vedrenne}, G.}, \bibinfo{author}{{Roques}, J.P.},
  \bibinfo{author}{{Sch{\"o}nfelder}, V.e.a.}, \bibinfo{year}{2003}.
\newblock \bibinfo{title}{{SPI: The spectrometer aboard INTEGRAL}}.
\newblock \bibinfo{journal}{A\&A} \bibinfo{volume}{411},
  \bibinfo{pages}{L63--L70}.
\newblock \DOIprefix\doi{10.1051/0004-6361:20031482}.
\bibitem[{{Ventura} and {D'Antona}(2011)}]{vent11}
\bibinfo{author}{{Ventura}, P.}, \bibinfo{author}{{D'Antona}, F.},
  \bibinfo{year}{2011}.
\newblock \bibinfo{title}{{Hot bottom burning in the envelope of super
  asymptotic giant branch stars}}.
\newblock \bibinfo{journal}{Mon. Not. R. Astron. Soc.} \bibinfo{volume}{410},
  \bibinfo{pages}{2760--2766}.
\newblock \DOIprefix\doi{10.1111/j.1365-2966.2010.17651.x}.
\bibitem[{{Vink}(2012)}]{vink12}
\bibinfo{author}{{Vink}, J.}, \bibinfo{year}{2012}.
\newblock \bibinfo{title}{{Supernova remnants: the X-ray perspective}}.
\newblock \bibinfo{journal}{Astron. Astrophys Rev} \bibinfo{volume}{20},
  \bibinfo{pages}{49}.
\newblock \DOIprefix\doi{10.1007/s00159-011-0049-1},
  \href{http://arxiv.org/abs/1112.0576}{\tt arXiv:1112.0576}.
\bibitem[{{Vink} et~al.(1996){Vink}, {Kaastra} and {Bleeker}}]{vink96}
\bibinfo{author}{{Vink}, J.}, \bibinfo{author}{{Kaastra}, J.S.},
  \bibinfo{author}{{Bleeker}, J.A.M.}, \bibinfo{year}{1996}.
\newblock \bibinfo{title}{{A new mass estimate and puzzling abundances of SNR
  Cassiopeia A.}}
\newblock \bibinfo{journal}{A\&A} \bibinfo{volume}{307},
  \bibinfo{pages}{L41--L44}.
\bibitem[{{Vink} et~al.(2001){Vink}, {Laming}, {Kaastra}, {Bleeker}, {Bloemen}
  and {Oberlack}}]{vink01}
\bibinfo{author}{{Vink}, J.}, \bibinfo{author}{{Laming}, J.M.},
  \bibinfo{author}{{Kaastra}, J.S.}, \bibinfo{author}{{Bleeker}, J.A.M.},
  \bibinfo{author}{{Bloemen}, H.}, \bibinfo{author}{{Oberlack}, U.},
  \bibinfo{year}{2001}.
\newblock \bibinfo{title}{{Detection of the 67.9 and 78.4 keV Lines Associated
  with the Radioactive Decay of $^{44}$Ti in Cassiopeia A}}.
\newblock \bibinfo{journal}{Astrophys. J. Lett.} \bibinfo{volume}{560},
  \bibinfo{pages}{L79--L82}.
\newblock \DOIprefix\doi{10.1086/324172}.
\bibitem[{{von Ballmoos} et~al.(2005){von Ballmoos}, {Halloin}, {Evrard},
  {Skinner}, {Abrosimov}, {Alvarez}, {Bastie}, {Hamelin}, {Hernanz}, {Jean},
  {Kn{\"o}dlseder} and {Smither}}]{ball05}
\bibinfo{author}{{von Ballmoos}, P.}, \bibinfo{author}{{Halloin}, H.},
  \bibinfo{author}{{Evrard}, J.}, \bibinfo{author}{{Skinner}, G.},
  \bibinfo{author}{{Abrosimov}, N.}, \bibinfo{author}{{Alvarez}, J.},
  \bibinfo{author}{{Bastie}, P.}, \bibinfo{author}{{Hamelin}, B.},
  \bibinfo{author}{{Hernanz}, M.}, \bibinfo{author}{{Jean}, P.},
  \bibinfo{author}{{Kn{\"o}dlseder}, J.}, \bibinfo{author}{{Smither}, B.},
  \bibinfo{year}{2005}.
\newblock \bibinfo{title}{{CLAIRE: First light for a gamma-ray lens}}.
\newblock \bibinfo{journal}{Experimental Astronomy} \bibinfo{volume}{20},
  \bibinfo{pages}{253--267}.
\newblock \DOIprefix\doi{10.1007/s10686-006-9071-0}.
\bibitem[{{Walker}(1954)}]{walk54}
\bibinfo{author}{{Walker}, M.F.}, \bibinfo{year}{1954}.
\newblock \bibinfo{title}{{Nova DQ Herculis (1934): an Eclipsing Binary with
  Very Short Period}}.
\newblock \bibinfo{journal}{Publ. Astron. Soc. Pac.} \bibinfo{volume}{66},
  \bibinfo{pages}{230}.
\newblock \DOIprefix\doi{10.1086/126703}.
\bibitem[{{Wanajo} et~al.(2009){Wanajo}, {Nomoto}, {Janka}, {Kitaura} and
  {M{\"u}ller}}]{wana09}
\bibinfo{author}{{Wanajo}, S.}, \bibinfo{author}{{Nomoto}, K.},
  \bibinfo{author}{{Janka}, H.T.}, \bibinfo{author}{{Kitaura}, F.S.},
  \bibinfo{author}{{M{\"u}ller}, B.}, \bibinfo{year}{2009}.
\newblock \bibinfo{title}{{Nucleosynthesis in Electron Capture Supernovae of
  Asymptotic Giant Branch Stars}}.
\newblock \bibinfo{journal}{Astrophys. J.} \bibinfo{volume}{695},
  \bibinfo{pages}{208--220}.
\newblock \DOIprefix\doi{10.1088/0004-637X/695/1/208}.
\bibitem[{{Warren} et~al.(2005){Warren}, {Hughes}, {Badenes}, {Ghavamian},
  {McKee}, {Moffett}, {Plucinsky}, {Rakowski}, {Reynoso} and {Slane}}]{warr05}
\bibinfo{author}{{Warren}, J.S.}, \bibinfo{author}{{Hughes}, J.P.},
  \bibinfo{author}{{Badenes}, C.}, \bibinfo{author}{{Ghavamian}, P.},
  \bibinfo{author}{{McKee}, C.F.}, \bibinfo{author}{{Moffett}, D.},
  \bibinfo{author}{{Plucinsky}, P.P.}, \bibinfo{author}{{Rakowski}, C.},
  \bibinfo{author}{{Reynoso}, E.}, \bibinfo{author}{{Slane}, P.},
  \bibinfo{year}{2005}.
\newblock \bibinfo{title}{{Cosmic-Ray Acceleration at the Forward Shock in
  Tycho's Supernova Remnant: Evidence from Chandra X-Ray Observations}}.
\newblock \bibinfo{journal}{Astrophys. J.} \bibinfo{volume}{634},
  \bibinfo{pages}{376--389}.
\newblock \DOIprefix\doi{10.1086/496941},
  \href{http://arxiv.org/abs/astro-ph/0507478}{\tt arXiv:astro-ph/0507478}.
\bibitem[{{Webbink}(1984)}]{webb84}
\bibinfo{author}{{Webbink}, R.F.}, \bibinfo{year}{1984}.
\newblock \bibinfo{title}{{Double white dwarfs as progenitors of R Coronae
  Borealis stars and Type I supernovae}}.
\newblock \bibinfo{journal}{Astrophys. J.} \bibinfo{volume}{277},
  \bibinfo{pages}{355--360}.
\newblock \DOIprefix\doi{10.1086/161701}.
\bibitem[{{Weil} et~al.(2020){Weil}, {Fesen}, {Patnaude}, {Raymond},
  {Chevalier}, {Milisavljevic} and {Gerardy}}]{weil20}
\bibinfo{author}{{Weil}, K.E.}, \bibinfo{author}{{Fesen}, R.A.},
  \bibinfo{author}{{Patnaude}, D.J.}, \bibinfo{author}{{Raymond}, J.C.},
  \bibinfo{author}{{Chevalier}, R.A.}, \bibinfo{author}{{Milisavljevic}, D.},
  \bibinfo{author}{{Gerardy}, C.L.}, \bibinfo{year}{2020}.
\newblock \bibinfo{title}{{Detection of the Red Supergiant Wind from the
  Progenitor of Cassiopeia A}}.
\newblock \bibinfo{journal}{Astrophys. J.} \bibinfo{volume}{891},
  \bibinfo{pages}{116}.
\newblock \DOIprefix\doi{10.3847/1538-4357/ab76bf},
  \href{http://arxiv.org/abs/2002.08442}{\tt arXiv:2002.08442}.
\bibitem[{{Weinberger} et~al.(2020){Weinberger}, {Diehl}, {Pleintinger},
  {Siegert} and {Greiner}}]{wein20}
\bibinfo{author}{{Weinberger}, C.}, \bibinfo{author}{{Diehl}, R.},
  \bibinfo{author}{{Pleintinger}, M.M.M.}, \bibinfo{author}{{Siegert}, T.},
  \bibinfo{author}{{Greiner}, J.}, \bibinfo{year}{2020}.
\newblock \bibinfo{title}{{$^{44}$Ti ejecta in young supernova remnants}}.
\newblock \bibinfo{journal}{Astronom. Astrophys.} \bibinfo{volume}{638},
  \bibinfo{pages}{A83}.
\newblock \DOIprefix\doi{10.1051/0004-6361/202037536},
  \href{http://arxiv.org/abs/2004.12688}{\tt arXiv:2004.12688}.
\bibitem[{{Whalen} et~al.(2014){Whalen}, {Smidt}, {Even}, {Woosley}, {Heger},
  {Stiavelli} and {Fryer}}]{whal14}
\bibinfo{author}{{Whalen}, D.J.}, \bibinfo{author}{{Smidt}, J.},
  \bibinfo{author}{{Even}, W.}, \bibinfo{author}{{Woosley}, S.E.},
  \bibinfo{author}{{Heger}, A.}, \bibinfo{author}{{Stiavelli}, M.},
  \bibinfo{author}{{Fryer}, C.L.}, \bibinfo{year}{2014}.
\newblock \bibinfo{title}{{Finding the First Cosmic Explosions. III.
  Pulsational Pair-instability Supernovae}}.
\newblock \bibinfo{journal}{Astrophys. J.} \bibinfo{volume}{781},
  \bibinfo{pages}{106}.
\bibitem[{{Wheeler} and {Harkness}(1990)}]{whee90}
\bibinfo{author}{{Wheeler}, J.C.}, \bibinfo{author}{{Harkness}, R.P.},
  \bibinfo{year}{1990}.
\newblock \bibinfo{title}{{Type I supernovae}}.
\newblock \bibinfo{journal}{Reports on Progress in Physics}
  \bibinfo{volume}{53}, \bibinfo{pages}{1467--1557}.
\newblock \DOIprefix\doi{10.1088/0034-4885/53/12/001}.
\bibitem[{{Wheeler} et~al.(2015){Wheeler}, {Johnson} and
  {Clocchiatti}}]{whee15}
\bibinfo{author}{{Wheeler}, J.C.}, \bibinfo{author}{{Johnson}, V.},
  \bibinfo{author}{{Clocchiatti}, A.}, \bibinfo{year}{2015}.
\newblock \bibinfo{title}{{Analysis of late-time light curves of Type IIb, Ib
  and Ic supernovae}}.
\newblock \bibinfo{journal}{Mon. Not. R. Astron. Soc.} \bibinfo{volume}{450},
  \bibinfo{pages}{1295--1307}.
\newblock \DOIprefix\doi{10.1093/mnras/stv650}.
\bibitem[{{Whelan} and {Iben}(1973)}]{whel73}
\bibinfo{author}{{Whelan}, J.}, \bibinfo{author}{{Iben}, Jr., I.},
  \bibinfo{year}{1973}.
\newblock \bibinfo{title}{{Binaries and Supernovae of Type I}}.
\newblock \bibinfo{journal}{Astrophys. J.} \bibinfo{volume}{186},
  \bibinfo{pages}{1007--1014}.
\newblock \DOIprefix\doi{10.1086/152565}.
\bibitem[{{White} and {Kasliwal}(2015)}]{whit15}
\bibinfo{author}{{White}, C.J.}, \bibinfo{author}{{Kasliwal}, M.M.e.a.},
  \bibinfo{year}{2015}.
\newblock \bibinfo{title}{{Slow-speed Supernovae from the Palomar Transient
  Factory: Two Channels}}.
\newblock \bibinfo{journal}{Astrophys. J.} \bibinfo{volume}{799},
  \bibinfo{pages}{52}.
\newblock \DOIprefix\doi{10.1088/0004-637X/799/1/52}.
\bibitem[{{Winkler} et~al.(2003){Winkler}, {Courvoisier} and {Di
  Cocco}}]{wink03}
\bibinfo{author}{{Winkler}, C.}, \bibinfo{author}{{Courvoisier}, T.J.L.},
  \bibinfo{author}{{Di Cocco}, G.e.a.}, \bibinfo{year}{2003}.
\newblock \bibinfo{title}{{The INTEGRAL mission}}.
\newblock \bibinfo{journal}{A\&A} \bibinfo{volume}{411},
  \bibinfo{pages}{L1--L6}.
\newblock \DOIprefix\doi{10.1051/0004-6361:20031288}.
\bibitem[{{Woltjer}(1972)}]{wolt72}
\bibinfo{author}{{Woltjer}, L.}, \bibinfo{year}{1972}.
\newblock \bibinfo{title}{{Supernova Remnants}}.
\newblock \bibinfo{journal}{Annu. Rev. Astron. Astrophys.}
  \bibinfo{volume}{10}, \bibinfo{pages}{129}.
\newblock \DOIprefix\doi{10.1146/annurev.aa.10.090172.001021}.
\bibitem[{{Woods} et~al.(2017){Woods}, {Ghavamian}, {Badenes} and
  {Gilfanov}}]{wood17}
\bibinfo{author}{{Woods}, T.E.}, \bibinfo{author}{{Ghavamian}, P.},
  \bibinfo{author}{{Badenes}, C.}, \bibinfo{author}{{Gilfanov}, M.},
  \bibinfo{year}{2017}.
\newblock \bibinfo{title}{{No hot and luminous progenitor for Tycho's
  supernova}}.
\newblock \bibinfo{journal}{Nature Astronomy} \bibinfo{volume}{1},
  \bibinfo{pages}{800--804}.
\newblock \DOIprefix\doi{10.1038/s41550-017-0263-5},
  \href{http://arxiv.org/abs/1709.09190}{\tt arXiv:1709.09190}.
\bibitem[{{Woosley}(2017)}]{woos17}
\bibinfo{author}{{Woosley}, S.E.}, \bibinfo{year}{2017}.
\newblock \bibinfo{title}{{Pulsational Pair-instability Supernovae}}.
\newblock \bibinfo{journal}{Astrophys. J.} \bibinfo{volume}{836},
  \bibinfo{pages}{244}.
\newblock \DOIprefix\doi{10.3847/1538-4357/836/2/244}.
\bibitem[{{Woosley} and {Heger}(2015)}]{woos15}
\bibinfo{author}{{Woosley}, S.E.}, \bibinfo{author}{{Heger}, A.},
  \bibinfo{year}{2015}.
\newblock \bibinfo{title}{{The Deaths of Very Massive Stars}}. volume
  \bibinfo{volume}{412} of \textit{\bibinfo{series}{Astrophysics and Space
  Science Library}}.
\newblock p. \bibinfo{pages}{199}.
\newblock \DOIprefix\doi{10.1007/978-3-319-09596-7_7}.
\bibitem[{{Woosley} et~al.(2002){Woosley}, {Heger} and {Weaver}}]{woos02}
\bibinfo{author}{{Woosley}, S.E.}, \bibinfo{author}{{Heger}, A.},
  \bibinfo{author}{{Weaver}, T.A.}, \bibinfo{year}{2002}.
\newblock \bibinfo{title}{{The evolution and explosion of massive stars}}.
\newblock \bibinfo{journal}{Reviews of Modern Physics} \bibinfo{volume}{74},
  \bibinfo{pages}{1015--1071}.
\newblock \DOIprefix\doi{10.1103/RevModPhys.74.1015}.
\bibitem[{{Woosley} and {Weaver}(1994)}]{woos94a}
\bibinfo{author}{{Woosley}, S.E.}, \bibinfo{author}{{Weaver}, T.A.},
  \bibinfo{year}{1994}.
\newblock \bibinfo{title}{{Sub-Chandrasekhar mass models for Type IA
  supernovae}}.
\newblock \bibinfo{journal}{Astrophys. J.} \bibinfo{volume}{423},
  \bibinfo{pages}{371--379}.
\newblock \DOIprefix\doi{10.1086/173813}.
\bibitem[{{Woosley} et~al.(1980){Woosley}, {Weaver} and {Taam}}]{woos80}
\bibinfo{author}{{Woosley}, S.E.}, \bibinfo{author}{{Weaver}, T.A.},
  \bibinfo{author}{{Taam}, R.E.}, \bibinfo{year}{1980}.
\newblock \bibinfo{title}{{Models for Type I supernovae}}, in:
  \bibinfo{editor}{{J.~C.~Wheeler}} (Ed.), \bibinfo{booktitle}{Texas Workshop
  on Type I Supernovae}, pp. \bibinfo{pages}{96--112}.
\bibitem[{{Xu} et~al.(1988){Xu}, {Sutherland}, {McCray} and {Ross}}]{xu88}
\bibinfo{author}{{Xu}, Y.}, \bibinfo{author}{{Sutherland}, P.},
  \bibinfo{author}{{McCray}, R.}, \bibinfo{author}{{Ross}, R.R.},
  \bibinfo{year}{1988}.
\newblock \bibinfo{title}{{X-Rays from Supernova 1987A}}.
\newblock \bibinfo{journal}{Astrophys. J.} \bibinfo{volume}{327},
  \bibinfo{pages}{197}.
\newblock \DOIprefix\doi{10.1086/166181}.
\bibitem[{{Yamaguchi} et~al.(2015){Yamaguchi}, {Badenes}, {Foster}, {Bravo},
  {Williams}, {Maeda}, {Nobukawa}, {Eriksen}, {Brickhouse}, {Petre} and
  {Koyama}}]{yama15}
\bibinfo{author}{{Yamaguchi}, H.}, \bibinfo{author}{{Badenes}, C.},
  \bibinfo{author}{{Foster}, A.R.}, \bibinfo{author}{{Bravo}, E.},
  \bibinfo{author}{{Williams}, B.J.}, \bibinfo{author}{{Maeda}, K.},
  \bibinfo{author}{{Nobukawa}, M.}, \bibinfo{author}{{Eriksen}, K.A.},
  \bibinfo{author}{{Brickhouse}, N.S.}, \bibinfo{author}{{Petre}, R.},
  \bibinfo{author}{{Koyama}, K.}, \bibinfo{year}{2015}.
\newblock \bibinfo{title}{{A Chandrasekhar Mass Progenitor for the Type Ia
  Supernova Remnant 3C 397 from the Enhanced Abundances of Nickel and
  Manganese}}.
\newblock \bibinfo{journal}{Astrophys. J. Lett.} \bibinfo{volume}{801},
  \bibinfo{pages}{L31}.
\newblock \DOIprefix\doi{10.1088/2041-8205/801/2/L31},
  \href{http://arxiv.org/abs/1502.04255}{\tt arXiv:1502.04255}.
\bibitem[{{Yoon} et~al.(2012){Yoon}, {Dierks} and {Langer}}]{yoon12}
\bibinfo{author}{{Yoon}, S.C.}, \bibinfo{author}{{Dierks}, A.},
  \bibinfo{author}{{Langer}, N.}, \bibinfo{year}{2012}.
\newblock \bibinfo{title}{{Evolution of massive Population III stars with
  rotation and magnetic fields}}.
\newblock \bibinfo{journal}{A\&A} \bibinfo{volume}{542}, \bibinfo{pages}{A113}.
\newblock \DOIprefix\doi{10.1051/0004-6361/201117769}.
\bibitem[{{Young} et~al.(2006){Young}, {Fryer}, {Hungerford}, {Arnett},
  {Rockefeller}, {Timmes}, {Voit}, {Meakin} and {Eriksen}}]{youn06}
\bibinfo{author}{{Young}, P.A.}, \bibinfo{author}{{Fryer}, C.L.},
  \bibinfo{author}{{Hungerford}, A.}, \bibinfo{author}{{Arnett}, D.},
  \bibinfo{author}{{Rockefeller}, G.}, \bibinfo{author}{{Timmes}, F.X.},
  \bibinfo{author}{{Voit}, B.}, \bibinfo{author}{{Meakin}, C.},
  \bibinfo{author}{{Eriksen}, K.A.}, \bibinfo{year}{2006}.
\newblock \bibinfo{title}{{Constraints on the Progenitor of Cassiopeia A}}.
\newblock \bibinfo{journal}{Astrophys. J.} \bibinfo{volume}{640},
  \bibinfo{pages}{891--900}.
\newblock \DOIprefix\doi{10.1086/500108}.
\bibitem[{{Zha} et~al.(2019){Zha}, {Leung}, {Suzuki} and {Nomoto}}]{zha19}
\bibinfo{author}{{Zha}, S.}, \bibinfo{author}{{Leung}, S.C.},
  \bibinfo{author}{{Suzuki}, T.}, \bibinfo{author}{{Nomoto}, K.},
  \bibinfo{year}{2019}.
\newblock \bibinfo{title}{{Evolution of ONeMg Core in Super-AGB Stars toward
  Electron-capture Supernovae: Effects of Updated Electron-capture Rate}}.
\newblock \bibinfo{journal}{Astrophys. J.} \bibinfo{volume}{886},
  \bibinfo{pages}{22}.
\newblock \DOIprefix\doi{10.3847/1538-4357/ab4b4b},
  \href{http://arxiv.org/abs/1907.04184}{\tt arXiv:1907.04184}.
\bibitem[{{Zheng} et~al.(2014){Zheng}, {Shivvers}, {Filippenko} and
  et~al.}]{zhen14}
\bibinfo{author}{{Zheng}, W.}, \bibinfo{author}{{Shivvers}, I.},
  \bibinfo{author}{{Filippenko}, A.V.}, \bibinfo{author}{et~al.},
  \bibinfo{year}{2014}.
\newblock \bibinfo{title}{{Estimating the First-light Time of the Type Ia
  Supernova 2014J in M82}}.
\newblock \bibinfo{journal}{Astrophys. J. Lett.} \bibinfo{volume}{783},
  \bibinfo{pages}{L24}.
\newblock \DOIprefix\doi{10.1088/2041-8205/783/1/L24}.
\bibitem[{{Zwicky}(1938)}]{zwic38}
\bibinfo{author}{{Zwicky}, F.}, \bibinfo{year}{1938}.
\newblock \bibinfo{title}{{On Collapsed Neutron Stars.}}
\newblock \bibinfo{journal}{Astrophys. J.} \bibinfo{volume}{88},
  \bibinfo{pages}{522--525}.
\newblock \DOIprefix\doi{10.1086/144003}.

\end{thebibliography}

\end{document}